\documentclass[acmsmall]{acmart} %

\allowdisplaybreaks

\usepackage{graphicx} %
\usepackage{thm-restate}
\usepackage{mathtools}
\usepackage{amsmath,amsthm}%
\usepackage{thmtools}
\usepackage{mathpartir}
\usepackage{libertine}
\usepackage{prooftree}
\usepackage{hyperref}
\usepackage[inline]{enumitem}
\usepackage{relsize}
\usepackage{turnstile}
\usepackage{mathrsfs}
\usepackage{tikz}
\usepackage{xfrac}
\usepackage{nicefrac}%
\usepackage{makecell}
\usepackage[dvipsnames]{xcolor}
\usepackage{threeparttable}
\usepackage{caption}
\usepackage{listings}
\usepackage{subcaption}
\usepackage{pifont}
\usepackage[dvipsnames]{xcolor}
\usepackage{wrapfig}
\usepackage{centernot}

\usepackage{algorithm}
\usepackage{algpseudocode}

\usetikzlibrary{automata, positioning, arrows, arrows.meta, calc, fit,
  shapes.misc, angles, quotes, decorations.pathreplacing, calligraphy}
\tikzset{
  >=stealth,
  node distance=2cm,
  every state/.style={thick, fill=gray!10},
  initial text=$ $,
}
\hypersetup{
    colorlinks=true,
    linkcolor=blue,
    filecolor=magenta,
    urlcolor=cyan,
    pdftitle={Subject Reduction for Vanilla Session Types},
    pdfpagemode=FullScreen,
    citecolor=cyan,
    }
\usepackage[capitalize,nameinlink,noabbrev]{cleveref} %
\crefname{notation}{Notation}{Notations}
\Crefname{notation}{Notation}{Notations}

\usepackage{xr}
\usepackage{pgffor}

\newcounter{myequation}
\makeatletter
\@addtoreset{equation}{myequation}
\makeatother

\makeatletter
\g@addto@macro{\proof}{\stepcounter{myequation}}
\makeatother

\makeatletter
\DeclareRobustCommand{\filename}[1]{%
 \begingroup
  \def\textendash{-}%
  \filename@parse{#1}%
  \edef\filename@base{\detokenize\expandafter{\filename@base}}%
  \textbf{\textsf{\filename@base.\filename@ext}}
 \endgroup
}
\makeatother

\lstdefinestyle{smjleo}{
    basicstyle=\ttfamily\footnotesize,
    showstringspaces=false,
    breaklines=true
}
\setcopyright{cc}
\setcctype{by}
\acmDOI{10.1145/3871757}
\acmYear{2027}
\acmJournal{PACMPL}
\acmVolume{11}
\acmNumber{POPL}
\acmArticle{55}
\acmMonth{1}
\acmSubmissionID{popl27main-p484-p}
\received{2026-07-09}
\received[accepted]{2026-10-05}

\begin{document}
\newcommand{\s}{\ensuremath{{\sf T}}}
\newcommand{\roleFmt}[1]{\ensuremath{{\boldsymbol{\roleCol{\mathtt{#1}}}}}}%
\newcommand{\pp}{\roleFmt{p}}
\newcommand{\ppp}{\roleFmt{p'}}
\newcommand{\ppi}[1]{\roleFmt{p_{#1}}}
\newcommand{\q}{\roleFmt{q}}
\newcommand{\qp}{\roleFmt{q'}}
\newcommand{\qi}[1]{\roleFmt{q_{#1}}}
\newcommand{\pq}{\roleFmt{q}}
\newcommand{\pr}{\roleFmt{r}}
\newcommand{\pri}[1]{\roleFmt{r_{#1}}}
\newcommand{\ps}{\roleFmt{s}}
\newcommand{\pu}{\roleFmt{u}}
\newcommand{\e}{\kf{e}}
\newcommand{\x}{x}
\newcommand{\y}{y}
\newcommand{\h}{h}
\newcommand{\val}{\kf{v}}
\newcommand{\bool}{\kf{b}}
\newcommand{\valn}{\kf{n}}
\newcommand{\valr}{\kf{i}}
\newcommand{\vali}{\valr}
\newcommand{\valb}{\kf{b}}
\newcommand{\DEF}{\ensuremath{::=}}
\newcommand{\SEP}{\ensuremath{~~\mathbf{|\!\!|}~~ }}
\newcommand{\kf}[1]{\ensuremath{\mathsf{#1}}}
\newcommand{\pc}{\ensuremath{~|~}}

\newcommand{\Tset}{\mathcal{T}}
\newcommand{\G}{\ensuremath{{\sf G}}}
\newcommand{\T}{\ensuremath{{\sf T}}}

\newcommand{\distP}{\dist_{\PP}}
\newcommand{\distN}{\dist_{\N}}
\newcommand{\distT}{\dist_{\T}}
\newcommand{\distD}{\dist_{\boldsymbol{\Delta}}}
\newcommand{\distDsub}[1]{\dist_{\boldsymbol{\Delta}_{#1}}}
\newcommand{\distNsub}[1]{\dist_{\N_{#1}}}
\newcommand{\config}{C}

\newcommand{\Basic}{\ensuremath{{\sf B}}}
\newcommand{\RedG}{\Longrightarrow}
\newcommand{\Gvti}[5]{\ensuremath{#1\to#2:\{#3_i({#4}_i). #5_i \}_{i \in I}}}
\newcommand{\Gvtj}[5]{\ensuremath{#1\to#2:\{#3_j({#4}_j). #5_j \}_{j \in J}}}
\newcommand{\Gvtk}[5]{\ensuremath{#1\to#2:\{#3_k({#4}_k). #5_k \}_{k \in K}}}
\newcommand{\GvtiSet}[6]{\ensuremath{#1\to#2:\{#3_i({#4}_i). #5_i \}_{i \in {#6}}}}
\newcommand{\GvtlSet}[6]{\ensuremath{#1\to#2:\{#3_l({#4}_l). #5_l \}_{l \in {#6}}}}
\newcommand{\GvtmSet}[6]{\ensuremath{#1\to#2:\{#3_m({#4}_m). #5_m \}_{m \in {#6}}}}
\newcommand{\Gvt}[5]{\ensuremath{#1\to#2: #3({#4}). #5 }}
\newcommand{\GvtIdx}[6]{\ensuremath{#1\to#2: \{#3({#4}). #5\}_{#6}}}
\newcommand{\ty}{\mathbf{t}}
\newcommand{\GvtPair}[3]{\ensuremath{#1\to#2: \{#3\}_{i\in I}}}

\newcommand{\Sy}{\ensuremath{\mathcal{S}}}
\newcommand{\N}{\ensuremath{M}}
\newcommand{\M}{\ensuremath{M}}

\newcommand{\pa}[2]{#1 \triangleleft  #2}
\newcommand{\set}[1]{\{#1\}}
\newcommand{\eval}[2]{#1 \downarrow #2}
\newcommand{\true}{\kf{true}}
\newcommand{\false}{\kf{false}}

\newcommand{\myrule}[3]{\begin{prooftree} #1 \justifies   #2   \using{\rln{#3}} \end{prooftree}}
\newcommand{\rln}[1]{\textsc{#1}}
\newcommand{\der}[3]{ #1 \vdash   #2  :#3}
\newcommand{\dercoup}[4]{ #1 \vdash_{#4}   #2  :#3}
\newcommand{\pro}[3]{ #1 \upharpoonright_#2 #3}
\newcommand{\CP}[1]{ {\mathcal P}\!\left({#1}\right)}
\newcommand{\CG}[2]{ {\mathcal G}(#1,#2)}
\newcommand{\CGZ}[3]{ {\mathcal G}_0(#1,#2,#3)}
\newcommand{\res}{\setminus}
\newcommand{\rem}{\bbslash}
\newcommand{\CC}[1]{ {\mathcal C}[#1]}
\newcommand*{\st}[0]{\ensuremath{:\,}}
\newcommand{\rpath}[0]{\ensuremath{\mathtt{r}}}
\newcommand{\Rpath}[0]{\ensuremath{\mathrm{R}}}
\newcommand{\Rpathf}[4]{\ensuremath{\mathrm{R}_{#1}(#2,#3,#4)}}
\newcommand{\tpath}[0]{\ensuremath{\xi}}
\newcommand{\tpathp}[0]{\ensuremath{\xi'}}
\newcommand{\Tpath}[0]{\ensuremath{\Xi}}
\newcommand{\inferT}[1]{\ensuremath{\mathit{infer}(#1)}}
\newcommand{\visited}[1]{\ensuremath{\mathsf{visited}(#1)}}
\newcommand{\normalize}[2]{#1\ensuremath{\leadsto}#2}
\newcommand{\normalizestar}[2]{#1\ensuremath{\leadsto^*}#2}
\newcommand{\norm}[1]{\ensuremath{\mathcal{N}{(#1)}}}
\newcommand{\ghostT}[2]{\ensuremath{{\tt{ghostT}}(#1).#2}}
\newcommand{\ghostB}[2]{\ensuremath{{\tt{ghostBool}}(#1).#2}}
\newcommand{\normn}[0]{\ensuremath{\mathcal{N}}}
\newcommand{\rteq}[2]{\ensuremath{#1\approx_{\tt{rt}}#2}}
\newcommand{\intaction}[2]{\ensuremath{\delta(#1,#2)}}

\newcommand{\ActD}{\mathsf{Act}}
\newcommand{\stepd}[2]{\mathsf{step}(#1,#2)}
\newcommand{\suppd}{\mathrm{supp}}
\newcommand{\ReachD}[1]{\ensuremath{\{\Delta' \mid \Delta \predcontstar{}{}\Delta' \}}}
\newcommand{\sbj}[1]{\mathsf{sbj}(#1)}
\newcommand{\Sclos}{S^{\ast}}
\newcommand{\Aclos}{A^{\ast}}
\newcommand{\GoodP}[1]{\sff{EC}\wasllpp{#1}} 
\newcommand{\Confine}{\mathsf{Conf}}

\newcommand{\inout}[1]{{\mathtt{actions}}(#1)}

\newcommand{\restrict}[1]{\ensuremath{\big|_{#1}}}
\newcommand{\clbound}[3]{\ensuremath{\mbox{}_{(#1)}[#3_{#2}}}
\newcommand{\crbound}[3]{\ensuremath{#3_{#2}]_{(#1)}}}
\newcommand{\lbound}[2]{\ensuremath{\mbox{}_{(#1)}[\N_{#2}}}
\newcommand{\rbound}[2]{\ensuremath{\N_{#2}]_{(#1)}}}
\newcommand{\dlbound}[2]{\ensuremath{\mbox{}_{(#1)}[\Delta_{#2}}}
\newcommand{\drbound}[2]{\ensuremath{\Delta_{#2}]_{(#1)}}}

\newcommand{\participant}[1]{\mathtt{pt}\{#1\}}
\newcommand{\pn}[1]{\mathtt{pn}(#1)}
\newcommand{\pnin}[1]{\mathtt{pn}_?(#1)}
\newcommand{\pnout}[1]{\mathtt{pn}_!(#1)}
\newcommand{\proj}[2]{ #1 \upharpoonright_{#2}}
\newcommand{\projt}[2]{{#1}{\upharpoonright}{#2}}
\newcommand{\sub}[2]{\set{#1/#2}}

\newcommand{\exprt}[2]{\mathtt{expression}(#1,#2)}
\newcommand{\valt}[1]{\mathtt{value}(#1)}

\newcommand{\MP}{M}
\newcommand{\new}[1]{(\nu#1)}
\newcommand{\procin}[3]{#1 ?  #2 .#3}
\newcommand{\procoutnl}[3]{#1 ! \labels{#2} . #3}
\newcommand{\procout}[4]{#1 ! \labels{#2} \langle #3 \rangle . #4}
\newcommand{\probprocout}[6]{\sum\nolimits_{#1} #2 : \pchan{#3} ! \labels{#4} \langle #5 \rangle . #6}

\newcommand{\procoutu}[3]{#1 ! \labels{#2} . #3}
\newcommand{\pprocout}[5]{#2 ! \, #1{:}#3\langle #4 \rangle . #5}
\newcommand{\pprocoutnew}[6]{ \sum_{#2}#1 ! \,#3 {:} #4\langle #5 \rangle . #6}
\newcommand{\pprocbranch}[5]{\pchan{#1} !\, #2 {:} \labels{#3}\langle #4 \rangle.#5}
\newcommand{\pprocouttytwo}[2]{\scalebox{1.3}{$\oplus$}\{#1 , #2\}}
\newcommand{\pprocoutty}[6]{ \scalebox{1.3}{$\oplus$}_{#2} \,#3{:}\,\pchan{#1} !
    \labels{#4}\langle #5 \rangle . #6}
\newcommand{\probpprocoutty}[7]{ {\textstyle\sum\nolimits_{#1}} \scalebox{1.3}{$\oplus$}_{#3} \, #4
  {:}\, \pchan{#2} ! \labels{#5}\langle #6 \rangle . #7}
\newcommand{\pselsingle}[4]{#1 {:} \labels{#2}\langle #3 \rangle
  . #4}
\newcommand{\psinglenoval}[3]{#1 {:} \labels{#2}. #3}
\newcommand{\psinglenovaltail}[2]{#1 {:} \labels{#2}}
\newcommand{\pselsinglenotail}[3]{#1 {:} \labels{#2}\langle #3
  \rangle}
\newcommand{\pbrasinglenotail}[3]{#1 {:} \labels{#2}(#3)}
\newcommand{\pbrasingle}[3]{\labels{#1}(#2).{#3}}
\newcommand{\pselsinglenovaltail}[3]{#1 {:} \labels{#2}\langle #3 \rangle}
\newcommand{\pprocinnew}[5]{  \scalebox{1.2}{\&}_{#2} \ \pchan{#1} ? \labels{#3} (#4) . #5}
\newcommand{\procdag}[3]{#1 \dagger  #2 .#3}
\newcommand{\procddag}[3]{#1 \dagger #2. #3}
\newcommand{\error}{\kf{error}}
\newcommand{\PP}{\ensuremath{P}}
\newcommand{\Q}{\ensuremath{Q}}
\newcommand{\cond}[3]{\kf{if}~ #1 ~\kf{then} ~#2 ~\kf{else}~#3}
\newcommand{\condc}[3]{\kf{if}~ #1
    {\begin{cases}
        \kf{then}\; #2\\
        \kf{else}\; #3
    \end{cases}}
}
\newcommand{\ccond}[2]{\kf{if}~ #1 ~\kf{then} ~#2 ~\kf{else}}
\newcommand{\labels}[1]{{\color{gtColor}\sff{#1}}}
\newcommand{\labelsell}{\labels{\ell}}
\newcommand{\labelselli}[1]{\labels{\ell_{#1}}}

\newcommand{\inact}{\ensuremath{\mathbf{0}}}
\newcommand{\internal}{\oplus}
\newcommand{\external}{+}
\newcommand{\co}[1]{\mathtt{coherent}\{#1\}}
\newcommand{\emptyqueue}{\epsilon}

\newcommand{\pC}{\pchan{c}}\newcommand{\pM}{\pchan{m}}
\newcommand{\lk}{\labelselli{k}}\newcommand{\llo}{\labelselli{a}}
\newcommand{\llt}{\labelselli{b}}\newcommand{\lstop}{\labels{stop}}
\newcommand{\rederror}{\mathsf{error}}
\newcommand{\isz}[1]{#1 = 0}
\newcommand{\prd}[1]{\mathsf{pred}(#1)}
\newcommand{\eqk}[2]{#1 = #2}
\newcommand{\Disp}{\mathsf{Disp}}
\newcommand{\enc}[1]{\langle\!\langle #1 \rangle\!\rangle}
\newcommand{\snd}[3]{\procout{\pM}{\lk}{#1}{\procout{\pM}{\llo}{#2}{\procout{\pM}{\llt}{#3}{X}}}}
\newcommand{\reds}{\longrightarrow}
\newcommand{\tend}{\mathtt{end}}
\newcommand{\tbool}{\mathtt{bool}}
\newcommand{\tpbool}{\mathtt{pbool}}
\newcommand{\tnat}{\mathtt{nat}}
\newcommand{\tint}{\mathtt{int}}
\newcommand{\tin}[3]{#1?#2(#3)}
\newcommand{\tout}[3]{#1!#2(#3)}
\newcommand{\tdag}[3]{#1\dagger #2(#3)}
\newcommand{\tddag}[3]{#1\ddagger #2 (#3)}
\newcommand{\tinternal}{\bigvee\limits_{i\in I}}
\newcommand{\texternal}{\bigwedge\limits_{i\in I}}
\newcommand{\prob}{\ensuremath{\textcolor{blue}{p}}}
\newcommand{\probp}{\ensuremath{\textcolor{blue}{p'}}}
\newcommand{\probpp}{\ensuremath{\textcolor{blue}{p''}}}
\newcommand{\probi}[1]{\ensuremath{\textcolor{blue}{p_{#1}}}}
\newcommand{\probip}[1]{\ensuremath{\textcolor{blue}{p'_{#1}}}}
\newcommand{\probipp}[1]{\ensuremath{\textcolor{blue}{p''_{#1}}}}
\newcommand{\probippp}[1]{\ensuremath{\textcolor{blue}{p'''_{#1}}}}
\newcommand{\probq}{\ensuremath{\textcolor{blue}{q}}}
\newcommand{\probqp}{\ensuremath{\textcolor{blue}{q'}}}
\newcommand{\probn}[1]{\ensuremath{\textcolor{blue}{#1}}}
\newcommand{\pone}{\probn{1}}
\newcommand{\trace}{\ensuremath{\rho}}
\newcommand{\pathcont}[1]{\ensuremath{\Xi(#1)}}
\newcommand{\ppaths}[1]{\ensuremath{#1(\xi)}}
\newcommand{\spcon}{\Rrightarrow}
\newcommand{\obs}[1]{\ensuremath{\mathsf{obs}(#1)}}

\newcommand{\lin}{part}
\newcommand{\lout}{lab}

\newcommand{\queue}{\mathtt{queue}}
\newcommand{\lts}[1]{\xrightarrow{#1}}
\newcommand{\redG}[4]{#1\setminus#2\xrightarrow#3#4}
\newcommand{\SortI}[2]{ {#1}_{#2}}
\newcommand{\context}{\mathcal{C}}
\newcommand{\Econtext}{\mathcal{E}}

\newcommand{\Gctx}[1][]{\ifempty{#1}{\textswab{G}}{\textswab{G}_{#1}}}%
\newcommand{\GctxHole}[1]{[\;]^{#1}}%
\newcommand{\GctxApp}[4]{{#1}[{#2}]^{#3}_{#4}}%

\newcommand{\synct}{\mathrel{\ltimes}}
\newcommand{\subt}{\leqslant}
\newcommand{\supt}{\geqslant}
\newcommand{\subta}{\mathrel{\leqslant_a}}
\newcommand{\subs}{\leq\vcentcolon}
\newcommand{\nsubs}{\not\leq\vcentcolon}
\newcommand{\red}{\longrightarrow}
\newcommand{\nsubt}{\not\trianglelefteq}
\newcommand{\someRel}[1][]{\mathrel{\ifempty{#1}{\mathcal{R}}{\mathcal{R}_{#1}}}}%
\newcommand{\someReli}[1][]{\mathrel{\ifempty{#1}{\mathcal{R}'}{\mathcal{R}'_{#1}}}}%
\newcommand{\dist}{\ensuremath{\mathcal{D}}}

\newcommand{\fsqrt}[1]{{\tt neg}(#1)}
\newcommand{\fneg}{\fsqrt}
\newcommand{\fsucc}[1]{{\tt succ}(#1)}
\newcommand{\flip}[3]{\textsf{flip}(\probn{#1}) \{\textsc{h}
  {\text{\small$\Rightarrow$}}\, #2 \,[\!\!]\,
  \textsc{t}{\text{\small$\Rightarrow$}}\, #3\}}
\newcommand{\flipc}[3]{\textsf{flip}(\probn{#1})
    {\begin{cases}
        \textsc{h}{\text{\small$\Rightarrow$}}\; #2 \\
        \textsc{t}{\text{\small$\Rightarrow$}}\; #3
    \end{cases}}
}
\newcommand{\pluss}[4]{\sum
    {\begin{cases}
        #1 \\
        #2 \\
        #3 \\
        #4
    \end{cases}}
}
\newcommand{\plusc}[2]{\sum
    {\begin{cases}
        #1 \\
        #2
    \end{cases}}
}
\newcommand{\multiset}[1]{\{\!\!\{\,#1\,\}\!\!\}}
\newcommand{\ftv}[1]{{\tt ftv}(#1)}
\newcommand{\fv}[1]{{\tt fv}(#1)}
\newcommand{\tof}[1]{{\tt tOf}(#1)}
\newcommand{\IH}{\mathtt{IH}}
\newcommand{\reach}[0]{\ensuremath{\Diamond}}
\newcommand{\stuck}[1]{{\tt stuck}\!\left({#1}\right)}
\newcommand{\dual}[1]{\overline #1}

\newcommand{\fpv}[1]{\mathit{fpv}(#1)}

\newcommand{\dom}[1]{\operatorname{dom}({#1})}
\newcommand{\ran}[1]{\operatorname{ran}({#1})}
\newcommand{\subst}[2]{\{\nicefrac{#1}{#2}\}}%

\newcommand{\contTag}{{\,\mathsf{C}}}%
\newcommand{\payloadTag}{{\,\mathsf{P}}}%
\newcommand{\ttreeSym}{\mathcal{T}}%
\newcommand{\ttree}[1]{\operatorname{\ttreeSym}\!\left({#1}\right)}%
\newcommand{\tsynt}[1]{\mathbb{#1}}%
\newcommand{\ts}{\tsynt}%

\newcommand{\AContext}{\mathcal{A}}
\newcommand{\BContext}{\mathcal{B}}
\newcommand{\AContextf}[1]{\AContext [ #1 ]}
\newcommand{\AContextfp}[1]{\AContext' [ #1 ]}
\newcommand{\AContextfi}[2]{\AContext_#1 [ #2 ]}
\newcommand{\AContextfip}[2]{\AContext'_#1 [ #2 ]}


\newcommand{\selfloop}{\textcolor{gtColor}{\textsf{loop}}}

\newcommand{\aleks}[1]{\textcolor{magenta}{Aleks: #1}}
\newcommand{\joe}[1]{\textcolor{blue}{Joe: #1}}
\newcommand{\leo}[1]{\textcolor{orange}{Leo: #1}}
\newcommand{\NY}[1]{\textcolor{violet}{NY: #1}}

\newcommand{\pchoicee}[1]{\oplus}
\newcommand{\probmerge}[1]{\sqcap_{#1}}
\newcommand{\pchoicep}[1]{+_{#1}}
\newcommand{\pchoicepbig}{\sum}
\newcommand{\peval}[3]{#1 \downarrow_{#3} #2}
\newcommand{\pred}[1]{\longrightarrow_{#1}}
\newcommand{\guardlab}[1]{{\ensuremath{\color{darkgray}{\scriptsize{[\,#1\,]}}}}}
\newcommand{\predcont}[2]{\xrightarrow[]{#1}_{\probn{#2}}}
\newcommand{\longtwoheadrightarrow}{\mathrel{\relbar\joinrel\twoheadrightarrow}}

\newcommand{\intred}[1]{\xRightarrow{#1}}
\newcommand{\intredstar}[1]{\xRightarrow{#1}\!{}^{*}}
\newcommand{\intrednf}[1]{\xRightarrow{#1}\!{}_{\mathtt{nf}}}
\newcommand{\intrednfsched}[2]{\xRightarrow{#1}\!{}_{\mathtt{nf}}^{#2}}
\newcommand{\commredsched}[2]{\xRightarrow{}\!{}_{#1}^{#2}}
\newcommand{\intnf}[1]{\sff{Comm_{NF}}}

\newcommand{\predblock}[2]{\mathrel{\xLongrightarrow[\;#2\;]{\;#1\;}}}  
\newcommand{\xtwoheadrightarrow}[2][]{%
  \xrightarrow{#2}\mathrel{\mkern-14mu}\rightarrow_{\mkern-2mu #1}%
}
\newcommand{\predcontdouble}[2]{\xtwoheadrightarrow[\probn{#2}]{#1}}
\newcommand{\predcontsched}[2]{\xrightarrow[]{#1}_{#2}}
\newcommand{\predcontschedstar}[2]{%
  \ensuremath{\mbox{$\xrightarrow{#1}\!{}_{#2}^{*}$}}%
}
\newcommand{\predcontstar}[2]{%
  \ensuremath{\mbox{$\xrightarrow{#1}\!{}_{\probn{#2}}^{*}$}}%
}
\newcommand{\predcontwc}[3]{\xrightarrow{#1}\!{}_{\probn{#2}}^{#3}}
\newcommand{\pathproj}[2]{{\mathit{\pi}}_{#1}(#2)}
\newcommand{\id}[0]{{\tt{Id}}}
\newcommand{\nd}{{\textcolor{violet}n}}
\newcommand{\predcontnd}[1]{\xrightarrow[]{#1}_{\nd}}
\newcommand{\ndstep}[0]{\mathrel{\rightarrow_{\nd}}}
\newcommand{\predcontndstar}[1]{\xrightarrow{#1}\!{}_{\nd}^{*}}
\newcommand{\predmixstar}[1]{\xrightarrow{}\!{}_{\probn{#1}}^{\nd*}}
\newcommand{\predmix}[1]{\xrightarrow{}\!{}_{\probn{#1}}^{\nd}}
\newcommand{\pqsend}{\pp:\q!\labelsell\langle\Basic\rangle}
\newcommand{\pqrecv}{\pp:\q?\labelsell(\Basic)}
\newcommand{\qprecv}{\q:\pp?\labelsell(\Basic)}
\newcommand{\potsend}{\ppi{1}:\ppi{2}!\labelsell\langle\Basic\rangle}
\newcommand{\potrecv}{\ppi{1}:\ppi{2}?\labelsell(\Basic)}
\newcommand{\ptorecv}{\ppi{2}:\ppi{1}?\labelsell(\Basic)}
\newcommand{\term}[1]{{\sf{term}}(#1)}
\newcommand{\termm}{{\sf{term}}}
\newcommand{\pterm}[2]{{\sf{probterm}}(#1,#2)}
\newcommand{\df}[1]{{\sf{df}} (#1)}
\newcommand{\dff}[0]{{\sf{df}}}
\newcommand{\sdf}[1]{{\sf{sdf}} (#1)}
\newcommand{\sterm}[1]{{\sf{sterm}} (#1)}
\newcommand{\pdf}[2]{{\sf{probdf}} (#1, #2)}
\newcommand{\pdfone}[2]{{\sf{normpdf1}} (#1, #2)}
\newcommand{\pdftwo}[2]{{\sf{normpdf2}} (#1, #2)}
\newcommand{\pdfthree}[2]{{\sf{normpdf}} (#1, #2)}
\newcommand{\pureterm}[1]{{\sf{asterm}} (#1)}
\newcommand{\puredf}[1]{{\sf{asdf}} (#1)}
\newcommand{\sff}[1]{\relax\ifmmode\mathsf{#1}\else\textsf{#1}\fi}
\newcommand{\safe}[1]{{\sff{safe}} (#1)}
\newcommand{\safee}[0]{{\sff{safe}}}
\newcommand{\deadlock}[1]{\mathsf{deadlock}(#1)}
\newcommand{\deadlockk}{\mathsf{deadlock}}
\newcommand{\sdeadlock}[1]{\mathsf{sdeadlock}(#1)}
\newcommand{\sdeadlockk}{\mathsf{sdeadlock}}
\newcommand{\live}[1]{{\sff{live}} (#1)}
\newcommand{\livee}{{\sff{live}}}
\newcommand{\fairr}{{\sff{fair}}}
\newcommand{\fair}[1]{{\sff{fair}}(#1)}
\newcommand{\plive}[2]{{\sff{prolive}} (#1,#2)}
\newcommand{\aslive}[1]{{\sff{aslive}} (#1)}
\newcommand{\waslive}[2]{{\sff{waslive}} (#1,#2)}
\newcommand{\aslivee}{{\sff{aslive}} }
\newcommand{\waslivee}{{\sff{waslive}} }
\newcommand{\slive}[1]{{\sf{slive}} (#1)}
\newcommand{\saslive}[1]{{\sff{saslive}} (#1)}
\newcommand{\swaslivee}{{\sff{swaslive}} }
\newcommand{\raslivee}{{\sff{raslive}} }
\newcommand{\raslive}[1]{{\sff{raslive}} (#1)}
\newcommand{\pend}[2]{{\sff{Pend}}_{#1}(#2)}
\newcommand{\pendd}[1]{{\sff{Pend}}_{#1}}
\newcommand{\blkp}[2]{{\sff{Blk}}_{#1}(#2)}
\newcommand{\blkpp}[1]{{\sff{Blk}}_{#1}}
\newcommand{\typesm}[1]{{\sff{Types}}_{{#1}}}
\newcommand{\wasllp}[2]{{\sff{WASlivelock}}_{#1}(#2)}
\newcommand{\wasll}[1]{{\sff{WASlivelock}}(#1)}
\newcommand{\wasllpp}[1]{{\sff{WASlivelock}}_{#1}}

\newcommand{\fpaths}[2]{\Tpath_{#1}^{\mathrm{nf}}(#2)}
\newcommand{\paths}[2]{\Tpath_{#1}(#2)}
\newcommand{\mdp}[1]{\ensuremath{\mathcal{M}_{#1}}}
\newcommand{\last}[1]{\mathit{last}(#1)}
\newcommand{\prefixlt}{\sqsubset}
\newcommand{\states}{S}                          
\newcommand{\acts}{\mathit{Act}}                 
\newcommand{\actsen}[1]{\acts(#1)}               
\newcommand{\transf}{\mathbf{P}}                 
\newcommand{\dirac}[1]{\delta_{#1}}              
\newcommand{\pmeasure}[2]{\mathrm{Pr}^{#1}\left(#2\right)}
\newcommand{\pmeasuret}[3]{\mathrm{Pr}^{#1}_{#2}\left(#3\right)}

\newcommand{\dfp}[1]{{\sf{df}} (#1)}
\newcommand{\sdfp}[1]{{\sf{safedf}} (#1)}
\newcommand{\pdfp}[2]{{\sf{prodf}} (#1, #2)}
\newcommand{\puredfp}[1]{{\sf{asdf}} (#1)}
\newcommand{\dtmc}[0]{\ensuremath{\mathsf{DTMC}}}

\newcommand{\probpath}[2]{\mathbb{P}^{#1}(#2)}
\newcommand{\probpathmeasure}[2]{\mathbb{P}^{#1}_{#2}}
\newcommand{\probpathsig}[2]{\mathbb{P}^{\phi}_{{#1}}(#2)}
\newcommand{\probpathsiglr}[2]{\mathbb{P}^{\phi}_{{#1}}\left(#2\right)}
\newcommand{\probpathsigp}[2]{\mathbb{P}^{\phi'}_{{#1}}(#2)}
\newcommand{\probpathmax}[2]{\mathbb{P}^{\mathtt{max}}_{{#1}}(#2)}
\newcommand{\probpathmin}[2]{\mathbb{P}^{\mathtt{min}}_{{#1}}(#2)}
\newcommand{\probdf}[1]{\mathbb{P}_\mathsf{df}(#1)}
\newcommand{\probndf}[1]{\mathbb{P}_\mathsf{ndf}(#1)}
\newcommand{\probterm}[1]{\mathbb{P}_\mathsf{term}(#1)}
\newcommand{\probnterm}[1]{\mathbb{P}_\mathsf{nterm}(#1)}

\newcommand{\cyl}[1]{{\sf{Cyl}} (#1)}
\newcommand{\cylsig}[1]{\cyl{#1}}
\newcommand{\cylsigp}[1]{\cyl{#1}}

\newcommand{\termmp}[1]{{\sf{term}} (#1)}
\newcommand{\ptermmp}[2]{{\sf{p\_term}} (#1,#2)}
\newcommand{\dfmp}[1]{{\sf{df}} (#1)}
\newcommand{\pdfmp}[2]{{\sf{p\_df}} (#1, #2)}
\newcommand{\puredfmp}[1]{{\sf{as\_df}} (#1)}
\newcommand{\safemp}[1]{{\sff{safe}}(#1)}
\newcommand{\livemp}[1]{{\sff{live}}(#1)}
\newcommand{\plivemp}[2]{{\sff{p\_live}} (#1,#2)}

\newcommand{\pchan}[1]{\roleFmt{#1}}
\newcommand{\psyntout}[6]{\textstyle\sum\nolimits_{#1}\procout{\pchan{#3}}{\, #2{:}\labels{\labelselli{#4}}}{ #5 }{#6}}
\newcommand{\pselectsingle}[5]{\procout{\pchan{#1}}{\, #2{:}\labels{#3}}{ #4 }{#5
}}
\newcommand{\pselectsinglelabel}[4]{\pchan{#1}!{\, #2{:}\labels{#3}}.{#4}}
\newcommand{\psyntin}[5]{\textstyle\sum\nolimits_{#1} \procin{\pchan{#2}}{\labels{\labelselli{#3}}(#4)}{#5}}
\newcommand{\psyntinnl}[5]{\Sigma_{#1} \procin{\pchan{#2}}{\labels{\labelselli{#3}}(#4)}{#5}} 
\newcommand{\psyntif}[3]{\kf{if}~ #1 ~\kf{then} ~#2 ~\kf{else}~#3}
\newcommand{\psyntrec}[2]{  \mu #1.#2   }
\newcommand{\plblout}[4]{def}
\newcommand{\plblin}[4]{def}
\newcommand{\psyntses}[2]{ \pchan{#1} \triangleleft #2 }
\newcommand{\Multi}{\ensuremath{M}}
\newcommand{\topprob}[1]{\ensuremath{{\sff{ tp}} (#1)}}
\newcommand{\unf}[1]{\ensuremath{{\sff{ unf}} (#1)}}

\newcommand{\procoutdistty}[1]{\scalebox{1.3}{$\oplus$} \ \{#1\}}
\newcommand{\procoutnddistty}[1]{\sum\scalebox{1.3}{$\oplus$} \ \{#1\}}
\newcommand{\procoutlab}[3]{#1 ! \labels{#2} . #3}
\newcommand{\psyntinlab}[4]{\sum\nolimits_{#1} \procin{\pchan{#2}}{\labels{#3}}{#4}}

\newcommand{\ints}{\mathtt{int}}
\newcommand{\bools}{\mathtt{bool}}
\newcommand{\float}{\mathtt{float}}
\newcommand{\nat}{\mathtt{nat}}
\newcommand{\real}{\mathtt{real}}
\newcommand{\unit}{\mathtt{unit}}
\newcommand{\pmultipi}{\text{\small{\sc{promt}}}}


\newcommand{\bnfsep}{\mathbin{\;\big|\;}}
\newcommand{\prismrole}[1]{{\fontfamily{qcr}\selectfont
    #1
  }}
\newcommand{\prismcommandgeneric}[5]{[#1]\,#2 \rightarrow \textstyle\sum_{#3}#4: #5}
\newcommand{\prismcommand}[3]{[#1]\,#2 \rightarrow #3}
\newcommand{\prismmodule}[3]{\prismrole{#1}\!: \{#2\}_{#3}}
\newcommand{\prismmodulecomp}[2]{#1 \ || \ #2}
\newcommand{\prismbigcomp}[3]{||_{#2}#3}
\newcommand{\prismreduce}[2]{\xrightarrow[]{#1}_{#2}}
\newcommand{\prismhas}{\rightsquigarrow}

\newcommand{\irulePRISMsync}{M-sync}%
\newcommand{\irulePRISMnosync}{M-move}%
\newcommand{\irulePRISMstmove}{S-move}%

\definecolor{mpColor}{rgb}{0, 0, 0}
\newcommand{\mpFmt}[1]{{\color{mpColor}#1}}%
\newcommand{\mpNil}{\mpFmt{\mathbf{0}}}%

\newcounter{sarrow}
\newcommand\PRISMmove[2]{%
  \stepcounter{sarrow}%
  \mathrel{\begin{tikzpicture}[baseline= {( $ (current bounding box.south) + (0,-0.5ex) $ )}]
      \node[inner sep=.5ex] (\thesarrow) {$\scriptstyle #1$};
      \path[draw,<-,decorate,
      decoration={zigzag,amplitude=0.7pt,segment length=1.2mm,pre=lineto,pre length=4pt}] 
      (\thesarrow.south east) -- (\thesarrow.south west);
      \node[inner sep=0pt, below right=-1pt of \thesarrow.south east] {$\scriptstyle #2$};
    \end{tikzpicture}}%
}

\newcommand{\closure}[1]{\textsf{closure}(#1)}
\newcommand{\cmp}{ \ || \ }

\newcommand{\OR}[1]{OR(#1)}%
\newcommand{\indexs}[5]{Idx_{(#1,#2,#3,#4)}(#5)}%
\newcommand{\CR}[1]{CR(#1)}%
\newcommand{\Role}[1]{R(#1)}%
\newcommand{\StateS}[1]{{\color{MidnightBlue}\textsf{SS}(}#1{\color{MidnightBlue})}}%
\newcommand{\tranone}[1]{{ [\! [ #1 ] \! ]}}%
\newcommand{\trantwo}[1]{{ (\! | #1 | \! )}}%
\newcommand{\tranthree}[2]{{ [\! [ #1 ] \! ]}^2_{#2}}%
\newcommand{\oxtrantwo}[5]{ \{\!|  #1 | \! \}_{(#2,#3,#4,#5)}}%
\newcommand{\oxtranthree}[7]{ (\!|  #1 | \! )_{(#2,#3,#4,#5,#6,#7)}^2}%
\newcommand{\prism}{PRISM}%
\newcommand{\prismend}[1]{End(#1) }%
\newcommand{\pctlP}[2]{Prob_{#1}(#2) }%
\newcommand{\pmeasuremin}[1]{\mathrm{Pr}^{min}\left(#1\right)}
\newcommand{\pmeasuremax}[1]{\mathrm{Pr}^{max}\left(#1\right)}
\newcommand{\pmin}[1]{\textsf{Pmin}\left(#1\right)}
\newcommand{\pmax}[1]{\textsf{Pmax}\left(#1\right)}
\newcommand{\always}{\square \ }
\newcommand{\until}{\;\mathbin{\mathrm{U}}\;}
\newcommand{\eventually}{\rotatebox[origin=c]{45}{$\square$} \ }
\newcommand{\nextstate}{\bigcirc}
\newcommand{\lblid}[1]{\mathtt{ID}_{(#1)} }%
\newcommand{\lblidcount}[2]{\mathtt{ID}_{(#1,#2)} }%
\newcommand{\lblidfun}{\mathtt{ID} }%
\newcommand{\lbls}[1]{\mathtt{Lbl}(#1) }%
\newcommand{\lblssend}[1]{\mathtt{Lbl}^+(#1) }%
\newcommand{\lblsrec}[2]{\mathtt{Lbl}^-(#1,#2) }%
\newcommand{\domid}[3]{\operatorname{domid}({#1},{#2},{#3})}
\newcommand{\ap}{a}
\newcommand{\trancount}[2]{\mathtt{Tcnt}(#1,#2)}
\newcommand{\mapz}{g_0}
\newcommand{\ReductionsSin}[4]{\mathtt{RS}^-(#1,#2,#3,#4)}
\newcommand{\ReductionsSTin}[5]{\mathtt{RS}^-(#1,#2,#3,#4,#5)}
\newcommand{\ReductionsSout}[4]{\mathtt{RS}^+(#1,#2,#3,#4)}
\newcommand{\ReductionsSTout}[5]{\mathtt{RS}^+(#1,#2,#3,#4,#5)}
\newcommand{\EOL}[4]{\mathtt{EOL}(#1,#2,#3,#4)}
\newcommand{\EOLT}[5]{\mathtt{EOL}'(#1,#2,#3,#4,#5)}
\newcommand{\EB}[3]{\mathtt{EB}(#1,#2,#3)}
\newcommand{\EBT}[4]{\mathtt{EB}'(#1,#2,#3,#4)}
\newcommand{\EBL}[4]{\mathtt{EBL}(#1,#2,#3,#4)}
\newcommand{\EBLT}[5]{\mathtt{EBL}'(#1,#2,#3,#4,#5)}
\newcommand{\lab}[2]{#1::#2::\mathit{label}}
\newcommand{\pqlab}{\lab{\pp}{\q}}
\newcommand{\qplab}{\lab{\q}{\pp}}
\newcommand{\rslab}{\lab{\pr}{\ps}}
\newcommand{\srlab}{\lab{\ps}{\pr}}
\newcommand{\fail}{\mathit{fail}}
\newcommand{\source}[3]{\mathtt{source}_{#1 : #2}(#3)}
\newcommand{\sourcenobrackets}[2]{\mathtt{source}_{#1 : #2}}
\newcommand{\emptytran}{\trantwo{\cdot}}
\newcommand{\actions}[2]{\textsf{actions}_{#1}(#2)}
\newcommand{\conc}{::}
\newcommand{\nexts}[1]{\mathtt{next}(#1)}
\newcommand{\sume}[1]{\mathtt{sel}(#1)}
\newcommand{\selsummand}[1]{\T^{\oplus}_{#1}}
\newcommand{\come}[1]{\mathtt{com}(#1)}
\newcommand{\comsel}[1]{\T^{!}_{#1}}

\newcommand{\mut}{\mu \ty . \T}
\newcommand{\mutprime}{\mu \ty . \T'}
\newcommand{\tunroll}{\T \subst{\mut}{\ty}}
\newcommand{\tprimeunroll}{\T' \subst{\mut}{\ty}}
\newcommand{\reducesto}{\predcont{\alpha}{\prob}}
\newcommand{\sconf}{\sqsubseteq_{\T, \ty}}
\newcommand{\tfail}{\mathtt{fail}}

\newcommand{\sourcep}[4]{\mathtt{source}'_{(#2, #3, #4)}(#1)}
\newcommand{\sourcepp}[1]{\sourcep{#1}{\pp}{n}{f}}

\newcommand{\pmut}{\pp : \mut}
\newcommand{\pmuttran}{\trantwo{\pmut}}
\newcommand{\ptunroll}{\pp : \tunroll}
\newcommand{\ptunrolltran}{\trantwo{\ptunroll}}
\newcommand{\mutsource}[1]{\source{\pp}{\mut}{#1}}
\newcommand{\tunrollsource}[1]{\source{\pp}{\tunroll}{#1}}

\newcommand{\pmutprime}{\pp : \mutprime}
\newcommand{\pmutprimetran}{\trantwo{\pmutprime}{\lblidfun}}
\newcommand{\ptprimeunroll}{\pp : \T' \subst{\mu \ty . \T'}{\ty}}
\newcommand{\ptprimeunrolltran}{\trantwo{\ptprimeunroll}}
\newcommand{\mutprimesource}[1]{\source{\mutprime}{#1}}
\newcommand{\tprimeunrollsource}[1]{\source{\tprimeunroll}{#1}}

\newcommand{\dtran}{\tranone{\Delta}}
\newcommand{\dptran}{\tranone{\Delta'}}

\newcommand{\dtrantwo}{\trantwo{\Delta}}
\newcommand{\dptrantwo}{\trantwo{\Delta'}}

\newcommand{\pttran}{\trantwo{\pp : \T}}
\newcommand{\ptptran}{\trantwo{\pp : \T'}}
\newcommand{\qttran}{\trantwo{\q : \T'}}


\newcommand{\myparagraph}[1]{\smallskip\noindent\emph{\bf #1}}
\newcommand*{\fullref}[1]{\hyperref[{#1}]{\cref*{#1}: \nameref*{#1}}}  
\newcommand*{\partref}[1]{\hyperref[{#1}]{\nameref*{#1} \ref*{#1}}}  
\newcommand{\rulename}[1]{\text{\textnormal{\footnotesize{\color{ruleColor}[\textsc{#1}]}}}}

\newcommand{\cinferruletop}[3][]{%
  \mprset{fraction={===},
  fractionaboveskip=0.2ex,
  fractionbelowskip=0.4ex}
  \inferrule[{\scriptsize #1}]{#2}{#3}
}
\newcommand{\cinferrule}[3][]{
  \mprset{fraction={===},
  fractionaboveskip=0.2ex,
  fractionbelowskip=0.4ex}
  \inferrule[#1]{#2}{#3}
}

\newcommand{\cinferruleleft}[3][]{
  \mprset{fraction={===},
  fractionaboveskip=0.2ex,
  fractionbelowskip=0.4ex}
  \inferrule*[left=#1]{#2}{#3}
}

\newcommand{\ccinferrule}[3][]{
  \mprset{fraction={===},
  fractionaboveskip=1.8ex,
  fractionbelowskip=0.4ex}
  \inferrule[#1]{#2}{#3}
}


\definecolor{ruleColor}{rgb}{0.1, 0.3, 0.1}
\definecolor{hlColor}{rgb}{0.65, 1.0, 0.65}
\definecolor{tyColorCustom}{rgb}{0.0, 0.0, 0.85}%
\definecolor{roleColor}{rgb}{0.5, 0.0, 0.0}
\definecolor{gtColor}{rgb}{0.43, 0.21, 0.1}
\definecolor{stColor}{rgb}{0.54, 0.17, 0.89}
\definecolor{mpColor}{rgb}{0.54, 0.17, 0.89}
\newcommand{\roleCol}{\color{magenta}}


\newcommand{\actionprocin}[4]{#1{:}\pchan{#2} ? \labels{#3}(#4)}
\newcommand{\actionprocout}[4]{#1{:}\pchan{#2} ! \labels{#3}\ENCan{#4}}
\newcommand{\actionprocoutnb}[3]{#1{:}\pchan{#2} ! \labels{#3}}
\newcommand{\actionnd}[1]{#1{:} \tau}
\newcommand{\actiontau}[3]{(#1,#2)\labels{#3}}
\newcommand{\actionprob}[1]{#1{:}p}
\newcommand{\actionunknown}[3]{#1{:}\pchan{#2} \dagger \labels{#3}}
\newcommand{\actionunknowndual}[3]{#1{:}\pchan{#2} \overline{\dagger} \labels{#3}}

\newcommand{\ENCan}[1]{\langle #1 \rangle}

\newcommand{\DeltaOAuth}{\Delta_{\text{oa}}}

\newcommand{\probexp}[3]{\ENCan{#1}_{{#2},{#3}}}

\theoremstyle{definition}
\newtheorem{remark}{Remark}[section]

\hypersetup{hypertexnames=false}
\newcommand{\proofreference}[1]{(\text{Proof in {\color{blue}#1}}).}
\AddToHook{cmd/appendix/before}{\renewcommand{\proofreference}[1]{\ignorespaces}}

\newcommand{\labelend}{\textit{``end''}}
\newcommand{\deadlocklabel}{\textit{``deadlock''}}
\newcommand{\livelocklabel}{\textit{``livelock''}}
\newcommand{\receiveanylabel}[2]{\textit{``receive $\conc#1\conc#2$"}}
\newcommand{\receivelabel}[3]{\textit{``receive $\conc#1\conc#2\conc#3$"}}
\newcommand{\sendlabel}[3]{\textit{``send $\conc#1\conc#2\conc#3$"}}
\newcommand{\upotl}{\up{\ppi{1}}{\ppi{2}}{\labelsell}}
\newcommand{\downot}{\down{\ppi{1}}{\ppi{2}}}

\newcommand{\case}[1]{\paragraph{Case #1}}
\newcommand{\ddptrantwo}{\trantwo{\Delta, \Delta'}}
\newcommand{\ptran}[1]{\trantwo{\ppi{#1} : \T_{#1}}}
\newcommand{\ptranprime}[1]{\trantwo{\ppi{#1} : \T'_{#1}}}
\newcommand{\deltaupto}[1]{\ppi{1} : \T_1, \cdots, \ppi{#1} : \T_{#1}}
\newcommand{\deltaprimeupto}[1]{\ppi{1} : \T'_1, \cdots,
  \ppi{#1} : \T'_{#1}}
\newcommand{\tranupto}[1]{\ptran{1} \cmp \dots \cmp \ptran{#1}}
\newcommand{\tranprimeupto}[1]{\ptranprime{1} \cmp  \dots \cmp \ptranprime{#1}}
\newcommand{\pipj}[2]{\ppi{#1}::\ppi{#2}}
\newcommand{\popt}{\pipj{1}{2}}
\newcommand{\pipjl}[2]{\pipj{#1}{#2}::\labelsell}
\newcommand{\poptl}{\pipjl{1}{2}}
\newcommand{\allstates}{S_1 \cup \cdots \cup S_n}
\newcommand{\allstatesp}{S'_1 \cup \cdots \cup S'_n}
\newcommand{\allstatest}{T_1 \cup \cdots \cup T_n}
\newcommand{\allstatespt}{T'_1 \cup \cdots \cup T'_n}
\newcommand{\allstatesu}{U_1 \cup \cdots \cup U_n}
\newcommand{\allstatespu}{U'_1 \cup \cdots \cup U'_n}
\newcommand{\subsetsim}{\mathrel{\ooalign{\raise.3ex\hbox{$\subset$}\cr$\raise-.7ex\hbox{$\sim$}$}}}
\newcommand{\prsim}{\sim^+}
\newcommand{\refines}[1]{\subsetsim_{#1}}
\newcommand{\deltaprimeuptotwo}[1]{\ppi{1} : \T'_1, \ppi{2} : \T'_2, \ppi{3}:
  \T_3, \cdots,
  \ppi{#1} : \T_{#1}}
\newcommand{\tranprimeuptotwo}[1]{\ptranprime{1} \cmp \ptranprime{2} \cmp
  \ptran{3} \cmp \dots \cmp \ptran{#1}}
\newcommand{\parttran}{\oxtrantwo{\T}{\pp}{n}{m}{f}}
\newcommand{\parttranp}{\oxtrantwo{\T'}{\pp}{n}{m}{f \cup \{\ty \mapsto n\}}}
\newcommand{\parttrani}{\oxtrantwo{\T'_i}{\pp}{\nexts{i}}{m}{f}}
\newcommand{\vars}[1]{{\sf vars}(#1)}
\newcommand{\abreduce}[1]{\prismreduce{*}{#1}}

\newcommand{\dpptran}{\tranone{\Delta''}}

\newcommand{\detran}{\Delta_\text{tran}}
\newcommand{\mtran}{M_\text{tran}}
\newcommand{\mtranp}{M_{\text{tran}, \pp}}
\newcommand{\mtranq}{M_{\text{tran}, \q}}
\newcommand{\scc}{S_{\pchan{c}}}
\newcommand{\spp}{S_\pp}
\newcommand{\sq}{S_\q}
\newcommand{\pqa}{\pp :: \q}
\newcommand{\pqli}[1]{\pp::\q::\labelselli{#1}}

\newcommand{\prose}{\textsc{Prose}}

\newcommand{\mapper}{\roleFmt{mapper}}
\newcommand{\worker}[1]{\roleFmt{ worker_{#1}}}
\newcommand{\reducer}{\roleFmt{reducer}}

\newcommand{\datum}{{\texttt{datum}}}
\newcommand{\resultl}{{\texttt{result}}}
\newcommand{\continuel}{{\texttt{continue}}}
\newcommand{\stopl}{{\texttt{stop}}}

\newcommand{\gol}{\texttt{go}}
\newcommand{\redol}{\texttt{redo}}
\newcommand{\donel}{\texttt{done}}
\newcommand{\dice}[1]{\roleFmt{d_{#1}}}
\newcommand{\dummyp}{\roleFmt{dummy}}

\newcommand{\car}{\roleFmt{car}}
\newcommand{\host}{\roleFmt{host}}
\newcommand{\player}{\roleFmt{player}}

\newcommand{\rel}{\mathcal{R}}

\Crefname{section}{\S\!}{\S\!}%
\Crefname{subsection}{\S\!}{\S\!}%
\Crefname{subsubsection}{\S\!}{\S\!}%
\Crefname{appendix}{Appendix\@}{Appendix\@}
\Crefname{definition}{Def.\@}{Defs.\@}%
\Crefname{figure}{Fig.\@}{Figs.\@}%
\Crefname{example}{Example\@}{Examples\@}%
\Crefname{corollary}{Corollary\@}{Corollaries\@}%
\Crefname{theorem}{Theorem\@}{Theorems\@}%
\Crefname{proposition}{Proposition\@}{Propositions\@}%
\Crefname{item}{Item\@}{Items\@}%
\Crefname{remark}{Remark\@}{Remarks\@}%
\Crefname{notation}{Not.\@}{Nots.\@}%

\crefname{section}{\S\!}{\S\!}%
\crefname{subsection}{\S\!}{\S\!}%
\crefname{subsubsection}{\S\!}{\S\!}%
\crefname{appendix}{Appendix \S\!}{Appendix \S\!}
\crefname{definition}{Def.\@}{Defs.\@}%
\crefname{figure}{Fig.\@}{Figs.\@}%
\crefname{example}{Ex.\@}{Exs.\@}%
\crefname{corollary}{Cor.\@}{Cors.\@}%
\crefname{theorem}{Thm.\@}{Thms.\@}%
\crefname{proposition}{Prop.\@}{Props.\@}%
\crefname{remark}{Rem.\@}{Rems.\@}%
\crefname{notation}{Not.\@}{Nots.\@}%

\providecommand{\ifnotsplit}[2]{#1} 

\newif\ifpldi
\pldifalse
\newcommand{\procoutbound}[5]{ #1 : #2 ! \labels{#3} \langle #4 \rangle . #5}
\newcommand{\probmergebig}[3]{\bigsqcap_{#1}({#2},{#3})}
\newcommand{\probmergebigg}[3]{\bigsqcap_{#1}\left({{#2},{#3}}\right)}
\newcommand{\probpathsigi}[3]{\mathbb{P}_{\phi_{#1}, {#2}}(#3)}

\newcommand{\gtick}{\textcolor{OliveGreen}{\ding{52}}}
\newcommand{\rcross}{\textcolor{red}{\ding{55}}}

\newcommand{\req}{\labels{\texttt{req}}}
\newcommand{\ress}{\labels{\texttt{res}}}
\newcommand{\err}{\labels{\texttt{err}}}
\newcommand{\wrk}[1]{\roleFmt{w_{#1}}}
\newcommand{\dum}{\roleFmt{d}}

\newcommand{\livelock}{\mathtt{llock}}
\newcommand{\probdfminplain}[0]{\mathbb{P}_\mathsf{df}^\mathsf{ min}}
\newcommand{\probliveminplain}[0]{\mathbb{P}_\mathsf{live}^\mathsf{ min}}
\newcommand{\probdfmin}[1]{\mathbb{P}_\mathsf{df}^\mathsf{ min}(#1)}
\newcommand{\probphimin}[1]{\mathbb{P}_\varphi^\mathsf{ min}(#1)}
\newcommand{\probdfmax}[1]{\mathbb{P}_\mathsf{df}^\mathsf{ max}(#1)}
\newcommand{\probtermmin}[1]{\mathbb{P}_\mathsf{term}^\mathsf{ min}(#1)}
\newcommand{\probtermmax}[1]{\mathbb{P}_\mathsf{term}^\mathsf{ max}(#1)}

\newcommand{\problive}[1]{\mathbb{P}_\mathsf{live}(#1)}
\newcommand{\problivemin}[1]{\mathbb{P}_\mathsf{live}^\mathsf{ min}(#1)}
\newcommand{\problivemax}[1]{\mathbb{P}_\mathsf{live}^\mathsf{ max}(#1)}
\newcommand{\waslll}{{\sff{WASlivelock}}}

\AtEndPreamble{%
\theoremstyle{acmdefinition}
\newtheorem{notation}[theorem]{Notation}}
\makeatletter
\@for\thm@env:=theorem,definition,example,lemma,proposition,corollary,conjecture,notation\do{%
  \expandafter\def\csname theH\thm@env\endcsname{\thesection.\arabic{theorem}}}
\def\theHremark{\thesection.\arabic{remark}}
\makeatother


\title{Model Checking Probabilistic Multiparty Session Types}

\author{Aleksander Junge}
\authornote{Both authors contributed equally to this research.}
\orcid{0009-0005-7572-880X}
\email{aleksander.junge@cs.ox.ac.uk}

\author{Leo Seojun Lee}
\authornotemark[2]
\email{leo@leol.ee}
\orcid{0009-0003-2157-4038}

\affiliation{%
    \institution{University of Oxford}
    \city{Oxford}
    \country{UK}
}

\author{Joseph Paulus}
\email{joe.paulus@cs.ox.ac.uk}
\orcid{0000-0002-1711-9361}

\author{Nobuko Yoshida}
\correspondingauthor
\email{nobuko.yoshida@cs.ox.ac.uk}
\orcid{0000-0002-3925-8557}

\author{David Parker}
\email{david.parker@cs.ox.ac.uk}
\orcid{0000-0003-4137-8862}

\affiliation{%
    \institution{University of Oxford}
    \city{Oxford}
    \country{UK}
}

\begin{abstract}
Distributed systems increasingly rely on randomisation;
yet, verification techniques for such systems remain separate from the
programming abstractions used to build them.
This paper proposes the first probabilistic extension of \emph{general
multiparty session types}
--- typing sessions bottom-up, with no global specification ---
deriving quantitative properties of a
multiparty session from its session types alone.
Our types combine \emph{probabilistic} choice with
\emph{nondeterministic} choice that abstracts data-dependent control,
and allow \emph{separated choices}, where each branch of a selection
or branching may involve a different participant --- a combination
beyond previous probabilistic multiparty session type systems.
To achieve this, we develop a novel probabilistic subtyping relation and
merge operator, using which we prove type soundness and session fidelity.
The semantics of typing contexts separates nondeterministic resolution,
probabilistic choice, and synchronisation, so that every typing context
induces a finite Markov decision process. 
On it, we derive sound \emph{lower bounds}, holding under every fair
adversarial scheduler, on the probability that a session is
deadlock-free and live --- the first behavioural property guarantees,
partial or otherwise, for a probabilistic session calculus.
To connect session types with probabilistic model checking, we
introduce the first translation from session types into the \prism{}
modelling language, and prove it sound and complete with respect to our 
operational semantics.
We implement the framework in the toolchain \prose{}, which translates
probabilistic multiparty session typing contexts into \prism{} and
checks safety, deadlock-freedom, and liveness.

\end{abstract}



\newif\ifshowcomments
\showcommentsfalse

\maketitle

\section{Introduction}
\label{sec:intro}
\myparagraph{Probabilistic model checking. }
Concurrent components have been deployed in many computing systems, from
multicore architectures to web services to autonomous 
systems. Ensuring their safe and reliable execution is an essential
requirement for the design and specification of system components, 
but it creates additional demand for verification, as components
operate in \emph{uncertain} environments.
\emph{Probabilistic model
checking} \cite{model-checking-book,pctlthesis} aims to
automatically check
whether components that exhibit uncertain (probabilistic) and
nondeterministic behaviours satisfy given requirements, usually
specified by probabilistic extensions of temporal
logics. 
Probability is used to quantify environmental uncertainty or randomisation, nondeterminism to represent decisions or concurrency. Over the last two decades, probabilistic model checking tools (e.g., 
\cite{prism,hensel2022probabilistic,Hartmanns14,ABB24})
have been actively
developed and used for quantitative verification
in a broad spectrum of application domains~\cite{KNP25}.
Our aim is to lay the first groundwork connecting \emph{probabilistic
multiparty session types} to probabilistic model checking, so that
session-typed systems benefit directly from advances in
model checking techniques. Through a \emph{precise} behavioural
correspondence between probabilistic session types and a
model checker's input language, we analyse a spectrum of probabilistic
properties of multiparty sessions, developing a static typing system
that yields sound bounds on the probability of ``good'' behaviour ---
such as deadlock-freedom and liveness.

\myparagraph{Multiparty Session Types (MPST)} \cite{Honda2008,HYC2016} are
an extension of binary session types \cite{Takeuchi1994,honda.vasconcelos.kubo:language-primitives}, 
facilitating specification and verification of complex many-party interacting components.
Historically, the MPST theory has been developed based on
the \emph{top-down approach} where the chief system architect 
first writes a multiparty protocol (called
a \emph{global type}) between participants. This protocol
is projected to a set of end-point (\emph{local}) types against which
conformance of each component (participant) is checked. 
A collection of typed components then automatically
satisfies \emph{qualitative properties} such as type-safety,
deadlock-freedom and liveness \cite{YH2024}.

More recently, Scalas and Yoshida \cite{Scalas2019} proposed
\emph{generalised multiparty session types} that do \emph{not} require 
global types, but can enforce a desired property $\varphi$
by directly checking that a set of local types 
satisfies $\varphi$.
The bottom-up approach types a \emph{strictly larger set} of processes:
protocols beyond the reach of any global type --- with asymmetric or
emergent interaction structures that projection cannot express --- as
well as type-safe processes that may deadlock, run forever, or fail to
progress. This ``partial'' nature is a natural fit with probabilistic model
checking: rather than demanding the outright absence of
deadlock or non-termination, tools such as PRISM and Storm
\emph{quantify} such behaviours --- e.g., computing the probability of
reaching a deadlock or of termination~\cite{ABB24,HoffmannIMNV16}. 
\subsection*{Overview}\label{sec:overview}
\newcommand{\introTextboxWidth}{5cm}
\newcommand{\introTextboxShift}{2.55cm}
\newcommand{\introBraceShift}{-2.9cm}

\begin{figure}[t]
\centering
\small
\begin{tabular}{cc}
  \subcaptionbox{Type checking \label{subfig:typechecking}}
{\begin{minipage}{0.5\textwidth}
  \begin{tikzpicture}[>=Latex, node distance=1.2cm]
    \node[draw=black, fill=gray!20, rectangle, rounded corners] (gt) {Probabilistic property $\mathbb{P}$};
    \node (tt1) [below of=gt, xshift=-1.0cm] {$[\![\T_1,$};
    \node (tt2) [below of=gt, xshift=-0.2cm] {$\T_2,$};
    \node (ttn) [below of=gt, xshift=0.9cm] {$\T_n]\!]$};
    \node (t1) [below of=tt1, yshift=0.3cm] {$\T_1$};
    \node (t2) [below of=tt2, yshift=0.3cm] {$\T_2$};
    \node (tn) [below of=ttn, yshift=0.3cm] {$\T_n$};
    \node (u1) [below of=t1, yshift=0.5cm] {$\T_{1}'$};
    \node (u2) [below of=t2, yshift=0.5cm] {$\T_{2}'$};
    \node (un) [below of=tn, yshift=0.5cm] {$\T_{n}'$};
    \node (p1) [below of=u1, yshift=0.3cm] {$\PP_1$};
    \node (p2) [below of=u2, yshift=0.3cm] {$\PP_2$};
    \node (pn) [below of=un, yshift=0.3cm] {$\PP_n$};
    
    \node[rotate=90] (s1) at ($(t1)!0.5!(u1)$) {{\large $\subt$}};
    \node[rotate=90] (s2) at ($(t2)!0.5!(u2)$) {{\large $\subt$}};
    \node[rotate=90] (sn) at ($(tn)!0.5!(un)$) {{\large $\subt$}};

    \draw[->] (t1) -- (tt1);
    \draw[->] (t2) -- (tt2);
    \draw[->] (tn) -- (ttn);

    \draw[->] (tt1) -- (gt);
    \draw[->] (tt2) -- (gt);
    \draw[->] (ttn) -- (gt);
    
    \draw[->] (p1) -- (u1);
    \draw[->] (p2) -- (u2);
    \draw[->] (pn) -- (un);

    \node (ttd) at ($(tt2)!0.5!(ttn)$) {$\cdots,$};
    \node (td) at ($(t2)!0.5!(tn)$) {$\cdots$};
    \node (ud) at ($(u2)!0.5!(un)$) {$\cdots$};
    \node (pd) at ($(p2)!0.5!(pn)$) {$\cdots$};

    \node[text width=\introTextboxWidth] (text1) [right of=ttn, xshift=\introTextboxShift, yshift=0.45cm]
    {};
    \node[text width=\introTextboxWidth] (text1') [yshift=0cm] at (text1)
    {\small \sff{(3)} \sff{model checking}};%
    \node[text width=\introTextboxWidth] (text2) [right of=tn, xshift=\introTextboxShift, yshift=0.45cm] 
    {\small \sff{(2) translation}}; %
    \node[text width=\introTextboxWidth] (text3) [right of=un, xshift=\introTextboxShift, yshift=0.35cm]
        {\small \sff{(1b) subtype check}};%
    \node[text width=\introTextboxWidth] (text4) [right of=pn, xshift=\introTextboxShift, yshift=0.40cm]
    {\small \sff{(1a) type inference}}; %

    \newcommand{\drawBrace}[2]{%
      \node (brace1-#1) at (#1) [xshift=\introBraceShift, yshift=-#2] {};
      \node (brace2-#1) at (#1) [xshift=\introBraceShift, yshift=#2] {};
      \draw[decorate,decoration={calligraphic brace,amplitude=5pt,mirror,raise=3pt},yshift=0pt,line width=1.25pt] (brace1-#1)--(brace2-#1);
    }
    \drawBrace{text1}{0.55cm}
    \drawBrace{text2}{0.52cm}
    \drawBrace{text3}{0.45cm}
    \drawBrace{text4}{0.55cm}

  \end{tikzpicture}
  \end{minipage}
}  
&\hspace{-0.5cm}
\subcaptionbox{Type Inference \label{subfig:typeinference}}
{\begin{minipage}{0.5\textwidth}
  \begin{tikzpicture}[>=Latex, node distance=1.2cm]
    \node[draw=black, fill=gray!20, rectangle, rounded corners] (gt) {Probabilistic property $\mathbb{P}$};
    \node (tt1) [below of=gt, xshift=-1.0cm] {$[\![\T_{1}$};
    \node (tt2) [below of=gt, xshift=-0.2cm]     {$\T_{2}$};
    \node (ttn) [below of=gt, xshift=0.9cm]      {$\T_{n}]\!]$};
    \node (t1) [below of=tt1, yshift=-0.05cm] {$\T_{1}$};
    \node (t2) [below of=tt2, yshift=-0.05cm] {$\T_{2}$};
    \node (tn) [below of=ttn, yshift=-0.05cm] {$\T_{n}$};
    \node (p1) [below of=t1, yshift=-0.05cm] {$\PP_1$};
    \node (p2) [below of=t2, yshift=-0.05cm] {$\PP_2$};
    \node (pn) [below of=tn, yshift=-0.05cm] {$\PP_n$};

    \draw[->] (t1) -- (tt1);
    \draw[->] (t2) -- (tt2);
    \draw[->] (tn) -- (ttn);

    \draw[->] (tt1) -- (gt);
    \draw[->] (tt2) -- (gt);
    \draw[->] (ttn) -- (gt);
    
    \draw[->] (p1) -- (t1);
    \draw[->] (p2) -- (t2);
    \draw[->] (pn) -- (tn);

    \node (ttd) at ($(tt2)!0.5!(ttn)$) {$\cdots,$};
    \node (td) at ($(t2)!0.5!(tn)$) {$\cdots$};
    \node (pd) at ($(p2)!0.5!(pn)$) {$\cdots$};

    \node[text width=\introTextboxWidth] (text1) [right of=ttn, xshift=\introTextboxShift, yshift=0.45cm]
    {};
    \node[text width=\introTextboxWidth] (text1') [yshift=0cm] at (text1)
    {\small \sff{(3)} \sff{model checking}};%
    \node[text width=\introTextboxWidth] (text2) [right of=tn, xshift=\introTextboxShift, yshift=0.58cm] 
    {\small \sff{(2) translation}}; %
    \node[text width=\introTextboxWidth] (text3) [right of=un, xshift=\introTextboxShift, yshift=-0.25cm]
    {\small \sff{(1a) type inference}};%

    \newcommand{\drawBrace}[2]{%
      \node (brace1-#1) at (#1) [xshift=\introBraceShift, yshift=-#2] {};
      \node (brace2-#1) at (#1) [xshift=\introBraceShift, yshift=#2] {};
      \draw[decorate,decoration={calligraphic brace,amplitude=5pt,mirror,raise=3pt},yshift=0pt,line width=1.25pt] (brace1-#1)--(brace2-#1);
    }
    \drawBrace{text1}{0.55cm}
    \drawBrace{text2}{0.66cm}
    \drawBrace{text3}{0.7cm}
  \end{tikzpicture}
  \end{minipage}
}  
\end{tabular}
\vspace{-3mm}
\caption{Overview of the probabilistic property verification
procedure. In (a), given $\T_i$ and $P_i$ ($i\in \{1..n\}$) (1a) infer
local type $\T_{i}'$ from process $P_i$; 
(1b) check $\T_{i}' \subt \T_i$; (2) translate
$\T_i$ to \prism{} and (3) model check whether $\{T_i\}_{i \in I}$
satisfy the desired probabilistic property $\mathbb{P}$.  
In (b), directly verify the inferred types. 
}
\label{fig:workflow}
\end{figure}
We give an overview of the workflow in \Cref{fig:workflow} and explain
each step below. 

\myparagraph{Step (1): Type checking. }%
Following the bottom-up approach~\cite{Scalas2019}, the first stage
of our framework abstracts each process as a finite probabilistic session type $\T$.
This abstraction is necessary: communication safety of untyped
session processes is undecidable, as we prove by encoding
two-counter Minsky machines into sessions (Prop.~\ref{prop:procsafeundec}).
After inferring local types our toolchain supports two modes; either model check the types directly (\cref{fig:workflow}b), or type check them against a type specification via subtyping before model checking (\cref{fig:workflow}a).

Fig.~\ref{fig:abstraction-intro} shows a session (bottom-left)
in which a worker $\pchan{u}$ drives a counter held by
$\pchan{s}$ up or down, switching mode depending on its state.
Following~\cite{FuDG25}, our calculus distinguishes
data-dependent conditionals ($\mathsf{if}$) from probabilistic
choices ($\mathsf{flip}$).
Switching down fails with probability $\probn{0.1}$,
resulting in a deadlock, while switching up terminates
successfully with probability $\probn{0.2}$.

The unbounded counter gives rise to an infinite state space (left, with one column per counter value), preventing direct finite-state model checking.
We instead infer local types that soundly over-approximate communication behaviour, following existing approaches that model check behavioural types as \emph{abstractions of processes}~\cite{Scalas2019} and \emph{programs}~\cite{DBLP:conf/icse/LangeNTY18,PLDI19Effpi,BSYZ2022}.

\begin{figure}
\centering
\begin{subfigure}[c]{0.54\textwidth}
\centering
\begin{tikzpicture}[
    >={Stealth[length=2.1mm,width=1.5mm]},
    every node/.style={inner sep=1.5pt},
    st/.style={circle, fill=black, inner sep=1.2pt},
    edge/.style={draw, ->, line width=0.45pt, shorten >=2pt, shorten <=2pt},
    lab/.style={font=\fontsize{6.5}{7.5}\selectfont},
    ann/.style={font=\fontsize{6.5}{7.5}\selectfont},
    val/.style={font=\fontsize{6.5}{7.5}\selectfont, text=gray},
]
\node[st] (tm) at (-2.3,0.85) {};
\node[st] (t0) at (0,1.3)     {};
\node[st] (t1) at (2.3,1.75)  {};
\node[st] (bm) at (-2.3,-0.45) {};
\node[st] (b0) at (0,0)       {};
\node[st] (b1) at (2.3,0.45)  {};
\node (tdotsL) at (-3.2,0.62) {$\cdots$};
\node (tdotsR) at (3.2,1.98)  {$\cdots$};
\node (bdotsL) at (-3.2,-0.68) {$\cdots$};
\node (bdotsR) at (3.4,0.7)  {$\cdots$};
\coordinate (bin) at (3.2,0.68);
\draw[edge,opacity=0.45] (bin) -- (b1);
\node[st] (dlm) at (-3.55,1.3)  {};
\node[st] (dl0) at (-1.25,1.75) {};
\node[st] (dl1) at (1.05,2.2)   {};
\node[st] (em)  at (-1.15,-0.95) {};
\node[st] (e0)  at (1.375,-0.605)  {};
\node[st] (e1)  at (3.45,-0.05) {};
\node[val,anchor=east] at (-1.4,1.75) {{\color{darkgray}deadlock}};
\node[val,anchor=west] at (1.3,-0.73)  {{\color{darkgray}end}};
\node[val] at (-1.98,1.23) {$\langle -1\rangle$};
\node[val] at (0.32,1.68)  {$\langle 0\rangle$};
\node[val] at (2.62,2.13)  {$\langle 1\rangle$};
\coordinate (in) at (0,2.15);
\draw[edge] (in) -- (t0);

\draw[edge,opacity=0.45] (tdotsL) -- (tm);
\draw[edge] (tm) -- (t0);
\draw[edge] (t0) -- node[lab,above,sloped,pos=0.32]{}%
                    node[lab,above,sloped,pos=0.62]{$\actiontau{\pchan{u}}{\pchan{s}}{\labels{incr}}$} (t1);
\draw[edge,opacity=0.45] (t1) -- (tdotsR);
\draw[edge,opacity=0.45] (b1) -- (b0);
\draw[edge] (b0) -- node[lab,below,sloped,pos=0.4]{$\actiontau{\pchan{u}}{\pchan{s}}{\labels{decr}}$} (bm);
\draw[edge] (bm) -- (bdotsL);
\draw[edge,opacity=0.7] (tm) to[bend right=35]
    node[ann,left,pos=0.85]{} (bm);
\draw[edge,opacity=0.7] (bm) to[bend right=35]
    node[ann,right,pos=0.85]{} (tm);
\draw[edge] (t0) to[bend right=35]
    node[ann,left,pos=0.85,xshift=-0.08cm]{$\probn{0.9}$}
    node[ann,left,pos=0.4]{}%
    node[lab,left,pos=0.65]{$\actiontau{\pchan{u}}{\pchan{s}}{\labels{switch}}$} (b0);
\draw[edge] (b0) to[bend right=35]
    node[ann,right,pos=0.85,xshift=0.08cm]{$\probn{0.8}$}
    node[ann,right,pos=0.65]{}%
    node[lab,right,pos=0.4]{$\actiontau{\pchan{u}}{\pchan{s}}{\labels{switch}}$} (t0);
\draw[edge,opacity=0.7] (t1) to[bend right=35]
    node[ann,left,pos=0.85]{} (b1);
\draw[edge,opacity=0.7] (b1) to[bend right=35]
    node[ann,right,pos=0.85]{} (t1);
\draw[edge,opacity=0.7] (tm) -- (dlm);
\draw[edge] (t0) -- node[lab,above,sloped,pos=0.45]{$\actiontau{\pchan{u}}{\pchan{s}}{\labels{fail}}$}
                    node[ann,below,sloped,pos=0.55]{$\probn{0.1}$} (dl0);
\draw[edge,opacity=0.7] (t1) -- (dl1);
\draw[edge,opacity=0.7] (bm) -- (em);
\draw[edge] (b0) -- node[lab,above,sloped,pos=0.55]{$\actiontau{\pchan{u}}{\pchan{s}}{\labels{done}}$}
                    node[ann,below,sloped,pos=0.85]{$\probn{0.2}$} (e0);
\draw[edge,opacity=0.7] (b1) -- (e1);
\end{tikzpicture}
\end{subfigure}%
\begin{subfigure}[c]{0.04\textwidth}
  \centering
  {\Large$\rightsquigarrow$}
\end{subfigure}%
\hfill
\begin{subfigure}[c]{0.40\textwidth}
\centering
\hspace*{-2mm}%
\begin{tikzpicture}[
    >={Stealth[length=2.1mm,width=1.5mm]},
    every node/.style={inner sep=1.5pt},
    edge/.style={draw, ->, line width=0.45pt, shorten >=2pt, shorten <=2pt},
    lab/.style={font=\fontsize{6.5}{7.5}\selectfont},
    ann/.style={font=\fontsize{6.5}{7.5}\selectfont},
    val/.style={font=\fontsize{6.5}{7.5}\selectfont, text=gray},
]
\node (D2) at (0,1.4)   {$\Delta_2$};
\node (D1) at (1.7,1.4) {$\Delta_1$};
\node (D3) at (3.6,1.4) {$\Delta_3$};
\node (D8) at (5.2,1.4) {$\Delta_8$};
\node (D7) at (0,0)   {$\Delta_7$};
\node (D6) at (1.7,0) {$\Delta_6$};
\node (D4) at (3.6,0) {$\Delta_4$};
\node (D5) at (5.2,0) {$\Delta_5$};
\coordinate (in) at (1.7,2.2);
\node[val] at (5.1,1.8)  {deadlock};
\node[val] at (0,-0.4)   {end};

\draw[edge] (in) -- (D1);
\draw[edge] (D1) to[bend right=22]
    node[lab,above,pos=0.5]{$\actionnd{\pchan{u}}$} (D2);
\draw[edge] (D2) to[bend right=22]
    node[lab,below,pos=0.5]{$\actiontau{\pchan{u}}{\pchan{s}}{\labels{incr}}$} (D1);
\draw[edge] (D1) -- node[lab,above,pos=0.45]{$\actionnd{\pchan{u}}$} (D3);
\draw[edge] (D3) -- node[lab,above,pos=0.45]{$\actiontau{\pchan{u}}{\pchan{s}}{\labels{fail}}$}
                    node[ann,below,pos=0.9,yshift=-0.03cm]{$\probn{0.1}$} (D8);
\draw[edge] (D3) -- node[lab,right,pos=0.5]{$\actiontau{\pchan{u}}{\pchan{s}}{\labels{switch}}$}
                    node[ann,left,pos=0.8,xshift=-0.05cm]{$\probn{0.9}$} (D4);
\draw[edge] (D4) to[bend left=22]
    node[lab,above,pos=0.5]{$\actionnd{\pchan{u}}$} (D5);
\draw[edge] (D5) to[bend left=22]
    node[lab,below,pos=0.5]{$\actiontau{\pchan{u}}{\pchan{s}}{\labels{decr}}$} (D4);
\draw[edge] (D4) -- node[lab,above,pos=0.45]{$\actionnd{\pchan{u}}$} (D6);
\draw[edge] (D6) -- node[lab,right,pos=0.55]{$\actiontau{\pchan{u}}{\pchan{s}}{\labels{switch}}$}
                    node[ann,right,pos=0.8,xshift=0.03cm]{$\probn{0.8}$} (D1);
\draw[edge] (D6) -- node[lab,above,pos=0.45]{$\actiontau{\pchan{u}}{\pchan{s}}{\labels{done}}$}
                    node[ann,below,pos=0.8,yshift=-0.03cm]{$\probn{0.2}$} (D7);
\end{tikzpicture}
\hspace*{2mm}\par
\end{subfigure}
\begin{subfigure}[l]{0.48\textwidth}
  \vspace{1mm}
{$
  \footnotesize
\hspace*{-5mm}
\begin{array}{l}
\PP_{\pchan{u}}  =
\psyntrec{X}{\,
  \cond{e}{\procoutnl{\ps}{\labels{incr}}{X}}\\
           \hspace{1.6cm}{\flip{\probn{0.9}}
             {\procoutnl{\ps}{\labels{switch}}{\PP'}}
             {\procoutnl{\ps}{\labels{fail}}{X}}}}
\\%
\PP'  =
\psyntrec{Y}{\,
  \cond{e'}{\procoutnl{\ps}{\labels{decr}}{Y}}\\
 \hspace{1.6cm}{\flip{\probn{0.8}}
              {\procoutnl{\ps}{\labels{switch}}{X}}
              {\procoutnl{\ps}{\labels{done}}{\inact}}}}
\\%
\PP_{\pchan{s}}  =
\psyntrec{X}{\big(
  \procin{\pu}{\labels{incr}}{X}
  + \procin{\pu}{\labels{decr}}{X}\\
  \hspace{1.5cm}+ \procin{\pu}{\labels{switch}}{X}
  + \procin{\pu}{\labels{done}}{\inact}
  + \procin{\pu}{\labels{fail}}{\inact}
\big)}
\end{array}
$}
\end{subfigure}%
\hspace*{-5mm}
\begin{subfigure}[c]{0.02\textwidth}
\centering
{\Large$\rightsquigarrow$}
\hspace*{-5mm}
\end{subfigure}%
\begin{subfigure}[l]{0.40\textwidth}
  \vspace{1mm}
{$
\footnotesize
\begin{array}{r@{}l}
 \pu : \T_{\pchan u} &\;=\;
  \mu\ty.\;
   {\sum}\big\{\,
      {\oplus\ps!\labels{incr}.\,\ty},\\
    & \hspace{1.2cm}{{\oplus}\{\,
         \probn{0.9}\!:\ps!\labels{switch}.\,\T',\;
         \probn{0.1}\!:\ps!\labels{fail}.\,\ty
      \,\}}%
   \,\big\}
 \\%
 \T' &\;=\; 
    \mu\ty'.\;
    {\sum}\big\{\,
      {\oplus\ps!\labels{decr}.\,\ty'},\\
   &\hspace{1.2cm} {\oplus}\{\,
         \probn{0.8}\!:\ps!\labels{switch}.\,\ty,\;
         \probn{0.2}\!:\ps!\labels{done}.\,\tend
      \,\}
    \,\big\}
 \\%
 \ps : \T_{\pchan s} & = 
 \mu\ty.\;\&\big\{\,
     \pu?\labels{incr}.\,\ty,\;
     \pu?\labels{decr}.\,\ty,\;\\
     &\hspace{1cm}\pu?\labels{switch}.\,\ty,\;
     \pu?\labels{done}.\,\tend,\;
     \pu?\labels{fail}.\,\tend
 \,\big\}
\end{array}
$}
\end{subfigure}
\vspace{-3mm}
\caption{
    An infinite-state session (left; one column per counter value)
 and its finite typing context $\Delta_i$ (right).
The data-dependent decision by $\mathsf{if}$ becomes a nondeterministic choice over its two alternatives, while the probabilities of $\mathsf{flip}$ are preserved as edge annotations.
Processes are shown bottom-left and their local types bottom-right.
``$\mu$'' denotes recursion, $!$/$?$ output/input,
and $\oplus/\&$ internal/external choice
(see \cref{sec:processes,sec:types}).}
\label{fig:abstraction-intro}
\end{figure}

Typing replaces data-dependent conditionals with nondeterministic choices, accounting for the possible outcomes of tests whose values the types do not track, while retaining probability information from flips.
In \Cref{fig:abstraction-intro}, this removes the need to track the unbounded counter: the types describe a finite model of the communication modes instead.

The resulting model is a \emph{finite Markov decision process} (\textsf{MDP}), combining probabilistic choices with nondeterminism arising from both abstracted data and participant scheduling.
We analyse it under the least favourable (adversarial) resolutions of nondeterminism, subject to our fairness assumptions.
This accounts for the information lost through abstraction and yields \emph{sound lower bounds} on the probabilities of properties of the original session.
Type checking is decidable (Thm.~\ref{thm:decidability}), and the finite abstraction enables probabilistic model checking.
For instance, this analysis establishes that the session in \Cref{fig:abstraction-intro} is deadlock-free with probability at least $\probn{\nicefrac{9}{14}} \approx \probn{0.64}$ under every fair scheduler.

\smallskip

\myparagraph{Step (2): Precise translation from session types into \prism{}.}
To verify type-level properties automatically, we employ the \prism{}~\cite{prism} probabilistic model checker,
a well-established tool which has been used to study a diverse range of applications~\cite{KNP25}.
Other popular probabilistic model checkers such as Storm~\cite{storm} and
the Modest Toolset~\cite{Hartmanns14} also support the PRISM language as input 
(the latter by converting first into the JANI~\cite{Jani} format).

To fill a gap between  
types and models, we \emph{encode} session types into the PRISM modelling language 
and prove a \emph{precise} (\emph{sound} and \emph{complete}) 
behavioural correspondence. 
Building on earlier approaches to model checking behavioural
types~\cite{Scalas2019,DBLP:conf/icse/LangeNTY18,PLDI19Effpi,BSYZ2022},
we verify the translation itself by developing new bisimulation
and operational correspondence relations for the \prism{}
semantics, precisely relating type-level behaviours to those
of the generated models.

\myparagraph{Step (3): Probabilistic property analysis and model checking.}
As the final step, we express the properties of typing contexts
(\Cref{sec:typingcontextproperties}) as specifications in Probabilistic Linear Temporal Logic (PLTL) and run \prism{} to verify them.
For deadlock-freedom and liveness this establishes lower bounds (\Cref{sec:typingcontextproperties}), while for safety we get a yes/no answer.

There are two main challenges, both rooted in the nondeterminism of
the type-level MDP.
First, the probability of a set of paths is well-defined only relative
to a \emph{scheduler}: both the resolution of the abstracted
data-dependent choices and the scheduling of parallel
roles can change the probability of satisfying a given property.
We define the probability of a set of paths under a fixed scheduler by
standard means, using \emph{cylinder sets} (\Cref{def:cylinder}), and
obtain lower bounds by taking the \emph{minimum}
over all fair schedulers (\Cref{def:scheduler}); these minima then
transfer to well-typed sessions
(\Cref{thm:proc:df,thm:liveness}).
The second challenge is that \emph{probabilistic deadlock-freedom} and  \emph{probabilistic liveness} are 
quantified over \emph{fair} schedulers only, while model checkers impose fairness on all
nondeterministic choices uniformly---including the internal
nondeterminism, which must remain unconstrained.
We therefore characterise probabilistic deadlock-freedom as a safety property over finite paths that 
\prism{} checks without any fairness constraint. Characterising liveness in this manner is more challenging: to do this,
we introduce the notion of a \emph{weak almost-sure livelocked} region, and prove that the 
maximum probability of reaching this region corresponds exactly to the minimum probability 
that the typing context is live~(\Cref{thm:wasll}).

\myparagraph{Contributions and Outline. }
To the best of our knowledge, this work is the first to propose
a probabilistic extension of generalised (bottom-up)
multiparty session types, to prove that a typing context yields
\emph{sound lower bounds} of probabilistic \emph{properties} of typed
session processes, and to translate session types
into a probabilistic modelling language with proofs of
\emph{soundness} and \emph{completeness} with respect to operational
semantics.
\textbf{\Cref{sec:processes}}
proposes a \emph{probabilistic multiparty session calculus} ($\pmultipi$)
as a conservative extension of synchronous multiparty session
processes in \cite{Ghilezan2019}, with a single probabilistic construct
$\flip{\prob}{\PP}{\Q}$, and proves that safety of sessions is
undecidable (\Cref{prop:procsafeundec}).
\textbf{\Cref{sec:types}}
proposes a \emph{probabilistic multiparty session typing system},
introducing \emph{probabilistic subtyping} (\Cref{def:pmpst_subtype}),
which widens data-dependent choices into nondeterministic sums,
and the \emph{probabilistic merging operator} (\Cref{def:mergerel}),
which reconciles the two outcomes of a flip. We further demonstrate that
type checking is decidable (\Cref{thm:decidability}).
\textbf{\Cref{sec:properties}}
proves \emph{subject reduction (type safety)} (\Cref{thm:subred})
and \emph{session fidelity} (\Cref{thm:sessionfidelitysmall}),
interprets typing context reductions as finite MDPs with path probabilities relative to schedulers (\Cref{def:scheduler}),
and proves the main results: a typing context induces
\emph{lower bounds} on the probabilities
of deadlock-freedom (\Cref{thm:proc:df}) and
liveness (\Cref{thm:liveness}) of the sessions it types,
with their non-probabilistic counterparts as special cases.
\textbf{\Cref{sec:translation}}
proposes a translation from session types to the \prism{} language and 
proves sound and complete operational correspondence (\Cref{thm:soundness,thm:completeness}).
\textbf{\Cref{sec:modelchecking}}
expresses safety, deadlock-freedom and liveness as PLTL
specifications, reducing liveness to reachability of weak
almost-sure livelocks (\Cref{thm:wasll}), benchmarks our toolchain \prose{} on
case studies from the \prism{} homepage and related literature,
and outlines a prototype implementation of steps (1a) and (1b) of Fig.~\ref{fig:workflow}.
\textbf{\Cref{sec:related}} concludes with related work.
Our artifact is explained
in \Cref{sec:datastatement} and available in \cite{prose}.

\section{Probabilistic Multiparty Processes and Sessions}
\label{sec:processes}
This section introduces the
syntax and semantics of a \emph{probabilistic multiparty session calculus} ($\pmultipi$-calculus).
Since our focus is \emph{probabilistic verification}, we
base our calculus on the simplest multiparty session calculus
\cite{Ghilezan2019}
extended with a branching construct (external choice) which receives a message
from different participants \cite{PPGSY2026} 
and add a single new construct, $\flip{\prob}{P}{Q}$.
\label{sec:mpst-syntax-terms}

\myparagraph{Probability and nondeterminism.}
The calculus separates two sources of branching behaviour.
\emph{Probability} is introduced operationally by $\flip{\prob}{P}{Q}$,
whose two continuations are taken with complementary probabilities $\prob$ and
$\probn{1} - \prob$; the reduction relation tracks these as
explicit weights.
\emph{Nondeterminism} arises at the process level from the scheduling of parallel roles. 
The typed account of how nondeterminism is abstracted is deferred
to \Cref{sec:types}. %

\myparagraph{Expressions} ($\e, \e', \dots$) are variables ($x,y,z,\dots$),
a value $\val$, which is either a natural number $\valn$, an integer
$\valr$, or a Boolean $\true$/$\false$,
or expressions built by applying operators such as
${\tt neg}$, $\neg$ and $+$.
\centerline{
$
\label{fig:expr}
\e ::= \true \SEP \false \SEP \valn \SEP \vali \SEP x
     \SEP \e \vee \e \SEP \neg \e \SEP \e + \e \SEP
\fsucc{\e} \SEP \fneg{\e} \SEP \e > \e'$}\\[1mm]
\myparagraph{Processes}
($\PP, \Q, R, \dots$) and \textbf{multiparty sessions} ($\N, \N', \dots$) are defined by the following syntax:\\[1mm]
\centerline{
$\begin{array}{rll}
\PP, \Q   ::= & \inact  \SEP 
\procout{\pp}{\ell}{\e}{\PP} 
\SEP
\psyntinnl{i\in I}{\pp_i}{i}{x_i}{\PP_i}
\SEP 
\cond{\e}{\PP}{\Q} \SEP 
& \text{\footnotesize (nil, selection, branching, conditional)}\\
 & \flip{p}{\PP}{\Q} \SEP \mu X. \PP \SEP X & \text{\footnotesize(flip, recursion, proc variable)}\\
\N, \N'  ::=   & \pa \pp \PP  \ \SEP \ \N \pc\N' \SEP \rederror & \text{\footnotesize (role, parallel composition, error)}
\end{array}
$}\\[1mm]
$\pp, \pq, \pr, \ps, \dots$ are \emph{participants} (\emph{roles}) from
a countably infinite set. The \emph{nil} $\inact$ denotes termination;
process $\procout{\pp}{\ell}{\e}{\PP}$
denotes the \emph{selection}
of label $\labels{\ell}$ and expression $\e$
communicating with participant $\pchan{p}$ and
continuing with behaviour $\PP$.
Process $\psyntinnl{i\in I}{\pp_i}{i}{x_i}{\PP_i}$
is \emph{branching} (\emph{external choice}): for any $i \in I \neq \emptyset$
it can accept label $\labels{\ell_i}$ with a value bound to $x_i$ from
participant $\pchan{\pp_i}$, continuing as $\PP_i$.
The sender $\pchan{\pp_i}$ may vary across branches, allowing a
single external choice to await messages from different participants.
This follows the external choice in \cite{PPGSY2026,SE2024}, which 
can represent 
protocols in which a server accepts
client updates in arbitrary order, as in session-typed models
of federated learning. 
On the sending side, as $\procout{\pp}{\ell}{\e}{\PP}$ names its own target
$\pp$, a role may also direct the branches of a $\textsf{flip}$ or $\textsf{if}$ statement 
to \emph{different} participants.
We often write $\procin{\pp}{\labels{\ell}(x)}{\PP}$ for a singleton branching, expand it as, e.g.,
$\procin{\pchan{\pp_1}}{\labels{\ell_1}(x_1)}{\PP_1}+
\procin{\pchan{\pp_2}}{\labels{\ell_2}(x_2)}{\PP_2}$, and omit the payloads, e.g., $\procin{\pchan{\pp_1}}{\labels{\ell_1}}{\PP_1}+
\procin{\pchan{\pp_2}}{\labels{\ell_2}}{\PP_2}$, when they are not needed.
\emph{Conditional processes} $\cond{\e}{\PP}{Q}$ and \emph{recursive processes}
$\mu X.\PP$ are standard. We assume recursive processes are \emph{guarded}, i.e.,
every recursive variable occurs under a prefix containing at least one input or
output. Thus e.g.\ $\mu X. X$ and $\mu X. \cond{\e}{X}{\Q}$ are not valid, whereas
$\mu X.\procin{\pp}{\labels{\ell}(x)}{X}$ is. 
We let $\prob,\probp,\probpp$ range over \emph{probabilities} and introduce the flip construct $\flip{\prob}{P}{Q}$ which continues as $\PP$ on heads and $\Q$ on tails; we syntactically constrain $\prob \in (\probn{0},\probn{1})$ to the \emph{open} unit interval, so that neither outcome of a $\mathsf{flip}$ is excluded. 

Process $\PP$ plays the participant $\pp$ (denoted $\pa\pp\PP$)
and can interact with other roles through parallel composition $\pc$, which we assume to be commutative
and associative; we write $\prod_{i \in I}\pa{\pp_i}{\PP_i}$ for
$\pa{\pp_1}{\PP_1}\pc\cdots\pc\pa{\pp_n}{\PP_n}$ with
$I=\set{1,\dots,n}$. $\rederror$ denotes an error session.

\begin{remark}[$\pmultipi$ probability constructs]
\label{rem:syntax}
The probabilistic behaviour of $\pmultipi$ only originates from 
$\flip{\prob}{\PP}{\Q}$, inspired by the
flip of \cite{Das2023} and close to the top-level probabilistic internal choice
$\PP \boxplus_{\prob} \Q$ of \cite{Inverso2020}. $\pmultipi$ has no primitive
probabilistic output choice as in \cite{AmanC22,Aman2019,varacca2007probabilistic},
but simulates one:
\\[1mm]
\centerline{
    $\PP\ = \ \flip{\prob}{\procout{\pchan{\pp_1}}{\labels{\ell_1}}{\e_1}{\PP_1}}
    {\procout{\pchan{\pp_2}}{\labels{\ell_2}}{\e_2}{\PP_2}}$
}\\[1mm]
emits $\labels{\ell_1}\langle\e_1\rangle$ on heads (probability $\prob$) and
$\labels{\ell_2}\langle\e_2\rangle$ on tails ($\probn{1}-\prob$), which in
\cite{AmanC22,Aman2019,varacca2007probabilistic} would be written as the explicit
probabilistic output choice
$\pprocout{\prob}{\pchan{\pp_1}}{\labelselli{1}}{\e_1}{\PP_1} \ \boxplus_{\prob}\
\pprocout{(\probn{1}-\prob)}{\pchan{\pp_2}}{\labelselli{2}}{\e_2}{\PP_2}$. 
Nesting flips extends this construction to probabilistic output choices over any finite number of alternatives.
\end{remark}

\begin{example}[Calculator service]\label{exam:running}
A user $\pchan{u}$ submits integers to a server $\pchan{s}$, which forwards 
the request to a calculator $\pchan{c}$; the calculator may
(depending on the size of the integer) occasionally produce an error. The three roles realise the session 
$\N \ = \ \psyntses{\pchan{u}}{P_{\pchan{u}}} \SEP
         \psyntses{\pchan{s}}{P_{\pchan{s}}} \SEP
         \psyntses{\pchan{c}}{P_{\pchan{c}}}$
where $\PP_\ps, \PP_{\pchan{c}}, \PP_\pu$ are:\\[1mm]
\centerline{
$
\small
\begin{array}{r@{\hspace{3pt}}l}
    P_{\pchan{u}} & = 
\procout{\pchan{s}}{\labels{incr}}{1}{
 \psyntrec{X}{\,
\procin{\pchan{c}}{\labels{res}(x)}{
\procout{\pchan{s}}{\labels{incr}}{\x}{X}}}}
\qquad
    P_{\pchan{s}}  = \psyntrec{X}{
{\big(\procin{\pchan{c}}{\labels{error}(x)}{\procout{\pchan{c}}{\labels{incr}}{x}{X}}}
\ + \ 
{\procin{\pu}{\labels{incr}(x)}{\procout{\pchan{c}}{\labels{incr}}{x}{X}}}\big)}
\\[2mm]
        P_{\pchan{c}} & = 
\psyntrec{X}{\procin{\ps}{\labels{incr}(x)}{\cond{x < 2^8}{\procout{\pu}{\labels{res}}{\fsucc{x}}{X}}{\flip{0.8}{\procout{\pu}{\labels{res}}{\fsucc{x}}{X}}{\procout{\ps}{\labels{error}}{x}{X}}}}}
\end{array}
$}\\[1mm]

The user loops, sending $\labels{incr}\langle e\rangle$ and then awaits a 
result $\labels{res}(x)$ from $\pchan{c}$, binding the result of the computation to $\x$ 
before recursing. 
The server waits for one of two external choices: (1) $\labels{incr}$ from $\pchan{u}$,
after which it sends $\labels{incr}\langle \e \rangle$ to $\pchan{c}$
and loops; or (2) an $\labels{error}(x)$ from the calculator $\pchan{c}$ on a previous computation
upon which it retries the calculation by sending $\labels{incr}\langle \x \rangle$ to $\pchan{c}$ again.
The calculator is accurate up to a certain integer size --- if the input from $\pchan{s}$ 
is smaller than $2^8$, $\pchan{c}$ successfully increments $\x$ and forwards it to $\pchan{u}$, on the other hand if the number is greater than or equal, it has a $\probn{0.2}$ probability of sending $\labels{error}$ to $\ps$ before recursing (and $\probn{0.8}$ probability of succeeding as before). %
\end{example}
\begin{figure}[t]
\small
         $\begin{array}{c}    
                \begin{array}{ll}

                              \rulename{r-head} &               \rulename{r-tail}\\ 
                                  \pa\pp{{\flip{\prob}{\PP}{\Q}}}\;\pc\; \Multi  
                                  \predcont{}{\prob} \pa\pp{\PP} \;\pc\; \Multi
                &
                                \pa\pp{{\flip{\prob}{\PP}{\Q}}}\;\pc\; \Multi  
                                \predcont{}{\probn{1-p}} \pa\pp{\Q} \;\pc\; \Multi
                \end{array}
                \\[3.5mm]
                \inferrule*[left=\rulename{r-t-if}]{
                 \peval{\e}{\true}{}}{
             \pa\pp{{\cond{\e}{\PP}{\Q}}}\;\pc\; \Multi  \predcont{}{\probn{1}} \pa\pp{\PP} \;\pc\; \N}
                \inferrule*[left=\rulename{r-f-if}]{
                 \peval{\e}{\false}{}}{
             \pa\pp{{\cond{\e}{\PP}{\Q}}}\;\pc\; \Multi  \predcont{}{\probn{1}} \pa\pp{\Q} \;\pc\; \N}
                  \\[1mm]
                \inferrule*[left=\rulename{r-comm}]
                 {
                   j \in I 
                   \\ 
                   \labels{\ell}=\labels{\ell_j}
                   \\
                    \pp = \pchan{\pp_j}
                   \\
                   \eval{\e}{\val}
                   }{
                    \pa\pp{\procout{\q}{\ell}{\e}{\PP}}
                    \pc
                    \pa\q{\psyntin{i\in I}{\pp_i}{i}{x_i}{\Q_i}}
                   \;\pc\; \N
                   \predcont{}{\probn{1}}
                   \pa\pp{{\PP}}\;\pc\;\pa\q{\Q_j\subst{\val}{\x_j}} 
                   \;\pc\; \N
                   }
                   \\[1mm]
                \inferrule*[left=\rulename{r-err}]
                 {
                     \exists j \in I \st \pp = \pchan{\pp_j},
                     \quad
                     \nexists i \in I : \ppi{i} = \pp \,\wedge\, \labels{\ell_i} = \labels{\ell}
                   }
                   {
                   \pa\pp{\procout{\q}{\ell}{\e}{\PP}}
                   \pc
                   \pa\q{\psyntin{i\in I}{\pp_i}{i}{x_i}{\Q_i}}
                   \;\pc\; \N
                   \pred{\probn{1}}
                   \rederror
                   }
               \inferrule*[left=\rulename{r-sc}]{
                   \N_1 \spcon \N'_1 \quad \N_1' \predcont{}{\probn{p}} \N_2 \quad \N_2 \spcon \N'_2
                }
                { 
                    \N_1 \predcont{}{\probn{p}} \N'_2
                }
\end{array}$
\vspace{-2mm}
\caption{Reduction rules for $\pmultipi$-calculus}%
            \label{fig:pmpstc_red}
\end{figure}

\myparagraph{Reduction relations.}
We write $\eval\e\val$ when fully evaluating a closed expression $\e$ (one with no
free variables) yields value $\val$; the rules are standard \cite[Table~1]{Ghilezan2019}.
The reduction relation for sessions is given in
\Cref{fig:pmpstc_red}; reductions $\pred{\prob}$ are annotated with a weight that
tracks probabilistic reduction paths. The probabilistic rules are
$\rulename{r-head}$ and $\rulename{r-tail}$, taken when a $\flip{\prob}{P}{Q}$
resolves---heads (weight $\prob$) continuing as $P$, tails ($\probn{1}-\prob$) as
$Q$; all other rules carry weight $\probn{1}$ and coincide with the
non-probabilistic MPST calculus of \cite{Ghilezan2019,UdomsrirungruangY25}. Rules
$\rulename{r-t-if}$ and $\rulename{r-f-if}$ fire once the guard $\e$ has evaluated
to $\true$ or $\false$. In $\rulename{r-comm}$, a participant $\pp$ selecting
label $\labels{\ell}$ to $\q$ synchronises with $\q$'s matching branch (the one
whose sender is $\pp$), binding $\val$ to $\x_j$ in the continuation $\Q_j$; the
weight is $\probn{1}$, as the communication surely happens. Rule \rulename{r-err} is enabled when
$\pp$ and $\q$ can communicate but there's a label \emph{mismatch}, i.e. $\pp$ sends a label $\labelsell$ not expected by $\q$. Rule $\rulename{r-sc}$
closes reduction under the \emph{structural precongruence} $\spcon$, defined by
$\pa\pp{\mu X.\PP} \pc \N \spcon \pa\pp{\PP\sub{\mu X.\PP}{X}} \pc \N$ and
$\prod_{i \in I}\pa{\pp_i}{\PP_i} \spcon \prod_{j \in J}\pa{\pp_j}{\PP_j}$ when
$I$ is a permutation of $J$; we write $\equiv$ for $(\spcon \cup \spcon^{-1})$.
We use $\predcontstar{}{\prob}$ for the \emph{reflexive} and \emph{transitive closure} of $\pred{\probp}$, where $\prob$ is the cumulative probability of that path, e.g. if $\N\pred{\probp}\pred{\probpp}\N'$ then $\N \ \predcontstar{}{\probp\cdot\probpp}\N'$.

\begin{example}[Reductions]\label{exam:runningreduction}
We trace a reduction of the session $\N$ from~\cref{exam:running}.
We abbreviate $\PP'_\pu = \psyntrec{X}{\procin{\pchan{c}}{\labels{res}(x)}{\procout{\pchan{s}}{\labels{incr}}{\x}{X}}}$ for the loop of $\pu$, and unfold recursion silently by $\rulename{r-sc}$:
\begin{equation*}
    \small
{\begin{array}{clr}
        \overset{\rulename{r-comm}}{\pred{\probn{1}}} & \pa\pu{
\procin{\pchan{c}}{\labels{res}(x)}{
\procout{\pchan{s}}{\labels{incr}}{\x}{\PP'_\pu}}} \pc \pa\ps{\procout{\pchan{c}}{\labels{incr}}{1}{\PP_{\pchan{s}}}} \pc \pa{\pchan{c}}{\PP_{\pchan{c}}} & (\pu\overset{\labels{incr}\langle 1\rangle}{\rightarrow} \ps) \\
\overset{\rulename{r-comm}}{\pred{\probn{1}}} & \pa\pu{\procin{\pchan{c}}{\labels{res}(x)}{\procout{\pchan{s}}{\labels{incr}}{\x}{\PP'_\pu}}} \pc \pa\ps{\PP_\ps} \pc \pa{\pchan{c}}{\cond{1 < 2^8}{\procout{\pu}{\labels{res}}{2}{\PP_{\pchan{c}}}}{\ldots}} & (\ps\overset{\labels{incr}\langle 1\rangle}{\rightarrow} \pchan{c}) \\
\overset{\rulename{r-t-if}}{\pred{\probn{1}}} & \pa\pu{\procin{\pchan{c}}{\labels{res}(x)}{\procout{\pchan{s}}{\labels{incr}}{\x}{\PP'_\pu}}} \pc \pa\ps{\PP_\ps} \pc \pa{\pchan{c}}{\procout{\pu}{\labels{res}}{2}{\PP_{\pchan{c}}}} & (\text{since }1 < 2^8)\\
\overset{\rulename{r-comm}}{\pred{\probn{1}}} & \pa{\pchan{u}}{\procout{\pchan s}{\labels{incr}}{2}{\PP'_\pu}} \pc \pa{\pchan{s}}{\PP_{\pchan s}} \pc \pa{\pchan{c}}{\PP_{\pchan c}} & (\pchan{c}\overset{\labels{res}\langle 2 \rangle}{\rightarrow}\pu)
\end{array}}
\end{equation*}
and the value loops back incremented. For larger inputs the calculator may error and the edges would carry a probability $\prob \neq \probn{1}$.
\end{example}

However, our processes are too expressive to analyse directly.
Indeed, the following result tells us that safety of (untyped) processes cannot be established in a sound and complete manner.
\begin{restatable}{proposition}{procsafeundec}
\label{prop:procsafeundec}
\proofreference{\cref{app:minskyencoding}}
It is undecidable to check safety of (untyped) sessions:   
no algorithm exists that, given
an arbitrary session $\N$,
can decide whether $\N\not\predcontstar{}{}\rederror$.
\end{restatable}
\begin{proof}
We reduce the halting problem to the problem under consideration. A two-counter Minsky machine $\mathcal M$ is
encoded as a session $\enc{\mathcal M}$ that simulates it
and reaches $\mathsf{error}$ exactly when $\mathcal M$ reaches its $\mathrm{HALT}$ instruction; 
hence $\enc{\mathcal M}$ is safe iff $\mathcal M$
does not halt. As halting is undecidable~\cite{Minsky1967}, so is safety.
\end{proof} 

This motivates a type system for $\pmultipi$: typing provides
a \emph{decidable}, \emph{sound} criterion for safety (well-typed
$\!\Rightarrow\!$ safe), and safety alone forces incompleteness (safe
$\not\Rightarrow$ well-typed), irrespective of any further properties
(deadlock-freedom and liveness) we wish to analyse.
The abstraction by types allows us to check safety and other probabilistic properties in a tractable manner.

\section{Probabilistic Multiparty Session Types and Typing System}
\label{sec:types}
\subsection{Probabilistic Multiparty Session Types}
\label{sec:mpsttypes}
We start by defining probabilistic multiparty session types, subtyping and merging. 
We then define their labelled transition relation, and properties of typing contexts.

\myparagraph{Probabilistic Multiparty Session Types} ($\T, \T', \dots$)
define a probabilistic communication specification for each participant.
\emph{Sorts} (or \emph{base types}) are ranged over by $\Basic, \Basic', \dots$\\[1mm]
\centerline{ 
$
\begin{array}{c}
\Basic  \ ::= \ \bools \SEP \nat \SEP \ints 
\qquad 
\T \ ::= \ \tend 
    \SEP 
\probpprocoutty{i \in I}{\pp_j}{j \in J_i}{\probi{i,j}}{\ell_j}{\Basic_j}{\T_{i,j}}
    \SEP  
    \pprocinnew{\pp_i}{i\in I}{\ell_i}{\Basic_i}{\T_i}
     \SEP 
     \mu \ty . \T
     \SEP \ty 
\end{array}
$}\\[1mm]
\emph{Sorts} are boolean, naturals and integers;
\emph{probabilistic multiparty session types} are: type $\tend$
denoting termination; 
type $\pprocinnew{\pp_i}{i\in I}{\ell_i}{\Basic_i}{\T_i}$ denoting
\emph{branching}  
(external choice) for each $i$ offering label $\labels{\ell_i}$ to $\ppi{i}$. 
For \emph{branching} to be well-formed, we require each $\ppi{i}?\labelselli{i}$ be distinct and $I \neq \emptyset$. %
Type $\probpprocoutty{i \in I}{\pp_j}{j \in J_i}{\probi{i,j}}{\ell_j}{\Basic_j}{\T_{i,j}}$
denotes \emph{selection} (internal choice) as a two-step resolution; the outer sum $\sum_{i \in I}$ represents a \emph{non-deterministic} choice over its summands $\pprocoutty{\pp_j}{j \in J_i}{\probi{i,j}}{\ell_j}{\Basic_j}{\T_{i,j}}$, that each denote a discrete output distribution with a fixed probability $\probi{i,j}\in (0,1]$ assigned to outputting $\labelselli{j}\langle \Basic_j \rangle$ to $\pchan{\pp_j}$ and continuing as $\T_{i,j}$. For \emph{selection} to be \emph{well-formed} we require $I\neq \emptyset$, each $\ppi{j}!\labelselli{j}\;j \in J_i$ be distinct and $\sum_{j \in J_i} \probi{i,j}=\probn{1}$ for all $i$.
We sometimes write $\pprocoutty{\pp_j}{j \in J_1}{\probi{1,j}}{\ell_j}{\Basic_j}{\T_{1,j}} + \probpprocoutty{i \in I\setminus \{1\}}{\pp_j}{j \in J_i}{\probi{i,j}}{\ell_j}{\Basic_j}{\T_{i,j}}$ for $\probpprocoutty{i \in I}{\pp_j}{j \in J_i}{\probi{i,j}}{\ell_j}{\Basic_j}{\T_{i,j}}$ and often omit $\sum_{i \in I}$ when $|I| = 1$.
$\mu \ty. \T$ and $\ty$ are \emph{recursive types} and \emph{type variables}, %
where $\T$ is \emph{guarded}: every type variable occurs under a branching or
selection prefix (so $\mu\ty.\ty$ is excluded), ensuring contractiveness.
We often omit $\tend$, base types $\Basic$ and the probability $\probn{1}$. %
We denote $\ftv{\T}$ as the set of \emph{free type variables} in $\T$,
and call $\T$ \emph{closed} if $\ftv{\T}=\emptyset$. Hereafter we assume all types are closed and well-formed.

\begin{example}[Calculator service continued]\label{exam:runningtype}
    Consider again the \emph{user-server-calculator} session from \Cref{exam:running}.
    Letting $\pp:\T$ denote that participant $\pp$ has type $\T$, the session is typed as:\\[2mm]
$
    \small
{\begin{array}{r@{\;=\;}l}
\pu : \T_{\pchan u} &
  \mu\ty.\;\oplus\pchan{s}!\labels{incr}\langle\ints\rangle.\;
  \&\pchan{c}?\labels{res}(\ints).\;\ty
\\[2mm]
\ps : \T_{\pchan s} &
\mu\ty.\;\&\big\{\,
    \pchan{c}?\labels{error}(\ints).\,\oplus\pchan{c}!\labels{incr}\langle\ints\rangle.\,\ty,\;
    \pchan{u}?\labels{incr}(\ints).\,\oplus\pchan{c}!\labels{incr}\langle\ints\rangle.\,\ty
\,\big\}
\\[2mm]
\pchan{c} : \T_{\pchan c} &
  \mu\ty.\;
  \&\pchan{s}?\labels{incr}(\ints).\;
  {\sum}\big\{\,
    \underbrace{\oplus\pchan{u}!\labels{res}\langle\ints\rangle.\,\ty}_{\text{$x<2^8$}}
    \;,\;
    \underbrace{{\oplus}\{\,
       \probn{0.8}\!:\pchan{u}!\labels{res}\langle\ints\rangle.\,\ty,\;
       \probn{0.2}\!:\pchan{s}!\labels{error}\langle\ints\rangle.\,\ty
\,\}}_{\text{else, }\flip{\probn{0.8}}{\ldots}{\ldots}}
  \,\big\}
\end{array}}
$\\[2mm]
\noindent Note that the condition in the \emph{if}-statement of
$\PP_{\pchan{c}}$, depending on the size of the input $x$, is
abstracted to a \emph{non-deterministic choice} over the two output
distributions --- rendering \emph{safety}, \emph{deadlock-freedom} and
\emph{liveness}, undecidable at the process level
(\cref{prop:procsafeundec}), decidable.
Observe also that the choice of $\pchan{c}$ is \emph{separated}: within
one distribution, the $\labels{res}$ branch answers $\pchan{u}$ while the
$\labels{error}$ branch targets $\pchan{s}$ --- expressible because our
selections fix a participant per branch, not per choice.
\end{example}

\myparagraph{Probabilistic subtyping.} 
Subtyping $\T\subt\T'$ allows a process typed by a smaller type $\T$ to stand in wherever one typed by $\T'$ is expected (the substitution principle~\cite{Liskov94}): it narrows selections (fewer nondeterministic options) and widens branchings (more offered inputs), as the rules make precise.
\begin{definition}[Probabilistic subtyping]
\label{def:pmpst_subtype}
\emph{Probabilistic subtyping} $\subt$ is the largest relation on session types (taken up to unfolding), coinductively defined by the rules below:\\[1mm]
\centerline{
$\small
\begin{array}{@{}c@{}}
\cinferruletop[\rulename{sub-$\oplus$}]
{
 \forall i \in I: \;
 \T_i \subt \T'_i
}
{
\pprocoutty{\pp_i}{{i \in I}}{\probi{i}}{\ell_i}{\Basic_i}{\T_i}
\subt
\pprocoutty{\pp_i}{{i \in I}}{\probi{i}}{\ell_i}{\Basic_i}{\T'_i}
}
\quad
\cinferruletop[\rulename{sub-$\&$}]
{
\forall i \in I: \quad
\T_i \subt \T'_i \qquad
\{\ppi{i}\}_{i \in I \cup J} = \{\ppi{i}\}_{i \in I}
}
{
\pprocinnew{\pp_i}{i \in I \cup J}{\ell_i}{\Basic_i}{\T_i} 
\subt
\pprocinnew{\pp_i}{i \in I}{\ell_i}{\Basic_i}{\T'_i}}
\\[5mm]
\cinferruletop[\rulename{sub-end}]
{-}
{\tend \subt \tend}
\quad
\cinferruletop[\rulename{sub-$\Sigma$}]
{
|I| \cdot |K| > 1
\quad
\forall i \in I: \exists k \in K \st \;
\pprocoutty{\pp_j}{{j \in J_i}}{\probi{i,j}}{\ell_j}{\Basic_j}{\T_{i,j}}
\subt
\pprocoutty{\pp_j}{{j \in J'_k}}{\probi{k,j}}{\ell_j}{\Basic_j}{\T'_{k,j}}
}
{
\probpprocoutty{i \in I}{\pp_j}{j \in J_i}{\probi{i,j}}{\ell_j}{\Basic_j}{\T_{i,j}}
\subt
\probpprocoutty{k \in K}{\pp_j}{j \in J'_k}{\probi{k,j}}{\ell_j}{\Basic_j}{\T'_{k,j}}
}
\end{array}
$
}
\end{definition}

The rules adapt the multiparty subtyping of \cite{Ghilezan2019} to fit our probabilistic types. \rulename{sub-end} relates $\tend$ only to itself.
\rulename{sub-$\&$} is contravariant: the subtype may offer \emph{more} inputs ($I \cup J$) than the supertype ($I$), as a process handling extra messages is safe where fewer are expected.
As in the input subtyping rule \rulename{s-in} in \cite[Def.~5]{PPGSY2026},
we require the set of participants $\{\ppi{i}\}_{i \in I}$ be
identical to $\{\ppi{i}\}_{i \in I \cup J}$ for downward-closure for
typing context properties (\cref{lemma:safeproperties,corollary:dfclosed}).
\rulename{sub-$\oplus$} relates two selection distributions only when
targets, labels, sorts, and probabilities coincide (changing
any would be observable and break
substitutability). Rule \rulename{sub-$\Sigma$} is less restrictive in comparison: every
summand of the subtype must embed (via \rulename{sub-$\oplus$}) into
some summand of the supertype, so the supertype's non-deterministic
options cover the subtype's; the premise $|I|\cdot|K| > 1$ prevents degenerate derivations (closing coinductive derivation after 1 step of \rulename{sub-$\Sigma$}).

\begin{restatable}[$\subt$-preorder]{lemma}{subtpreorder}\label{lem:subtpreorder}
\proofreference{\cref{app:sec3proofs}}
$\subt$ is reflexive and transitive.
\end{restatable}

\begin{example}[Subtyping]\label{exam:subtyping}
    Consider the following subprocess of some larger session:\\[1mm]
    $\PP_{cond} = \cond{x > 3}{\procout{\pp}{\labels{hi}}{\true}{\inact}}{\procout{\q}{\labels{bye}}{1}{\inact}}$,
    where the types for the ``\textsf{then}'' and ``\textsf{else}'' branch are $\T_{\mathsf{then}} = \oplus\pp!\labels{hi}\langle\tbool\rangle.\,\tend$ and $\T_{\mathsf{else}} =\oplus\q!\labels{bye}\langle\ints\rangle.\,\tend$.
    Then by \rulename{sub-$\Sigma$}, we have $\T_{\mathsf{then}} \subt \sum \big\{\T_{\mathsf{then}}, \T_{\mathsf{else}}\big\}$ and $\T_{\mathsf{else}} \subt \sum \big\{\T_{\mathsf{then}}, \T_{\mathsf{else}}\big\}$.
    Thus, by \rulename{t-sub} and \rulename{t-cond} (both introduced later), we are able to type $\PP_{cond}$ with the common supertype $\T_{cond} = \sum \big\{\T_{\mathsf{then}}, \T_{\mathsf{else}}\big\}$.
\end{example}

\begin{restatable}[$\subt$-decidable]{proposition}{decidablesubtype}
\label{prop:decidablesubtype}
\proofreference{\cref{app:sec3proofs}}
Given $\T_1$ and $\T_2$, checking $\T_1 \subt \T_2$ is decidable.
\end{restatable}

\begin{definition}[$\subt$-least upper-bound]
$\cdot \sqcup \cdot = \cdot$ is a partial relation on session types with $\T_1 \sqcup \T_2 = \T$ defined exactly when $\T_1 \subt \T$ and $\T_2 \subt \T$ and $\forall \T'\st (\T_1 \subt \T' \land \T_2 \subt \T') \Rightarrow \T \subt \T'$
\end{definition}

We write $\T_1 \sqcup \T_2$ for any $\T$ such that $\T_1 \sqcup \T_2 = \T$ and note that $\T$ is unique up to mutual subtyping.

\begin{definition}[Unfolding]
    Define $\unf{\cdot}$ by $\unf{\mu\ty .\T}=\unf{\T\subst{\mu\ty.\T}{\ty}}$ and $\unf{\T} = \T$ otherwise.
\end{definition}

\myparagraph{Probabilistic merging.} 
Subtyping allows us to type \textsf{if}-statements by widening the type sufficiently to 
accommodate both possible behaviours. Analogously,
\emph{merging} will allow us to reconcile 
the behaviour of both \textsf{head} and \textsf{tail} branches of a $\flip{\prob}{\ldots}{\ldots}$.

\begin{definition}[Probabilistic merge]\label{def:mergerel}
    The merge is the largest relation on session types, such that whenever $\T_1 \probmerge{\prob} \T_2 = \T$ the clauses below hold, with other cases being undefined: 
\begin{itemize}
    \item[\rulename{m-end}] {\small If $\unf{\T_1}=\unf{\T_2}=\tend$ then $\unf{\T} = \tend$}
    \item[\rulename{m-$\&$}]
        {\small If $\unf{\T_1}=\pprocinnew{\q_i}{i\in I}{\ell_i}{\Basic_i}{\T_i}$ and $\unf{\T_2}=\pprocinnew{\q_j}{j\in J}{\ell_j}{\Basic_j}{\T_j'}$ with $K = \{k \in I \cap J | \exists \T : \T_k\sqcup \T_k' = \T \}$
            and $\{\qi{i}\}_{i \in I} = \{\qi{j}\}_{j \in J} = \{\qi{k}\}_{k \in K}$ then $\unf{\T} = \pprocinnew{\q_k}{k\in K}{\ell_k}{\Basic_k}{(\T_k\sqcup \T_k')}$
    }
    \item[\rulename{m-$\oplus$}]
        {\small If $\unf{T_1}=\pprocoutty{\q_i}{i\in I}{\probi{i}}{\ell_i}{\Basic_i}{\T_i}$ and $\unf{T_2}=\pprocoutty{\q_j}{j\in J}{\probip{j}}{\ell_j}{\Basic_j}{\T_j'}$ with $\forall k \in I \cap J: \exists \T : \T_k \sqcup\T_k' = \T$
    then $\unf{T}=\pprocoutty{\q_k}{k\in I \cup J}{\probipp{k}}{\ell_k}{\Basic_k}{\T_k''}$ and for every $k\in I \cup J$:
$$
\begin{array}{rl}
    k\in I\setminus J\implies & \probipp{k} = \prob \cdot \probi{k} \land \T_k'' = \T_k\\[0pt]
     k\in J\setminus I\implies & \probipp{k} = (\probn{1}-\prob)\cdot \probip{k} \land T_k'' = \T_k'\\[0pt]
     k\in I\cap J\implies      & \probipp{k} = \prob\cdot \probi{k} + (1-\prob)\cdot \probip{k} \land \T_k'' = \T_k \sqcup \T_k'
\end{array}
$$
}
\item[\rulename{m-$\Sigma$}] {\small If $\unf{\T_1} = \sum_{i \in I}\T_i$ and $\unf{\T_2} = \sum_{j \in J}\T_j'$ with $|I| \cdot |J| > 1$ then $\unf{\T} = \sum_{(i,j) \in I \times J} \T_{i,j}$ where $\T_i \probmerge{\prob} \T_j' = \T_{i,j}$ for all $(i,j) \in I \times J$
}
\end{itemize}
\end{definition}

The above merging operator is analogous to the binary session operator $\T_1 \boxplus_{\prob} \T_2$
of \cite{Inverso2020}, but extended to multiparty session types. 
By \rulename{m-$\tend$} type $\tend$ is mergeable only with $\tend$. 
Merging of external choice \rulename{m-$\&$} must preserve the participant sets and drops any branches indexed by $\ppi{i}?\labelselli{i}$ when $i \not\in I \cap J$ (analogous to the contravariance of \rulename{sub-$\&$}), or whose continuations $\T_i$ and $\T_i'$ have no common supertype --- continuations are then typed by their least supertype $\T_k \sqcup \T_k'$.
Rule \rulename{m-$\oplus$} merges two selections, assigning to $\ppi{i}!\labelselli{i}\langle\Basic_i\rangle.\T_i$ in $i \in I\setminus J$ (resp. $i \in J \setminus I$) probability $\prob \cdot \probi{i}$ (resp. $(\probn{1}-\prob) \cdot \probip{i}$), while the probability for $i \in I \cap J$ is given by the convex combination $\prob\cdot \probi{i} + (\probn{1}-\prob)\cdot \probip{i}$ with continuations typed by the least upper bound $\T_k \sqcup \T_k'$.
Finally, the rule \rulename{m-$\Sigma$} merges every selection combination, the product $I \times J$ ensures every possible resolution of non-determinism is preserved; the side-condition $|I|\cdot |J| > 1$ prevents degenerate derivations.

\begin{notation}[Multi-way merge]
\label{sessfed}
We define $\probmergebig{i \in I}{\T_i}{\probi{i}}$ inductively as:
$\T_i$ if $\probi{i} = \probn{1}$ and $|I| = 1$; $\probmergebig{i \in I}{\T_i}{\probi{i}}= \T_j  \probmerge{\probi{j}}   \probmergebig{i \in I\setminus \{j\}} {\T_i }{\frac{\probi{i}}{\probn{1}-\probi{j}}}$ if $j \in I \land \probi{j} \neq 0$;
undefined otherwise.
\end{notation}

 \begin{example}[Probabilistic merging]
 \label{exam:probmerge}  
 The non-deterministic choice in the type $\T_{\pchan{c}}$ of the calculator (\cref{exam:runningtype}) contains the selection distribution
 $\T_{\pchan{c}-\mathsf{else}} = {\oplus}\{\,
       \probn{0.8}\!:\pchan{u}!\labels{res}\langle\ints\rangle.\,\ty,\;
       \probn{0.2}\!:\pchan{s}!\labels{error}\langle\ints\rangle.\,\ty
\,\}$, which is the merge $\T_1 \probmerge{\probn{0.8}} \T_2$ of $\T_1 = \oplus\probn{1}\!:\pchan{u}!\labels{res}\langle\ints\rangle.\,\ty$ and $\T_2 = \oplus\probn{1}\!:\pchan{s}!\labels{error}\langle\ints\rangle.\,\ty$ (with $\ty$ of type $\T_{\pchan{c}}$).
Since the label-role pairs $\pu!\labels{res}$ and $\ps!\labels{error}$ are disjoint, a single application of \rulename{m-$\oplus$} suffices (as $\probn{0.8} = \probn{0.8} \cdot \probn{1}$, $\probn{0.2} = \probn{0.2} \cdot \probn{1}$ and $\ty = \ty$, by the disjoint case of the statement).
 \end{example}

\begin{restatable}[Decidability of merge]{proposition}{mergedecidable}
\label{prop:mergedecidable}
\proofreference{\cref{app:sec3proofs}}
For $\T_1,\T_2$ and $\prob\in(0,1)$ it is decidable whether $\T_1\probmerge{\prob}\T_2$ is defined
in time $\mathcal O(|\T_1||\T_2|)$, and when defined the witness is computable in time\\[1mm]
$
 \mathcal O\!\left(
   (\max\{2,d\})^{2|\T_1||\T_2|}
   \log(|\T_1||\T_2|+1)
 \right)
$ where $d$ is the max out-degree of the type graphs of $\T_1, \T_2$.
\end{restatable}

\myparagraph{A typing context} ($\Delta, \Delta', \dots$) 
is a mapping from a participant to its type, representing 
behaviours of a multiparty session $\M$. 
Its behaviour (semantics) are defined by  
a labelled transition relation labelled with \emph{actions}  
($\alpha, \alpha', \dots$). Typing contexts and actions are 
defined as:\\[1mm]
\centerline{
$\Delta ::= \pp : \T \SEP \Delta , \Delta'$ 
\quad \quad 
$\alpha ::= 
\actionprocout{\pp}{\q}{\ell}{\Basic} 
\SEP 
\actionprocin{\pp}{\q}{\ell}{\Basic} 
\SEP 
\actiontau{\pp}{\q}{\ell}
\SEP 
\actionnd{\pp}
\SEP 
\pp {:} p
$}\\[1mm]
Action $\actionprocout{\pp}{\q}{\ell}{\Basic}$ is a \emph{selection} from 
participant $\pp$ to $\q$ with label $\ell$ and type $\Basic$; 
action $\actionprocin{\pp}{\q}{\ell}{\Basic}$ is \emph{branching} at 
participant $\pp$ from $\q$ with label $\ell$ and type $\Basic$; 
and action $\actiontau{\pp}{\q}{\ell}$ is a \emph{communication} 
between dual actions. $\actionnd{\pp}$ and $\pp{:}p$ denote \emph{nondeterministic} and \emph{probabilistic} internal choice respectively.
We write $\Delta_1 \subt \Delta_2$ if $\dom{\Delta_1}=\dom{\Delta_2}$ and 
$\forall\pp \in
\dom{\Delta_1} : \Delta_1(\pp) \subt
\Delta_2(\pp)$.

\myparagraph{Semantics (Labelled Transition System) of typing contexts}
is defined in \Cref{fig:pmpst_tct}.
We note that emitting a message decomposes into up to \emph{three}
steps: \rulename{ct-nd} resolves an internal \emph{non-deterministic}
choice among the summands of a $\Sigma$; \rulename{ct-prob} is an
internal \emph{probabilistic} reduction, resolving a participant $\pp$
at a \emph{proper} output selection
$\pp : \pprocoutty{\q_i}{i \in I}{\probi{i}}{\ell_i}{\Basic_i}{\T_i}$
to the singleton selection of branch $k$, with probability $\probi{k}$;
this enables \rulename{ct-out}, emitting the message to $\pchan{q_k}$
with label $\labels{\ell_k}$, sort $\Basic_k$, and probability
$\probn{1}$. Rule \rulename{ct-in} inputs label $\labels{\ell_k}$ with
sort $\Basic_k$ at participant $\pp$ with probability $\probn{1}$; rule
\rulename{ct-$\tau$} synchronises a singleton selection with its dual branching;
\rulename{ct-rec} is the standard rule for recursive types; and 
\rulename{ct-sc} is a context rule. We write: 

\begin{itemize}[leftmargin=*]
    \item $
        \Delta \predcontdouble{\actiontau{\pp}{\q}{\ell}}{\prob} \Delta'
        \;\text{iff}\;
        (\prob=\probn{1}
            \land
            \Delta\predcont{\actiontau{\pp}{\q}{\ell}}{\probn{1}}\Delta')
        $ or $
        \exists\Delta'':
        \Delta\predcont{\actionprob{\pp}}{\prob}\Delta''
        \land
        \Delta''\predcont{\actiontau{\pp}{\q}{\ell}}{\probn{1}}\Delta'.
        $
        \item
    $
    \Delta\predcont{}{\color{black}t}\Delta'
        \;\text{iff}\;
        \exists\alpha\in \{\actionnd{\pp},\actionprob{\pp},
    \actiontau{\pp}{\q}{\ell}\mid\pp,\q,\ell\}:
    \Delta\predcont{\alpha}{\color{black}t}\Delta'$;
        and
    $
        \Delta\predcont{}{}\Delta'
        \;\text{iff}\;
        \exists t\in(\probn{0};\probn{1}]\cup\{\nd\}:
        \Delta\predcont{}{\color{black}t}\Delta'
    $;\\
 and $\Delta\predcont{\alpha}{\color{black}t}$ iff $\exists\Delta':\Delta\predcont{\alpha}{\color{black}t}\Delta'$ (same for $\predcont{}{\color{black}t}$ and $\predcont{}{}$).
    These conventions also apply to
    $\predcontdouble{\alpha}{\prob}$.

    \item $\Delta_1 \pred{\probn p}^{+} \Delta_m$ for the
    \emph{transitive closure}: a sequence
    $\Delta_1 \cdots \pred{\probn{p_{m-1}}} \Delta_m$
    with $m \geq 2$ and
    $\probn p = \prod_{k=1}^{m-1}\probn{p_k}$.
    The \emph{reflexive} and \emph{transitive closure}
    $\pred{\probn p}^{\ast}$ additionally admits the empty
    path ($\Delta \pred{\probn 1}^{\ast} \Delta$).
\end{itemize}

\begin{figure}[t]
\small
\begin{mathpar}
\inferrule[\rulename{ct-rec}]
{
   \pp: \T \subst{\mu \ty . \T}{\ty} \predcont{\alpha}{{\color{black}t}} \pp : \T'
   \quad
   t \in (\probn{0};\probn{1}] \cup \{\nd\}
}
{
    \pp: \mu \ty . \T \predcont{\alpha}{{\color{black}t}} \pp : \T'
}
\quad\;\;
\inferrule[\rulename{ct-nd}]
{
  k \in I 
  \quad 
  |I| > 1
}
{
    \pp :  \probpprocoutty{i \in I}{\q_j}{j \in J_i}{\probi{i,j}}{\ell_j}{\Basic_j}{\T_{i,j}}
  \predcontnd{\actionnd{\pp}}{
      {\pp}:{\pprocoutty{\q_j}{j \in J_k}{\probi{k,j}}{\ell_j}{\Basic_j}{\T_{k,j}}}
  }
}
\\
\inferrule*[left=\rulename{ct-prob}]
{
    k \in I \\ \probi{k} \in (\probn{0};\probn{1})
}
{\pp :  \pprocoutty{\q_i}{i \in I}{\probi{i}}{\ell_i}{\Basic_i}{\T_i}
    \predcont{\actionprob{\pp}}{\probi{k}}
\pp :  \pprocoutty{\q_k}{}{\probn{1}}{\ell_k}{\Basic_k}{\T_k}
}
\;\;
\inferrule*[left=\rulename{ct-out}]
{
    \mbox{}
}
{
\pp :  \pprocoutty{\q}{}{\probn{1}}{\ell}{\Basic}{\T}
\predcont{\actionprocout{\pp}{\q}{\ell}{\Basic}}{\probn{1}}
\pp :  \T
}
\\
\inferrule[\rulename{ct-in}]
{
  k \in I
}
{
\pp : \pprocinnew{\q_i}{i \in I}{\ell_i}{\Basic_i}{\T_i}
\predcont{\actionprocin{\pp}{\q_k}{\ell_k}{\Basic_k}}{1} {\pp : \T_k}
}
\quad\;
\inferrule[\rulename{ct-$\tau$}]
{
  \Delta_1 \predcont{\actionprocout{\pp}{\q}{\ell}{\Basic}}{\probn{1}} \Delta'_1
  \quad
  \Delta_2 \predcont{\actionprocin{\q}{\pp}{\ell}{\Basic}}{1}
  \Delta'_2 
}
{
\Delta_1 , \Delta_2 
\predcont{\actiontau{\pp}{\q}{\ell}}{\probn{1}} \Delta'_1 , \Delta'_2 
}
\quad
\;
\inferrule[\rulename{ct-sc}]
{
    \Delta \predcont{\alpha}{\color{black}t} \Delta'
  \quad
  t \in (\probn{0};\probn{1}] \cup \{\nd\}
}
{
    \Delta , \pp : \T \predcont{\alpha}{\color{black}t} \Delta' , \pp : \T 
}
\end{mathpar}
\vspace{-2mm} 
\caption{Labelled transition system for typing contexts}
\label{fig:pmpst_tct}
\end{figure}

\myparagraph{Safety property. } In a general typing system \cite{Scalas2019}, 
we type-check that a multiparty session $\M$ satisfies
a \emph{safety property}: whether its typing context
only exchanges \emph{compatible} labels. 
\begin{definition}[Safety] \cite[Definition~4.1]{Scalas2019} 
\label{def:safety} 
$\varphi$ is a \emph{safety property} of $\Delta$ iff, 
whenever $\varphi(\Delta)$:   
\begin{enumerate*}
\item $\varphi(\emptyset)$;
\item\label{def:safety1} 
$\Delta \predcont{\actionprocout{\pp}{\q}{\ell}{\Basic}}{\probn{1}}$   
and 
$\Delta \predcont{\actionprocin{\q}{\pp}{{\ell'}}{\Basic'}}{\probn{1}}$
imply $\Delta \predcont{\actiontau{\pp}{\q}{\ell}}{\probn{1}}$;
\item\label{def:safety3} 
    $\Delta \predcont{}{} \Delta'$ implies $\varphi(\Delta')$
\end{enumerate*}
We write $\safemp{\Delta}$ if $\varphi(\Delta)$ holds for some safety 
property $\varphi$. 
\end{definition}
\noindent \Cref{def:safety}(1) above states
the empty context is considered safe,
(2) defines that if $\Delta$ contains the entries for 
selection at $\pp$ to $\q$ and branching at $\q$ from $\pp$, then 
branching at $\q$ supports \emph{all} of $\pp$'s labels. 
\Cref{def:safety}(3)
states that a safety property must be preserved under reductions.

\begin{example}[Semantics of types and safety]
\label{exam:typingcontextreductions}
Consider the typing context $\Delta = \pu : \T_{\pu}, \ps : \T_\ps, \pchan{c} : \T_{\pchan{c}}$ of \Cref{exam:runningtype}. We consider two runs starting from $\Delta$:\\[1mm]
\centerline{$\footnotesize
\begin{array}{l}
\Delta
   \predcont{\actiontau{\pchan u}{\pchan s}{\labels{incr}}}{1} \cdot
   \predcont{\actiontau{\pchan s}{\pchan c}{\labels{incr}}}{1} \cdot
   \predcontnd{\actionnd{\pchan c}}{} \cdot
   \predcontdouble{\actiontau{\pchan c}{\pchan s}{\labels{error}}}{0.2} \cdot
   \predcont{\actiontau{\pchan s}{\pchan c}{\labels{incr}}}{1} \cdot
   \predcontnd{\actionnd{\pchan c}}{} \cdot
   \predcont{\actiontau{\pchan c}{\pchan u}{\labels{res}}}{1} \Delta
   \cdots
\\[1.5mm]
\Delta
   \predcont{\actiontau{\pchan u}{\pchan s}{\labels{incr}}}{1} \cdot
   \predcont{\actiontau{\pchan s}{\pchan c}{\labels{incr}}}{1} \cdot
   \predcontnd{\actionnd{\pchan c}}{} \cdot
   \predcontdouble{\actiontau{\pchan c}{\pchan s}{\labels{error}}}{0.2} \cdot
   \predcont{\actiontau{\pchan s}{\pchan c}{\labels{incr}}}{1} \cdot
   \predcontnd{\actionnd{\pchan c}}{} \cdot
   \predcontdouble{\actiontau{\pchan c}{\pchan s}{\labels{error}}}{0.2} \cdot
   \predcont{\actiontau{\pchan s}{\pchan c}{\labels{incr}}}{1} \cdot
   \predcontnd{\actionnd{\pchan c}}{} \cdot
   \predcontdouble{\actiontau{\pchan c}{\pchan s}{\labels{error}}}{0.2} \cdot
   \cdots
\end{array}
$}\\[1mm]
The first run fails once then succeeds when the $\rulename{ct-nd}$ step commits $\pchan{c}$'s
non-deterministic choice to the selection sending $\labels{res}$ to $\pu$ with probability $\probn{1}$. 
The second fails indefinitely, a run of probability $\lim_{n\to\infty}\probn{0.2}^{\,n}=\probn{0}$ (\Cref{sec:properties}).
Note the $\predcontnd{\actionnd{\pchan c}}{}$ steps carry no probability, and instead abstract over the
$x<2^8$ test in \Cref{exam:runningtype}. 
\end{example}

\begin{figure}[t]
$\small
\begin{array}{c}
\quad
      \inferrule*[left=\rulename{t-rec}]
      {\der{\Gamma,X:\T}{\PP}{\T}}
      {\der\Gamma{\mu X.\PP}\T}
\quad 
      \inferrule*[left=\rulename{t-var}]{ }
            {\der{\Gamma,X:\T}{X}{\T}}
\quad 
      \inferrule*[left=\rulename{t-sub}]
      {
        \der{\Gamma}{P}{\T} \quad  \T \subt \T'
      }
      {
        \der{\Gamma}{P}{\T'}
      }
\\[2mm]
      \inferrule*[left=\rulename{t-$\inact$}]{ }
           {\der\Gamma{\inact}\tend}
           \quad
      \inferrule*[left=\rulename{t-flip}]
      {
          \der{\Gamma}{\PP_i}{\T_i} \quad i \in \{1,2\}
        }
        {\der{\Gamma}{\flip{\prob}{\PP_1}{\PP_2}} 
        {\T_1\probmerge{\prob} \T_2}}
\\[2mm]
      \inferrule*[left=\rulename{t-sel}]
      {
        \der{\Gamma}{\e}{\Basic}
        \quad 
        \der{\Gamma}{\PP}{\T}
        }
        {\der{\Gamma}{\procout{\pp}{\ell}{\e}{\PP}} 
        {\pprocoutty{\pp}{}{\probn{1}}{\ell}{\Basic}{\T}}
}\quad 
      \inferrule*[left=\rulename{t-bra}]
{\forall i\in I \quad\der{\Gamma,x_i:\Basic_i}{\PP_i}\T_i}
{\der{\Gamma}{\psyntinnl{i\in I}{\pp_i}{i}{x_i}{\PP_i}}
{\pprocinnew{\pp_i}{i\in I}{\ell_i}{\Basic_i}{\T_i}}
}
\\[2mm]
      \inferrule*[left=\rulename{t-cond}]
      {
     \der\Gamma\e\tbool\quad  
     \der{\Gamma}{\PP_1}{\T}\quad
     \der{\Gamma}{\PP_2}{\T}\quad 
     }{\der\Gamma{{\cond{\e}{\PP_1}{\PP_2}}}{\T}}
     \quad
      \inferrule*[left=\rulename{b-sess}]
      {
          \safemp{\{\ppi{i}: \T_i\}_{i \in I}} \quad 
          \forall i \in I: \; \der{}{\PP_i}{\T_i}
      }
      {
          \der{}{\textstyle \prod_{i \in I}{\pa{\ppi{i}}{\PP_i}}}{\{\ppi{i}: \T_i\}_{i \in I}}
      }
\end{array}
$
\vspace{-2mm}
  \caption{Probabilistic multiparty session typing system}
  \label{fig:pmpst_typrules}
\end{figure}

\subsection{Probabilistic Multiparty Session Typing System}
\label{sec:typingsystem}
We introduce the probabilistic multiparty session typing system.  
We define \emph{typing environments} ($\Gamma, \Gamma', \dots$) 
as a mapping from a variable to a sort for typing expressions 
and a process variable to a session type for typing recursive processes:  
$\Gamma :: = \varnothing \SEP \Gamma, x:\Basic \SEP \Gamma, X:\T$. 
The judgements for typing expressions, processes and sessions 
are written as $\der{\Gamma}{\e}{\Basic}$, $\der{\Gamma}{\PP}{\T}$, and 
$\der{}{\N}{\Delta}$, respectively.

We omit the typing rules for expressions as they are standard and 
identical with the rules in \cite{Ghilezan2019}. 
\Cref{fig:pmpst_typrules} defines the typing rules 
for processes and a session.
With the exception of
\rulename{t-flip},
\rulename{t-sel} and  
\rulename{b-sess}, 
the rules are standard and identical with 
the corresponding rules in \cite{Ghilezan2019}.
Rule
\rulename{t-flip} is the main rule for assigning probabilities
with respect to $\prob$ to types using the probabilistic merging  
operator in \Cref{def:mergerel}. 
Rule \rulename{t-sel} types a single 
selection with probability $\probn{1}$. 
Rule \rulename{b-sess} is identical with 
the bottom-up rule \rulename{b-sess} 
in \cite[Figure~5]{UdomsrirungruangY25} where 
session $\M$ is typed by a safe typing context  
(note: \cite{Ghilezan2019} uses the top-down typing system).

\begin{restatable}[Decidability]{theorem}{decidability}
\label{thm:decidability}
Given session $\N$ and typing context $\Delta$, $\der{}{\N}\Delta$ is decidable.  
\end{restatable}
\begin{proof}
    The proof is similar to~\cite[Thm. 4.11]{Scalas2019} with $\subt$, $\probmerge{\prob}$, and $\safe{\Delta}$ decidable by~\cref{prop:decidablesubtype,prop:mergedecidable} and~\cref{prop:decidablesafe}. Full proof in \cref{app:sec3proofs}.
\end{proof}

\section{Properties of Probabilistic Multiparty Sessions}
\label{sec:properties}
This section studies a range of probabilistic properties of
probabilistic multiparty session types and processes, after 
proving \emph{subject reduction} and \emph{session fidelity}. %

\subsection{Subject Reduction and Session Fidelity Theorems}
\label{sec:sr}
Without probabilities, subject reduction states that each process reduction is matched by a context transition.
Flips complicate this: contexts have no flip transitions, since $\probmerge{\prob}$ 
compiled each flip into new distributions at typing time, so a single probabilistic 
step cannot preserve typing alone. Instead, we look at the weighted \emph{pair} of outcomes, 
recombined through the merge (case~(3) below). 

\begin{restatable}[Subject Reduction]{theorem}{subred}
\label{thm:subred}
\proofreference{\cref{app:sr}}
Suppose $\vdash  \N: \Delta$. If $ \N \pred{\prob} \N_1$, then depending on the top-level rule used to derive $\N \pred{\prob} \N_1$ we have:
\begin{enumerate}[leftmargin=*]
    \item \rulename{r-t-if} or \rulename{r-f-if} implies: $\der{}{\N_1}{\Delta}$
    \item \rulename{r-comm} implies: $\exists \Delta_1$ such that $\vdash \N_1 : \Delta_1$ and either $\Delta \predcont{}{\probn{1}}\Delta_1$ or $\Delta \predcontnd{}\cdot\predcont{}{\probn{1}}\Delta_1$.
    \item \rulename{r-head} or \rulename{r-tail} implies $\exists \pp, \T, \{\T_i, \N_i\}_{i \in \{1,2\}},\Delta'$ such that $\Delta = \pp : \T, \Delta'$ and $\N \predcont{}{\probn{1}-\prob}\N_2$ with $\der{}{\N_i}{(\pp : \T_i,\Delta')}$ for $i \in \{1,2\}$ and $(\pp:\T_1 \probmerge{\prob} \T_2, \Delta') \subt \Delta$.
\end{enumerate} 
\end{restatable}
\begin{proof}
    By induction on the derivation of $\N \predcont{}{\prob} \N_1$.
\end{proof}

Case~(1) covers data-dependent control-flow steps.
In case~(2) the context matches the communication, possibly after first committing its nondeterministic 
choice via $\predcontnd{}$. 
In case~(3), as no context transition matches a \textsf{flip}, the statement instead mentions 
\emph{both} outcomes, whose residual typings recombine into $\T_1 \probmerge{\prob} \T_2 \subt \T$. 
Case (3) is illustrated in \Cref{fig:frontier}: merging the typings at each frontier of a flip tree yields 
a context that is a subtype of $\Delta$ (the composed chain follows by monotonicity of 
merging~\cref{lemma:monotonicmerge}).

\begin{corollary}[Communication Safety]
If $\vdash  \N: \Delta$, then $\N\nrightarrow^\ast\rederror$.
\end{corollary} 

\begin{figure}[t]
\begin{minipage}[c]{0.45\linewidth}
\centering
\begin{tikzpicture}[
  >={Stealth[length=4.5pt]},
  every node/.style={inner sep=1.2pt},
  sess/.style={font=\small},
  ann/.style={font=\scriptsize, text=black!55},
  plab/.style={font=\scriptsize, inner sep=0.8pt},
  alab/.style={font=\scriptsize, inner sep=0.8pt},
  hl/.style={line width=1pt},   %
  dedge/.style={-{Stealth[length=4.5pt,sep=-2.4pt]Stealth[length=4.5pt]}},
]
\node[sess, label={[ann, label distance=-3.5pt]50:{$:\Delta$}}]
  (M)  at (0,-0.05)    {$\N$};
\node[sess, label={[ann, label distance=-3.5pt]50:{$:\Delta_1$}}]
  (M1) at (1.0,0.75)   {$\N_1$};
\node[sess, label={[ann, label distance=-3.5pt]50:{$:\Delta_2$}}]
  (M2) at (1.0,-0.85)  {$\N_2$};
\node[sess, label={[ann, label distance=-3.5pt]50:{$:\Delta_3$}}]
  (M3) at (2.6,1.15)   {$\N_3$};
\node[sess, label={[ann, label distance=-3.5pt]50:{$:\Delta_4$}}]
  (M4) at (2.6,0.35)   {$\N_4$};
\node[sess, label={[ann, label distance=-3.5pt]50:{$:\Delta_4$}}]
  (M5) at (2.6,-0.45)  {$\N_5$};
\node[sess, label={[ann, label distance=-3.5pt]50:{$:\Delta_5$}}]
  (M6) at (2.6,-1.25)  {$\N_6$};
\node[sess, label={[ann, label distance=-3.5pt]50:{$:\Delta_6$}}]
  (M7) at (5,1.15)   {$\N_7$};
\node[sess, label={[ann, label distance=-3.5pt]50:{$:\Delta_7$}}]
  (M8) at (5,0.35)  {$\N_8$};
\node[sess, label={[ann, label distance=-3.5pt]50:{$:\Delta_7$}}]
  (M9) at (5,-0.45)  {$\N_9$};
\node[sess, label={[ann, label distance=-3.5pt]50:{$:\Delta_8$}}]
  (M10) at (5,-1.25)  {$\N_{10}$};
\draw[->,hl] (M)  -- node[plab, pos=0.75, below,yshift=-0.03cm,xshift=0.04cm] {$\probn{0.8}$} (M1);
\draw[->,hl] (M)  -- node[plab, pos=0.75, below,yshift=-0.03cm,xshift=-0.04cm] {$\probn{0.2}$} (M2);
\draw[->]    (M1) -- node[plab, pos=0.72, below] {$\probn{0.5}$} (M3);
\draw[->,hl] (M1) -- node[plab, pos=0.72, below] {$\probn{0.5}$} (M4);
\draw[->,hl] (M2) -- node[plab, pos=0.72, below] {$\probn{0.5}$} (M5);
\draw[->]    (M2) -- node[plab, pos=0.72, below] {$\probn{0.5}$} (M6);
\draw[->] (M3) -- node[alab, pos=0.42, above]
    {$(\pchan{p},\pchan{r})\labels{\ell_{1}}$}
  node[plab, pos=0.9, below] {$\probn{1}$} (M7);
\draw[->,hl] (M4) -- node[alab, pos=0.42, above, sloped]
    {$(\pchan{p},\pchan{q})\labels{\ell_2}$}
  node[plab, pos=0.9, below, sloped] {$\probn{1}$} (M8.west);
\draw[->,hl] (M5) -- node[alab, pos=0.42, above, sloped]
    {$(\pchan{p},\pchan{q})\labels{\ell_2}$}
  node[plab, pos=0.9, below, sloped] {$\probn{1}$} (M9.west);
\draw[->] (M6) -- node[alab, pos=0.42, above]
    {$(\pchan{p},\pchan{r})\labels{\ell_3}$}
  node[plab, pos=0.94, below] {$\probn{1}$} (M10);
\node[font=\small] (E0) at (0,-2.45)    {$\Delta$};
\node[font=\small]      at (0.4,-2.45) {$\geqslant$};
\node[font=\small, align=center, inner sep=1pt] (E1) at (0.8,-2.45)
  {$\Delta_1$\\[-2pt]$\probmerge{\probn{0.8}}$\\[-2pt]$\Delta_2$};
\node[font=\small]      at (1.45,-2.45) {$\geqslant$};
\node[font=\small, align=center, inner sep=1pt] (E2) at (2.4,-2.45)
    {$\big(\Delta_3 \probmerge{\probn{0.5}}\Delta_4\big)$\\[-2pt]
   $\probmerge{\probn{0.8}} $\\[-2pt]$\big(\Delta_4 \probmerge{\probn{0.5}} \Delta_5\big)$};
\node[font=\small] (D7) at (5,-1.65)  {$\Delta_6$};
\node[font=\small] (D8) at (5,-2.45)  {$\Delta_7$};
\node[font=\small] (D9) at (5,-3.25)  {$\Delta_8$};
\draw[dedge] ($(E2.east)+(0,3pt)$) -- node[alab, pos=0.5, above, sloped]
    {$(\pchan{p},\pchan{r})\labels{\ell_1}$}
  node[plab, pos=0.87, below, sloped,yshift=-0.03cm] {$\probn{0.4}$} (D7.west);
\draw[dedge,hl] (E2.east) -- node[alab, pos=0.5, above]
    {$(\pchan{p},\pchan{q})\labels{\ell_2}$}
  node[plab, pos=0.87, below,yshift=-0.03cm] {$\probn{0.5}$} (D8.west);
\draw[dedge] ($(E2.east)+(0,-3pt)$) -- node[alab, pos=0.5, above, sloped]
    {$(\pchan{p},\pchan{r})\labels{\ell_3}$}
  node[plab, pos=0.87, below, sloped,yshift=-0.03cm] {$\probn{0.1}$} (D9.west);
\end{tikzpicture}
\end{minipage}\hfill
\begin{minipage}[c]{0.55\linewidth}
\caption{Subject reduction: The session $\N$ reduces probabilistically to one of $\N_3,\dots,\N_6$ before $\pp$ sends a message to $\q$ or $\pr$. Both $\N_4,\N_5$ (and likewise $\N_8,\N_9$) are typed by the same context.
\cref{thm:subred}(3) states that merging the typings $\T_1,\T_2$ of a participant $\pp$ 
across both outcomes of its \textsf{flip} always produces a subtype of its type in 
$\Delta$ ($\T_1 \probmerge{\prob} \T_2 \subt \T$). 
Under the merged context (bottom), $\pp$ holds a distribution over its
sends.\\[1mm]
Session fidelity (\cref{thm:sessionfidelitysmall}(3)) recovers the resolving
step's probability as the total weight of the flip paths exposing that
send: for $(\pp,\q)\ell_2$, $\probn{0.8} \cdot \probn{0.5} + \probn{0.2} \cdot \probn{0.5} = \probn{0.5}$ (bold
edges). The resolved context mirrors $\Delta_4$, which types both
$\N_4$ and $\N_5$.
}
\label{fig:frontier}
\end{minipage}
\end{figure}

Session fidelity states the converse: typing context reductions must be matched by the process.

\begin{notation}[Annotated session reductions]
    \label{not:annotated}
We write $\N \predcont{\actiontau{\pp}{\q}{\ell}}{\probn{1}} \N'$ when
$\N \predcont{}{\probn{1}} \N'$ is derived by top-level rule \rulename{r-comm} exchanging
label $\labelsell$ from $\pp$ to $\q$.%
\end{notation}
\begin{restatable}[Session Fidelity]{theorem}{sessionfidelitysmall}
\label{thm:sessionfidelitysmall}
\proofreference{\cref{app:sr}}
    \mbox{}
    \begin{enumerate}
        \item If $\der{}{\N}{\Delta}$ then
  $\Delta \predcont{\actiontau{\pp}{\q}{\ell}}{\probn{1}} \Delta'$ implies
  $\exists I \st \{\N \predcontstar{}{\probi{i}} \N_i
     \predcont{\actiontau{\pp}{\q}{\ell}}{\probn{1}}\}_{i\in I}$
     with $\sum_{i \in I} \probi{i} = 1$ 
 \item If $\der{}{\PP}{\T}$ then $\pp : \T \predcontnd{\actionnd{\pp}}$ implies $\exists \T'\st \pp : \T \predcontnd{\actionnd{\pp}} \pp : \T'$ and $\exists I \st \{\pa\pp\PP \predcontstar{}{\probi{i}} \pa\pp\PP_i\mid \pa\pp\PP_i \not\rightarrow\}_{i \in I}$ with $\sum_{i \in I}\probi{i} = \probn{1}$
            and a set of types $\{\T_i \mid \der{\,}{\PP_i}{\T_i}\}_{i \in I}$ such that $\probmergebig{i \in I}{\T_i}{\probi{i}} \subt \T'$ %
\item If $\der{}{\PP}{\T}$ then
  $\pp : \T \predcont{\actionprob{\pp}}{\prob}
     \pp : \pprocoutty{\q}{}{\probn{1}}{\ell}{\Basic}{\T'}$ implies
  $\exists I \st \{\pa\pp\PP \predcontstar{}{\probi{i}}
     \pa\pp{\procout{\q}{\ell}{\e_i}{\PP_i}}\}_{i \in I}$
 with $\sum_{i \in I} \probi{i} = \prob$ and $\der{}{\PP_i}{\T'}$ for all $i \in I$
    \end{enumerate}
\end{restatable}
(2) and (3) are easily lifted to $\der{}\N\Delta$ through \rulename{ct-sc}.
\begin{proof}
    Corollary of a strengthened statement, proved by induction.
\end{proof}

Case~(1) states that a context communication of probability $\probn{1}$ is realised by weighted paths $\N \ \predcontstar{}{\probi{i}} \N_i$, each arriving where the communication fires surely, with $\sum_{i \in I} \probi{i} = \probn{1}$. 
Case~(2) states that when the context commits its nondeterministic choice, the process resolves its flips down to leaves $\{\PP_i\}_{i \in I}$ that reduce no further, 
and merging the leaf types at their path weights, $\probmergebig{i \in I}{\T_i}{\probi{i}}$, recovers a subtype of the committed type. Analogously, case~(3) guarantees that $\pp$'s probabilistic transition to $ \pp : \pprocoutty{\q}{}{\probn{1}}{\ell}{\Basic}{\T'}$ in the typing context corresponds to the probability of reducing to a leaf in $\{\pa\pp{\procout{\q}{\ell}{\e_i}{\PP_i}}\}_{i \in I}$, and that the continuation $\PP_i$ is typed by $\T'$.
This is illustrated in \Cref{fig:frontier}, with the bold paths representing $(\pp,\q)\labels{\ell_2}$ with weights $\probn{0.4} + \probn{0.1} = \probn{0.5}$.

\subsection{Paths and Probabilities}
\label{sec:path}
To take probabilities into account, we formalise the notion of
\emph{paths} and the probability of obtaining a set of paths.
Since typing contexts combine probabilistic choice with
nondeterminism, such probabilities are well-defined only relative to a
\emph{scheduler} that resolves the nondeterminism. We therefore interpret
a typing context as a finite Markov Decision Process (\textsf{MDP}), define
the probability of a set of paths under a fixed scheduler by standard
means~\cite{KSK76}, and obtain lower bounds that transfer to well-typed sessions
by taking the \emph{minimum} over all (fair) schedulers~\cref{sec:processproperties}.

\begin{wrapfigure}{l}{0.3\textwidth}
    \centering
    \begin{tikzpicture}[
        >={Stealth[length=2.1mm,width=1.5mm]},
        every node/.style={inner sep=1.5pt},
        edge/.style={draw, ->, line width=0.45pt, shorten >=2pt, shorten <=2pt},
        dedge/.style={draw, line width=0.45pt, shorten >=2pt, shorten <=2pt,
            -{Stealth[length=2.1mm,width=1.5mm,sep=-0.6mm]
              Stealth[length=2.1mm,width=1.5mm]}},
        lab/.style={font=\footnotesize},
        ann/.style={font=\footnotesize},
    ]
    \clip (-1.45,-1.9) rectangle (1.85,0.72);
    \node (I) at (0.2,0.0) {$\Delta_{1}$};
    \node (K) at (0.2,-1.2) {$\Delta_{2}$};
    \node at ($(K)+(0,-0.35)$) {\footnotesize (deadlock)};
    \draw[dedge] (I) to[out=270,in=90,looseness=1.0]
        node[lab,left,pos=0.6,xshift=-0.03cm]{$\actiontau{\pr}{\pp}{\labels{stop}}$}
        node[ann,right,pos=0.92]{$\probn{0.5}$} (K);
    \draw[dedge] (I) to[out=210,in=170,looseness=18] (I);
    \node[ann] at (-0.2, 0.28) {$\probn{0.5}$};
    \node[lab] at (-0.9, 0.27) {$\actiontau{\pchan{\pr}}{\pchan{\pp}}{\labels{go}}$};
    \draw[edge] (I) to[out=-30,in=10,looseness=18] (I);
    \node[ann] at (0.42, 0.22) {$\probn{1}$};
    \node[lab] at (1.25, 0.25) {$\actiontau{\pchan{\q}}{\pchan{\pp}}{\labels{go}}$};
    \draw[edge] (K.east) -- ++(0.65,0);
    \draw[line width=0.8pt] ($(K.east)+(0.26,-0.16)$) -- ($(K.east)+(0.44,0.16)$);
    \end{tikzpicture}
    \caption{A deadlock dependent\\[1mm]\hphantom{xxxxxxx}on scheduling.}
    \label{fig:scheduling}
\vspace{-4mm} %
\end{wrapfigure}

A typing context can be seen as inducing a finite \textsf{MDP} --- an LTS with \emph{nondeterministic} and \emph{probabilistic} state transitions. We distinguish between \emph{two} sources of nondeterminism: (1); internal $\Sigma$ resolutions $\pp{:}\sum_{i \in I}\T_i \predcontnd{\actionnd{\pp}}\pp{:}\T_i$, and (2); scheduling of parallel enabled roles --- as \emph{separated choice} means that a participant $\pp$ may be simultaneously waiting to receive from both $\q$ and $\pr$, and both being enabled, the scheduler must choose which pair gets to communicate, consider for instance~\cref{fig:scheduling} --- if $\pr$ is scheduled to send there is a $\probn{0.5}$ chance of \emph{deadlocking} ($\pr$ sends $\labels{stop}$ to $\pp$ and both quit, leaving $\q$ hanging), scheduling $\q$ on the other hand sends $\labels{go}$ to $\pp$ and recurses. Thus scheduling of concurrently enabled participants may affect the probability of satisfying a property. Later, we will work with a restricted class of \emph{fair} schedulers that cannot starve persistently enabled actions.

\begin{definition}[Path]
\label{def:path}
A path $\xi$ is a (potentially infinite) sequence of contexts
$\Delta_1; \Delta_2; \cdots$ such that for all $i<|\xi|$ we have
$\Delta_i \predcont{}{\prob} \Delta_{i+1}$ or
$\Delta_i \predcont{}{\nd} \Delta_{i+1}$. A path $\tpath$
is \emph{complete} if it is either infinite or
$\Delta_{|\xi|} \not\rightarrow$. We write $\pathcont{\Delta}$ for the
set of all complete paths starting from $\Delta$.
\end{definition}

\begin{definition}[Scheduler]
\label{def:scheduler}
A scheduler $\phi \in \Phi^\Delta$ for typing context $\Delta$ is a function mapping
each finite path $\Delta_1; \cdots; \Delta_n$ of $\Delta$ with
$\Delta_n \!\rightarrow$ to a pair $(\pp, \Delta)$, of a participant and an optional
typing context, such that:
(1) if $\phi(\Delta_1; \cdots; \Delta_n) = (\pp, \emptyset)$, then either
$\Delta_n \predcont{\actionprob{\pp}}{\probn{p}}$ for some $\probn{p}$,
or $\Delta_n \predcont{\actiontau{\pp}{\pchan{q}}{\ell}}{\probn{1}}$ for
some $\q$ and $\labelsell$; and
(2) if $\phi(\Delta_1; \cdots; \Delta_n) = (\pp, \Delta_{n+1})$, then
$\Delta_n \predcontnd{\actionnd{\pp}} \Delta_{n+1}$.
\end{definition}

Fixing a scheduler $\phi \in \Phi^\Delta$ resolves all nondeterminism in the \textsf{MDP} of $\Delta$, inducing a Discrete-time Markov chain (\textsf{DTMC}) in which every transition carries a probability (the scheduler induced transitions can be seen to carry probability $\probn{1}$).
The probability of a \emph{finite} path is then the product of its transition probabilities; we write $\Tpath_\phi(\Delta)$ for the set of complete paths induced by $\phi$ from $\Delta$. To lift finite-path probabilities to a measure on complete paths, we follow~\cite{KSK76} and take cylinder sets as our basic events.

\begin{definition}[Cylinder set]
\label{def:cylinder}
The \emph{cylinder set} $\cylsig{\xi}$ of a finite path
$\xi = \Delta_1; \cdots; \Delta_n$ of the \textsf{DTMC} induced by a
scheduler $\phi$ is the set of all complete paths in
$\Tpath_\phi(\Delta_1)$ with prefix $\xi$; in particular, if $\xi$ is
itself complete, then $\cylsig{\xi} = \{\xi\}$. We write
$\probpathmeasure{\phi}{\Delta_1}$ for the unique probability measure on
the $\sigma$-algebra generated by the cylinder sets with
$\probpathsig{\Delta_1}{\cylsig{\xi}} = \prod_{i=1}^{n-1} \probi{i}$,
where each $\Delta_i \predcont{}{\probi{i}}
\Delta_{i+1}$; measurable sets of paths are \emph{events}.
\end{definition}

For a predicate $f$ on paths, we write $\probpathmeasure{\phi}{\Delta}(f)=\probpathmeasure{\phi}{\Delta}(\{\tpath \in \Tpath_\phi(\Delta) \SEP f(\tpath)\})$ for the probability of obtaining a path satisfying $f$ under $\phi$ from $\Delta$.
For state predicate $g$ we define $\eventually g$ to be the event ``eventually $g$'', where $(\eventually g)(\tpath) \iff \exists i \leq |\tpath|: g(\tpath[i])$.

\subsection{Probabilistic Deadlock-Freedom and Liveness for Typing Contexts}
\label{sec:typingcontextproperties}
In this section we define \emph{two} properties of typing contexts, \emph{deadlock-freedom} and the slightly stronger property of \emph{liveness}.
We first define the non-probabilistic version of each before introducing their probabilistic counterparts.   
A context $\Delta$ is \emph{deadlocked} if it has no transitions and some participant has a pending action.
A context is \emph{deadlock-free} if no \emph{deadlocked} state is reachable.

\begin{definition}[Deadlock-freedom]
\label{def:termdf}
\quad 
\begin{enumerate}
\item 
\label{item:deadlocked}
A context $\Delta$ is \emph{deadlocked}, written $\deadlock{\Delta}$, if
$\Delta \not{\rightarrow}$ and $\exists \pp : \Delta(\pp) \neq \tend$.
\item
\label{item:df}
A context $\Delta$ is \emph{deadlock-free}, written $\df{\Delta}$, if $\Delta
\rightarrow^* \Delta'$ implies $\neg \deadlock{\Delta'}$.
\end{enumerate}
\end{definition}

Before introducing the concept of \emph{liveness}, we define what it means for a path to be \emph{fair}, as our notion of liveness depends upon the scheduler being fair.

\begin{definition}[Fair path]
    \label{def:fairpath}
    A complete path $\tpath$ is \emph{fair}, written $\fair{\tpath}$, if the following hold:
  \begin{enumerate}
      \item $\forall \pp, \q, \labelsell:$ if
      $\forall i\, \exists j \geq i:$
      $\Delta_{j} \predcont{\actiontau{\pp}{\q}{\ell}}{\probn{1}}$
      then
      $\forall i\, \exists n \geq i:$
      $\Delta_{n} \predcont{\actiontau{\pp}{\q}{\labelsell}}{\probn{1}} \Delta_{n+1}$
  \item $\forall \pp\,\forall i:$ if
    $\Delta_{i} \predcontnd{\actionnd{\pp}}{}$ then 
    $\exists n \geq i:$ 
    $\Delta_{n} \predcontnd{\actionnd{\pp}} \Delta_{n+1}$
  \item $\forall \pp\,\forall i:$ if
    $\Delta_{i} \predcont{\actionprob{\pp}}{}$ then 
    $\exists n \geq i, \probn{p}:$ 
    $\Delta_{n} \predcont{\actionprob{\pp}}{\probn{p}} \Delta_{n+1}$
  \end{enumerate}
\end{definition}
Intuitively, a path is \emph{fair} if a participant $\pp$ cannot be denied sending an \emph{enabled} communication action indefinitely; and if $\pp$ has an internal step to take then it will eventually do so.

A path is then \emph{live} if any $\pp$ with a pending \emph{input} or \emph{output} action along it eventually communicates --- note that this does not require the communication partner to be immediately available.

\begin{definition}[Live path]
    \label{def:livepath}
    A path $\xi$ is \textit{live}, written $\live{\tpath}$, if for every $\Delta_i \in \xi$:
  \begin{enumerate}
    \item if $\Delta_i \predcont{\actionprocout{\pp}{\q}{\ell}{\Basic}}{\probn{1}}$, then $\exists n \geq i:$ $\Delta_n \predcont{\actiontau{\pp}{\q}{\ell}}{\probn{1}}{ \Delta_{n+1}}$
  \item if $\Delta_i \predcont{\actionprocin{\q}{\pp}{\ell}{\Basic}}{\probn{1}}$, then $\exists n \geq i$, $\pchan{r}, \labelsell':$
      $\Delta_n \predcont{\actiontau{\pchan{r}}{\q}{\ell'}}{\probn{1}}{ \Delta_{n+1}}$
  \end{enumerate}
\end{definition}

\begin{definition}[Live contexts]
\label{def:live}
Given a context $\Delta$, $\Delta$ is \emph{live}, written $\live{\Delta}$, if every
complete path $\xi \in \Xi(\Delta)$ that is fair is also live.
\end{definition}

\begin{example}[Non-liveness of the calculator service]
    \label{exam:runninglive}
    Consider again the calculator service of~\cref{exam:running}, whose types are given in~\cref{exam:runningtype}. The typing context formed by $\Delta = \pu{:}\T_\pu, \ps{:}\T_\ps, \pchan{c}{:}\T_{\pchan{c}}$ is \emph{not live} according to~\cref{def:live}. To see this, note that there is an \emph{infinite} reduction path\\[1mm]
\centerline{$\footnotesize
\begin{array}{l}
\Delta
   \predcont{\actiontau{\pchan u}{\pchan s}{\labels{incr}}}{1} \cdot
   \predcont{\actiontau{\pchan s}{\pchan c}{\labels{incr}}}{1} \cdot
   \predcontnd{\actionnd{\pchan c}}{} \cdot
   \predcontdouble{\actiontau{\pchan c}{\pchan s}{\labels{error}}}{0.2} \cdot
   \predcont{\actiontau{\pchan s}{\pchan c}{\labels{incr}}}{1} \cdot
   \predcontnd{\actionnd{\pchan c}}{} \cdot
   \predcontdouble{\actiontau{\pchan c}{\pchan s}{\labels{error}}}{0.2} \cdot
   \predcont{\actiontau{\pchan s}{\pchan c}{\labels{incr}}}{1} \cdot
   \predcontnd{\actionnd{\pchan c}}{} \cdot
   \predcontdouble{\actiontau{\pchan c}{\pchan s}{\labels{error}}}{0.2} \cdot
   \cdots
\end{array}
$}\\[1mm]
where after $\pu$'s initial message, $\pu$ is waiting at $\pu{}\predcont{\actionprocin{\pu}{\pchan{c}}{\labels{res}}{\ints}}{\probn{1}}$ while $\ps$ and $\pchan{c}$ are communicating indefinitely, yet the probability of seeing this path is $\lim_{n\to\infty}\probn{0.2}^{\,n}=\probn{0}$.
\end{example}

\cref{exam:runninglive} illustrates that for \emph{probabilistic}
programs, it would be desirable to have more permissive notions of the
properties, admitting \emph{non-live} (or deadlocking) paths so long as
they have measure $0$ --- or indeed any quantifiable measure. 
We therefore measure both properties, asking: what probability of
deadlock-freedom, and of liveness, can be \emph{guaranteed} however the
nondeterminism is resolved? 
Since the probability of an event is
well-defined only relative to a scheduler
(\cref{def:scheduler,def:cylinder}), we quantify over those schedulers that induce \emph{fair} paths with probability 1; 
\begin{definition}[Fair schedulers]
\label{def:fairscheduler}
A scheduler $\phi$ over $\Delta$ is \emph{fair} if $\phi \in \Phi_F^{\Delta}= \{ \phi \in \Phi^\Delta \;|\; \probpathsig{\Delta}{\fairr} = 1 \}$
\end{definition}

\begin{definition}[Probabilistic deadlock-freedom and liveness]
\label{def:context-probprop}
For a context $\Delta$, we define the \emph{minimum probability} of
deadlock-freedom and of liveness as
\[
  \probdfmin{\Delta} = \probn{1} - \max_{\phi \in \Phi^{\Delta}_F} \probpathsig{\Delta}{\eventually \deadlockk}
  \qquad\qquad
  \problivemin{\Delta} = \min_{\phi \in \Phi_F^\Delta}
    \probpathsig{\Delta}{\livee}%
\]
\end{definition}

\begin{definition}[Almost-sure properties]
\label{def:context-asprop}
$\Delta$ is \emph{almost-surely deadlock-free},
written $\puredf{\Delta}$, if $\probdfmin{\Delta} = 1$, and
\emph{almost-surely live}, written $\aslive{\Delta}$, if
$\problivemin{\Delta} = 1$. 
\end{definition}

\begin{figure}
\vspace{-2mm}      
\definecolor{lightred}{RGB}{233,120,110}
\definecolor{mgnta}{RGB}{201,31,150}
\definecolor{lblue}{RGB}{72,146,205}
\centering
\begin{tabular}{@{}cc}
\begin{minipage}{0.62\textwidth}
\vspace{-2mm}
\begin{tikzpicture}[
    >={Stealth[length=2.1mm,width=1.5mm]},
    every node/.style={inner sep=1.5pt},
    edge/.style={draw, ->, line width=0.45pt, shorten >=2pt, shorten <=2pt},
    dedge/.style={draw, line width=0.45pt, shorten >=2pt, shorten <=2pt,
        -{Stealth[length=2.1mm,width=1.5mm,sep=-0.6mm]
          Stealth[length=2.1mm,width=1.5mm]}},
    lab/.style={font=\footnotesize},
    ann/.style={font=\footnotesize},
]

\node (A) at (1.7,6.5) {$\Delta_1$};
\node (C) at (4.0,6.5) {$\Delta_2$};
\node[text=lblue]    (F) at (6.9,6.5) {$\Delta_6$};
\node[text=lblue]    (G) at (8.5,6.5) {$\Delta_7$};
\node (B) at (1.7,5.1) {$\Delta_3$};
\node (D) at (4.0,5.1) {$\Delta_4$};
\node[text=lblue]    (H) at (7.7,5.0) {$\Delta_8$};
\node (E) at (4.0,3.9) {$\Delta_5$};
\node[text=mgnta]    (I) at (6.4,3.9) {$\Delta_{10}$};
\node[text=lightred] (J) at (1.8,3.9) {$\Delta_{9}$};
\node[text=mgnta]    (K) at (8.2,4.4) {$\Delta_{11}$};
\node[text=mgnta]    (L) at (6.4,5.0) {$\Delta_{12}$};
\coordinate (in) at (0.3,6.5);

\draw[edge] (in) -- (A);                                                    %

\draw[edge] (A) -- node[lab,above]{\actiontau{\pchan{u}}{\pchan{s}}{\labels{incr}}}
                   node[ann,below,pos=0.94]{} (C);

\draw[dedge] (C) -- node[lab,left,pos=0.5]{\actiontau{\pchan{s}}{\pchan{c}}{\labels{incr}}}
                   node[ann,right,pos=0.85]{\probn{0.9}} (D);

\draw[dedge,draw=lblue] (C) -- node[lab,above]{\actiontau{\pchan{s}}{\pchan{c}}{\labels{upgrade}}}
                   node[ann,below,pos=0.93]{\probn{0.1}} (F);

\draw[edge] (D) -- node[lab,above,pos=0.55]{$\actionnd{\pchan{c}}$}
                   node[ann,below,pos=0.9]{$\nd$} (B);

\draw[edge] (D) -- node[lab,left,pos=0.4]{$\actionnd{\pchan{c}}$}
                   node[ann,left,pos=0.88]{$\nd$} (E);

\draw[edge] (B) -- node[lab,right,pos=0.45]{\actiontau{\pchan{c}}{\pchan{u}}{\labels{res}}}
                   node[ann,right,pos=0.9]{} (A);

\draw[dedge] (E) to[out=170,in=230,looseness=1.2]
                   node[lab,left,pos=0.73]{\actiontau{\pchan{c}}{\pchan{u}}{\labels{res}}}
                   node[ann,left,pos=0.93,xshift=-0.1cm]{\probn{0.8}} (A);

\draw[dedge] (E) to[out=40,in=292,looseness=1.5]
                   node[lab,right,pos=0.52]{\actiontau{\pchan{c}}{\pchan{s}}{\labels{error}}}
                   node[ann,right,pos=0.88]{\probn{0.1}} (C);

\draw[dedge,draw=mgnta] (E) -- node[lab,above,pos=0.56]{\actiontau{\pchan{c}}{\pchan{s}}{\labels{disconn}}}
                   node[ann,below,pos=0.9]{\probn{0.05}} (I);

\draw[dedge,draw=lightred] (E) -- node[lab,below right,pos=0.6]{\actiontau{\pchan{c}}{\pchan{s}}{\labels{crit}}}
                   node[ann,below,pos=0.9]{\probn{0.05}} (J);

\draw[edge,draw=mgnta] (I) to[out=30,in=182,looseness=1.4] 
                   node[lab,above,pos=0.4]{$\actionnd{\pchan{s}}$}
                   node[ann,above,pos=0.88]{$\nd$} (K);

\draw[edge,draw=mgnta] (K) to[out=220,in=0,looseness=1.3] 
                   node[lab,below,pos=0.3]{$\actiontau{\pchan{s}}{\pchan{c}}{\labels{retry}}$}
                   node[ann,below,pos=0.94]{} (I);

\draw[edge,draw=mgnta] (I) -- node[lab,left,pos=0.4]{$\actionnd{\pchan{s}}$}
                              node[ann,left,pos=0.88]{$\nd$} (L);

\draw[edge,draw=mgnta] (L) -- node[lab,right,pos=0.65]{$\actiontau{\pchan{s}}{\pchan{c}}{\labels{conn}}$}
                              node[ann,above,pos=0.85]{} (C);

\draw[edge,draw=lblue] (F) to node[lab,above,pos=0.5]{\actiontau{\pchan{s}}{\pchan{c}}{\labels{incr}}}
                   node[ann,below,pos=0.9]{} (G);
\draw[edge,draw=lblue] (G) to node[lab,right,pos=0.5]{\actiontau{\pchan{c}}{\pchan{u}}{\labels{res}}}
                   node[ann,below,pos=0.9]{} (H);
\draw[edge,draw=lblue] (H) to node[lab,left,pos=0.5]{\actiontau{\pchan{u}}{\pchan{s}}{\labels{incr}}}
                   node[ann,right,pos=0.9]{} (F);

\end{tikzpicture}
  \end{minipage}
  \begin{minipage}{0.38\textwidth}
    \centering
    \small
    \begin{tabular}{|l|l|l|l|l|l|}
      \hline
      & $\probliveminplain\!\!$ & $\probdfminplain\!\!$ & & $\probliveminplain\!\!$ &  $\probdfminplain\!\!$ \\
      \hline
        $\Delta_1$ & \probn{0.52} &\probn{0.69} &$\Delta_7$   &\probn{1} &\probn{1} \\
      \hline                                   
        $\Delta_2$ & \probn{0.52} &\probn{0.69} &$\Delta_8$   &\probn{1} &\probn{1} \\
      \hline                                      
        $\Delta_3$ & \probn{0.52} &\probn{0.69} &$\Delta_9$   &\probn{0} &\probn{0} \\
      \hline
        $\Delta_4$ & \probn{0.47} & \probn{0.65} &$\Delta_{10}$&\probn{0} &\probn{0.69} \\
      \hline
        $\Delta_5$ & \probn{0.47} & \probn{0.65} &$\Delta_{11}$&\probn{0} &\probn{0.69} \\
      \hline
        $\Delta_6$ & \probn{1} & \probn{1} &$\Delta_{12}$&\probn{0.52} &\probn{0.69} \\
      \hline
    \end{tabular}
  \end{minipage}
  \end{tabular}
  \vspace{-2mm}
  \caption{A refined calculator service~\cref{exam:mdp}.\label{fig:mdp}}
\end{figure}

\begin{example}[Calculator service extended]
    \label{exam:mdp}
    Consider a refinement of the \emph{calculator service} from~\cref{exam:runningtype} whose labelled transition system is given in~\cref{fig:mdp}. $\Delta_{1-5}$ correspond to the original session from~\cref{exam:runningtype}, but the $\probn{0.2}$ probability of an $\labels{error}$ occurring has been distributed into \emph{two} additional events now, (1); with probability $\probn{0.05}$ the error is \emph{critical} and $\pchan{c}$ informs $\ps$ after which both quit, leaving $\pu$ hanging (a deadlock, $\Delta_9$), (2); with probability $\probn{0.05}$, $\pchan{c}$ loses its connection ($\labels{disconn}$) to $\ps$, entering a $\labels{retry}$ loop ($\Delta_{10-11}$), eventually it \emph{may} reconnect ($\Delta_{12}$) but this is \emph{not} guaranteed. Finally, $\ps$ knows $\pchan{c}$ is faulty, and will with probability $\probn{0.1}$ \labels{upgrade} to a non-faulty mode, after which the 3 participants can communicate and remain \emph{live} ($\Delta_{6-8}$). The table on the right side of~\cref{fig:mdp} gives $\problivemin{\Delta}$ and $\probdfmin{\Delta}$ for each $\Delta$. Worthy of note is the probability of \emph{liveness} is \probn{0} in $\Delta_{10-11}$, as there exists a \emph{fair} path ($\labels{retry}$ indefinitely) that starves $\pu$ --- a fair adversarial scheduler can always pick the action $\Delta_{10} \predcontnd{\actionnd{\ps}} \Delta_{11}$, keeping the region closed. The full type of $\Delta_1$ is given in~\cref{exam:mdpfull}.
\end{example}

\begin{example}[Separating examples]
    ~\Cref{fig:property-relationships} summarises the relationships between the properties of this section by exhibiting a few examples that fall into different categories of either \emph{partially} or \emph{fully} satisfying a property. The examples are numbered to match their regions in the diagram. 
\end{example}

The properties are stable along \emph{subtyping}: replacing a context by any subtype 
--- as the verification workflow of \cref{fig:workflow} does when checking an inferred context against a declared one --- 
preserves every property of this section, qualitatively and in probability:

\begin{restatable}[$\subt$-monotonicity]{proposition}{submono}
\label{prop:sub-monotone}
\proofreference{\cref{app:path}}
Let $\Delta \subt \Delta'$ and $\safe{\Delta'}$, then:\\
\begin{enumerate*}
    \item $\varphi(\Delta') \implies \varphi(\Delta)$ if
    $\varphi \in \{\safee,\dff,\livee\}$; and
    \item $\probdfmin{\Delta'} \leq \probdfmin{\Delta}$ and $\;\problivemin{\Delta'} \leq \problivemin{\Delta}$.
\end{enumerate*}
\end{restatable}

\noindent
Thus, properties model checked for $\Delta'$ transfer to every subtype and, by \rulename{t-sub}, to every session typed below it. On the other hand, \emph{widening} to a supertype may weaken the bounds by admitting additional behaviours: the guarantees remain sound but need not be tight.

\begin{example}[Bounds]\label{exam:bounds}
To show how $\subt$ may weaken bounds, consider first the session:\\[1mm]
\centerline{$ 
    \der{}{(\pa\q{\flip{\probn{0.5}}{\pp!\labels{m}}{\pr!\labels{m}}} \pc \pa\pp{\q?\labels{m}})}{\q:{\oplus\{{\probn{0.5}:\pp!\labels{m},}{\probn{0.5}:\pr!\labels{m}}\}},\;\pp:{\q?\labels{m}}}
$}
whose type is partially deadlock-free and live as shown in bullet $4$ of~\cref{fig:property-relationships}. In this case the session and typing context LTS are isomorphic, and indeed their lower bounds are identical. Consider now what happens when $\q$ has an always-true \emph{if}-statement over action $\pp!\labels{m}$ and its previous \emph{coinflip}:\\[1mm]
\centerline{$ 
    \begin{array}{ll}
        \vdash & {\pa\q{\cond{\true}{\pp!\labels{m}}{\flip{\probn{0.5}}{\pp!\labels{m}}{\pr!\labels{m}}}} \pc \pa\pp{\q?\labels{m}}}\\
        :  & {\q:{\Sigma\{\oplus\{\pp!\labels{m}\},\;\oplus\{{\probn{0.5}:\pp!\labels{m},}{\probn{0.5}:\pr!\labels{m}}\}\}},\;\pp:{\q?\labels{m}}}
    \end{array}
$}
By \rulename{sub-$\Sigma$} the two branches have been widened to a nondeterministic choice (abstracting away the conditional used to make the choice), but the cost is obvious; while the session is clearly \emph{live} the type still reports probability $\probn{0.5}$ of satisfying \emph{liveness}, as an adversarial scheduler may always choose to pick the coinflip branch.
\end{example}

Finally, we note merging is also compatible with refinement:

\begin{restatable}[$\probmerge{}$-monotonicity]{lemma}{monotonicmerge}
\label{lemma:monotonicmerge}
\proofreference{\cref{app:path}}
Suppose $\T_i \subt \T_i'$ for $i \in \{1,2\}$ and $\T_1' \probmerge{\prob}\T_2'$ is defined, then $\T_1 \probmerge{\prob} \T_2$ is defined and $\T_1 \probmerge{\prob} \T_2 \subt \T_1' \probmerge{\prob} \T_2'$.
\end{restatable}

\noindent Hence, replacing its operands by subtypes can only strengthen the resulting bounds.

\begin{figure}[tb]
\centering
\begin{tikzpicture}[
    x=0.75cm, y=0.72cm,
    font=\small,
    set box/.style={
        draw=black!75,
        line width=0.5pt,
        rounded corners=2pt
    },
    set label/.style={
        anchor=north west,
        inner sep=4pt
    },
    number/.style={
        circle,
        draw=black!70,
        fill=gray!8,
        line width=0.4pt,
        minimum size=4mm,
        inner sep=0pt,
        font=\scriptsize
    },
    table line/.style={
        draw=black!75,
        line width=0.5pt
    }
]

\definecolor{safeFill}{HTML}{E2F5DE}
\definecolor{pdfFill}{HTML}{BFF3B5}
\definecolor{pliveFill}{HTML}{C6E8F1}
\definecolor{dfFill}{HTML}{F7BCBC}
\definecolor{overlapFill}{HTML}{E8CDD9}
\definecolor{asliveFill}{HTML}{FFF18A}
\definecolor{liveFill}{HTML}{FFB45C}

\path[set box, fill=safeFill]
    (0,0) rectangle (7,4.8);

\path[set box, fill=pdfFill]
    (0.14,0.14) rectangle (6.86,4.02);

\path[rounded corners=2pt, fill=pliveFill]
    (2.95,0.28) rectangle (6.72,3.92);

\path[rounded corners=2pt, fill=dfFill]
    (0.28,0.28) rectangle (6.58,2.65);

\begin{scope}
    \clip[rounded corners=2pt]
        (2.95,0.28) rectangle (6.72,3.92);
    \path[rounded corners=2pt, fill=overlapFill]
        (0.28,0.28) rectangle (6.58,2.65);
\end{scope}

\draw[set box]
    (2.95,0.28) rectangle (6.72,3.92);
\draw[set box]
    (0.28,0.28) rectangle (6.58,2.65);

\path[set box, fill=asliveFill]
    (3.09,0.42) rectangle (6.44,1.85);

\path[set box, fill=liveFill]
    (4.88,0.56) rectangle (6.30,1.72);

\node[set label] at (0,4.8)
    {$\safe{\Delta}$};

\node[set label] at (0.14,4.02)
    {$\probdfmin{\Delta} > 0$};

\node[set label] at (2.95,4.02)
    {$\problivemin{\Delta} > 0$};

\node[set label] at (0.28,2.65)
    {$\df{\Delta}$};

\node[set label] at (3.09,1.85)
    {$\aslive{\Delta}$};

\node[set label] at (4.88,1.72)
    {$\live{\Delta}$};

\node[number] at (6.00,4.43) {1};
\node[number] at (1.50,3.05) {2};
\node[number] at (1.50,0.95) {3};
\node[number] at (6.00,3.25) {4};
\node[number] at (6.00,2.25) {5};
\node[number] at (3.95,0.88) {6};
\node[number] at (5.60,0.88) {7};

\begin{scope}
\newcommand{\examplemark}[1]{%
    \tikz[baseline=(n.base)]{%
        \node[number,minimum size=3.7mm] (n) {#1};
    }%
}

\node[anchor=north west,inner sep=0pt] at  (7.05,4.8) {%
    \scriptsize
    \setlength{\tabcolsep}{2pt}
    \renewcommand{\arraystretch}{1.1}
    \begin{tabular}{|c|p{7.7cm}|}
    \hline
    & \centering Examples \tabularnewline
    \hline

    \examplemark{1} &
    $\begin{array}[t]{@{}l@{}}
        \pchan{p}:\oplus\pchan{q}!\labels{m},
        \quad
        \pchan{q}:\oplus\pchan{p}!\labels{m}
    \end{array}$
    \tabularnewline[2pt]
    \hline

    \examplemark{2} &
    $\begin{array}[t]{@{}l@{}}
        \pchan{p}:\mu\ty.\&\{
            \pchan{q}?\labels{m}.\ty
        \},
        \;\;
        \pchan{r}:\oplus\pchan{q}!\labels{m},
        \;\;
        \pchan{q}:\oplus\{
            \probn{0.5}:\pchan{r}!\labels{m},
            \;
            \probn{0.5}:\pchan{p}!\labels{m}.
                \mu\ty.\oplus\pchan{p}!\labels{m}.\ty
        \}
    \end{array}$
    \tabularnewline[2pt]
    \hline

    \examplemark{3} &
    $\begin{array}[t]{@{}l@{}}
        \pchan{p}:\mu\ty.\&\{
            \pchan{q}?\labels{m}.\ty
        \},
        \quad
        \pchan{q}:\mu\ty.\oplus\pchan{p}!\labels{m}.\ty,
        \quad
        \pchan{r}:\&\pchan{p}?\labels{m}
    \end{array}$
    \tabularnewline[2pt]
    \hline

    \examplemark{4} &
    $\begin{array}[t]{@{}l@{}}
        \pchan{p}:\&\{
            \pchan{q}?\labels{m}
        \},
        \quad
        \pchan{q}:\oplus\{
            \probn{0.5}:\pchan{p}!\labels{m},
            \;
            \probn{0.5}:\pchan{r}!\labels{m}
        \}
    \end{array}$
    \tabularnewline[2pt]
    \hline

    \examplemark{5} &
    $\begin{array}[t]{@{}l@{}}
        \pchan{p}:\mu\ty.\&\{
            \pchan{q}?\labels{m}.\ty,
            \;
            \pchan{r}?\labels{m}.\ty
        \},
        \;\;
        \pchan{q}:\mu\ty.\oplus\pchan{p}!\labels{m}.\ty,
        \;\;
        \pchan{r}:\oplus\{
            \probn{0.5}:\pchan{p}!\labels{m},
            \;
            \probn{0.5}:\pchan{q}!\labels{m}
        \}
    \end{array}$
    \tabularnewline[2pt]
    \hline

    \examplemark{6} &
    $\begin{array}[t]{@{}l@{}}
        \pchan{p}:\mu\ty.\oplus\{
            \probn{0.5}:\pchan{q}!\labels{m}.\ty,
            \;
            \probn{0.5}:\pchan{r}!\labels{m}.\ty
        \},
        \quad
        \pchan{q}:\mu\ty.\&\{
            \pchan{p}?\labels{m}.\ty
        \},
        \quad
        \pchan{r}:\mu\ty.\&\{
            \pchan{p}?\labels{m}.\ty
        \}
    \end{array}$
    \tabularnewline[2pt]
    \hline

    \examplemark{7} &
    $\begin{array}[t]{@{}l@{}}
        \pchan{p}:\oplus\pchan{q}!\labels{m},
        \quad
        \pchan{q}:\&\pchan{p}?\labels{m}
    \end{array}$
    \tabularnewline[2pt]
    \hline

    \end{tabular}%
};
\end{scope}

\end{tikzpicture}
\caption{Relationships between the properties. Numbers identify examples in the right-hand panel which separate properties. For example, the typing context in $2$ is safe and partially deadlock-free, but not live nor completely deadlock-free.}
\label{fig:property-relationships}
\end{figure}

\subsection{Probabilistic Deadlock-Freedom and Liveness for Multiparty Processes}
\label{sec:processproperties}
In this section we transfer the bounds obtained in the previous section on $\Delta$ back to sessions $\Multi$ typed as $\der{}{\Multi}{\Delta}$. We first adapt the definitions of \cref{sec:path,sec:typingcontextproperties} from typing contexts to sessions, and then show that $\probdfmin{\Delta}$ and $\problivemin{\Delta}$ bound the corresponding properties of $\Multi$.

\ifpldi
\aleks{move these definitions to section 2's LTS?}
\fi
\begin{definition}[Session paths]
\label{def:proc:path}
A (session) path $\xi$ is a sequence $\M_1; \M_2; \cdots$ with
$\M_i \pred{\probi{i}} \M_{i+1}$ for all $i \in [1,|\xi|)$;
\emph{complete} if infinite or $\M_{|\xi|} \not\rightarrow$. 
\end{definition}

\begin{definition}[Session scheduler]
\label{def:sessscheduler}
A scheduler $\phi \in \Phi^\N$ for session $\N$ is a function mapping
each finite path $\N_1; \cdots; \N_n$ of $\N$ with
$\N_n \!\rightarrow$ to a participant $\pp$ such that
if $\phi(\N_1; \cdots; \N_n) = \pp$, then either
$\N_n \spcon \pa\pp\PP \pc \N'$ and $\pa\pp\PP\predcont{}{\prob} \pa\pp\PP'$
for some $\probn{p}$,
or $\N_n \predcont{\actiontau{\pp}{\pchan{q}}{\ell}}{\probn{1}}$ for
some $\q$ and $\labelsell$.
\end{definition}

Complete paths $\pathcont{\M}$ and $\Xi_\phi(\M)$, and probability measure $\probpathmeasure{\phi}{\M}$ are as defined in~\cref{def:scheduler,def:cylinder}. 

\begin{definition}[Fair session paths]
\label{def:proc:fairpath}
A session path $\xi$ is \emph{fair}, written $\fairr(\tpath)$ if:
\begin{enumerate}
\item $\forall \pp, \q$: if $\forall n\, \exists m \geq n, \labelsell$
    with $\M_m \predcont{\actiontau{\pp}{\q}{\labelsell}}{}$, then
  $\forall n\, \exists m \geq n$ with
  $\M_m \predcont{\actiontau{\pp}{\q}{\labelsell}}{} \M_{m+1}$
\item $\forall \pp\, \forall n$: if
  $\M_n \spcon \pa{\pp}{\flip{\prob}{\PP}{\Q}} \pc \M'$ or
  $\M_n \spcon \pa{\pp}{\cond{\e}{\PP}{\Q}} \pc \M'$, then
  $\exists m > n$ with
  $\M_m \spcon \pa{\pp}{\PP} \pc \M''$ or
  $\M_m \spcon \pa{\pp}{\Q} \pc \M''$
\end{enumerate}
\end{definition}

\begin{definition}[Live session paths]
\label{def:proc:live}
A session path $\xi$ is \emph{live}, written $\live{\tpath}$ if for every $\M_n \in \xi$:
\begin{enumerate}
    \item if $\M_n \spcon \pa{\pp}{\procout{\q}{\ell}{\e}{\PP}} \pc \M'$, then $\exists m \geq n$ with $\M_m \predcont{\actiontau{\pp}{\q}{\labelsell}}{} \M_{m+1}$
    \item if $\M_n \spcon \pa{\q}{\psyntin{i \in I}{\pp_i}{i}{x_i}{\Q_i}} \pc \M'$, then $\exists m \geq n, j \in I$ with $\M_m \predcont{\actiontau{\ppi{j}}{\q}{\labelselli{j}}}{} \M_{m+1}$
\end{enumerate}
\end{definition}

A session $\N$ is \emph{live}, written $\live{\N}$, if every fair
complete $\xi \in \pathcont{\N}$ is live;
$\N$ is $\deadlock{\N}$, if $\N \not\rightarrow$ and $\exists \pp, \PP \st \N \spcon \pa{\pp}{\PP} \pc \N'$ with $\PP \neq \inact$; and $\df{\N}$ if $\N \predcontstar{}{} \N'$ implies $\neg \deadlock{\N'}$. %

Analogously to~\cref{def:context-probprop}, we define 
the \emph{lower bounds} on these properties over all fair schedulers:
\ifpldi
\aleks{instead of defining these one by one as $\problivemin{\Delta}$ maybe just define $\probpathmin{\Phi_F}{\Delta \models P} = \inf_{\phi \in \Phi_F}(\probpathsig{}{\Delta \models P})$and give the final theorems instead of 10 definitions. } 
\fi

\begin{definition}[Probabilistic properties of sessions]
\label{def:session-probprop}
\[
  \probdfmin{\N} = \probn{1} - \max_{\phi \in \Phi^{\N}_F} \probpathsig{\N}{\eventually \deadlockk}
  \qquad
  \problivemin{\Multi} = \min_{\phi \in \Phi_F^{\Multi}}
    \probpathsig{\Multi}{\livee}
\]
with $\puredf{\Multi}$ and $\aslive{\Multi}$ as defined in \cref{def:context-asprop}.
\end{definition}

The main theorems of this paper follow; each property transfers from $\Delta$ to $\Multi$, qualitatively and in probability.

\begin{restatable}{theorem}{procdf}
\label{thm:proc:df}
\proofreference{\cref{app:sec44proofs}}
If $\der{}{\Multi}{\Delta}$, then
\begin{enumerate*}
    \item $\df{\Delta}$ implies $\df{\N}$; and 
    \item $\probdfmin{\Delta} \leq \probdfmin{\Multi}$
\end{enumerate*}
\end{restatable}

\begin{restatable}[Liveness]{theorem}{livenss}
\label{thm:liveness}
\proofreference{\cref{app:sec44proofs}}
If $\der{}{\Multi}{\Delta}$, then
\begin{enumerate*}
    \item $\live{\Delta}$ implies $\live{\Multi}$; 
    \item $\problivemin{\Delta} \leq \problivemin{\Multi}$.
\end{enumerate*}
\end{restatable}

\begin{corollary}
\label{cor:as-transfer}
If $\der{}{\N}{\Delta}$, then $\puredf{\Delta}$ implies $\puredf{\Multi}$ and $\aslive{\Delta}$ implies $\aslive{\Multi}$.
\end{corollary}

 \section{Translating Probabilistic Multiparty Session Types into \prism{}}
 \label{sec:translation}
 This section presents the translation from probabilistic multiparty session types
to the \prism{} modelling language. We first introduce the syntax and semantics of the
\prism{} language (\Cref{sec:prism}), before describing the translation
function (\Cref{sec:translating}). Finally, we show the correctness of
this encoding with 
respect to the operational semantics of session types and of \prism{}
(\Cref{sec:correctness}).

\subsection{\prism{} Syntax and Operational Semantics}
\label{sec:prism}
To enable a precise discussion, we first formally define the relevant subset of the \prism{} language. We closely follow \cite{prismdoc,Carbone2024}, with modifications to syntax and a more precise discussion of states.

\begin{definition}[\prism{} Syntax]\label{def:prismsyntax}
The syntax of \prism{} \textit{models} is given by
\begin{alignat*}{4}
        M, N \ &&\DEF \ &\{ C_i \}_{i \in I} \bnfsep \prismmodulecomp{M}{N} 
& \text{\small (module, composition)} \\
        C \ &&\DEF \
        &\prismcommandgeneric{\alpha}{g}{i \in I}{\probi{i}}{u_i}
        & \hspace{-5cm}\text{\small (command, with $\textstyle\sum_{i \in I}\probi{i} = 1$ and $\probi{i} \in [0, 1]$)} \\
        g, g' \ &&\DEF \ & (x = v) \bnfsep \neg g \bnfsep g \wedge
        g' \bnfsep g \vee g' \qquad
 u, u' \ \DEF \  (x' = v) \bnfsep u \wedge u'
& \text{\small (guard, update)} \\
v, v' \ &&\DEF \ &v + v' \bnfsep v - v' \bnfsep v \times v' \bnfsep
        v \div v' \bnfsep  x \bnfsep n \in \mathbb{Z}
        & \hspace{-4cm}\text{\small (operations,
        variable, constant integer)} 
\end{alignat*}
\end{definition}

\prism{} models are a \emph{composition} of \emph{modules}, each containing
\emph{commands} $C_i$. Each command has an \emph{action} $\alpha$ (potentially
empty, denoted as $\alpha = \varepsilon$), \emph{guard}
$g$, and updates $u_i$, each with probability $\probi{i}$. Guards and updates
operate on \emph{states} throughout the execution of the model.

\begin{definition}
\label{def:prismstate}
A \textbf{state} $S$ is a mapping $\{x_i \mapsto n_i\}_{i \in I}$ from
variables $x_i$ to constants $n_i$. \textbf{Updates} override mappings: $S[(x'
= n)] = (S \setminus \{x \mapsto S(x)\}) \cup \{x \mapsto n\}$.
\prism{} distinguishes equality from updates by denoting them as $(x = v)$
and $(x' = v)$, respectively: in the latter, we update $x$, not $x'$.
A state $S$ is \textbf{a state for} model $M$ if every $x \in \dom{S}$ appears
in some guard or update in $M$. We call $(M, S)$ a \textbf{model-state pair}.
\end{definition}

\begin{notation}
For tidiness, we treat zero- and \false-valued variables differently:
\begin{enumerate*}
  \item for $x \notin \dom{S}$, we treat $S(x) = 0 = \false$; and
  \item $S \cup \{x \mapsto 0\} \equiv S \cup \{x \mapsto \false\}
    \equiv S$.
\end{enumerate*}
Hence we succinctly describe the state $\{x \mapsto 42, y \mapsto 0\}$ as $\{x
\mapsto 42\}$. 
Here, we initialise integer variables to 0 and Booleans to false, so the 
\emph{initial state} of any model is $\varnothing.$ 
We write $M$ as a shorthand for $(M, \varnothing)$.
\end{notation}

\begin{figure}
  \centering
\small
$  
  \begin{array}{c}
        \inferrule*[left=\rulename{m-module}]{ 
            \prismcommandgeneric{\alpha}{g}{i \in I}{\probi{i}}{u_i} \in \{C_j\}_{j \in J}
        }{
            \{C_j\}_{j \in J} \prismhas
            \prismcommandgeneric{\alpha}{g}{i \in I}{\probi{i}}{u_i}
        }
        \
        \inferrule*[left=\rulename{m-$\varepsilon$}]{
            M \prismhas \prismcommandgeneric{}{g}{i \in I}{\probi{i}}{u_i}
        }{
            \prismmodulecomp{M}{N} \prismhas
            \prismcommandgeneric{}{g}{i \in I}{\probi{i}}{u_i}
        }
\\[3mm]
        \inferrule*[left=\rulename{m-move}]{
            \alpha \neq \varepsilon \\
            M \prismhas \prismcommandgeneric{\alpha}{g}{i \in I}{\probi{i}}{u_i}
            \\
            \forall J, \probip{j}, u'_j, g' : N \ \not \prismhas \  \prismcommandgeneric{\alpha}{g'}{j \in J}{\probip{j}}{u'_j}
        }{
            \prismmodulecomp{M}{N} \prismhas
            \prismcommandgeneric{\alpha}{g}{i \in I}{\probi{i}}{u_i}
        }
\\[1mm]        
        \inferrule[\rulename{m-sync}]{
            \alpha \neq \varepsilon \quad
            M \prismhas \prismcommandgeneric{\alpha}{g}{i \in I}{\probi{i}}{u_i} \quad
            N \prismhas \prismcommandgeneric{\alpha}{g'}{j \in J}{\probip{j}}{u'_j}
        }{
            \prismmodulecomp{M}{N} \prismhas
            \prismcommandgeneric{\alpha}{g \land g'}{i,j}{\probi{i} \times \probip{j}}{u_i \land u'_j}
        }
        \quad 
        \inferrule[\rulename{s-step}]{
            M \prismhas \prismcommandgeneric{\alpha}{g}{i \in I}{\probi{i}}{u_i} \quad
            S \vdash g 
        }{
            (M, S) \prismreduce{\alpha}{\probi{i}} (M, S[u_i])
        }
    \end{array}
$
    \vspace{-3mm}
    \caption{Rules for $\prismhas$ (\rulename{m-module, $\varepsilon$,
        move, sync}) and for model-state pairs (\rulename{s-step})}
    \label{fig:prism-has}
    \label{fig:prism-transition}
\end{figure}

We now define the semantics of the
\prism{} language. Following \cite{Carbone2024}, we first
introduce a relation $M \prismhas C$ (pronounced \emph{``$M$ has command $C$''})
to capture the effects of composition. The rules for $\prismhas$ are shown in
\Cref{fig:prism-has}. \rulename{m-module}
says a model with one module inherits its commands, while
\rulename{m-$\varepsilon$} ensures that commands with empty actions are
preserved.
If $M$ has action $\alpha$ and $N$ does not, \rulename{m-move} keeps
$M$\emph{'}s command in their composition $\prismmodulecomp{M}{N}$. If both
contain $\alpha$, \rulename{m-sync} synchronises them, executing both commands
together.
Based on this, we define the transition relation $\prismreduce{\alpha}{\prob}$
between model–state pairs (\Cref{fig:prism-transition}). Its sole rule
\rulename{s-step} does what we expect: if $M$ has a command guarded by $g$, and
state $S$ satisfies $g$ ($S \vdash g$), then with probability $\probi{i}$, we
apply the update $u_i$. Since each probabilistic branch becomes a separate transition, we
assume that no two enabled commands with more than one branch share an action, so that the distributions remain recoverable from
$\prismreduce{\alpha}{\prob}$. This is guaranteed by our encoding (\Cref{sec:typeencoding}).
Similarly to typing contexts, we write
$\prismreduce{}{\prob}^*$ for the (probabilistic) reflexive, transitive closure.

\subsection{Translating Multiparty Session Type Semantics}\label{sec:translating}
\subsubsection{Dealing with States}
The main complexity for the encoding arises from the fact that
session types are regarded as a term
rewriting system, whereas \prism{} is \emph{state-based}.
Thus, we must first determine how to map terms into states. To facilitate this, we will be translating each
participant in the typing context into its own module, with variable $S_\pp$
tracking $\pp$'s current position in its type. We then compose these modules
into a single model.

We inductively define %
the \emph{state space} of a type $\T$, written $\StateS{\T}$, as:
$\StateS{\tend} = 0$;
$\StateS{\mu \ty . \T'} = \StateS{\T'}$;
$\StateS{\ty} = 0$;
$\StateS{\pprocinnew{\pp_i}{i \in I}{\labelselli{i}}{\Basic_i}{\T_i}}
                      = 1 + \textstyle\sum_{i \in I} \StateS{\T_i}$;
$\StateS{\pprocoutty{\q}{}{\probn{1}}{\labelsell}{\Basic}{\T'}}
= 1 + \StateS{\T'}$ for a \emph{committed} selection (a single branch of
probability $\probn{1}$); for a proper selection ($|J| \geq 2$),
$\StateS{\pprocoutty{\pp_j}{j \in J}{\probi{j}}{\labelselli{j}}{\Basic_j}{\T_{j}}}
= 1 + \textstyle\sum_{j \in J} \StateS{\comsel{j}}$, where
$\comsel{j} = \pprocoutty{\pp_j}{}{\probn{1}}{\labelselli{j}}{\Basic_j}{\T_{j}}$
denotes the committed selection for branch $j$;
                      and, for a proper sum ($|I| \geq 2$),
$\StateS{\probpprocoutty{i \in I}{\pp_j}{j \in J_i}{\probi{(i,j)}}{\labelselli{j}}{\Basic_j}{\T_{(i,j)}}}
= 1 + \textstyle\sum_{i \in I} \StateS{\selsummand{i}}$, where
$\selsummand{i} = \pprocoutty{\pp_j}{j \in
  J_i}{\probi{(i,j)}}{\labelselli{j}}{\Basic_j}{\T_{(i,j)}}$ denotes the $i$-th
summand of the sum (a singleton selection).
    Each variable $S_\pp$ in the encoding of $\Delta = \pp : \T, \Delta'$ has a
    value between 0 and $\StateS{\T}$, inclusive, where $S_\pp =
    \StateS{\T}$ represents the end state.

  The counts mirror the type-level semantics, which resolves the
  nondeterministic choice between summands, the probabilistic
  choice between branches, and the communication itself by separate
  reductions.
  A branching needs only an entry state, from which the participant
  synchronises with the chosen sender. A committed selection similarly needs
  only an entry state, from which the participant synchronises with the
  recipient. A proper selection contributes an entry state, from which the
  participant commits to a branch probabilistically, moving to the entry
  state of the corresponding committed selection $\comsel{j}$. A proper
  sum contributes an entry state from which a summand is picked
  nondeterministically. Recursion contributes no states, as we simply loop
  back to the state of the corresponding $\mu$-binding. Similarly,
  $\tend$ also contributes no states, since all terminated participants share
  the single end state. The transitions between these states are made precise
  by the encoding below.

\subsubsection{Encoding}
We define an encoding from typing contexts into
\prism{}. We first present an \emph{inner} encoding
$\trantwo{\cdot}$, and then extend this with the \emph{closure module} to give
the final encoding $\tranone{\cdot}$.

\myparagraph{Inner top-level encoding $\trantwo{\cdot}$.}
A typing context $\Delta$ is translated as: 
$\trantwo{\varnothing}  = \varnothing$; 
$\trantwo{\Delta, \Delta'} = \trantwo{\Delta} \cmp
\trantwo{\Delta'}$, and 
$\trantwo{\pp : \T} = \oxtrantwo{\T}{\pp}{0}{\StateS{\T}}{\varnothing}$
with the type-level encoding $\oxtrantwo{\T}{\pp}{n}{m}{f}$,
which takes a type $\T$ being translated, participant $\pp$, current state $n$,
end state $m$ and a mapping $f$ from variables to their state value.

\myparagraph{Inner type-level encoding
$\oxtrantwo{\cdot}{\pp}{n}{m}{f}$.} \label{sec:typeencoding}
The inner type-level encoding is presented in Fig.~\ref{fig:innertranslation}.
Each reduction of the typing context LTS maps to exactly \emph{one} \prism{}
transition, distinguished by its action label. For a proper sum ($|I| \geq 2$), the nondeterministic reduction
$\predcontnd{\actionnd{\pp}}$ maps to an internal transition
$\prismreduce{\pp\conc\tau}{\probn{1}}$ (mirroring the type-level action
$\actionnd{\pp}$) that picks a
summand $i \in I$ and moves to the entry state of the encoding of the singleton
selection $\selsummand{i}$. For a proper selection, the probabilistic reduction
$\predcont{\actionprob{\pp}}{\prob}$ (rule \rulename{ct-prob}) maps to an internal
transition $\prismreduce{\pp\conc p}{\prob}$ in which $\pp$ probabilistically picks a
branch $j \in J$ and moves to the entry state of the encoding of the committed
selection $\comsel{j}$. For a committed selection, the communication
$\predcont{(\pp, \q) \labelsell}{\probn{1}}$ maps to the synchronisation
$\prismreduce{\pp\conc\q\conc\labelsell}{\probn{1}}$
that informs $\q$ of $\pp$'s choice. Since participant names are unique, no other
module has an action labelled $\pp\conc\tau$ or $\pp\conc p$, so these transitions cannot
synchronise and are therefore internal. For a branching, the receiver synchronises
directly with the chosen
sender in a single $\prismreduce{\q\conc\pp\conc\labelsell}{\probn{1}}$ transition.
For recursion $\mu \ty . \T$, we
extend the mapping with $\ty \mapsto n$ and continue translating $\T$
(using $f$ if a continuation is a variable).
\begin{figure}[th]
{$\small
\begin{array}{c}
   \oxtrantwo{\tend}{\pp}{n}{m}{f} = \varnothing \qquad 
   \oxtrantwo{\ty}{\pp}{n}{m}{f} = \varnothing \qquad
   \oxtrantwo{\mu \ty . \T}{\pp}{n}{m}{f} =
   \oxtrantwo{\T}{\pp}{n}{m}{f \cup \{\ty \mapsto n\}}
\end{array}$
}\\[1mm]
\centerline{
\small
$\begin{array}{lll}
\oxtrantwo{\textstyle\probpprocoutty{i \in I}{\pp_j}{j \in J_i}{\probi{(i,j)}}{\ell_j}{\Basic_j}{\T_{(i,j)}}}{\pp}{n}{m}{f}\\[1mm]
\qquad = \textstyle\bigcup_{i \in I} \ \prismcommand{\pp\conc\tau}{(S_{\pp} = n)}{\probn{1}: (S'_\pp = \sume{i})} 
    \ \cup \ \textstyle\bigcup_{i \in I}
    \oxtrantwo{\selsummand{i}}{\pp}{\sume{i}}{m}{f} \\[2mm]
\hspace{3cm}\text{where} \quad |I| \geq 2\quad 
\selsummand{i} = \pprocoutty{\pp_j}{j \in
    J_i}{\probi{(i,j)}}{\ell_j}{\Basic_j}{\T_{(i,j)}} \quad   
    \sume{i} = n + 1 + \textstyle \sum_{i' < i} \StateS{\selsummand{i'}} \\[1mm]
\oxtrantwo{\pprocoutty{\pp_j}{j \in J}{\probi{j}}{\ell_j}{\Basic_j}{\T_{j}}}{\pp}{n}{m}{f}
\qquad (|J| \geq 2)\\[2mm]
\qquad = \prismcommand{\pp\conc p}{(S_{\pp} = n)}{\textstyle\sum_{j \in J}(\probn{p_{j}}: (S'_\pp = \come{j}))}
     \ \cup \ \textstyle\bigcup_{j \in J}
     \oxtrantwo{\comsel{j}}{\pp}{\come{j}}{m}{f} \\[1mm]
\hspace{3cm}\text{where} \quad
     \comsel{j} = \pprocoutty{\pp_j}{}{\probn{1}}{\ell_j}{\Basic_j}{\T_{j}} \quad
     \come{j} = n + 1 + \textstyle \sum_{j' <
     j} \StateS{\comsel{j'}} \\[1mm]
\oxtrantwo{\pprocoutty{\q}{}{\probn{1}}{\ell}{\Basic}{\T'}}{\pp}{n}{m}{f}\\%
\qquad = \prismcommand{\pp \conc \q \conc \ell}{(S_\pp = n)}{
       \probn{1} : \begin{cases}
         S'_{\pp} = m & \T' = \tend \\
         S'_{\pp} = f(\ty) & \T' = \ty \\
         S'_{\pp} = n + 1 & \text{otherwise}
       \end{cases}
     }
     \cup \ \oxtrantwo{\T'}{\pp}{n + 1}{m}{f} \\[1mm]
     \oxtrantwo{\pprocinnew{\pp_i}{i\in I}{\ell_i}{\Basic_i}{\T_i}}{\pp}{n}{m}{f}\\%
\qquad     = \textstyle \bigcup_{i \in I}
     \prismcommand{\ppi{i} :: \pp :: \ell_i}{(S_\pp = n)}{\probn{1} :
       \begin{cases}
         S'_\pp = m & \T_i = \tend \\
         S'_\pp = f(\ty) & \T_i = \ty \\
         S'_\pp = \nexts{i} & \text{otherwise}
       \end{cases}
     } 
     \cup \textstyle \bigcup_{i \in I} \oxtrantwo{\T_i}{\pp}{\nexts{i}}{m}{f} \\[2mm]
\hspace{3cm}\text{where} \quad          \nexts{i} = n + 1 + \textstyle \sum_{i' < i} \StateS{\T_{i'}}
\end{array}
$}
\vspace{-2mm}
\caption{Inner Translation}
\label{fig:innertranslation}
\end{figure}
We refer to the two kinds of internal transitions produced by
the encoding as follows: a \emph{summand step} is a transition labelled
$\pp\conc\tau$, introduced for
a proper sum, moving $S_\pp$ from the entry state of the sum to the entry
state of one of its summands $\selsummand{i}$ with probability $\probn{1}$;
a \emph{commitment step} is a transition labelled $\pp\conc p$, introduced
for a proper selection, moving
$S_\pp$ from the entry state of the selection to the entry state of one of
its committed selections $\comsel{j}$ with probability $\probi{j}$.
Summand steps resolve the nondeterministic choice
(rule \rulename{ct-nd}), commitment
steps resolve the probabilistic choice (rule \rulename{ct-prob}), and
synchronisations perform the communication (rule \rulename{ct-$\tau$}).
Commitment steps are the only commands with more than one branch, and at most one is enabled in
any state per module, as each is guarded by a distinct entry state and never synchronises.

The inner encoding has slightly more transitions
than needed since we treat each $\prismreduce{\pp\conc\q\conc\labelsell}{}$
transition as a communication, implicitly
assuming both $\pp$ and $\q$ participate. However, in some cases, only one of
$\pp$ or $\q$ contains this action, and \rulename{m-move} permits such
transitions to occur without synchronisation.
Thus, we introduce a \emph{closure module} that blocks these spurious
transitions.

\begin{definition}
    \label{def:closure}
    Let $\actions{\Delta}{\pp}$ be the set of actions in $\trantwo{\pp :
      \Delta(\pp)}$ where $\actions{\Delta}{\pp} = \varnothing$ if $\pp
    \notin \dom{\Delta}$. Define $\actions{}{\Delta} =
    \bigcup_{\pp \in \dom{\Delta}} \actions{\Delta}{\pp}$.
    The \textbf{closure module} for $\Delta$ is given by\\[1mm]
\centerline{$    
        \closure{\Delta} =
        \{ \prismcommand{\pp \conc \q \conc \labelsell}{\false}{\probn{1}
        :
        ()} \SEP \pp\conc\q\conc\labelsell \in \actions{\Delta}{\pp} \oplus \actions{\Delta}{\q}\},
$}\\[1mm]        
    where $\oplus$ is the \emph{symmetric difference}, and the $()$ denotes a
    dummy update that will never be applied.
    The \textbf{final encoding} $\tranone{\cdot}$ of contexts
    into \prism{} models is defined by 
        $\tranone{\Delta} = \trantwo{\Delta} \cmp \closure{\Delta}$.
\end{definition}

\begin{example}\label{exam:state-mapping}
Consider the calculator $\pchan{c} : \T_{\pchan c}$ from
\Cref{exam:runningtype}, for which $\StateS{\T_{\pchan c}} = 6$.
\Cref{fig:state-mapping} shows its state encoding.
Each typing context reduction corresponds to exactly \emph{one} \prism{}
transition. The recursion contributes no state, so the branching
$\&\pchan{s}?\labels{incr}$ is translated at state $0$ with $f = \{\ty \mapsto
0\}$. Then, receiving $\labels{incr}$ from $\ps$ moves $S_{\pchan c}$ to state $1$,
which is the entry state of the nondeterministic sum. 
From there, a summand step picks
one of the two summands nondeterministically, moving either to state $2$, the
singleton selection $\oplus\pchan{u}!\labels{res}.\ty$, or to state $3$, the
entry state of the probabilistic selection.
From state $3$, a commitment step
moves to state $4$ with probability \probn{0.8} or to state $5$ with
probability \probn{0.2}, the entry states of the two singleton selections
$\oplus\pchan{u}!\labels{res}.\ty$ and $\oplus\pchan{s}!\labels{error}.\ty$.
Each singleton selection then synchronises with its recipient. 
As every continuation is the variable $\ty$,
these synchronisations loop back to state $f(\ty) = 0$. 
The end state $6$ is
never reached, since the calculator never terminates.

The remaining participants $\pu$ and $\ps$ are encoded in a similar manner, yielding modules over $S_\pu$ and $S_\ps$. The final encoding is then: $\tranone{\Delta} = \trantwo{\pchan{c} : \T_{\pchan{c}}} \cmp \trantwo{\pchan{u} : \T_{\pchan{u}}} \cmp \trantwo{\ps : \T_\ps} \cmp \closure{\Delta}$

The resulting modules synchronise on
the four actions $\pu\conc\ps\conc\labels{incr}$, $\ps\conc\pchan{c}\conc\labels{incr}$,
$\pchan{c}\conc\pu\conc\labels{res}$ and $\pchan{c}\conc\ps\conc\labels{error}$.
Every action appears in both the sender's and the receiver's module, so the closure module
is empty. Tracing part of an execution, we could have:\\[1mm]
\centerline{$\small
    \begin{array}{l}
        (\tranone{\Delta}, \{\scc \mapsto 0,\ldots\}) \predcont{\ps ::\pchan{c}::\labels{incr}}{\probn{1}} (\tranone{\Delta}, \{\scc \mapsto 1, \ldots\}) \predcont{\pchan{c}::\tau}{\probn{1}}(\tranone{\Delta}, \{\scc \mapsto 3, \ldots\})\predcont{\pchan{c}::p}{\probn{0.8}}
        (\tranone{\Delta}, \{\scc \mapsto 4,\ldots\})%
    \end{array}
$}
\end{example}

\begin{figure}[t]
  \centering
\small
\begin{tabular}{@{}l@{}l@{}}
\begin{minipage}{0.42\textwidth}
\[
\begin{array}{l}
\trantwo{\pchan{c}:\T_{\pchan{c}}}=\\
\;\;\;\left\{ \!\!\!
\begin{array}{l}
    \prismcommand{\ps::\pchan{c}::\labels{incr}}{(\scc = 0)}{\probn{1} : (\scc '= 1)}\\
    \prismcommand{\pchan{c}::\tau}{(\scc = 1)}{\probn{1} : (\scc '= 2)}, \\
    \prismcommand{\pchan{c}::\tau}{(\scc = 1)}{\probn{1} : (\scc '= 3)}, \\
    \prismcommand{\pchan{c}::\pu::\labels{res}}{(\scc = 2)}{\probn{1} : (\scc '= 0)}, \\
    {\begin{array}{rl}
      \prismcommand{\pchan{c}::p}{(\scc = 3)} & \probn{0.8} : (\scc '= 4)\\
                                            + & \probn{0.2} : (\scc '= 5),
      \end{array}}\\ 
      \prismcommand{\pchan{c}::\pu::\labels{res}}{(\scc = 4)}{\probn{1} : (\scc '= 0)}, \\
  \prismcommand{\pchan{c}::\ps::\labels{error}}{(\scc = 5)}{\probn{1} : (\scc' = 0)} \ \}
\end{array}\right.
\end{array}
\]
\end{minipage}
&
\begin{minipage}{0.58\textwidth}
    \begin{tikzpicture}[x=1.55cm,scale=0.8]
        \tikzset{>={Stealth[width=1.5mm,length=1.5mm]},
                 lbl/.style={font=\smaller},
                 tlbl/.style={font=\fontsize{6}{7}\selectfont}}

        \draw[dashed] (1.55,-0.85) rectangle (2.45,0.75);
        \node[tlbl, anchor=south] at (2.1,0.8) {$\oplus\pchan{u}!\labels{res}\langle\ints\rangle.\,\ty$};
        \draw[dashed] (2.8,-0.85) rectangle (5.25,0.75);
        \node[tlbl, align=left, inner sep=1pt, anchor=north west] at (4.4,-0.9)
          {$\oplus\{\,\probn{0.8}\!:\pchan{u}!\labels{res}\langle\ints\rangle.\,\ty,$\\
           $\phantom{\oplus\{\,}\probn{0.2}\!:\pchan{s}!\labels{error}\langle\ints\rangle.\,\ty\,\}$};

        \draw (0, 0) -- (6,0);
        \foreach \x in {0, 1, ..., 6} \draw (\x,0.1) -- (\x,-0.1);
        \foreach \x in {0, 1, ..., 6} \node at (\x, -0.32) {\smaller \x};

        \draw[->] (0,0) to[bend left=40] coordinate[pos=0.5] (t1) (1,0);
        \draw[->] (1,0) to[bend left=40] coordinate[pos=0.2] (t2) (2,0);
        \draw[->] (1,0) to[bend left=40] coordinate[pos=0.25] (t3) (3,0);
        \draw[->] (3,0) to[bend left=40] coordinate[pos=0.5] (t4) (4,0);
        \draw[->] (3,0) to[bend left=40] coordinate[pos=0.75] (t5) (5,0);

        \draw[dashed] (t1) -- (t1 |- 0,0.9) node[above, lbl] {$\ps\conc\pchan{c}\conc\labels{incr}$};
        \draw[dashed] (t2) -- (t2 |- 0,1.3) node[above, lbl] {$\pchan{c}\conc\tau$};
        \draw[dashed] (t3) -- (t2 |- 0,1.3);
        \draw[dashed] (t4) -- (t4 |- 0,0.9) node[above, lbl] {$\pchan{c}\conc p$, \probn{0.8}};
        \draw[dashed] (t5) -- (t5 |- 0,1.3) node[above, lbl] {$\pchan{c}\conc p$, \probn{0.2}};

        \draw[->] (2,0) to[bend left=40, looseness=1.3] coordinate[pos=0.4] (b1) (0,0);
        \draw[->] (4,0) to[bend left=40, looseness=1.1] coordinate[pos=0.37] (b2) (0,0);
        \draw[->] (5,0) to[bend left=40, looseness=1.1] coordinate[pos=0.2] (b3) (0,0);

        \draw[dashed] (b1) -- (b1 |- 0,-1.75) node[below, lbl] {$\pchan{c}\conc\pu\conc\labels{res}$};
        \draw[dashed] (b2) -- (b2 |- 0,-1.75) node[below, lbl] {$\pchan{c}\conc\pu\conc\labels{res}$};
        \draw[dashed] (b3) -- (b3 |- 0,-1.75) node[below, lbl] {$\pchan{c}\conc\ps\conc\labels{error}$};
    \end{tikzpicture}
\end{minipage}
\end{tabular}
\vspace{-2mm}
\caption{States in the encoding of $\pchan{c} : \T_{\pchan c}$
  (\Cref{exam:runningtype}). Points on the number line represent values of
  $S_{\pchan c}$, and arrows represent transitions, labelled with
  their action and probability. 
  Unlabelled transitions have $\prob=\probn{1}$.
}
    \label{fig:state-mapping}
\end{figure}

\subsection{Correctness}
\label{sec:correctness}
We now show that the encoding is \emph{sound} (\Cref{thm:soundness}): every typing context reduction corresponds to
a transition of the encoded model; and \emph{complete} (\Cref{thm:completeness}): every \prism{} transition corresponds
to a typing context reduction. Detailed definitions are given in \cref{app:translation}.

\subsubsection{Soundness}
For soundness, we must show that 
$\emptytran$ preserves behaviour under
recursion unfolding: $\trantwo{\pp : \mu \ty . \T}$ should be \emph{equivalent}
to $\trantwo{\pp : \T \subst{\mu \ty . \T}{\ty}}$. %
Though the two modules' internal states and commands may 
differ, they should exhibit the same observable transition behaviour. This
motivates use of \emph{bisimulation} %
on model-state pairs 
$(M_1, I_1) \sim (M_2, I_2)$
to define equivalence.

\begin{restatable}[Translation soundness]{theorem}{soundness}
  \label{thm:soundness}
  \proofreference{\cref{app:sec5proofs}}
  Let $\Delta, \Delta'$ be contexts. %
  \begin{enumerate}
    \item If $\Delta \predcontnd{\actionnd{\pp}} \Delta'$, then there
    exists a state $S$ such
      that $\tranone{\Delta} \prismreduce{\pp\conc\tau}{\probn{1}} (\tranone{\Delta}, S)$
      and $\tranone{\Delta'} \sim (\tranone{\Delta}, S)$.
    \item If $\Delta \predcont{\actionprob{\pp}}{\prob} \Delta'$, %
      then exists a state $S$ such that $\tranone{\Delta}
      \prismreduce{\pp\conc p}{\prob} (\tranone{\Delta}, S)$ and $\tranone{\Delta'} \sim
      (\tranone{\Delta}, S)$.
    \item If $\Delta \predcont{\actiontau{\pp}{\q}{\labelsell}}{\probn{1}} \Delta'$,
      then exists a state $S$ such that $\tranone{\Delta}
      \prismreduce{\pp\conc\q\conc\labelsell}{\probn{1}} (\tranone{\Delta}, S)$
      and $\tranone{\Delta'} \sim (\tranone{\Delta}, S)$.
  \end{enumerate}
\end{restatable}

\subsubsection{Completeness}
Completeness is the dual of soundness: every transition of the encoded
model corresponds directly to a reduction of the typing context.
\begin{restatable}[Translation completeness]{theorem}{completeness}
  \label{thm:completeness}
  \proofreference{\cref{app:sec5proofs}}
  For every context $\Delta$:
  \begin{enumerate}
    \item if $\dtran \prismreduce{\pp\conc\tau}{\probn{1}} (\dtran, S)$,
      then there exists $\Delta'$ such that $\Delta \predcontnd{\actionnd{\pp}}
      \Delta'$ and $(\dtran, S) \sim \dptran$;
    \item if $\dtran \prismreduce{\pp\conc p}{\prob} (\dtran, S)$,
      then there exists $\Delta'$ such that
      $\Delta \predcont{\actionprob{\pp}}{\prob} \Delta'$ and $(\dtran, S) \sim \dptran$;
    \item if $\dtran \prismreduce{\pp::\q::\labelsell}{\pone} (\dtran, S)$,
      then there exists $\Delta'$ such that
      $\Delta \predcont{\actiontau{\pp}{\q}{\labelsell}}{\pone} \Delta'$ and $(\dtran, S) \sim \dptran$.
  \end{enumerate}
\end{restatable}

 \section{Model-Checking Probabilistic Multiparty Session Types in \prism{}}
 \label{sec:modelchecking}
 With the encoding in place, we must now express the properties of
\Cref{sec:typingcontextproperties} as logical formulae to enable
model checking. 
We use Probabilistic LTL (PLTL), the linear-time fragment of PCTL*, to specify and model check properties.

\myparagraph{A brief introduction to PLTL}.
Due to space constraints, we opt for a brief presentation; a more principled
discussion can be found in \cite{model-checking-book,BdA95}.
The syntax of our fragment of PLTL is given by\\[1mm]
\centerline{
$\begin{array}{rll}
\Phi \DEF &\pmin{\varphi} \bnfsep \pmax{\varphi} & \text{\footnotesize (lower- and upper-bound probability constructs)} \\
\varphi, \psi \DEF & a \bnfsep
\neg \varphi \bnfsep \varphi \wedge \psi \bnfsep \varphi \vee \psi
& \text{\footnotesize (atomic proposition,
negation, conjunction, disjunction)}\\
\bnfsep & \always \varphi \bnfsep \eventually \varphi &
\text{\footnotesize (always, eventually)}
\end{array}
$}\\[1mm]
We write $\varphi \implies \psi$ as a shorthand for $\neg \varphi \vee \psi$. We omit the \emph{next} ($\mathsf{X}$) and \emph{until} ($\mathsf{U}$) operators of PLTL, as they are not needed to express our properties.

\emph{Path} $\pi$ of a \prism{} model $M$ is a (potentially infinite) sequence of states $S_1; S_2; \cdots$ such that $\forall i \in [1, |\pi|) : (M, S_i) \prismreduce{}{} (M, S_{i+1})$.
We call $\varphi$ a \emph{path formula}. Path formulae act on paths $\pi$ of a
model $M$, and we write $\pi \models \varphi$ if $\pi$ satisfies $\varphi$. The
semantics of path formulae are given in \Cref{def:properties:pctl*}.

Given a model $M$, $\pmin{\varphi}$ (resp.\ $\pmax{\varphi}$) tells us the
\textbf{minimum} (resp.\ \textbf{maximum}) probability of
obtaining a path $\pi$ of $M$ such that $\pi \models \varphi$, across all possible
schedulers. Since our \prism{} models are MDPs, they contain
both probabilistic choice (within the same action) and nondeterminism (over
which action to fire, when multiple are available).

\begin{definition}[Semantics of a path formula]
\label{def:properties:pctl*}
Let $\pi = S_1; S_2; \cdots$ be a path, and denote by $\pi[i..]$ the subpath $S_i; S_{i+1}; \cdots$.
The \emph{semantics of a path formula} $\varphi$ is given inductively:\\[1mm]
\centerline{
$\begin{array}{rlcrlcrl}
\pi \models a &\hspace*{-3mm}\iff S_1 \vdash a
& &                                                                             \pi \models \varphi \land \psi
        &\hspace*{-3mm}\iff \pi \models \varphi \land \pi \models \psi
& &   \pi \models \varphi \vee \psi
        &\hspace*{-3mm}\iff \pi \models \varphi \vee \pi \models \psi \\
     \pi \models \neg \varphi&\hspace*{-3mm}\iff  \pi \ \not\models \ \varphi
        & &
        \pi \models \always \varphi & \hspace*{-3mm}\iff \forall
        i \cdot \pi[i..] \models \varphi & &
      \pi \models \eventually \varphi &\hspace*{-3mm}\iff \exists i \cdot \pi[i..] \models \varphi
    \end{array}
$}
\end{definition}
The meaning of $S_1 \vdash a$ will be explained in \Cref{sec:spec:pctl}.
The \emph{always} and \emph{eventually} operators originate from \emph{linear
 temporal logic}; informally, $\always \varphi$ is true if $\varphi$ is always
true throughout the path, and $\eventually \varphi$ is true if $\varphi$
eventually becomes true. Other operators are standard.

\subsection{Defining Specifications}
\label{sec:spec:pctl}
We define several \prism{} \emph{labels} -- boolean expressions over state variables
-- to act as atomic propositions. For
conciseness, we only give informal meanings here, though it is straightforward
to express them as conjunctions or disjunctions of state variables.

\begin{definition}
\label{def:labels}
Given an encoding $\tranone{\Delta}$, we define the \emph{labels} (their names denoted with \textit{``quotes"}):
\begin{itemize}
        \item $S \vdash \labelend \iff \forall \pp \in \dom{\Delta} : S(S_\pp) = \StateS{\Delta(\pp)}$
        \item $\forall \pp\conc\q\conc\labelsell \in \actions{\Delta}{\pp} \cup \actions{\Delta}{\q} : \
        S \vdash \sendlabel{\pp}{\q}{\labelsell} \iff (\trantwo{\Delta(\pp)}, S) \prismreduce{\pp\conc\q\conc\labelsell}{}$
        \item $\forall \pp, \q \in \dom{\Delta} : S \vdash
          \receiveanylabel{\pp}{\q} \iff
        \exists \labelsell : (\trantwo{\Delta(\q)}, S) \prismreduce{\pp\conc\q\conc\labelsell}{}$
        \item $\forall \pp\conc\q\conc\labelsell \in \actions{\Delta}{\pp} \cup
          \actions{\Delta}{\q} : \
        S \vdash \receivelabel{\pp}{\q}{\labelsell} \iff (\trantwo{\Delta(\q)}, S) \prismreduce{\pp\conc\q\conc\labelsell}{}$
\end{itemize}
Additionally, \prism{} has a built-in label $\deadlocklabel$ such that $S
  \vdash \deadlocklabel \iff (\tranone{\Delta}, S) \not\rightarrow$.
\end{definition}
\noindent The label $\labelend$ is true iff all participants are in the end state.
$\sendlabel{\pp}{\q}{\labelsell}$ is true if $\pp$ is able to send $\labelsell$
to $\q$, i.e.\ $\pp$ is at the entry state of a committed selection and offers
the synchronisation $\pp\conc\q\conc\labelsell$; and
$\receivelabel{\pp}{\q}{\labelsell}$ is true if $\q$ is able to receive $\labelsell$ from
$\pp$. Finally, $\receiveanylabel{\pp}{\q}$ is true if $\q$ is able to receive
\emph{any} label from $\pp$.

\myparagraph{Fairness.}
The properties of \Cref{def:context-probprop} are quantified over \emph{fair}
schedulers, whereas $\textsf{Pmin}$ and $\textsf{Pmax}$ quantify over
\emph{all} schedulers of the model. This restriction cannot be imposed
directly, as \prism{} applies fairness to all nondeterministic choices
uniformly, including the internal nondeterminism $\Sigma_{i \in
  I}\T_i\predcontnd{}\T_k$, which must remain unconstrained. As we will soon
see, we instead express deadlock-freedom and liveness through the reachability of their
negations, for which the restriction to fair schedulers is unnecessary.

\myparagraph{Safety and deadlock-freedom.}
Using the labels, we write the following \emph{property
specifications}:\\[1mm]
\centerline{$
\begin{array}{l}
\begin{array}{rl}
    \safe{\Delta} \iff \textsf{Pmin}\big(\always\big(\bigwedge_{\pp\conc\q\conc\labelsell \in \actions{}{\Delta}} \big(\sendlabel{\pp}{\q}{\labelsell} \ \land \ \receiveanylabel{\pp}{\q}\big)& \\
     \implies \receivelabel{\pp}{\q}{\labelsell} &\big)\big) = \probn{1}\\[1mm]
\end{array}\\[1mm]
\begin{array}{rlcrl}
    \probdfmin{\Delta} = & \probn{1} - \pmax{\eventually (\deadlocklabel \wedge \neg\labelend)} &\qquad&
    \puredf{\Delta} \iff & \probdfmin{\Delta} = \probn{1}
\end{array}
\end{array}
$}\\[1mm]
In a safe context (\Cref{def:safety}), if $\pp$ has committed to sending
$\labelsell$ to $\q$ and $\q$ is ready to receive from $\pp$, then $\q$ must
accept $\labelsell$.
For deadlock-freedom, recall from \Cref{def:termdf} that a deadlocked context
is one that cannot reduce even though some participant has not ended. The path
formula $\eventually (\deadlocklabel \wedge \neg\labelend)$ holds exactly on
the paths that reach such a state, so the specification computes $\probn{1}$
minus the maximal probability of deadlock across all schedulers. The following
theorem shows that this maximum agrees with the maximum over fair schedulers
of \Cref{def:context-probprop}.

\begin{restatable}[]{theorem}{dffair}
\label{thm:dffair}
\proofreference{\cref{app:spec:pctl}}
$\max_{\phi \in \Phi_F^{\Delta}} \probpathsig{\Delta}{\eventually \deadlockk} = \max_{\phi \in \Phi^{\Delta}}  \probpathsig{\Delta}{\eventually \deadlockk}$
\end{restatable}

Intuitively, every deadlocking path is finite, and every finite complete path is fair.
Admitting unfair schedulers therefore does not change the maximal probability of
deadlock.

\myparagraph{Liveness} is trickier, as it is a property of infinite paths whose negation is not
immediately a reachability event.
We note that a context fails to be almost-surely live precisely when it can reach a region
of states in which some participant can be kept pending forever while
scheduling remains fair. We call this a \emph{weak almost-sure livelock}, 
weak in that an actual livelock may only surface when the nondeterminism is resolved adversarially.

To define the \emph{weak almost-sure livelock} region, we first define $\pend{\pp}{\Delta}$ to be the set of states reachable from $\Delta$ where $\pp$ is nonterminated but has no enabled communication:\\[1mm]
\centerline{
$\pend{\pp}{\Delta} = \{ \Delta' \in (\Delta \predcontstar{}{}) \ | \Delta'(\pp)\neq \tend,
\forall \q, \ell, \prob \st 
\Delta' \centernot{\predcontdouble{\actiontau{\pp}{\q}{\ell}}{\prob}} \wedge \ 
\Delta' \centernot{\predcontdouble{\actiontau{\q}{\pp}{\ell}}{\prob}} \}.
$}\\[1mm]
Writing $\pendd{\pp}$ for membership in $\pend{\pp}{\Delta}$, a fair path $\pi$ is non-live iff
$\pi \models \bigvee_{\pp \in \dom{\Delta}} \eventually\always\pendd{\pp}$.
Indeed, fairness resolves any remaining internal choices on such a pending suffix, leaving the participant permanently stuck.

Writing $\visited{\tpath} = \{\Delta_i \;|\; \exists i \leq |\tpath| : \tpath[i] = \Delta_i\}$ for the set of states visited along $\tpath$, we define:
\begin{definition}[Weak almost-sure livelock]
\label{def:wasll}
The \emph{weak almost-surely livelocked region} for
$\pp$:\\[1mm]
\centerline{$\wasllp{\pp}{\Delta} \;=\; \{\, \Delta' \in
\pend{\pp}{\Delta} \SEP
\exists \phi \in \Phi_F^{\Delta'}: \forall \tpath \in \Tpath_\phi(\Delta'): \visited{\tpath} \subseteq \pend{\pp}{\Delta}
\}$}\\[1mm]
and $\wasll{\Delta} = \bigcup_{\pp \in \dom{\Delta}}
\wasllp{\pp}{\Delta}$.
\end{definition}

For a \emph{limit property} $\varphi$, the standard end-component reduction~\cite[Theorem~10.122(b)]{model-checking-book} gives ${\small\pmin{\varphi} = \probn{1} - \pmax{\eventually(\bigcup_{i \in I} S_i)}}$, where $I$ indexes all end components $(S_i,A_i)$ whose recurrent behaviour violates $\varphi$. Applying this reduction with fairness incorporated yields:

\begin{restatable}{theorem}{wasllthm}
\label{thm:wasll}
\proofreference{\cref{app:spec:pctl}}
$\problivemin{\Delta} =
\min_{\phi \in \Phi_F^{\Delta}}
    \probpathsig{\Delta}{\livee} =
    \probn{1} - \max_{\phi \in \Phi^{\Delta}}
    \probpathsig{\Delta}{\eventually \waslll}$
\end{restatable}

Then, membership of the livelocked region becomes an atomic proposition. We define the label\\[1mm]
\centerline{
$S \vdash \livelocklabel \iff \exists \Delta' \in \wasll{\Delta} : (\tranone{\Delta}, S) \sim \tranone{\Delta'}$
}\\[1mm]
which, like the labels of \Cref{def:labels}, is a finite disjunction over
valuations of the state variables $S_\pp$. 
The implementation precomputes $\wasll{\Delta}$ by an end-component analysis of the context MDP (details in~\cref{app:wasll-alg}).
By \Cref{thm:wasll}, we then obtain the liveness specification:\\[1mm]
\centerline{
$\problivemin{\Delta} = \probn{1} - \pmax{\eventually \livelocklabel}$
}\\[1mm]
Since the right-hand side is a reachability query maximised over all
schedulers, no fairness constraint needs to be imposed on \prism{}, and
$\Delta$ is almost-surely live iff $\pmax{\eventually \livelocklabel} =
\probn{0}$.

\myparagraph{Complexity.}
The runtime cost of our verification approach is dominated by the weak 
almost-sure 
livelock precomputation as well as the PRISM model-checking.
Writing $|\Delta|$ for the total number of nodes in the syntax trees of 
the types in
$\Delta$, we have:

\begin{restatable}{theorem}{complexitythm}
\label{thm:complexity}
\proofreference{\cref{app:complexity}}
Let $N$ be the number of reachable states of $\tranone{\Delta}$ and $R =
|\ReachD{\Delta}|$. Then $\safe{\Delta}$, $\probdfmin{\Delta}$ and
$\problivemin{\Delta}$ can be computed in time polynomial in $N$, $R$ and
$|\Delta|$.
\end{restatable}

\begin{restatable}{corollary}{complexitycor}
\label{cor:complexity}
\proofreference{\cref{app:complexity}}
Let $k = |\dom{\Delta}|$. Then $\safe{\Delta}$, $\probdfmin{\Delta}$ and
$\problivemin{\Delta}$ can be computed in time $|\Delta|^{O(k)}$, i.e.\
exponential in the number of participants.
\end{restatable}

\noindent In practice, synchronous communication often tightly couples the
reachable set, as can be seen in our empirical evaluation. For instance, the
product of the state spaces of the ten participants of Multiparty Workers
exceeds a million states, yet only 1222 of them are reachable
(\Cref{tab:benchmarks}).

\begin{example}
\label{exam:endtoend}
We consider the calculator service in \Cref{exam:runningtype} for
the final time, with the encoding given in \Cref{fig:state-mapping}.
The label $\labelend$ requires all three state variables $S_\pu$, $S_\ps$ and $S_{\pchan c}$ to be at their end
states, and, for instance, $\sendlabel{\pchan{c}}{\ps}{\labels{error}}$ holds
exactly when $S_{\pchan c} = 5$ in \Cref{fig:state-mapping}, while
$\receiveanylabel{\pchan{c}}{\ps}$ holds when $S_\ps$ is at the entry state of
the branching of $\T_\ps$.
\prism{} reports that $\Delta$ is safe and that no reachable state satisfies
$\deadlocklabel$, since the context never stops, and so
$\probdfmin{\Delta} = \probn{1}$. The precomputed region $\wasll{\Delta}$ is
empty: $\pu$ may be left pending while $\ps$ and $\pchan{c}$ exchange
$\labels{error}$ and $\labels{incr}$ (\Cref{exam:runninglive}), but no fair
scheduler can keep all paths in this region, as every commitment step of
$\pchan{c}$ leaves it with probability $\probn{0.8}$. Hence
$\livelocklabel$ is the constant $\false$ and $\problivemin{\Delta} = \probn{1}$.
The calculator is thus almost-surely deadlock-free and live.

For a negative case, consider the extended context $\Delta_1$ of
\Cref{exam:mdp} (\Cref{fig:mdp}). Here \prism{} finds exactly one state
satisfying $\deadlocklabel \wedge \neg\labelend$, namely the encoding of
$\Delta_9$, in which $S_\ps$ and $S_{\pchan c}$ are at their end states while
$S_\pu$ is at the entry state of the branching $\&\pchan{c}?\labels{res}$. The
label $\livelocklabel$ is the disjunction of the valuations encoding $\Delta_9$,
where $\pu$ is pending forever because other participants have ended, and
$\Delta_{10}$ and $\Delta_{11}$, where $\pu$ is pending while $\ps$ and
$\pchan{c}$ exchange $\labels{retry}$. \prism{} returns $\probn{0.31}$ and
$\probn{0.474}$ for the two \textsf{Pmax} queries, giving $\probdfmin{\Delta_1} = \probn{0.69}$
and $\problivemin{\Delta_1} = \probn{0.526}$.
\end{example}

\begin{table}[]
    \centering
    \caption{Benchmark results for \prose{}.
      \textsf{WASL} is the precomputation of the weak almost-sure livelocked region (\Cref{def:wasll}). All benchmarked processes can be found in the \texttt{/examples/} folder of the artifact~\cite{prose}.}
    \vspace{-1mm}
    \scriptsize
    \begin{threeparttable}
    \setlength{\tabcolsep}{1.5pt}
    \begin{tabular}{@{}l|c|c|c|c|c|c|c||r|r@{}}
    & \multicolumn{7}{c||}{
        \textsf{\textbf{Mean runtime (in ms) of 30 runs after 1 warmup, $\pm$ relative standard error (\%)}}}
    & \multicolumn{2}{c}{
        \textsf{\textbf{Model Size}}} \\
    \cline{2-10}
    \textsf{\textbf{Typing context}} &
    \textsf{\textbf{Inference}} &
    \textsf{\textbf{Translation}} &
    \textsf{\textbf{WASL}} &
    \textsf{\textbf{Safety}} &
    \textsf{\textbf{DF}} &
    \textsf{\textbf{Liveness}} &
    \textsf{\textbf{End-to-end}} &
    \textsf{\textbf{States}} &
    \textsf{\textbf{Trans.}}\\
    \hline
    OAuth Protocol\tnote{1}
      & $21.64{\footnotesize\pm 1.5\%}$ & $0.42{\footnotesize\pm 4.8\%}$ & $0.03{\footnotesize\pm 3.9\%}$ & $373.16{\footnotesize\pm 2.7\%}$ & $402.53{\footnotesize\pm 2.8\%}$ & $409.50{\footnotesize\pm 3.0\%}$ & $606.70{\footnotesize\pm 2.3\%}$
      & 19 & 23 \\
    Knuth-Yao Dice\tnote{2}
      & $23.16{\footnotesize\pm 0.6\%}$ & $0.60{\footnotesize\pm 8.3\%}$ & $0.39{\footnotesize\pm 2.6\%}$ & $421.52{\footnotesize\pm 3.9\%}$ & $443.06{\footnotesize\pm 2.4\%}$ & $436.38{\footnotesize\pm 2.2\%}$ & $679.95{\footnotesize\pm 0.7\%}$
      & 74 & 83 \\
      Dining Phils.~\cref{exam:diningphilo}\tnote{2}
      & $22.14{\footnotesize\pm 0.3\%}$ & $0.83{\footnotesize\pm 4.8\%}$ & $107.95{\footnotesize\pm 0.7\%}$ & $551.27{\footnotesize\pm 3.2\%}$ & $542.88{\footnotesize\pm 2.9\%}$ & $931.20{\footnotesize\pm 1.9\%}$ & $1262.56{\footnotesize\pm 0.6\%}$
      & 19579 & 58308 \\
    Synch. Leader Election\tnote{2}
      & $22.58{\footnotesize\pm 0.5\%}$ & $0.41{\footnotesize\pm 4.9\%}$ & $0.02{\footnotesize\pm 4.3\%}$ & $374.92{\footnotesize\pm 2.5\%}$ & $404.71{\footnotesize\pm 2.7\%}$ & $407.08{\footnotesize\pm 2.8\%}$ & $619.37{\footnotesize\pm 0.7\%}$
      & 14 & 17 \\
    Monty Hall, Change
      & $23.06{\footnotesize\pm 0.3\%}$ & $0.43{\footnotesize\pm 4.7\%}$ & $0.03{\footnotesize\pm 2.4\%}$ & $380.95{\footnotesize\pm 2.7\%}$ & $413.72{\footnotesize\pm 2.5\%}$ & $418.95{\footnotesize\pm 2.1\%}$ & $623.16{\footnotesize\pm 0.7\%}$
      & 20 & 23 \\
    Monty Hall, Stay
      & $23.31{\footnotesize\pm 1.1\%}$ & $0.44{\footnotesize\pm 4.5\%}$ & $0.03{\footnotesize\pm 1.1\%}$ & $380.74{\footnotesize\pm 2.4\%}$ & $411.81{\footnotesize\pm 2.7\%}$ & $412.80{\footnotesize\pm 2.7\%}$ & $627.97{\footnotesize\pm 1.1\%}$
      & 20 & 23 \\
    Multiparty Workers\tnote{1}
      & $20.63{\footnotesize\pm 1.6\%}$ & $0.53{\footnotesize\pm 5.7\%}$ & $3.26{\footnotesize\pm 0.3\%}$ & $408.86{\footnotesize\pm 3.9\%}$ & $415.49{\footnotesize\pm 3.1\%}$ & $424.30{\footnotesize\pm 3.5\%}$ & $638.86{\footnotesize\pm 0.9\%}$
      & 1222 & 3643 \\
    Gambler's Ruin
      & $23.30{\footnotesize\pm 0.3\%}$ & $0.42{\footnotesize\pm 4.8\%}$ & $0.02{\footnotesize\pm 3.3\%}$ & $376.44{\footnotesize\pm 2.0\%}$ & $407.10{\footnotesize\pm 2.2\%}$ & $418.77{\footnotesize\pm 2.5\%}$ & $612.53{\footnotesize\pm 1.7\%}$
      & 16 & 19 \\
    Recursive Map-Reduce\tnote{1}
      & $23.40{\footnotesize\pm 0.9\%}$ & $0.46{\footnotesize\pm 6.5\%}$ & $0.08{\footnotesize\pm 1.3\%}$ & $379.64{\footnotesize\pm 2.9\%}$ & $401.59{\footnotesize\pm 2.8\%}$ & $400.31{\footnotesize\pm 2.5\%}$ & $625.37{\footnotesize\pm 0.8\%}$
      & 68 & 119 \\
    Recursive Two Buyers\tnote{1}
      & $24.82{\footnotesize\pm 4.6\%}$ & $0.42{\footnotesize\pm 2.4\%}$ & $0.02{\footnotesize\pm 3.0\%}$ & $375.84{\footnotesize\pm 2.1\%}$ & $406.04{\footnotesize\pm 2.4\%}$ & $414.30{\footnotesize\pm 2.2\%}$ & $607.63{\footnotesize\pm 1.1\%}$
      & 15 & 17 \\
    Example \ref{exam:mdp}
      & $24.51{\footnotesize\pm 6.1\%}$ & $0.57{\footnotesize\pm 7.0\%}$ & $0.05{\footnotesize\pm 1.7\%}$ & $409.89{\footnotesize\pm 4.0\%}$ & $414.10{\footnotesize\pm 1.8\%}$ & $421.18{\footnotesize\pm 2.1\%}$ & $625.69{\footnotesize\pm 0.8\%}$
      & 43 & 51 \\
    Von-Neumann Coin\tnote{2}
      & $22.57{\footnotesize\pm 0.4\%}$ & $0.39{\footnotesize\pm 2.6\%}$ & $0.02{\footnotesize\pm 3.8\%}$ & $371.26{\footnotesize\pm 2.0\%}$ & $414.59{\footnotesize\pm 1.8\%}$ & $412.39{\footnotesize\pm 2.2\%}$ & $627.61{\footnotesize\pm 0.3\%}$
      & 16 & 19 \\
    \end{tabular}
    \begin{tablenotes}
    \item[1] Probabilistic variant of an example from~\cite{Scalas2019}.
    \item[2] Adapted from the \prism{} case studies~\cite{prismcase}.
    \end{tablenotes}
    \end{threeparttable}
    \label{tab:benchmarks}
\end{table}

\subsection{Implementation and Performance Evaluation}

\label{sec:implementation}\label{sec:performance}
We implement our encoding and property specifications in \prose{}, a tool for
the verification of \textbf{pro}babilistic multiparty \textbf{se}ssion
types.
\prose{} has been implemented in roughly 2800 lines of OCaml. It accepts a
typing context as its input, and outputs the probabilities of properties
in \Cref{sec:typingcontextproperties}.
\prose{} is designed to be a self-contained, standalone
toolkit for verifying types as abstractions of programs.
\prose{} aims to be used in conjunction
with real programs
(such as Go or Scala programs) following the line of
\cite{DBLP:conf/icse/LangeNTY18,PLDI19Effpi,BSYZ2022}
which have applied model checkers for types
to real programs.

We evaluate the efficiency of \prose{} through performance benchmarks. The 
experiments were run on a laptop with an Apple M4 processor and 16GB of RAM, 
with OCaml 5.2.1 and \prism{} v4.10.1. 
We discard one warmup run, then repeat each measurement 30 times and report its mean and relative standard error.
For each benchmark process, we record the type-inference time and use the inferred typing context for subsequent measurements: translation into a \prism{} model, precomputation of the weak almost-sure livelocked region (\Cref{def:wasll}), and separate \prism{} invocations for the three properties of \Cref{sec:spec:pctl} (safety, deadlock-freedom, and liveness).
We also measure the end-to-end runtime, including inference and translation, with a single \prism{} invocation checking all three properties. As the translation function executes very quickly, each of its 30 data points is the mean of 100 consecutive calls (totalling 3,000 function calls), to reduce the effect of measurement overhead.

\Cref{tab:benchmarks} presents the results, along with the size (number of states and transitions) of each context MDP. The benchmark processes and their type specifications are provided in the 
\texttt{/examples/} folder of the accompanying artifact~\cite{prose}; measurements use freshly inferred contexts. 
The results of the verification process can be found in~\cref{app:proseresults}.
Most contexts are verified end-to-end in around $0.6s$, and Dining Phils.\ (DP) in around $1.3s$,
demonstrating that \prose{} is suitable for practical
use. Translation cost is negligible (<$1$ms), and the livelock-region precomputation is
cheap (<$6$ms) on all contexts but DP, as the algorithm enumerates its large context state space. This makes \prism{} model-checking the dominating cost.
Model checking the same \prism{} encoding with Storm~\cite{storm} yielded
similar results.
End-to-end runtimes are close to those for individual
properties, as \prism{}'s internal model construction, performed once per invocation, accounts for most of each runtime.

\begin{figure}
\centering
\begin{subfigure}{0.49\textwidth}
  \centering
  \includegraphics[width=\linewidth]{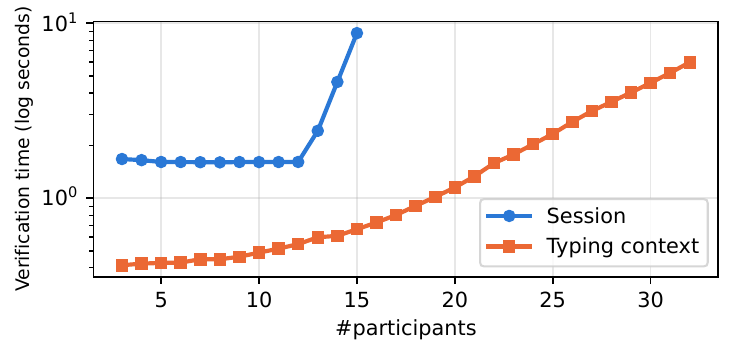}
\vspace*{-6mm}
\caption{Verification times}
\label{fig:factorial:a}
\vspace{-2mm}  
\end{subfigure}
\hfill
\begin{subfigure}{0.49\textwidth}
  \centering
  \includegraphics[width=\linewidth]{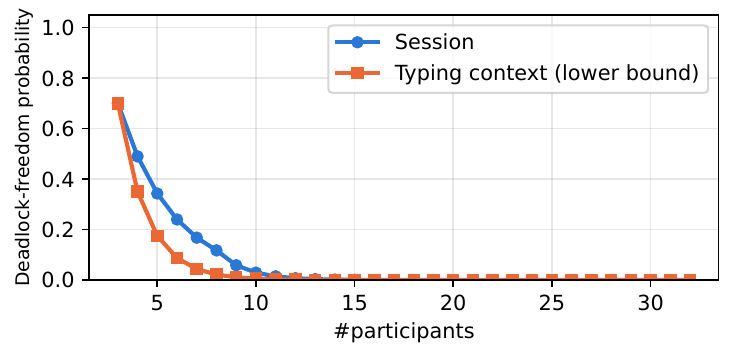}
\vspace*{-6mm}  
  \caption{Deadlock-freedom probabilities}
  \label{fig:factorial:b}
\vspace{-2mm}  
\end{subfigure}
\caption{Verification times and deadlock-freedom probabilities of the
  factorial session $\Multi_n$ which probabilistically produces an
  error; and its typing context $\Delta_n$,
  with varying values of $n$. The verification times
are a mean of 10 consecutive runs. 
For $n \geq 14$, the \prism{} invocation for session $\Multi_n$ 
ran out of memory. }
\label{fig:factorial}
\end{figure}

\myparagraph{Comparison with model checking processes.}
To compare scalability of verifying session types and 
processes, we have measured verification times and deadlock-freedom probabilities
of a simple factorial multiparty session $\M_i$ and its typing context $\Delta_i$ (for each $1 \leq i \leq 30$).  
While verifying sessions is undecidable in general (\Cref{prop:procsafeundec}),
this particular session can be translated into a finite-state PRISM module with
one variable keeping track of its current state and one
integer variable per received variable and per sent payload.
\Cref{fig:factorial:a} shows for $n \geq 14$, the
\prism{} invocations for session verification exhausted the allocated Java heap 
space (the default allocation is 1 GB), and
consequently did not finish
execution.

Moreover, \Cref{fig:factorial:b} shows that the deadlock-freedom probability obtained from $\Delta_n$
is a lower bound of the precise deadlock-freedom probability of $\M_n$, with both tending to $\probn{0}$ as $n$ grows.
As we describe in \Cref{exam:bounds}, the loss of precision comes from the conditional expressions 
in our calculus, which are typed as nondeterministic choice. 
See \cref{app:modelcheckingprocesses} for a more detailed explanation, as well as the factorial processes and types.

\subsection{Prototype implementation of Type Inference \& Checking}
\label{sec:typecheckimplementation}
We implement a prototype of the type system, with roughly 1,500 lines
of Haskell (GHC~9.6.7) (\emph{type inference} and \emph{subtype check}
in Fig.~\ref{fig:workflow}). The tool takes as input a program: a list
of named $\pmultipi$ processes in a plain-text surface syntax
mirroring~\cref{sec:processes}, each optionally paired with a
specified local type $\T_{\mathsf{spec}}$.  The main challenge to
inference is recursion: \rulename{t-rec} assumes the very type being
inferred, while deriving a type for \emph{conditionals} and \emph{flips} requires knowledge of the type constructor of their branch types $\T_1$ and $\T_2$ (\emph{conditionals} must compute the least upper bound of $\T_1$ and $\T_2$ with respect to $\subt$, and \emph{flips} the merge $\T_1 \probmerge{\prob} \T_2$).

Consider $\PP=\mu X.\procoutnl{\q}{m}{F}$ where   $F=\flip{\prob}{X}{\procoutnl{\pr}{m}{\inact}}$. To type $F$ requires the type of $X$, but finding that type requires typing $F$. We handle this circularity by recording type constraints and solving them on a finite graph. For example let $\T_X^?$ denote the \textit{unknown} type of $\PP$ (and $X$), and $\T_F^?$ the \textit{unknown} type of $F$. Using the same type unknown $\T_X^?$ for the recursive process and its bound occurrences of $X$, the prototype seeks a solution of the equations: $\T_X^?=\oplus\{1:\procoutnl{\q}{m}{\T_F^?}\}$ and $\T_F^?=\T_X^?\probmerge{\prob}\oplus\{1:\procoutnl{\pr}{m}{\tend}\}$.

The first equation determines an output node for $\T_X^?$ with an edge to the node reserved for $\T_F^?$, whose outgoing edges are still unknown. To infer them, the solver follows $X$ to its binder and exposes the leading $\q!m$, retaining the reference to $\T_F^?$ beneath that send, yielding: $\T_F^?=\oplus\{1:\procoutnl{\q}{m}{\T_F^?}\}\probmerge{\prob} \oplus\{1:\procoutnl{\pr}{m}{\tend}\}$. As both constructors under $\probmerge{\prob}$ are revealed, we can merge to get:
$\T_F^?=\oplus\{
   \prob:\procoutnl{\q}{m}{\T_F^?},\;
   (1-\prob):\procoutnl{\pr}{m}{\tend}\}$.
The $\q!m$ continuation reuses the recorded node for $F$ instead of
unfolding the definition again, closing the cycle.
Reading this graph as a recursive type yields
$\T_{\mathsf{inf}}=
 \oplus\{1:\procoutnl{\q}{m}{
   \mu\ty.\oplus\{
     \prob:\procoutnl{\q}{m}{\ty},\;
     (1-\prob):\procoutnl{\pr}{m}{\tend}\}}\}$.

The tool outputs, per process, the inferred local type $\T_{\mathsf{inf}}$, and optionally for each type specification a verdict of the subtype check $\T_{\mathsf{inf}}\subt\T_{\mathsf{spec}}$, displaying both types on failure; \cref{fig:workflow} highlights the two workflows. Inferred types can also be exported to the input format of \prose{}, our types to model checker translation~\cref{sec:translation}, connecting inference directly to the verification pipeline.

The graph construction and the Cartesian products in \rulename{m-$\Sigma$} can both incur exponential growth, in the number of graph nodes and output summands respectively. This is consistent with the $\mathcal{O}(n2^n)$ worst-case behaviour reported for bottom-up type inference in~\cite[Thm.~5.26]{UdomsrirungruangY25}, where $n=|\PP|$; the left-most column of~\cref{tab:benchmarks} reports measured inference times for our benchmark programs.

 \section{Related Work and Conclusion}
 \label{sec:related}
 
\myparagraph{Probabilistic linear and binary session types. }
While many probabilistic concurrent models have been studied in
process algebra (e.g., CCS~\cite{DBLP:conf/ecrts/Hansson92},
CSP~\cite{DBLP:journals/tcs/Lowe95,DBLP:conf/mmb/GeorgievskaA12}
and mobile process calculi~\cite{Herescu00FOSSACS,Norman07QEST}),
few works use \emph{behavioural types} to control and analyse
probabilistic communications.
The work in \cite{varacca2007probabilistic} 
applies linear types developed in \cite{DBLP:journals/iandc/YoshidaBH04,BHY}
to give a sound and complete correspondence between 
probabilistic processes and 
confusion-free event structures. 
Dal Lago and Giusti \cite{LagoG22} develop
a probabilistic binary
(2-party)
session calculus based on
bounded linear logic, where typed processes execute
within probabilistic polynomial time, enabling reasoning
about cryptographic protocols.

Inverso et al.~\cite{Inverso2020}
propose a probabilistic type system
which statically estimates an upper bound
on the probability of typed processes reaching successful states,
using the top-level
probabilistic choice $P \boxplus_{\prob} Q$
for processes and types.
Dal Lago and Padovani \cite{LagoP24} study almost-sure termination of
a binary session calculus with the same process-level probabilistic
choice; their types carry no probability annotations, but real-valued
\emph{heights} bounding the expected number of reduction steps.
This property is qualitative, whereas we
derive quantitative bounds.
Both calculi~\cite{Inverso2020,LagoP24} are built
on the linear logic-based binary session type
systems~\cite{CairesPfenningToninho16},
expressing only tree-like network topologies and enforcing \emph{total} (100\%)
deadlock-freedom, and every choice is probabilistic: there is
no nondeterministic choice 
corresponding to our 
sums.
Neither work provides an implementation.

Das et al.~\cite{Das2023} propose probabilistic resource-aware
session types extended from intuitionistic linear logic-based
deadlock-free session types \cite{CairesPfenningToninho16}, implemented as
the programming language NomosPro, which verifies representative
probabilistic distributed algorithms with \emph{exact probability}
and \emph{resource bounds}.
Like our work, they provide both nondeterministic
and probabilistic choice, and our \sff{flip} construct and merging
operator for types are inspired by NomosPro.

Our merge $\T_1 \probmerge{\prob} \T_2$ (\Cref{def:mergerel}) takes the convex combination of the outermost probabilities and takes the least upper bounds under $\subt$ for continuations that have a common prefix. Where the weighted sums of \cite{Das2023,FuDG25} require these continuations to coincide, our merge accepts differing continuations whenever their join exists. E.g., consider a server that is equally likely to be healthy or degraded, acknowledging a request with probability $\probn{0.9}$ or $\probn{0.5}$, and then signalling $\labels{hi}$ or $\labels{lo}$:\\[1mm]
\centerline{\small
$\T_{\mathsf{H}} \ = \ 
\oplus\{\,
  \probn{0.9}\colon \pchan{q}!\labels{ok}.\,
    \oplus\,\pchan{q}!\labels{hi}.\,\tend,\;\;
  \probn{0.1}\colon \pchan{q}!\labels{err}.\,\tend
\,\},\ 
\T_{\mathsf{T}} \, = \,
\oplus\{\,
  \probn{0.5}\colon \pchan{q}!\labels{ok}.\,
    \oplus\,\pchan{q}!\labels{lo}.\,\tend,\;\;
  \probn{0.5}\colon \pchan{q}!\labels{err}.\,\tend
\,\}.$}\\[1mm]
Then we have:\\[1mm]
\centerline{
$\T_{\mathsf{H}} \probmerge{\probn{0.5}} \T_{\mathsf{T}}
=
\oplus\left\{
\probn{0.7}\colon \pchan{q}!\labels{ok}.\,
  \Sigma\{\,
    \oplus\,\pchan{q}!\labels{hi}.\,\tend,\;
    \oplus\,\pchan{q}!\labels{lo}.\,\tend
  \},\;
\probn{0.3}\colon \pchan{q}!\labels{err}.\,\tend
\right\}$,}\\[1mm]
while in \cite{Das2023,FuDG25}, this is undefined.  
The server remains typable despite its differing continuations: the probability of $\labels{ok}$ is retained exactly, while the subsequent choice between $\labels{hi}$ and $\labels{lo}$ is abstracted by nondeterminism. 
The work in \cite{Das2023} is extended to probabilistic refinement
session types \cite{FuDG25}, augmented by symbolic probabilities
indexed by refinement variables, enabling parameterised probabilistic
reasoning and analyses of expected
resource costs. Their implementation reconstructs the intermediate types of channels
with an SMT solver, verifying \emph{declared} types,
whereas type-checking in our system is decidable
(\Cref{thm:decidability}).
Their implementation of Lehmann and Rabin's randomised dining philosophers~\cite{Lehmann1981} uses a coordinating process and an indexed ring, with refinement predicates tracking fork availability and ownership.
Our multiparty formulation (\cref{exam:diningphilo}) expresses contention directly through separated external choice:
each fork accepts requests from either neighbour, then awaits its holder's release while rejecting competing requests.
Ownership is captured by the session-type continuations, without refinement indices or a central process mediating fork acquisition. 

None of the above calculi study probabilistic deadlock-freedom or liveness; processes with only partial deadlock-freedom are untypable.

\myparagraph{Probabilistic multiparty session types (MPST) and
choreographies.}
Aman and Ciobanu~\cite{Aman2019,AmanC22} extend
MPST~\cite{Honda2008,HYC2016} with probability intervals
on selections in global and local types.
Their top-down approach checks conformance to
interval-annotated global protocols.
Our bottom-up approach~\cite{UdomsrirungruangY25} instead analyses locally typed sessions and establishes lower bounds on their probabilities of deadlock-freedom and liveness, including sessions for which these properties do not hold on every fair execution.
Blechschmidt et al.~\cite{Blechschmidt2025} propose
a probabilistic mixed choice multiparty session $\pi$-calculus.
Their refinement subtyping permits
replacement of a channel by several channels implementing the same
behaviour, differing from the standard subtyping
relation defined between two local types.
The works in \cite{Aman2019,AmanC22,Blechschmidt2025} provide no implementation.  
Carbone and Veschetti~\cite{Carbone2024} propose a
probabilistic deadlock-free choreographic calculus with
choices explicitly annotated by probabilities (akin to
global types in \cite{Aman2019,AmanC22}), prove the correctness of
its translation into \prism{}, and compare
outcomes of probabilistic measurements
against the original \prism{} case studies.
Their language more closely resembles the \prism{} language than
our local types: a choreographic sending-receiving construct
describes both endpoints of a communication at once, mapping
directly to one action in \prism{}, whereas we translate each
participant of a typing context into its own \prism{} module
(\Cref{sec:typeencoding}), with a communication arising as a
synchronisation between the sender's and receiver's modules. Proving
that each reduction of the typing context corresponds to exactly one
transition of the composed model (\cref{thm:soundness,thm:completeness})
requires a closure module blocking spurious unsynchronised transitions
and a bisimulation relating models under recursion unfolding,
making our correctness proofs much more involved. No works above
analyse probabilistic bounds for properties such as deadlock-freedom and liveness.

\myparagraph{Model checking behavioural types and programs.}
We translate \emph{types} rather than processes to the \prism{}
language because:
(1) verification of session types is guaranteed to terminate, whereas
even safety of processes is undecidable (\Cref{prop:procsafeundec});
and (2) translating processes results in a far larger state space,
as exact values of every variable must be tracked, making verification of
types more efficient and scalable, as shown
in \Cref{fig:factorial} and \cref{app:modelcheckingprocesses}.

Model checking real-world programs has a long history~\cite{Visser03}.   
Chaki et al.~\cite{POPL02ModelChecking} are the first to propose
model checking behavioural types (an abstraction of $\pi$-processes)
developed in \cite{igarashi.kobayashi04:gentypPi}. 
More recently, \cite{Scalas2019} applied mCRL2 \cite{mcrl2book}
to verify multiparty session types, an approach subsequently extended 
to Go \cite{DBLP:conf/icse/LangeNTY18}
and Scala \cite{PLDI19Effpi} programs, 
and to distributed failures \cite{BSYZ2022}.
None of the above are probabilistic.

Beyond behavioural types, Enea et al.~\cite{Enea26} verify
\emph{fair almost-sure termination} of randomized distributed
algorithms such as Ben-Or's consensus, using a proof rule
combining martingale-based termination arguments with weak fairness,
discharged by an SMT solver. Their deductive approach works directly
on processes whose state spaces may grow unboundedly,
beyond the reach of finite-state model checking; conversely, they
establish termination with probability one, whereas we derive
quantitative lower bounds on deadlock-freedom and liveness.

\myparagraph{Conclusion. }
We have presented the first probabilistic
multiparty session type theory for deriving probabilistic properties 
(probabilistic deadlock-freedom and liveness)   
of typed multiparty session processes. 
Our nondeterministic sums abstract data-dependent control,
inducing lower bounds on probabilistic properties
that hold under every fair adversarial scheduler.
We have implemented 
the tool-chain \prose{} based on
the precise translation from session types to \prism{}, and evaluated it on several case studies.
Our future work includes the extension 
of the calculus with 
session delegations \cite{Scalas2019,HYC2016}, dynamically joining 
participants \cite{DBLP:conf/popl/DenielouY11,CY2023} and
asynchronous sessions \cite{PPGSY2026}.
Our subtyping system is \emph{not} precise in the sense of
\cite{Ghilezan2019,MY2027,GPPSY2023}. For example, $
    \oplus\{\probn{0.5}:\pp!\labels{m},
             \probn{0.5}:\q!\labels{m}\}\not\subt
    \Sigma\{\oplus\{\pp!\labels{m}\},
             \oplus\{\q!\labels{m}\}\},
$ although the distribution on the left is a convex combination
of the alternatives on the right. Investigating probabilistic
bisimulations~\cite{SegalaL95} to refine subtyping and minimise
session types is another direction for future work.
We also plan to 
analyse more advanced properties such as real-time session type
specifications 
\cite{hou_et_al:LIPIcs.ECOOP.2024.19,DBLP:journals/pacmpl/YaoICBZ25},
taking advantage of PRISM's support for probabilistic timed automata.

\subsubsection*{Acknowledgements.}
We thank the POPL reviewers for their careful comments,  
encouragement and detailed suggestions. We thank Jake Masters for discussions
and Kai Pischke for testing our artifact.
We thank Iona Kuhn, Wei Qing Neo, Kashishu Raimalani and Litu Zou
(in alphabetical order), who worked on probablisitic session types for
their MSc projects in the Department of Computer Science at the
University of Oxford. 
This work is supported by
ARIA MSAI-PR01-P020 and MSAI-PR01-P029; and
EPSRC
EP/T006544/2,
EP/T014709/2,
EP/Z533749/1, 
EP/Y005244/1,
EP/V000462/1 and
EP/X015955/1; 
Horizon EU TaRDIS 101093006 (UKRI number 10066667);
a grant from Trinity College Oxford;
and a grant from the Simons Foundation.

\section*{Data Availability Statement}
\label{sec:datastatement}
 The implementation of \prose{}, as well as scripts and instructions 
to reproduce the experimental results of this paper as well as the prototype for type-checking
and type-inference are included
as an artifact available on Zenodo~\cite{prose}.
\ifshowcomments
\aleks{remember to put DOIs and check refs}
\fi

\section*{Disclosure of Use of AI in Research}
\label{sec:aidisclosure}
We used OpenAI's GPT-6 Astra~\cite{openai-astra}, accessed through Codex, to assist with the implementation of the type-inference prototype $\pmultipi$, and to generate regression tests and explore counterexamples. 
We additionally used Anthropic's Claude Fable~\cite{fable}, accessed through Claude Code, to assist with the implementation of \prose{} and to generate benchmarking scripts.
The authors directed this work and take responsibility for the implementation and reported results.

 \bibliography{SessionTypes.bib}

\clearpage
\appendix

\section{Appendix for Section \ref{sec:processes}}
\label{app:processes}
\subsection{Proofs for~\cref{sec:processes}}
\label{app:minskyencoding}

\paragraph{Derived expressions.}
From the grammar in~\cref{fig:expr} we abbreviate, for a natural $e$ and a numeral $j$:
\[
  \prd{e}\triangleq e+\mathsf{neg}(1)\qquad
  \isz{e}\triangleq \neg\,(e>0)\qquad
  \eqk{e}{j}\triangleq \neg\big((e>j)\vee(j>e)\big),
\]
the decrement, the zero-test, and comparison against a fixed numeral.

\paragraph{Two-counter machines.}
A \emph{Minsky machine} $\mathcal M$ is a finite instruction list $I_0,\dots,I_K$ over counters $c_1,c_2$, each instruction of the form "$\mathrm{INC}_{c_r};\mathrm{goto}\,j'$", "$\mathrm{DECJZ}_{c_r}(j_z,j_{nz})$" or "$\mathrm{HALT}$". %
Configurations $(k,a,b)\in\{0,\dots,K\}\times\mathbb N^2$ step by a standard reduction relation $\hookrightarrow_{\mathcal M}$. The instruction pointer $k$ is bounded, so the control is finite; the Turing power resides entirely in the unbounded $\nat$ counters $a,b$. $\mathcal M$ \emph{halts} if $(0,0,0)\hookrightarrow_{\mathcal
M}^{*}(k,\cdot,\cdot)$ and $I_k = \mathrm{HALT}$. Determining whether $\mathcal M$ halts is \emph{undecidable}~\cite{Minsky1967}.

\paragraph{The encoding.}
We show how to encode a Minsky machine $\mathcal M$ into a multiparty session $\enc{\mathcal M}$.
A controller $\pC$ computes; a memory $\pM$ echoes the live state
$(k,a,b)$ back each round (three $\nat$ messages, labels $\lk,\llo,\llt$):
\[
\begin{aligned}
  Mem &\triangleq
  \procout{\pC}{\lk}{0}{\procout{\pC}{\llo}{0}{\procout{\pC}{\llt}{0}{}}}
  \mu Y.\,\pC?\lk(k).\pC?\llo(a).\pC?\llt(b).
  \procout{\pC}{\lk}{k}{\procout{\pC}{\llo}{a}{\procout{\pC}{\llt}{b}{Y}}}\\
  C &\triangleq \mu X.\,\pM?\lk(k).\pM?\llo(a).\pM?\llt(b).\,\Disp\\
  \Disp &\triangleq
  \cond{\eqk{k}{0}}{A_0}{(\cdots\,\cond{\eqk{k}{K-1}}{A_{K-1}}{A_K})}
\end{aligned}
\]
where, the branch $A_j$ realises instruction $I_j$ for $j \leq K$:
$$
\small
\begin{array}{ll}
  \enc{\mathrm{INC}_{c_1};\mathrm{goto}\,j'}: & A_j\triangleq\snd{j'}{\mathsf{succ}(a)}{b}\\
  \enc{\mathrm{INC}_{c_2};\mathrm{goto}\,j'}: & A_j\triangleq\snd{j'}{a}{\mathsf{succ}(b)}\\
  \enc{\mathrm{DECJZ}_{c_1}(j_z,j_{nz})}: & A_j\triangleq\cond{\isz{a}}{\snd{j_z}{a}{b}}{\snd{j_{nz}}{\prd{a}}{b}}\\
  \enc{\mathrm{DECJZ}_{c_2}(j_z,j_{nz})}: & A_j\triangleq\cond{\isz{b}}{\snd{j_z}{a}{b}}{\snd{j_{nz}}{a}{\prd{b}}}\\
  \enc{\mathrm{HALT}}: & A_{j}\triangleq\procout{\pM}{\lstop}{0}{\mathbf 0}
\end{array}
$$
The session is $\enc{\mathcal M}\triangleq\pa{\pC}{C}\;\pc\;\pa{\pM}{Mem}$. Its communication skeleton ranges over the finite label set $\{\lk,\llo,\llt,\lstop\}$ and all unboundedness sits in the $\nat$ payloads, so each role is typable with a finite-$\mu$ local type.

\noindent The encoding faithfully simulates $\mathcal M$ (\cref{lem:minskysim}), and the sole label that can ever mismatch is $\lstop$, emitted only by the $\mathrm{HALT}$ arm; so an error is reachable iff $\mathcal M$ reaches the $(k,\cdot,\cdot)$ for which $I_k = \mathrm{HALT}$ (shown below).

\paragraph{Dispatch configurations.} For $\gamma=(k,a,b)$ write $D(\gamma)$ for
the configuration with $\pC$ poised at $\Disp$ under $\{k,a,b\}$ and $\pM$ at its
loop head awaiting the next triple from $\pC$:
\[
  D(\gamma)\triangleq
  \pa{\pC}{\Disp\{k,a,b\}}\;\pc\;
  \pa{\pM}{\pC?\lk(k').\pC?\llo(a').\pC?\llt(b').
           \procout{\pC}{\lk}{k'}{\procout{\pC}{\llo}{a'}{\procout{\pC}{\llt}{b'}{Y}}}}.
\]

\begin{lemma}[Simulation]\label{lem:minskysim}
Let $\gamma_0\hookrightarrow_{\mathcal M}\gamma_1\hookrightarrow_{\mathcal M}\cdots$
be the run of $\mathcal M$ from $\gamma_0=(0,0,0)$. Then:
\begin{enumerate}
  \item $\enc{\mathcal M}\predcontwc{}{\probn1}{+}D(\gamma_0)$;
  \item if $\gamma_n=(k,a,b)$ with $I_k\neq\mathrm{HALT}$ and
        $\gamma_n\hookrightarrow_{\mathcal M}\gamma_{n+1}$, then
        $D(\gamma_n)\predcontwc{}{\probn1}{+}D(\gamma_{n+1})$ with $D(\gamma_{n+1}) \neq \rederror$
  \item if $\gamma_n=(k,a,b)$ with $I_k=\mathrm{HALT}$, then
        $D(\gamma_n)\predcontwc{}{\probn1}{+}\rederror$, and every maximal reduction
        from $D(\gamma_n)$ reaches $\rederror$.
\end{enumerate}
\end{lemma}
\begin{proof}
\emph{(1)} The three initial outputs of $Mem$ ($\lk\langle0\rangle,\llo\langle0\rangle, \llt\langle0\rangle$, all addressed to $\pC$) meet the matching inputs at the head of $C$ by \rulename{r-comm}; $C$ then stands at $\Disp \{0,0,0\}$, and unfolding $\mu Y$ once leaves $Mem$ at its loop head, together giving $D(\gamma_0)$.

\emph{(2)} Reachable configurations have $k\le K$, so the cascade selects $A_k$. That arm computes $\gamma_{n+1}$ from $\gamma_n$ using either $\enc{\mathrm{INC}_{c_r}\ldots}$ or $\enc{\mathrm{DECJZ}_{c_r}\ldots}$ and sends the result to $\pM$, which echoes it back to $\pC$, yielding $D(\gamma_{n+1})$. Every message uses a label in $\{\lk,\llo,\llt\}$ that its partner's input offers, so \rulename{r-err} never applies on the segment.

\emph{(3)} The cascade selects $A_k=\procout{\pM}{\lstop}{0}{\mathbf 0}$, so $D(\gamma_n)$ reduces internally to $\pa{\pC}{\procout{\pM}{\lstop}{0}{\mathbf 0}}\pc\pa{\pM}{\pC?\lk(k').\cdots}$. The memory offers a single input branch, from $\pC$ on $\lk$: thus $\pC$ is a partner ($\exists j\st\pp=\pchan{\pp_j}$) but no branch carries $\lstop$ ($\nexists i:\pchan{\pp_i}=\pC\wedge\labels{\ell_i}=\labels{\lstop}$), so \rulename{r-err} fires and the configuration reduces to $\rederror$.
No other reduction is available (both prefixes need a partner, and \rulename{r-comm} is blocked by the label clash), and the preceding dispatch is deterministic, so every maximal reduction from $D(\gamma_n)$ reaches $\rederror$.
\end{proof}

\procsafeundec*
\begin{proof}
Via the encoding of \cref{app:minskyencoding} and \cref{lem:minskysim}. We show $\enc{\mathcal M}$ is safe $\iff$ $\mathcal M$ does not halt.

\emph{($\mathcal M$ does not halt)}
I.e. its run is infinite and no
$\gamma_n=(k,\cdot,\cdot)$ has $I_k=\mathrm{HALT}$. By \cref{lem:minskysim}(1)--(2) and induction, every configuration reachable from $\enc{\mathcal M}$ lies on a segment $D(\gamma_n)\predcontwc{}{\probn1}{+} D(\gamma_{n+1})$, none of which reduces to $\rederror$ (the only label that could trigger \rulename{r-err} is $\lstop$, emitted solely by a $\mathrm{HALT}$ arm, never selected). Hence $\enc{\mathcal M}\not\predcontstar{}{}\rederror$: it is safe.

\emph{($\mathcal M$ halts)}
Some reachable $\gamma_n=(k,\cdot,\cdot)$ has $I_k=\mathrm{HALT}$. By \cref{lem:minskysim}(1)--(2), $\enc{\mathcal M}\predcontstar{}{\probn1}D(\gamma_n)$, and by~(3) $D(\gamma_n)\predcontwc{}{\probn1}{+}\rederror$. Hence
$\enc{\mathcal M}\predcontstar{}{}\rederror$: it is unsafe.

Thus $\enc{\mathcal M}\predcontstar{}{}\rederror$ iff $\mathcal M$ halts.
As $\mathcal M$ halting is undecidable, so is deciding $\enc{\N}\predcontstar{}{}\rederror$, and hence safety.
\end{proof}

\section{Appendix for Section \ref{sec:types}}
\label{app:types}

\subsection{Derivation rules for merging}
\label{app:mergeder}
For the proofs of the appendix, we give an equivalent rule-based presentation of \Cref{def:mergerel}. The rules are read up to unfolding. Only \rulename{m-$\Sigma$} has merge premises, each relating two selection distributions and hence discharged by \rulename{m-$\oplus$}. Thus merge derivations are \emph{finite}; continuation joins $(\sqcup)$ refer to the independently defined least-upper-bound relation.
{\small
\[
\begin{array}{@{}c@{}}
\inferrule*[left=\rulename{m-end}]
{
-
}
{
\tend \probmerge{\prob} \tend = \tend
}
\quad
\inferrule*[left=\rulename{m-$\&$}]
{
K = \{\,k \in I \cap J
    \mid \exists \T:\; \T_k \sqcup \T_k' = \T\,\}
\quad
\{\qi{i}\}_{i \in I}
=
\{\qi{j}\}_{j \in J}
=
\{\qi{k}\}_{k \in K}
}
{
\pprocinnew{\q_i}{i \in I}{\ell_i}{\Basic_i}{\T_i}
\probmerge{\prob}
\pprocinnew{\q_j}{j \in J}{\ell_j}{\Basic_j}{\T_j'}
=
\pprocinnew{\q_k}{k \in K}{\ell_k}{\Basic_k}
            {(\T_k \sqcup \T_k')}
}
\\[2mm]
{
\inferrule*[left=\rulename{m-$\oplus$}]
{
    {\begin{array}{ll}
\forall k \in I \setminus J:\quad
  \probipp{k} = \prob \cdot \probi{k}
  \;\land\;
  \T_k'' = \T_k & \\
\forall k \in J \setminus I:\quad
  \probipp{k} = (\probn{1}-\prob)\cdot\probip{k}
  \;\land\;
  \T_k'' = \T_k' &
\forall k \in I \cap J:\quad
  \probipp{k}
  = \prob\cdot\probi{k}
    +(\probn{1}-\prob)\cdot\probip{k}
  \;\land\;
  \T_k \sqcup \T_k' = \T_k''
    \end{array}}
}
{
\pprocoutty{\q_i}{i \in I}{\probi{i}}
           {\ell_i}{\Basic_i}{\T_i}
\probmerge{\prob}
\pprocoutty{\q_j}{j \in J}{\probip{j}}
           {\ell_j}{\Basic_j}{\T_j'}
           =
\pprocoutty{\q_k}{k \in I \cup J}{\probipp{k}}
           {\ell_k}{\Basic_k}{\T_k''}
}
}
\\[2mm]
\inferrule*[left=\rulename{m-$\Sigma$}]
{
|I|\cdot|J|>1
\qquad
\forall(i,j)\in I\times J:\; \T_i \probmerge{\prob} \T_j' = \T_{i,j}
}
{
\sum_{i\in I}\T_i
\probmerge{\prob}
\sum_{j\in J}\T_j'
=
\sum_{(i,j)\in I\times J}\T_{i,j}
}
\end{array}
\]
}

With other cases being undefined.

\subsection{Full dining philosophers example}

\begin{example}[Dining philosophers]
    \label{exam:diningphilo}
We consider the \emph{dining philosophers} example with $n$ philosophers,
indices $i \in \mathbb{Z}_{n}$ (all index arithmetic \emph{mod} $n$). Each
philosopher $\pchan{p_i}$ shares fork $\pchan{f_i}$ with its left neighbour and
$\pchan{f_{i+1}}$ with its right neighbour, and reports to a shared
$\pchan{food}$ role when it is ready to eat.
\vspace{-1mm}
\begin{equation*}
    \small
    \begin{array}{lr}
    \PP_i =
    \mu X. \flipc{p}
    {\mathtt{P}(\pchan{f_i},\pchan{f_{i+1}})}
    {\mathtt{P}(\pchan{f_{i+1}},\pchan{f_i})}
    &
    {\mathtt{P}}(\pchan{a},\pchan{b}) = \mu Y.\procoutlab{\pchan{a}}{pick}{\plusc
    {
        \procin{\pchan{a}}{\labels{free}}{ 
            \procoutlab{\pchan{b}}{pick}{\plusc
                    {
                        \procin{\pchan{b}}{\labels{notFree}}{
                            \procoutlab{\pchan{a}}{drop}{X}
                        }
                    }
                    {
                    \procin{\pchan{b}}{\labels{free}}{
                        \procoutlab{\pchan{food}}{\labels{eat}}{
                            \procoutlab{\pchan{a}}{drop}{
                                \procoutlab{\pchan{b}}{drop}{ X }
                            }
                        }
                    }
                }
            }
        } 
    }
    {
        \procin{\pchan{a}}{\labels{notFree}}{ Y }
    }
}
    \end{array}
\end{equation*}

\noindent
Philosopher $\PP_i$ flips a coin to decide which fork to reach for first: on
heads it runs $\mathtt P(\pchan{f_i},\pchan{f_{i+1}})$, requesting $\pchan{f_i}$
before $\pchan{f_{i+1}}$, and on tails the two are swapped---this randomisation
breaks the symmetry that would otherwise deadlock~\cite{Lehmann1981}.
The routine $\mathtt P(\pchan a,\pchan b)$ requests fork $\pchan a$; on $\labels{notFree}$ it
retries the \emph{same} fork, and on $\labels{free}$ it requests the second fork
$\pchan b$. If $\pchan b$ answers $\labels{notFree}$ it drops $\pchan a$ and
starts over; otherwise it holds both forks, signals $\pchan{food}$, and releases
them.
\vspace{-1mm}
\begin{equation*}
    \small
    \begin{array}{lcr}
        F_i = \mu X.\plusc 
        {\mathtt{Q}(\pchan{p_i},\pchan{p_{i-1}})}
        {\mathtt{Q}(\pchan{p_{i-1}},\pchan{p_i})}
    &
    {\mathtt{Q}}(\pchan{a}, \pchan{b})=
            {\procin{\pchan{a}}{\labels{ pick}}{
                    \pchan{a} !\, \labels{free} . 
                    \mu Y. \plusc {\procin{\pchan{a}}{ \labels{drop}}{X}}
                    {
                        \procin{\pchan{b}}{ \labels{pick}}{
                              \pchan{b} !\, \labels{notFree} . Y
                        }
                    }
            }}
    & 
    \text{Food} = \mu X . \displaystyle{\sum_{i \in \mathbb{Z}_{n}}}\procin{\pchan{p_i}}{eat}{X}
    \end{array}
\end{equation*}

\noindent
Once a philosopher holds both $\pchan{f_i}$ and $\pchan{f_{i+1}}$ he
sends $\labels{eat}$ to $\pchan{food}$, then puts both forks down and starts over.

\medskip
\noindent\emph{Session types.}\;
We now give the type of each participant. The only bulky fragment is the
continuation reached once the first fork has been granted, so we abbreviate it as
$\T_{\mathtt E}$ (the \emph{eat-attempt}) and let $\T_{\mathtt P}$ name the type of
the grab routine $\mathtt P(\pchan a,\pchan b)$:
\[
\small
\begin{array}{r@{\;\;=\;\;}l}
\T_{\mathtt E}(\pchan a,\pchan b) &
  \oplus\,\pchan b!\labels{pick}.\,
  \&\big\{\,
    \pchan b?\labels{free}.\,
      \oplus\,\pchan{food}!\labels{eat}.\,
      \oplus\,\pchan a!\labels{drop}.\,
      \oplus\,\pchan b!\labels{drop}.\,\ty,\;\;
    \pchan b?\labels{notFree}.\,
      \oplus\,\pchan a!\labels{drop}.\,\ty
  \,\big\}
\\[2.5mm]
\T_{\mathtt P}(\pchan a,\pchan b) &
  \mu\ty'.\,\oplus\,\pchan a!\labels{pick}.\,
  \&\big\{\,
    \pchan a?\labels{free}.\,\T_{\mathtt E}(\pchan a,\pchan b),\;\;
    \pchan a?\labels{notFree}.\,\ty'
  \,\big\}
\end{array}
\]
Here $\ty$ is the philosopher's outer loop, bound by the $\mu\ty$ below 
, while $\ty'$ in $\T_{\mathtt P}$ is the local retry of the same first fork.
The philosopher type is then the merge of the two flip orientations;\\[1mm]
$\T_{\ppi{i}} = \mu \ty . \big(\T_{\mathtt P}(\pchan{f_i},\pchan{f_{i+1}}) \probmerge{\prob} \T_{\mathtt P}(\pchan{f_{i+1}},\pchan{f_i})\big)$, which collapses to a single tagged selection: %
\[
\small
\T_{\pchan{p_i}} \;\;=\;\;
\mu\ty.\;\oplus\left\{
\begin{array}{@{}l@{}}
\probn{p}\;:\;\pchan{f_i}!\labels{pick}.\,
  \&\big\{\,
    \pchan{f_i}?\labels{free}.\,\T_{\mathtt E}(\pchan{f_i},\pchan{f_{i+1}}),\;\;
    \pchan{f_i}?\labels{notFree}.\,\T_{\mathtt P}(\pchan{f_i},\pchan{f_{i+1}})
  \,\big\},\\[1.5mm]
\probn{1{-}p}\;:\;\pchan{f_{i+1}}!\labels{pick}.\,
  \&\big\{\,
    \pchan{f_{i+1}}?\labels{free}.\,\T_{\mathtt E}(\pchan{f_{i+1}},\pchan{f_i}),\;\;
    \pchan{f_{i+1}}?\labels{notFree}.\,\T_{\mathtt P}(\pchan{f_{i+1}},\pchan{f_i})
  \,\big\}
\end{array}
\right\}
\]
The two summands differ only by swapping $\pchan{f_i}\leftrightarrow\pchan{f_{i+1}}$. Dually, fork $\pchan{f_i}$ offers $\labels{pick}$ to either neighbour and, once granted, refuses the other until the holder drops it; $\pchan{food}$ simply awaits "$\labels{eat}$" from every philosopher:
\[
\small
\begin{array}{r@{\;\;=\;\;}l}
\T_{\mathtt Q}(\pchan a,\pchan b) &
  \oplus\,\pchan a!\labels{free}.\,
  \mu\ty'.\,
  \&\big\{\,
    \pchan a?\labels{drop}.\,\ty,\;\;
    \pchan b?\labels{pick}.\,\oplus\,\pchan b!\labels{notFree}.\,\ty'
  \,\big\}
\\[2.5mm]
\T_{\pchan{f_i}} &
  \mu\ty.\;
  \&\big\{\,
    \pchan{p_i}?\labels{pick}.\,\T_{\mathtt Q}(\pchan{p_i},\pchan{p_{i-1}}),\;\;
    \pchan{p_{i-1}}?\labels{pick}.\,\T_{\mathtt Q}(\pchan{p_{i-1}},\pchan{p_i})
  \,\big\}
\\[2.5mm]
\T_{\pchan{food}} &
  \mu\ty.\;\&_{i\in\mathbb Z_n}\;\pchan{p_i}?\labels{eat}.\,\ty
\end{array}
\]

\end{example}

\subsection{Proofs of statements in~\cref{sec:types}}
\label{app:sec3proofs}

\subtpreorder*
\begin{proof}
    \mbox{}
    \begin{enumerate}
        \item (\emph{Reflexivity}) For any $\T$, we have $\T \subt \T$:\\[2mm]
            The identity relation $\mathrm{Id} = \{(\T,\T)\}$ is closed under the rules of
            ~\cref{def:pmpst_subtype}: the \rulename{sub-end}, \rulename{sub-$\oplus$} and \rulename{sub-$\&$} cases are
            immediate, each continuation obligation being again in $\mathrm{Id}$; and for
            \rulename{sub-$\Sigma$} (when $|I|>1$) the diagonal matching $i \mapsto i$ discharges every
            summand via \rulename{sub-$\oplus$}. By coinduction $\mathrm{Id} \subseteq {\subt}$, so
            $\T \subt \T$.

        \item (\emph{transitivity}) Suppose $\T_1 \subt \T_2$ and $\T_2 \subt \T_3$ then $\T_1 \subt \T_3$:\\[2mm]
            We proceed by case analysis on the rule used to derive $\T_1 \subt \T_2$:
        \begin{itemize}
            \item \rulename{sub-end}:
                $\T_1 \subt \T_2$ by \rulename{sub-end} implies $\T_1 = \T_2 = \tend$. Thus $\T_2 \subt \T_3$ must be derived as $\tend \subt \T_3$, implying $\T_3 = \tend$. As $\tend \subt \tend$, $\T_1 \subt \T_3$ follows immediately.
            \item \rulename{sub-$\oplus$}:
                If $\T_1 \subt \T_2$ by:
                $$
                \cinferruleleft[\rulename{sub-$\oplus$}]
                {
                \forall i \in I:\;
                \T_{1,i} \subt \T_{2,i}\quad
                }
                {
                \pprocoutty{\pp_i}{{i \in I}}{\probi{i}}{\ell_i}{\Basic_i}{\T_{1,i}}
                \subt
                \pprocoutty{\pp_i}{{i \in I}}{\probi{i}}{\ell_i}{\Basic_i}{\T_{2,i}}
                }
                $$
                i.e. $\T_2 = \pprocoutty{\pp_i}{{i \in I}}{\probi{i}}{\ell_i}{\Basic_i}{\T_{2,i}}$, there's two cases to consider for the derivation of $\T_2 \subt \T_3$:
                \rulename{sub-$\oplus$} or \rulename{sub-$\Sigma$} with $|I| = 1$:
                \begin{itemize}
                    \item $\T_2 \subt \T_3$ by \rulename{sub-$\oplus$}:
                        $$
                        \cinferruleleft[\rulename{sub-$\oplus$}]
                        {
                        \forall i \in I:\;
                        \T_{2,i} \subt \T_{3,i}\quad
                        }
                        {
                        \pprocoutty{\pp_i}{{i \in I}}{\probi{i}}{\ell_i}{\Basic_i}{\T_{2,i}}
                        \subt
                        \pprocoutty{\pp_i}{{i \in I}}{\probi{i}}{\ell_i}{\Basic_i}{\T_{3,i}}
                        }
                        $$
                        now we can apply $\IH$ on $\T_{1,i}\subt\T_{2,i}$ and $\T_{2,i}\subt \T_{3,i}$ to get $\T_{1,i} \subt \T_{3,i}$ for all $i \in I$ and thus:
                        $$
                        \cinferruleleft[\rulename{sub-$\oplus$}]
                        {
                        \forall i \in I:\;
                        \T_{1,i} \subt \T_{3,i}\quad
                        }
                        {
                        \pprocoutty{\pp_i}{{i \in I}}{\probi{i}}{\ell_i}{\Basic_i}{\T_{1,i}}
                        \subt
                        \pprocoutty{\pp_i}{{i \in I}}{\probi{i}}{\ell_i}{\Basic_i}{\T_{3,i}}
                        }
                        $$
                        i.e. $\T_1 \subt \T_3$ as desired.

                    \item $\T_2 \subt \T_3$ by \rulename{sub-$\Sigma$}:
                        $$
                        \cinferruleleft[\rulename{sub-$\Sigma$}]
                        {
                            \cinferruleleft[\rulename{sub-$\oplus$}]
                            {
                            \forall i \in I:\;
                            \T_{2,i} \subt \T_{3,i}\quad
                            }
                            {
                            \pprocoutty{\pp_i}{{i \in I}}{\probi{i}}{\ell_i}{\Basic_i}{\T_{2,i}}
                            \subt
                            \pprocoutty{\pp_i}{{i \in I}}{\probi{i}}{\ell_i}{\Basic_i}{\T_{3,i}}
                            }
                            \quad
                            |L| \geq 1
                        }
                        {
                            \pprocoutty{\pp_i}{{i \in I}}{\probi{i}}{\ell_i}{\Basic_i}{\T_{2,i}}
                            \subt
                            \pprocoutty{\pp_i}{{i \in I}}{\probi{i}}{\ell_i}{\Basic_i}{\T_{3,i}}
                            +
                            \sum_{l \in L}\T_l'
                        }
                        $$
                        To get $\T_1 \subt \T_3$ we apply $\IH$ on $\T_{1,i} \subt \T_{2,i}$ and $\T_{2,i} \subt \T_{3,i}$ to get for all $i \in I: \T_{1,i} \subt \T_{3,i}$ and use this to derive $\T_1 \subt \T_3$:
                        $$
                        \cinferruleleft[\rulename{sub-$\Sigma$}]
                        {
                            \cinferruleleft[\rulename{sub-$\oplus$}]
                            {
                            \forall i \in I:\;
                            \T_{1,i} \subt \T_{3,i}\quad
                            }
                            {
                            \pprocoutty{\pp_i}{{i \in I}}{\probi{i}}{\ell_i}{\Basic_i}{\T_{1,i}}
                            \subt
                            \pprocoutty{\pp_i}{{i \in I}}{\probi{i}}{\ell_i}{\Basic_i}{\T_{3,i}}
                            }
                            \quad
                            |L| \geq 1
                        }
                        {
                            \pprocoutty{\pp_i}{{i \in I}}{\probi{i}}{\ell_i}{\Basic_i}{\T_{1,i}}
                            \subt
                            \pprocoutty{\pp_i}{{i \in I}}{\probi{i}}{\ell_i}{\Basic_i}{\T_{3,i}}
                            +
                            \sum_{l \in L} \T_l'
                        }
                        $$
                \end{itemize}

            \item \rulename{sub-$\&$}, $\T_1 \subt \T_2$ by:
                $$
                \cinferruleleft[\rulename{sub-$\&$}]
                {
                \forall i \in I \cup J:\quad
                \T_{1,i} \subt \T_{2,i}\qquad
                \{\ppi{i}\}_{i \in I \cup J} = \{\ppi{i}\}_{i \in I \cup J \cup K}
                }
                {
                \pprocinnew{\pp_i}{i \in I \cup J \cup K}{\ell_i}{\Basic_i}{\T_{1,i}}
                \subt
                \pprocinnew{\pp_i}{i \in I \cup J}{\ell_i}{\Basic_i}{\T_{2,i}}
                }
                $$
                i.e. $\T_2 =\pprocinnew{\pp_i}{i \in I \cup J}{\ell_i}{\Basic_i}{\T_{2,i}}$, then $\T_2 \subt \T_3$ could only be derived through \rulename{sub-$\&$}:
                $$
                \cinferruleleft[\rulename{sub-$\&$}]
                {
                \forall i \in I:\quad
                \T_{2,i} \subt \T_{3,i} \qquad
                \{\ppi{i}\}_{i \in I} = \{\ppi{i}\}_{i \in I \cup J}
                }
                {
                \pprocinnew{\pp_i}{i \in I \cup J}{\ell_i}{\Basic_i}{\T_{2,i}}
                \subt
                \pprocinnew{\pp_i}{i \in I}{\ell_i}{\Basic_i}{\T_{3,i}}
                }
                $$
                i.e. $\T_3 = \pprocinnew{\pp_i}{i \in I}{\ell_i}{\Basic_i}{\T_{3,i}}$, we can apply $\IH$ on $\T_{1,i} \subt \T_{2,i}$ and $\T_{2,i} \subt \T_{3,i}$ to get $\T_{1,i} \subt \T_{3,i}$ for all $i \in I$ and use this to derive $\T_1 \subt \T_3$ as:
                $$
                \cinferruleleft[\rulename{sub-$\&$}]
                {
                \forall i \in I:\quad
                \T_{1,i} \subt \T_{3,i} \qquad
                \{\ppi{i}\}_{i \in I} = \{\ppi{i}\}_{i \in I \cup J \cup K}
                }
                {
                \pprocinnew{\pp_i}{i \in I \cup J \cup K}{\ell_i}{\Basic_i}{\T_{1,i}}
                \subt
                \pprocinnew{\pp_i}{i \in I}{\ell_i}{\Basic_i}{\T_{3,i}}
                }
                $$
                where $\{\ppi{i}\}_{i \in I} = \{\ppi{i}\}_{i \in I \cup J \cup K}$ follows transitively from the above.

            \item \rulename{sub-$\sum$}: Suppose $\T_1 \subt \T_2$ by
                $$
                \cinferruleleft[\rulename{sub-$\sum$}]
                {
                |I| \cdot |J| > 1\;\;
                \forall i \in I:\exists j \in J \st
                \T_{1,i} \subt \T_{2,j}
                }
                {
                \sum_{i\in I} \T_{1,i}
                \subt
                \sum_{j\in J} \T_{2,j}
                }
                $$
                Let $\T_2 \subt \T_3$ be derived by
                $$
                \cinferruleleft[\rulename{sub-$\sum$}]
                {
                |J| \cdot |K| > 1\;\;
                \forall j \in J:\exists k \in K \st
                \T_{2,j} \subt \T_{3,k}
                }
                {
                \sum_{j\in J} \T_{2,j}
                \subt
                \sum_{k\in K} \T_{3,k}
                }.
                $$

                Fix any $i\in I$. By the first premise there exists $j\in J$ such that
                $\T_{1,i} \subt \T_{2,j}$. By the second premise there exists $k\in K$ such that
                $\T_{2,j} \subt \T_{3,k}$. By the induction hypothesis,
                $\T_{1,i} \subt \T_{3,k}$. Since $i$ was arbitrary, we have
                \[
                \forall i\in I:\exists k\in K \st \T_{1,i} \subt \T_{3,k}.
                \]

                Hence, by another application of \rulename{sub-$\sum$},
                $$
                \cinferruleleft[\rulename{sub-$\sum$}]
                {
                |I| \cdot |K| > 1\;\;
                \forall i \in I:\exists k \in K \st
                \T_{1,i} \subt \T_{3,k}
                }
                {
                \sum_{i\in I} \T_{1,i}
                \subt
                \sum_{k\in K} \T_{3,k}
                }
                $$
                and therefore $\T_1 \subt \T_3$.
        \end{itemize}
    \end{enumerate}
\end{proof}

\begin{restatable}[]{proposition}{decidablesafety}
    \label{prop:decidablesafe}
    Given a type context $\Delta$ then $\safe{\Delta}$ is decidable.
\end{restatable}
\begin{proof}
    By~\cref{def:safety}, $\safe{\Delta}$ holds iff every context reachable from $\Delta$ is a safe state (cf.\ \cite[Lemma~4.4]{Scalas2019}). The reachable set is finite: by inspection of Fig.~\ref{fig:pmpst_tct}, every transition---including the $\predcontnd{}$ of \rulename{ct-nd} resolving non-determinism and the probability-annotated \rulename{ct-out}---sends each component $\T_\pp$ to one of its $\le |\T_\pp|$ subterms, so $|\{\Delta' \mid \Delta \predcontstar{}{} \Delta'\}| \le \prod_{\pp \in \dom{\Delta}} |\T_\pp|$.
Whether a single $\Delta'$ is a safe state is decidable, being a finite check of label compatibility over its offered actions (~\cref{def:safety}). Enumerating the reachable set and checking each is therefore effective, so $\safe{\Delta}$ is decidable.
\end{proof}

\decidablesubtype*
\begin{proof}
Inspired by~\cite{UdomsrirungruangY25}, we build a finite graph $G(T_1,T_2)$ and decide $T_1 \leqslant T_2$ by search.
The nodes are pairs of sub-terms $\mathrm{Sub}(T_1)\times\mathrm{Sub}(T_2)$ with root $(T_1,T_2)$;
since $T_1,T_2$ are regular, this set is of size $\mathcal{O}(|T_1||T_2|)$.
From a node $(U,V)$, unfold both components to expose
their top constructors. The rules of~\cref{def:pmpst_subtype} are constructor-directed,
so at most one applies, and its premises fix the successors: none for \rulename{sub-$\tend$};
one continuation pair $(T_i,T_i')$ per branch for \rulename{sub-$\oplus$}, \rulename{sub-$\&$}; and,
for \rulename{sub-$\Sigma$}, for each subtype summand $i\in I$ a matching supertype summand
$k\in K$ and its continuation pairs $(\T_{k,j},\T_{k,j}')$. Every successor is a pair of
subterms of the endpoints, so edges stay within the finite node set.

Call $(U,V)$ \emph{inconsistent} if, after unfolding, no rule applies: the top
constructors clash or a side condition fails---for \rulename{sub-$\Sigma$}, some $i\in I$
has no matching $k\in K$. Inconsistency of one node is decidable (a constructor
comparison, plus for \rulename{sub-$\Sigma$} a finite search over $I\times K$).

The procedure computes the nodes reachable from $(T_1,T_2)$ and reports
$T_1\leqslant T_2$ iff none is inconsistent; it terminates as $G$ is finite.
For correctness, note that a reachability-closed set of nodes with no inconsistent
member is exactly a relation closed under the rules, hence contained in $\leqslant$;
so an inconsistency-free reachable set witnesses $T_1\leqslant T_2$. Conversely,
a reachable inconsistent $(U,V)$ admits no rule relating $U$ and $V$, and its
obligations descend from $(T_1,T_2)$, so $(T_1,T_2)\notin{\leqslant}$.
\end{proof}

\begin{restatable}[$\T_1\sqcup\T_2$ is decidable]{lemma}{supdecidable}
\label{lem:supdecidable}
Let $\T_1,\T_2$ be types and let $d$ be the maximum out-degree of their finite type graphs. It is decidable whether $\T_1\sqcup\T_2$ is defined in time $\mathcal O(|\T_1||\T_2|)$. When defined, a witness is computable in time
\[
 \mathcal O\!\left(
   (\max\{2,d\})^{2|\T_1||\T_2|}
   \log(|\T_1||\T_2|+1)
 \right).
\]
\end{restatable}

\begin{proof}
    Here $|\T|$ counts all syntax-tree nodes. We begin by presenting the \textbf{\emph{Algorithm}}.
\[
 \mathsf{TypeGraph}(\T_i)=\mathbb G(\T_i)=(V_i,E_i,r_i),
 \qquad i\in\{1,2\}.
\]
The vertices $V_i$ are the occurrences of $\tend$, branching constructors, output distributions and non-singleton outer sums in the syntax of $\T_i$ (distinct occurrences of the same subterm remain distinct vertices). The set $E_i$ contains edge occurrences, and $r_i\in V_i$ is the root node of $\T_i$.

We follow the type-graph notation of~\cite{UdomsrirungruangY25}. Recursion binders and variables contribute no separate vertices: the memoised function ``$\mathsf{head}$'' resolves them to their constructor vertices. Each prefix is recorded on an edge to its continuation's head; $\tend$ has no outgoing edges.

We note that a proper sum has two graph levels:
\[
\begin{gathered}
 u=\sum_{i\in I}\T'_i,\qquad
 \T'_i=\pprocoutty{\pp_j}{j\in J_i}{\probi{i,j}}
       {\ell_j}{\Basic_j}{u_{i,j}},\qquad |I|>1,\\
 u\xrightarrow{i}\T'_i
   \xrightarrow{\probi{i,j}:\pp_j!\ell_j\langle\Basic_j\rangle}u_{i,j},
 \qquad i\in I,\ j\in J_i.
\end{gathered}
\]
Thus $V_i$ represents constructor subterm occurrences, rather than every node counted by $|\T_i|$. Write $\operatorname{od}(u)$ for the out-degree of a vertex $u$. For simplicity of presentation we identify a vertex with the type it represents. Equalities such as $u=\tend$ specify the constructor, while ancestor lookup uses vertex identities.

Align branching indices by participant, label and payload sort: shared indices denote the same prefix, with possibly different continuation vertices. For input vertices
\[
 u=\pprocinnew{\pp_i}{i\in I}{\ell_i}{\Basic_i}{u_i},
 \qquad
 v=\pprocinnew{\pp_j}{j\in J}{\ell_j}{\Basic_j}{v_j},
\]
define
\[
\begin{aligned}
 \mathsf{Pts}(u)&=\{\pp_i\mid i\in I\},
 \qquad \mathsf{Pts}(v)=\{\pp_j\mid j\in J\},\\
 M(u,v)&=\{(\pp_k,\ell_k,\Basic_k,(u_k,v_k))\mid k\in I\cap J\}.
\end{aligned}
\]
Well-formed inputs have nonempty sender sets and distinct participant--label pairs, so $M(u,v)$ retains each matching prefix.
Set $M(u,v)=\emptyset$ for other constructor pairs. The join dependency graph $\mathbb G_{\sqcup}(\T_1,\T_2)$ has vertex set $V_1\times V_2$ and edges
\[
 (u,v)\xrightarrow{\pp?\ell(\Basic)}(u',v')
 \quad\Longleftrightarrow\quad
 (\pp,\ell,\Basic,(u',v'))\in M(u,v).
\]
For $R\subseteq V_1\times V_2$, let $\mathsf{Pts}_R(u,v)$ contain the senders of matching prefixes whose continuation pairs belong to $R$:
\[
 \mathsf{Pts}_R(u,v)=
 \{\pp\mid(\pp,\ell,\Basic,(u',v'))\in M(u,v),\ (u',v')\in R\}.
\]
An input pair $(u,v)$ is \emph{consistent with respect to $R$}
exactly when
\[
 \mathsf{Pts}(u)=\mathsf{Pts}(v)=\mathsf{Pts}_R(u,v)\neq\emptyset.
\]
Selection pairs and $(\tend,\tend)$ are also consistent; all other constructor pairs are inconsistent.

Algorithm~\ref{alg:join} computes and stores $M(u,v)$ once per input pair. Its counter $c[(u,v),\pp]$ records the number of surviving matches from $\pp$; a sender disappears only when its last match is removed. The list $\mathsf{pred}[(u',v')]$ contains one occurrence of $((u,v),\pp)$ for each matching prefix from $\pp$ whose
continuation pair is $(u',v')$.
The worklist $\mathcal Q$ contains pairs awaiting deletion, and $Z$ records those already enqueued. Let $R_*$ denote the relation $R$ remaining after the deletion loop terminates, the fixed point reached by pruning inconsistent pairs.
$\mathsf{Pop}$ removes and returns one queued pair, and $\cdot$ denotes concatenation of predecessor lists.

For a relation $R\subseteq V_1\times V_2$ whose pairs are consistent with respect to $R$, introduce one new vertex $\T_{u,v}$ for each $(u,v)\in R$. Construct the auxiliary graph $\widehat{\mathcal H}_R$ on the disjoint vertex sets
\[
 V_1\uplus V_2\uplus\{\T_{u,v}\mid(u,v)\in R\},
\]
retaining $E_1,E_2$. Define its result vertices by constructor:
\[
 u=v=\tend\quad\Longrightarrow\quad\T_{u,v}=\tend.
\]
For selections, we use the tagged disjoint union
$I\uplus J=(\{1\}\times I)\cup(\{2\}\times J)$:
\[
 u=\sum_{i\in I}\T'_{1,i},\quad
 v=\sum_{j\in J}\T'_{2,j},\quad
 \T_{u,v}=\sum_{(a,k)\in I\uplus J}\T'_{a,k}
\]
For input pairs, put
\[
 \T_{u,v}=\pprocinnew{\pp}
      {\substack{(\pp,\ell,\Basic,(u',v'))\in M(u,v)\\(u',v')\in R}}
      {\ell}{\Basic}{\T_{u',v'}}.
\]
For $X\subseteq V(\mathcal G)$, let $\mathcal G[X]$ denote
the induced subgraph, and put
\[
 \operatorname{Reach}_{\mathcal G}(r)
   =\{v\mid r\longrightarrow_{\mathcal G}^{*}v\}.
\]
When $(r_1,r_2) \in R$, define the result graph $\mathcal H_R$ by $\mathsf{Build}(R,\mathbb G(\T_1),\mathbb G(\T_2))$ where
\[
 \mathcal H_R=
 \widehat{\mathcal H}_R
   [\operatorname{Reach}_{\widehat{\mathcal H}_R}(\T_{r_1,r_2})],
 \qquad r_{\mathcal H_R}=\T_{r_1,r_2}
\]
Thus the result contains only reachable result vertices and the original vertices needed by retained summands.

Given a result graph $\mathcal H_R$, we use the procedure $\mathsf{Read}$ below to read out the type $\T$ it denotes.

The ancestor map $A$ associates type vertices $u,v$  with fresh typevariables $\ty_{\text{fresh}}$. We note that for a non-singleton selection, the selection clause prints the sum vertex and its distribution vertices together, then recurses into continuations.
\[
\begin{array}{l}
\mathsf{Read}(u,A)=\\
\;\begin{gathered}
 \begin{cases}
   A(u),&u\in\operatorname{dom}(A),\\[2pt]
   \tend,&u\notin\operatorname{dom}(A),\ u=\tend,\\[5pt]
   \mu\ty_{\text{fresh}}.\pprocinnew{\pp_k}{k\in K}{\ell_k}{\Basic_k}
   {\mathsf{Read}(u_k,A[u \mapsto \ty_{\text{fresh}}])},
        &\begin{array}{l}
         u\notin\operatorname{dom}(A),\\
         u=\pprocinnew{\pp_k}{k\in K}{\ell_k}{\Basic_k}{u_k},
       \end{array}\\[8pt]
       \mu\ty_{\text{fresh}}.\probpprocoutty{i\in I}{\pp_j}{j\in J_i}{\probi{i,j}}
   {\ell_j}{\Basic_j}{\mathsf{Read}(u_{i,j},A[u \mapsto \ty_{\text{fresh}}])},
      &\begin{array}{l}
         u\notin\operatorname{dom}(A),\\
         u=\probpprocoutty{i\in I}{\pp_j}{j\in J_i}{\probi{i,j}}
            {\ell_j}{\Basic_j}{u_{i,j}}.
       \end{array}
 \end{cases}
 \\[3pt]
\end{gathered}
\end{array}
\]
The uniform construction may introduce unused binders.
They can be erased without changing the unfolding, since
\[
 \ty\notin\ftv{\T}
 \quad\Longrightarrow\quad
 \unf{\mu\ty.\T}=\T.
\]

\begin{algorithm}[tp]
\caption{Computing $\T_1\sqcup\T_2$; $\emptyset$ denotes failure.}
\label{alg:join}
\small
\begin{algorithmic}[1]
\Function{Join}{$\T_1,\T_2$}
  \State $(V_i,E_i,r_i)\gets\Call{TypeGraph}{\T_i}$, $i\in\{1,2\}$
  \State $R\gets V_1\times V_2$; $\mathcal Q\gets\emptyset$
  \State $\mathsf{pred}[(u,v)]\gets\langle\rangle$ \quad $((u,v)\in V_1\times V_2)$
  \ForAll{$(u,v)\in V_1\times V_2$}
    \If{$u=v=\tend$ \textbf{or} $(u=\sum_{i\in I}\T'_{1,i}\ \land\ v=\sum_{j\in J}\T'_{2,j})$}
      \State \textbf{continue}
    \ElsIf{$u=\pprocinnew{\pp_i}{i\in I}{\ell_i}{\Basic_i}{u_i}$ \textbf{and}
           $v=\pprocinnew{\pp_j}{j\in J}{\ell_j}{\Basic_j}{v_j}$}
      \State $c[(u,v),\pp]\gets 0$ \quad $(\pp\in\mathsf{Pts}(u))$
      \ForAll{$(\pp,\ell,\Basic,(u',v'))\in M(u,v)$}
        \State $c[(u,v),\pp]\gets c[(u,v),\pp]+1$
        \State $\mathsf{pred}[(u',v')]\gets\mathsf{pred}[(u',v')]\cdot\langle((u,v),\pp)\rangle$
      \EndFor
      \If{$\mathsf{Pts}(u)\neq\mathsf{Pts}(v)\ \lor\ \mathsf{Pts}_R(u,v)\neq\mathsf{Pts}(u)$}
        \State $\mathcal Q\gets \mathcal Q\cup\{(u,v)\}$
      \EndIf
    \Else
      \State $\mathcal Q\gets \mathcal Q\cup\{(u,v)\}$
    \EndIf
  \EndFor
  \State $Z\gets \mathcal Q$
  \While{$\mathcal Q\neq\emptyset$}
    \State $(u',v')\gets\Call{Pop}{\mathcal Q}$; $R\gets R\setminus\{(u',v')\}$
    \ForAll{$((u,v),\pp)\in\mathsf{pred}[(u',v')]:(u,v)\in R$}
      \State $c[(u,v),\pp]\gets c[(u,v),\pp]-1$
      \If{$c[(u,v),\pp]=0\ \land\ (u,v)\notin Z$}
        \State $\mathcal Q\gets \mathcal Q\cup\{(u,v)\}$; $Z\gets Z\cup\{(u,v)\}$
      \EndIf
    \EndFor
  \EndWhile
  \If{$(r_1,r_2)\notin R$}
    \State \Return $\emptyset$
  \EndIf
  \State $\mathcal H_R\gets\Call{Build}{R,\mathbb G(\T_1),\mathbb G(\T_2)}$
  \State \Return $\mathsf{Read}_{\mathcal H_R}(\T_{r_1,r_2},\emptyset)$
\EndFunction
\end{algorithmic}
\end{algorithm}

\par\noindent\emph{Correctness.}
Let
\[
 C=\{(u,v)\in V_1\times V_2\mid\exists\T:\ u\leq\T\land v\leq\T\}.
\]
All constructors are read up to head unfolding. For $(u,v)\in C$, inversion gives compatible heads. If both are inputs, the branches of a common upper bound cover their common sender set and have continuations
bounding both operand continuations. Thus
\[
 \mathsf{Pts}(u)=\mathsf{Pts}(v)=\mathsf{Pts}_C(u,v)\neq\emptyset.
\]
Whenever $C\subseteq R$, no member of $C$ is inconsistent with respect to $R$. After initialisation and at deletion-loop boundaries, the reverse-dependency updates maintain,
for every input pair $(u,v)\in R$,
\[
 \mathsf{Pts}_R(u,v)=
 \{\pp\in\mathsf{Pts}(u)\mid c[(u,v),\pp]>0\}.
\]
Thus only inconsistent pairs are enqueued, and deletion preserves $C\subseteq R$.
Initially $R=V_1\times V_2$, so $C\subseteq R_*$.

Every pair in $R_*$ is consistent with respect to $R_*$. Coinduction on the auxiliary graph
$\widehat{\mathcal H}_{R_*}$ gives
\[
 (u,v)\in R_*
 \Longrightarrow u\leq\T_{u,v}\land v\leq\T_{u,v},
\]
where each $\T_{u,v}$ denotes the type rooted at that vertex in $\widehat{\mathcal H}_{R_*}$.
Selections retain all operand summands; inputs retain matching branches with surviving continuation pairs and preserve the sender set. Hence $R_*=C$.

For leastness, let $u\leq\T$ and $v\leq\T$. For selections, write
\[
 u=\sum_{i\in I}\T'_{1,i},\qquad
 v=\sum_{j\in J}\T'_{2,j},\qquad
 \T=\sum_{k\in K}\T''_k.
\]
Inversion gives
\[
 \forall(a,i)\in I\uplus J\;\exists k\in K:\quad\T'_{a,i}\leq\T''_k.
\]
For inputs, write
\[
 \T=\pprocinnew{\pp_k}{k\in K}{\ell_k}{\Basic_k}{\T'_k}.
\]
Inversion gives
\[
\begin{gathered}
 \forall k\in K\;\exists(u',v')\in V_1\times V_2:\\
 (\pp_k,\ell_k,\Basic_k,(u',v'))\in M(u,v)
 \ \land\ u'\leq\T'_k\ \land\ v'\leq\T'_k.
\end{gathered}
\]
Hence each such $(u',v')$ lies in $C=R_*$, and every input
prefix of $\T$ occurs in $\T_{u,v}$.
By coinduction
\[
 u\leq\T\land v\leq\T\Longrightarrow\T_{u,v}\leq\T.
\]
Restricting to vertices reachable from $\T_{r_1,r_2}$ preserves its unfolding, as does $\mathsf{Read}$. The algorithm therefore returns a least upper bound exactly when one exists.

\textbf{Complexity.}
By the graph construction, $|V_i|\leq|\T_i|$ and $|E_i|=\mathcal O(|\T_i|)$. For each pair $(u,v)$, matching branches, initialising sender counters, and constructing the result vertex takes $\mathcal O((1+\operatorname{od}(u))(1+\operatorname{od}(v)))$. Summing over the entire product graph gives
\[
 \sum_{u\in V_1}\sum_{v\in V_2}
   \mathcal O\bigl((1+\operatorname{od}(u))(1+\operatorname{od}(v))\bigr)
 \quad=\quad
 \mathcal O\bigl((|V_1|+|E_1|)(|V_2|+|E_2|)\bigr)
 =\mathcal O(|\T_1||\T_2|).
\]
Each pair is deleted at most once and each reverse dependency is processed at most once. Restricting to the root-reachable graph is linear in its size. Hence definedness and the shared witness graph are computed in $\mathcal O(|\T_1||\T_2|)$ time.

To read out the witness type, the terminal case is immediate. Otherwise each operand contains a branch node not represented in $V_i$, so $|V_i|\leq|\T_i|-1$. Hence
\[
 |V(\widehat{\mathcal H}_{R_*})|
 \leq \;|V_1||V_2|+|V_1|+|V_2|\;
 \leq |\T_1||\T_2|-1.
\]
Along a readout path, each vertex contributes at most two constructor occurrences: a distribution may be emitted once under a selection join and once in its original context; ancestor lookup prevents further expansion. Thus the number of constructor occurrences along a path is at most
\[
 2|V(\widehat{\mathcal H}_{R_*})|
 \leq 2|\T_1||\T_2|.
\]
All out-degrees are at most $d$, except a selection-join vertex, whose out-degree is at most $2d$. At most one such vertex occurs per path, contributing only a constant factor. Combining this branching bound with the depth bound, the readout has
\[
 \mathcal O\!\left((\max\{2,d\})^{2|\T_1||\T_2|}\right)
\]
syntax nodes. The balanced ancestor map has size at most $2|\T_1||\T_2|$, so each lookup or update costs $\mathcal O(\log(|\T_1||\T_2|+1))$. All remaining work is linear in the emitted syntax, giving the stated witness-construction bound.
\end{proof}

\mergedecidable*
\begin{proof}
Use Algorithm~\ref{alg:merge}, with \textsc{Join} implemented
by \cref{lem:supdecidable}.

\textbf{Termination:}
Guardedness makes head unfolding effective, and every \textsc{Join} call terminates. Only \rulename{m-$\Sigma$} calls \textsc{Merge} recursively; its arguments are output distributions, so these calls use \rulename{m-$\oplus$} and cannot invoke \rulename{m-$\Sigma$} again. All loops are finite.

\textbf{Correctness:}
Inspect the unfolded constructors, aligning prefix indices
by participant and label.
\begin{itemize}
  \item[\rulename{m-end}]
  For $(\tend,\tend)$, the algorithm returns $\tend$.

  \item[\rulename{m-$\&$}]
  By \cref{lem:supdecidable}, the algorithm computes exactly
  \[
    K=\{k\in I\cap J\mid
      \Basic_k=\Basic'_k
      \land \T_k\sqcup\T'_k\text{ is defined}\},
  \]
  with $\T''_k=\T_k\sqcup\T'_k$ for every $k\in K$.
  It succeeds precisely when
  \[
    \{\q_i\}_{i\in I}
    =\{\q_j\}_{j\in J}
    =\{\q_k\}_{k\in K},
  \]
  returning
  $\pprocinnew{\q_k}{k\in K}{\ell_k}{\Basic_k}{\T''_k}$.

  \item[\rulename{m-$\oplus$}]
  The algorithm succeeds precisely when
  \[
    \forall k\in I\cap J:\quad
    \Basic_k=\Basic'_k
    \land \T_k\sqcup\T'_k\text{ is defined}.
  \]
  It returns
  $\pprocoutty{\q_k}{k\in I\cup J}{\probipp{k}}
    {\ell_k}{\Basic''_k}{\T''_k}$,
  with the weights and continuations prescribed by the rule.
  Since $\prob\in(0,1)$, every $\probipp{k}>0$ and
  \[
    \sum_{k\in I\cup J}\probipp{k}
    =\prob\sum_{i\in I}\probi{i}
      +(\probn{1}-\prob)\sum_{j\in J}\probip{j}
    =\probn{1}.
  \]

  \item[\rulename{m-$\Sigma$}]
  For $\T_1=\sum_{i\in I}\T_i$ and
  $\T_2=\sum_{j\in J}\T'_j$, with $|I||J|>1$,
  each recursive call is covered by \rulename{m-$\oplus$}.
  Thus success is equivalent to
  \[
    \forall(i,j)\in I\times J:\quad
    \T_i\probmerge{\prob}\T'_j\text{ is defined},
  \]
  with result
  $\sum_{(i,j)\in I\times J}(\T_i\probmerge{\prob}\T'_j)$.
\end{itemize}
All other constructor pairs are rejected. Hence the algorithm
returns $\emptyset$ exactly when the merge is undefined,
and otherwise returns a witness.

\textbf{Complexity results:}
Represent intermediate types by finite graphs, performing syntax readout only at the end. For the operand vertex sets $V_1,V_2$, compute $R_*$ and the full auxiliary graph $\widehat{\mathcal H}_{R_*}$ once, as in \cref{lem:supdecidable}, in $\mathcal O(|\T_1||\T_2|)$ time.
Each continuation join is then a constant-time membership test and witness-vertex lookup.
Each pair of summand distributions is processed once, so the remaining graph construction costs at most
\[
 \sum_{u\in V_1}\sum_{v\in V_2}
 \mathcal O\bigl((1+\operatorname{od}(u))(1+\operatorname{od}(v))\bigr)
 =\mathcal O(|\T_1||\T_2|).
\]
This proves the definedness bound.

For syntax readout, the input and terminal cases inherit
\cref{lem:supdecidable}. For selection operands,
$|V_i|\leq|\T_i|-1$, hence
\[
 |V(\widehat{\mathcal H}_{R_*})|
 \leq |V_1||V_2|+|V_1|+|V_2|
 \leq |\T_1||\T_2|-1.
\]
The LUB readout argument bounds each continuation path by $|V(\widehat{\mathcal H}_{R_*})|+1$ constructor occurrences:
only the first distribution reached from a join may be emitted twice.
Merge adds at most two initial vertices, of degrees at most $d^2$ for the sum and $2d$ for the merged distribution.
The latter contributes a constant factor; the former adds one further factor of $\max\{2,d\}$. The readout therefore has $\mathcal O((\max\{2,d\})^{|\T_1||\T_2|+3})$ nodes, with
\[
 |\T_1||\T_2|+3\leq2|\T_1||\T_2|
 \qquad(|\T_1|,|\T_2|\geq2).
\]
Ancestor-map operations contribute the same $\mathcal O(\log(|\T_1||\T_2|+1))$ factor as in \cref{lem:supdecidable}, giving the stated witness bound.
\end{proof}

\begin{algorithm}[h]
\caption{Merge of $\T_1,\T_2$ at probability $\prob\in(0,1)$;
returns a merged type, or $\emptyset$ if undefined.}
\label{alg:merge}
\small
\begin{algorithmic}[1]
\Function{Merge}{$\T_1,\T_2,\prob$}
  \State $\T_1\gets\unf{\T_1}$;
         $\T_2\gets\unf{\T_2}$

  \If{$\T_1=\tend$ \textbf{and} $\T_2=\tend$}
      \State \Return $\tend$
        \Comment{\rulename{m-end}}

  \ElsIf{$\T_1=
      \pprocinnew{\q_i}{i\in I}{\ell_i}{\Basic_i}{\T_i}$
      \textbf{and} $\T_2=
      \pprocinnew{\q_j}{j\in J}{\ell_j}{\Basic'_j}{\T'_j}$}
      \Comment{\rulename{m-$\&$}}
      \State $K\gets\emptyset$
      \ForAll{$k\in I\cap J$ with $\Basic_k=\Basic'_k$}
          \State $\T''_k\gets\Call{Join}{\T_k,\T'_k}$
          \If{$\T''_k\neq\emptyset$}
              \State $K\gets K\cup\{k\}$
          \EndIf
      \EndFor
      \If{$\{\q_i\}_{i\in I}\neq\{\q_j\}_{j\in J}$
          \textbf{or}
          $\{\q_k\}_{k\in K}\neq\{\q_i\}_{i\in I}$}
          \State \Return $\emptyset$
      \EndIf
      \State \Return
          $\pprocinnew{\q_k}{k\in K}{\ell_k}{\Basic_k}{\T''_k}$

  \ElsIf{$\T_1=
      \pprocoutty{\q_i}{i\in I}{\probi{i}}
          {\ell_i}{\Basic_i}{\T_i}$
      \textbf{and} $\T_2=
      \pprocoutty{\q_j}{j\in J}{\probip{j}}
          {\ell_j}{\Basic'_j}{\T'_j}$}
      \Comment{\rulename{m-$\oplus$}}
      \ForAll{$k\in I\cap J$}
          \If{$\Basic_k\neq\Basic'_k$}
              \State \Return $\emptyset$
          \EndIf
          \State $\T''_k\gets\Call{Join}{\T_k,\T'_k}$
          \If{$\T''_k=\emptyset$}
              \State \Return $\emptyset$
          \EndIf
          \State $(\probipp{k},\Basic''_k)\gets
              (\prob\cdot\probi{k}
               +(\probn{1}-\prob)\cdot\probip{k},
               \Basic_k)$
      \EndFor
      \ForAll{$i\in I\setminus J$}
          \State $(\probipp{i},\Basic''_i,\T''_i)\gets
              (\prob\cdot\probi{i},\Basic_i,\T_i)$
      \EndFor
      \ForAll{$j\in J\setminus I$}
          \State $(\probipp{j},\Basic''_j,\T''_j)\gets
              ((\probn{1}-\prob)\cdot\probip{j},
               \Basic'_j,\T'_j)$
      \EndFor
      \State \Return
          $\pprocoutty{\q_k}{k\in I\cup J}{\probipp{k}}
              {\ell_k}{\Basic''_k}{\T''_k}$

  \ElsIf{$\T_1=\sum_{i\in I}\T_i$
      \textbf{and} $\T_2=\sum_{j\in J}\T'_j$
      \textbf{and} $|I|\cdot|J|>1$}
      \Comment{\rulename{m-$\Sigma$}}
      \ForAll{$(i,j)\in I\times J$}
          \State $\T''_{i,j}\gets
              \Call{Merge}{\T_i,\T'_j,\prob}$
          \If{$\T''_{i,j}=\emptyset$}
              \State \Return $\emptyset$
          \EndIf
      \EndFor
      \State \Return $\sum_{(i,j)\in I\times J}\T''_{i,j}$

  \Else
      \State \Return $\emptyset$
  \EndIf
\EndFunction
\end{algorithmic}
\end{algorithm}

\decidability*

\begin{proof}
$\der{}{\N}{\Delta}$ is decidable for a \emph{given} typing context $\Delta$, with supplied receive-sort and recursion-type annotations, by a single bidirectional traversal of $\N$. Recursive-variable types come from these annotations: at \rulename{t-rec}, $\mu X.\PP$ synthesizes its annotation $\T_X$ after checking $\PP$ against $\T_X$ under $\Gamma,X:\T_X$; the matching \rulename{t-var} reads $\T_X$ back from $\Gamma$.

The remaining rules other than \rulename{t-sub} are syntax-directed on $\N$ (\cref{fig:pmpst_typrules}), and the traversal runs in two modes. In \emph{checking} mode, a demanded type is supplied by a recursion annotation or session boundary. In \emph{synthesis} mode, a type is produced from the process itself; this is used at leaves, prefixes, and the operands of conditionals and flips. Synthesis either returns a type or rejects; recursion variables are looked up in $\Gamma$, never unfolded.

For \rulename{t-cond}, synthesize $\T_1,\T_2$ and compute $\T_1\sqcup\T_2$ using \cref{lem:supdecidable}. At $\flip{\prob}{\PP_1}{\PP_2}$, the merge is not invertible, so the operands are likewise \emph{synthesised}: the sub-traversals return $\T_1,\T_2$, and a witness of $\T_1\probmerge{\prob}\T_2$ is computable by \cref{prop:mergedecidable}. Undefined joins or merges cause rejection.
Rule \rulename{t-sub} is discharged by a single subtyping check (\cref{prop:decidablesubtype}) wherever a synthesized type meets a demanded one. The side-condition $\safe{\Delta}$ of \rulename{b-sess} is decidable by \cref{prop:decidablesafe}.

Soundness follows from the rules; completeness for fixed annotations and sort choices follows by structural induction: the synthesized type lies below every derivable type, using the least-upper-bound property and \cref{lemma:monotonicmerge}.
Each traversal visits every syntactic node at most once and performs decidable operations; any remaining basic-sort choices range over a finite set. Hence typing respecting the supplied annotations is decidable (cf.\ the analogous argument in \cite[Thm.~4.11]{Scalas2019}).
\end{proof}

\section{Appendix for Section \ref{sec:properties}}
\label{app:properties}

\subsection{ Examples for ~\cref{sec:properties}}
\begin{example}
    \label{exam:mdpfull}
    The roles of~\cref{exam:mdp} are typed as follows (payloads $\ints$; sub-types named for the repeated
$Q$ and the larger continuations, each within the scope of the $\mu\ty$ of the type
that uses it):
\[
    \small
\begin{array}{r@{\;\;=\;\;}l}
\T_{\pchan u} &
  \oplus\,\pchan{s}!\labels{incr}\langle\ints\rangle.\,
  \mu\ty.\,
  \&\,\pchan{c}?\labels{res}(\ints).\,
  \oplus\,\pchan{s}!\labels{incr}\langle\ints\rangle.\,\ty
\\[2.5mm]
\T_{\pchan s} &
  \mu\ty.\;\&\Big\{\,
    \pchan{c}?\labels{crit}.\tend,\;\;
    \pchan{c}?\labels{disconn}.\,\mu\ty'.\,
      {\textstyle\sum}\big\{\,
        \pchan{c}!\labels{retry}.\,\ty',\;
        \pchan{c}!\labels{conn}.\,\T_Q
    \,\big\},\;\;
    \pchan{c}?\labels{err}(\ints).\,\T_Q,\;\;
    \pchan{u}?\labels{incr}(\ints).\,\T_Q
  \,\Big\}
\\[2.5mm]
\T_{\pchan c} &
  \mu\ty.\;\&\Big\{\,
    \pchan{s}?\labels{upgrade}.\,\T_{\mathsf{up}},\;\;
    \pchan{s}?\labels{incr}(\ints).\,
      {\textstyle\sum}\big\{\, \T_{\mathsf{flip}}\;,\;
        \pchan{u}!\labels{res}\langle\ints\rangle.\,\ty \,\big\}
  \,\Big\}
\end{array}
\]
where
\[
    \small
\begin{array}{r@{\;\;=\;\;}l}
\T_Q &
  \oplus\big\{\,
    \probn{0.9}\!:\pchan{c}!\labels{incr}\langle\ints\rangle.\,\ty,\;\;
    \probn{0.1}\!:\pchan{c}!\labels{upgrade}.\,\mu\ty''.\,
      \oplus\pchan{c}!\labels{incr}\langle\ints\rangle.\,
      \&\pchan{u}?\labels{incr}(\ints).\,\ty''
  \,\big\}
\\[2.5mm]
\T_{\mathsf{up}} &
  \mu\ty'.\;\&\pchan{s}?\labels{incr}(\ints).\,
  \oplus\pchan{u}!\labels{res}\langle\ints\rangle.\,\ty'
\\[2.5mm]
\T_{\mathsf{flip}} &
  \oplus\big\{
    \probn{0.8}\!:\pchan{u}!\labels{res}\langle\ints\rangle.\,\ty,
    \probn{0.1}\!:\pchan{s}!\labels{err}\langle\ints\rangle.\,\ty,
    \probn{0.05}\!:\pchan{s}!\labels{crit}.\tend,
    \probn{0.05}\!:\pchan{s}!\labels{disconn}.\,\mu\ty''.
      \&\big\{
        \pchan{s}?\labels{retry}.\ty'',
        \pchan{s}?\labels{conn}.\ty
      \big\}
  \big\}
\end{array}
\]
\end{example}

\subsection{Proofs for \Cref{sec:sr}}
\label{app:sr}

\begin{restatable}[Safety]{lemma}{safeproperties}
\label{lemma:safeproperties}
The following properties hold:
\begin{enumerate}
    \item\label{lem:item0:safeproperties} If $\safe{\Delta', \pp : \T}$ and $\T = \T_1 \probmerge{\prob} \T_2$ then $\safe{\Delta', \pp : \T_1}$
\item If $\safe{\Delta}$ and $\Delta = \Delta', \Delta''$, then $\safe{\Delta'}$.
\item\label{lem:item3:safeproperties} If $\safe{\Delta}$ and $\Delta' \subt \Delta$, then $\safe{\Delta'}$.
\end{enumerate}
\end{restatable}

\begin{proof}
  We prove each case individually:
  \begin{enumerate}
      \item
We use \cref{lem:item3:safeproperties}, established independently,
and induct on the finite derivation of
$\T_1\probmerge{\prob}\T_2=\T$.
All types are read up to unfolding.

First, the context reduction rules and the closure requirement of \cref{def:safety} give
\[
\begin{aligned}
    \safe{\Delta',\pp:\sum_{i\in I}\T_i}
    &\iff
    \forall i\in I:\;
        \safe{\Delta',\pp:\T_i},
    \\
    \safe{\Delta',\pp:
        \pprocoutty{\q_i}{i\in I}{\probi{i}}
            {\ell_i}{\Basic_i}{\T_i}}
    &\iff
    \forall i\in I:\;
        \safe{\Delta',\pp:
            \oplus\probn{1}:\q_i!\ell_i
                \langle\Basic_i\rangle.\T_i},
\end{aligned}
\]
where each $\T_i$ is a selection distribution. We proceed by case analysis on the last merge rule.
\begin{itemize}
    \item[\rulename{m-end}]
    $\T_1=\T=\tend$, so the claim is immediate.

    \item[\rulename{m-$\&$}]
    Write
    \[
    \begin{aligned}
        \T_1
        &=\pprocinnew{\q_i}{i\in I}
            {\ell_i}{\Basic_i}{\T_{1,i}},
        \\
        \T
        &=\pprocinnew{\q_k}{k\in K}
            {\ell_k}{\Basic_k}
            {(\T_{1,k}\sqcup\T_{2,k})}.
    \end{aligned}
    \]
    The rule gives
    \[
        K\subseteq I,
        \qquad
        \{\q_i\}_{i\in I}=\{\q_k\}_{k\in K},
        \qquad
        \forall k\in K:\;
        \T_{1,k}\subt\T_{1,k}\sqcup\T_{2,k}.
    \]
    Hence $\T_1\subt\T$ by \rulename{sub-$\&$}.
    Applying \cref{lem:item3:safeproperties} to
    \[
        (\Delta',\pp:\T_1)\subt(\Delta',\pp:\T)
    \]
    yields $\safe{\Delta',\pp:\T_1}$.

    \item[\rulename{m-$\oplus$}]
    Write
    \[
    \begin{aligned}
        \T_1
        &=\pprocoutty{\q_i}{i\in I}{\probi{i}}
            {\ell_i}{\Basic_i}{\T_{1,i}},
        \\
        \T
        &=\pprocoutty{\q_k}{k\in I\cup J}{\probipp{k}}
            {\ell_k}{\Basic_k}{\T''_k}.
    \end{aligned}
    \]
    For every $i\in I$, the rule gives
    \[
        \probipp{i}\geq\prob\cdot\probi{i}>0,
        \qquad
        \T_{1,i}\subt\T''_i,
    \]
    since $\T''_i$ is either $\T_{1,i}$ or
    $\T_{1,i}\sqcup\T_{2,i}$.
    By the second safety equivalence above,
    \[
        \safe{\Delta',\pp:
            \oplus\probn{1}:\q_i!\ell_i
                \langle\Basic_i\rangle.\T''_i}.
    \]
    Moreover, \rulename{sub-$\oplus$} gives
    \[
        \oplus\probn{1}:\q_i!\ell_i
            \langle\Basic_i\rangle.\T_{1,i}
        \subt
        \oplus\probn{1}:\q_i!\ell_i
            \langle\Basic_i\rangle.\T''_i.
    \]
    By \cref{lem:item3:safeproperties}, every committed
    alternative of $\T_1$ is safe. The same equivalence
    therefore yields $\safe{\Delta',\pp:\T_1}$.

    \item[\rulename{m-$\Sigma$}]
    Here
    \[
        \T_1=\sum_{i\in I}\T_{1,i},
        \qquad
        \T_2=\sum_{j\in J}\T_{2,j},
        \qquad
        \T=\sum_{(i,j)\in I\times J}\T''_{i,j},
    \]
    with
    $\T_{1,i}\probmerge{\prob}\T_{2,j}=\T''_{i,j}$.
    By the first safety equivalence,
    \[
        \forall(i,j)\in I\times J:\;
        \safe{\Delta',\pp:\T''_{i,j}}.
    \]
    Fix any $j\in J$.
    Applying the induction hypothesis to each premise gives
    \[
        \forall i\in I:\;
        \safe{\Delta',\pp:\T_{1,i}}.
    \]
    The first equivalence then gives
    $\safe{\Delta',\pp:\T_1}$.
\end{itemize}

\item Write $\Delta = \Delta', \Delta''$ and suppose, towards a contradiction, $\neg\safe{\Delta'}$.
    Since $\safe{\cdot}$ is the largest safety property, and clause~(3) of \cref{def:safety} close a safety property under all transitions, $\safe{\Delta_0}$ holds iff every context reachable from $\Delta_0$ satisfies the compatibility clause~(2); hence there is a finite sequence $\Delta' \predcont{}{}^{*} \Delta_1'$ with $\Delta_1'$ violating clause~(2), i.e.\ $\Delta_1' \predcont{\actionprocout{\pp}{\q}{\labelsell}{\Basic}}{\probn{1}}$ and $\Delta_1' \predcont{\actionprocin{\q}{\pp}{\labelsell'}{\Basic'}}{\probn{1}}$ but $\Delta_1' \not\predcont{\actiontau{\pp}{\q}{\labelsell}}{\probn{1}}$.
    Framing each step by \rulename{ct-sc} (induction on the length of the sequence, $\Delta''$ unchanged throughout) lifts it to $\Delta', \Delta'' \predcont{}{}^{*} \Delta_1', \Delta''$, and likewise lifts the two enabled actions to $\Delta_1', \Delta''$.
    By $\safe{\Delta', \Delta''}$ and closure, $\safe{\Delta_1', \Delta''}$, so clause~(2) gives $\Delta_1', \Delta'' \predcont{\actiontau{\pp}{\q}{\labelsell}}{\probn{1}}$. But its subjects $\pp, \q$ lie in $\dom{\Delta_1'}$, and every rule of \cref{fig:pmpst_tct} mentions only the entries of its acting participants, so the synchronisation restricts to $\Delta_1' \predcont{\actiontau{\pp}{\q}{\labelsell}}{\probn{1}}$ --- a contradiction.

    \item Let
    \[
      R \;=\; \{\,\Delta' \mid \exists \Delta \st \Delta' \subt \Delta \text{ and } \safe{\Delta}\,\}.
    \]
    By reflexivity of $\subt$ (\cref{lem:subtpreorder}), $R$ contains every safe context and $\emptyset$, giving condition~(1) of \cref{def:safety}; we show $R$ is a safety property, whence $R \subseteq \safe{\cdot}$ and the claim follows. So let $\Delta' \in R$, witnessed by $\Delta' \subt \Delta$ with $\safe{\Delta}$.
    By synchronisation (\cref{lemma:sync}), $\Delta \predcontndstar{}{} \hat{\Delta}$ with $\Delta' \synct \hat{\Delta}$; note $\safe{\hat{\Delta}}$ by closure of $\safe{\Delta}$, and $\hat{\Delta}$ equally witnesses $\Delta' \in R$.

    \emph{Compatibility~(2).} Suppose $\Delta' \predcont{\actionprocout{\pp}{\q}{\labelsell}{\Basic}}{\probn{1}}$ and $\Delta' \predcont{\actionprocin{\q}{\pp}{\labelsell'}{\Basic'}}{\probn{1}}$.
    By \rulename{ct-out} the former forces $\unf{\Delta'(\pp)}$ to be a singleton selection, so $\Delta'(\pp) \not\ndstep$ and, by \cref{def:synct}, $\Delta'(\pp) \subt \hat{\Delta}(\pp)$ is by \rulename{sub-$\oplus$}: $\unf{\hat{\Delta}(\pp)}$ is the \emph{same} singleton, and the output action is enabled at $\hat{\Delta}$.
    Moreover $\Delta'(\q) \subt \hat{\Delta}(\q)$ is by \rulename{sub-$\&$}, so $\unf{\hat{\Delta}(\q)}$ is a branching from $\pp$ --- nonempty by well-formedness --- enabling some input at $\q$ from $\pp$.
    Clause~(2) of $\safe{\hat{\Delta}}$ then gives $\hat{\Delta} \predcont{\actiontau{\pp}{\q}{\labelsell}}{\probn{1}}$; in particular $\hat{\Delta}(\q)$ offers $(\labelsell, \Basic)$, and by contravariance of \rulename{sub-$\&$} so does $\Delta'(\q)$, whence $\Delta' \predcont{\actiontau{\pp}{\q}{\labelsell}}{\probn{1}}$.

    \emph{Closure~(3).} If $\Delta' \predcont{\alpha}{t} \Delta_1'$, then by one-step preservation (\cref{lemma:syncstep}) there is $\hat{\Delta}_1$ with $\Delta_1' \synct \hat{\Delta}_1$, reached from $\hat{\Delta}$ by finitely many transitions; each preserves safety by closure, so $\safe{\hat{\Delta}_1}$, and $\hat{\Delta}_1$ witnesses $\Delta_1' \in R$.
\end{enumerate}
\end{proof}

\begin{lemma}[Substitution Lemma]\label{lemma:subst}
    \mbox{}
\begin{enumerate}
\item If $\der{\Gamma,x{:}\Basic}{\PP}\T$ then
$\der{\Gamma, y{:}\Basic}{{\PP}\subst{y}{x}}\T$;
and
\item if $\der{\Gamma,x{:}\Basic}{\PP}\T$ and $\der{\Gamma}{\val}{\Basic}$, then
$\der{\Gamma}{{\PP}\subst{\val}{x}}\T$.
and
\item if $\der{\Gamma,X{:}\T'}{\PP}\T$ and $\der{\Gamma}{\Q}{\T'}$, then
$\der{\Gamma}{{\PP}\subst{\Q}{X}}\T$.
\end{enumerate}
\end{lemma}
\begin{proof}
Parts (1) and (2) are by straightforward induction on the typing derivation; the cases
for the standard constructors follow \cite{Ghilezan2019} (see also \cite{Scalas2019,YH2024}).
We show the one case specific to our calculus, \rulename{t-flip}.

\emph{Case \rulename{t-flip}.} By inversion, $\der{\Gamma,x{:}\Basic}{\flip{\prob}{\PP_1}{\PP_2}}{\T}$
gives $\T_1,\T_2$ with $\der{\Gamma,x{:}\Basic}{\PP_i}{\T_i}$ and
$\T = \T_1 \probmerge{\prob}\T_2$. By the $\IH$,
$\der{\Gamma}{\PP_i\subst{\val}{x}}{\T_i}$ for $i=1,2$, so by \rulename{t-flip}
$\der{\Gamma}{\flip{\prob}{\PP_1\subst{\val}{x}}{\PP_2\subst{\val}{x}}}{\T}$,
i.e.\ $\der{\Gamma}{(\flip{\prob}{\PP_1}{\PP_2})\subst{\val}{x}}{\T}$.

Part (3) is analogous, replacing value substitution by process substitution
$\subst{\Q}{X}$ throughout.
\end{proof}

\begin{lemma}[Subject Precongruence]\label{lemma:precong}
Suppose $\der{}{\N}{\Delta}$ and $\N \spcon \N'$. Then
$\der{}{\N'}{\Delta}$.
\end{lemma}
\begin{proof}
By induction on the derivation of $\N \spcon \N'$.

\begin{itemize}
    \item \emph{Case} $\pa\pp{\mu X.\PP} \pc \N'' \spcon \pa\pp{\PP\subst{\mu X.\PP}{X}} \pc \N''$
        (recursion unfolding). By inversion of \rulename{b-sess}, $\der{}{\pa\pp{\mu X.\PP \pc \N''}}{\Delta}$
gives $\Delta = \pp{:}\T, \Delta'$ with $\der{}{\mu X.\PP}{\T}$ and $\der{}{\N''}{\Delta'}$.
Inverting \rulename{t-rec}, $\der{\, X{:}\T}{\PP}{\T}$. Since also $\der{}{\mu X.\PP}{\T}$,
\cref{lemma:subst}(3) yields $\der{}{\PP\subst{\mu X.\PP}{X}}{\T}$. Recomposing with
$\der{}{\N''}{\Delta'}$ by \rulename{b-sess} gives $\der{}{\pa\pp{\PP\subst{\mu X.\PP}{X} \pc \N''}}{\Delta}$.

\item \emph{Case} $\prod_{i\in I}\pa\ppi{i}{\PP_i} \spcon \prod_{j\in J}\pa\ppi{j}{\PP_j}$
    with $I$ a permutation of $J$ (parallel permutation). Immediate, as \rulename{b-sess}
    types a session from the (unordered) set of its components, so both sides are typed by
    the same $\Delta$.
\end{itemize}
The remaining closure cases are immediate from the IH.
\end{proof}

\begin{theorem}[Type Inversion]
    \label{lemma:typeinversion}
The following statements hold:
  \begin{enumerate}
    \item $\der\Gamma{\mu X.\PP}\T$ implies $\exists \T' \st \der{\Gamma,X:\T'}{\PP}{\T'}$ with $ \T' \subt \T $.

    \item $\der{\Gamma}{\flip{\prob}{\PP_1}{\PP_2}}{\T}$ implies $\exists \T_1, \T_2\;.$ $\der{\Gamma}{\PP_1}{\T_1}$ and $\der{\Gamma}{\PP_2}{\T_2}$ with $ \T_1 \probmerge{\prob} \T_2 \subt \T$.

    \item $\der{\Gamma}{\procout{\pp}{\ell}{\e}{\PP}}{\T}$ implies $\der{\Gamma}{\e}{\Basic}$ and $\der{\Gamma}{\PP}{\T'}$ with $\pprocoutty{\pp}{}{\probn{1}}{\ell}{\Basic}{\T'} \subt \T$  . %

    \item $\der{\Gamma}{\psyntin{i\in I}{\q}{i}{x_i}{\PP_i}}{\T}$ implies $\forall i \in I : \exists \T_i \;.\;$ $\der{\Gamma,x_i:\Basic_i}{\PP_i}\T_i$ with $\pprocinnew{\pchan{q_i}}{i\in I}{\ell_i}{\Basic_i}{\T_i} \subt \T$

    \item $\der\Gamma{{\cond{\e}{\PP_1}{\PP_2}}}{\T}$ implies $\der\Gamma\e\tbool$ , $\der{\Gamma}{\PP_1}{\T'}$ and $\der{\Gamma}{\PP_2}{\T'}$ with $ \T' \subt \T $.

    \item $\der{}{\prod\nolimits_{i\in I}\pa {\pp_i}\PP_i}\{ \pp_i : \T_i \}_{i \in I}$ implies $\safemp{\set{ \pp_i : \T_i}_{i \in I}}$ and $\forall i\in I$ we have $\der{}{\PP_i}{\T_i} $ %

  \end{enumerate}
\end{theorem}

\begin{proof}
  The proof follows as standard where subtyping is introduced by the application of $\rulename{t-sub}$ rule.
    \begin{enumerate}
      \item Follows from $\rulename{t-rec}$ .

      \item Follows from $\rulename{t-flip}$.

      \item Follows from $\rulename{t-sel}$.

      \item Follows from $\rulename{t-bra}$.

      \item Follows from $\rulename{t-cond}$.

      \item Follows from $\rulename{b-sess}$.
    \end{enumerate}
  \end{proof}

\subred*

\begin{proof}
By induction on the derivation $ \N \pred{\prob}  \N_1$.
\begin{itemize}
    \item \rulename{r-head}:

        As $\N \predcont{}{\prob}\N_1$ by \rulename{r-head} we want to show case (3) of the statement.
        \setcounter{equation}{0}
        By inversion of \rulename{r-head} we have:
            \begin{align}
            &\N = \pa\pp{\flip{\prob}{\PP_1}{\PP_2}}\pc\N'\\
            &\N_1 = \pa\pp\PP_1\pc \N'\\
            \text{(and }\exists \N_2 \st &\N_2 = \pa\pp \PP_2 \pc \N'\quad \text{ by }\rulename{r-tail})
            \end{align}
              let $\Delta = (\pp : \T, \Delta')$ then:
              \begin{equation}
              \der{\varnothing}{\flip{\prob}{\PP_1}{\PP_2}}{\T}
              \end{equation}
              since $\der{}{\N}{\Delta}$ and applying case (6) of type inversion (\cref{lemma:typeinversion})

        To recover the typing of $\pp$ we apply case (2) of \cref{lemma:typeinversion}:
              \begin{align}
              \exists\T_1, \T_2\st \;\;&\der{}{\PP_1}{\T_1}\\
              &\der{}{\PP_2}{\T_2}\\
              &\T_1 \probmerge{\prob}\T_2 \subt \T
              \end{align}

              since $\safe{\pp:\T,\Delta'}$ and by case (1) of \cref{lemma:safeproperties} we have:
        \begin{align}
        &\der{}{\N_1}{(\pp : \T_1,\Delta')}\\
        &\der{}{\N_2}{(\pp : \T_2,\Delta')}
        \end{align}

        Finally, since $\T_1 \probmerge{\prob} \T_2 \subt \T$, we get:
        \begin{equation}
            (\pp : \T_1 \probmerge{\prob} \T_2, \Delta') \subt \Delta
        \end{equation}

    \item \rulename{r-tail}: analogous to \rulename{r-head}.

    \item \rulename{r-comm}:
    By inversion of \rulename{r-comm} we have:
            \setcounter{equation}{0}
            \begin{align}
            \label{eq:21}
            &\N =
                \pa\pp{\procout{\q}{\ell}{\e}{\PP}}
                \pc
                \pa\q{\psyntin{i\in I}{\pp_i}{i}{x_i}{\Q_i}}
                \;\pc\; \N',\\
                \exists j \in I :\exists v\st \eval{\e}{v} \land \pp = \ppi{j}\land\labelselli{} = \labelselli{j}\land\; &\N_1 =
            \pa\pp{{\PP}}\;\pc\;\pa\q{\Q_j\subst{\val}{\x_j}}
            \;\pc\; \N'
            \end{align}
        Let $\Delta = \pp : \T_{\pp}, \q : \T_{\q}, \Delta'$, then by case (6) of type inversion (\cref{lemma:typeinversion}):

        \begin{equation}
        \label{eq:22}
      \inferrule*[left=\rulename{b-sess}]
      {
      \der{}{{\procout{\q}{\ell}{\e}{\PP}}}{\T_{\pp}}
      \quad
      \der{}{{\psyntin{i\in I}{\pp_i}{i}{x_i}{\Q_i}}}{\T_{\q}}
      \quad
      \forall (\pr:\T_{\pr}) \in \Delta': \;\der{}{\PP_{\pr}}{\T_{\pr}}
      \quad
      \safe{\Delta}
      }
      {\der{}{\N}\Delta
      }
      \end{equation}

    To recover the typing of $\q$, we apply case (4) of type inversion (\cref{lemma:typeinversion}):

        \begin{equation}
        \label{eq:23}
        \inferrule*[left=\rulename{t-sub}]
        {
            \inferrule*[left=\rulename{t-bra}]
            {
            \forall i \in I
            \quad
            \der{\x_i : \Basic_i}{\Q_i}{\T_{\q, i}}
            }
            {
                \der{}{\pa\q{\psyntin{i\in I}{\pp_i}{i}{x_i}{\Q_i}}}{\pprocinnew{\ppi{i}}{i\in I}{\ell_i}{\Basic_i}{\T_{\q,i}}}
            }
            \quad
            \pprocinnew{\ppi{i}}{i\in I}{\ell_i}{\Basic_i}{\T_{\q,i}} \subt \T_{\q}
        }
        {
        \der{}{\pa\q{\psyntin{i\in I}{\pp_i}{i}{x_i}{\Q_i}}}{\T_{\q}}
        }
        \end{equation}
        where $\T_{\q} =\pprocinnew{\pp_k}{k\in K}{\ell_k}{\Basic_k}{\T_{\q,k}'}$ with $K \subseteq I$
        and $\T_{\q, k} \subt \T_{\q, k}'$ for all $k\in K$ (by \rulename{sub-$\&$}).

         To recover the typing of $\pp$, we apply case (3) of type inversion (\cref{lemma:typeinversion}):

        \begin{equation}
        \label{eq:25}
        \inferrule*[left=\rulename{t-sub}]
        {
            \inferrule*[left=\rulename{t-sel}]
            {
            \der{}{\PP}{\T_{\pp, j}}
            }
            {
                \der{}{\pa\pp{\procout{\q}{\labelselli{j}}{\e}{\PP}}}{\pprocoutty{\q}{}{\probn{1}}{\labelselli{j}}{\Basic}{\T_{\pp,j}}}
            }
            \quad
            \pprocoutty{\q}{}{\probn{1}}{\labelselli{j}}{\Basic}{\T_{\pp,j}} \subt \T_{\pp}
        }
        {
            \der{}{\pa\pp{\procout{\q}{\labelselli{j}}{\e}{\PP}}}{\T_{\pp}}
        }
        \end{equation}
        where $\T_\pp$ is a supertype of $\pprocoutty{\q}{}{\probn{1}}{\labelselli{j}}{\Basic}{\T_{\pp,j}}$, by either \rulename{sub-$\Sigma$} or \rulename{sub-$\oplus$}:
        \begin{equation}
            \label{eq:sr1}
        \T_{\pp} =
        \pprocoutty{\q}{}{\probn{1}}{\labelselli{j}}{\Basic}{\T_{\pp,j}'}
        +
    \probpprocoutty{m \in M}{\pp_m}{n \in N_m}{\probi{m,n}}{\ell_m}{\Basic_m}{\T_{m,n}} \qquad \text{by \rulename{sub-$\Sigma$}}
        \end{equation}
        or
        \begin{equation}
            \label{eq:sr2}
        \T_{\pp} = \pprocoutty{\q}{}{\probn{1}}{\labelselli{j}}{\Basic}{\T_{\pp,j}'}\qquad \text{by \rulename{sub-$\oplus$}}
        \end{equation}
        in both cases $\T_{\pp,j} \subt \T_{\pp,j}'$; note that in both shapes the $j$-branch is the singleton selection at $\probn{1}$, \rulename{sub-$\oplus$} preserving probabilities exactly --- in particular $\T_\pp$ is always \emph{dirac} (point mass $\probn{1}$) so no \rulename{ct-prob} step arises.

        We will now argue that the type context can either step as:
        \begin{equation}
        \Delta \predcontnd{\actionnd{\pp}} (\pp : \pprocoutty{\q}{}{\probn{1}}{\labelselli{j}}{\Basic}{\T_{\pp,j}'}, \q : \T_\q, \Delta') \predcont{\actiontau{\pp}{\q}{\labelselli{j}}}{\probn{1}} \Delta_1
        \end{equation}
        (or directly):
        \begin{equation}
        \Delta \predcont{\actiontau{\pp}{\q}{\labelselli{j}}}{\probn{1}} \Delta_1
        \end{equation}
        Firstly if $\T_\pp =
        \pprocoutty{\q}{}{\probn{1}}{\labelselli{j}}{\Basic}{\T_{\pp,j}'}
        +
        \probpprocoutty{m \in M}{\pp_m}{n \in N_m}{\probi{m,n}}{\ell_m}{\Basic_m}{\T_{m,n}}$ then\\[1mm]
        $\Delta \predcontnd{\actionnd{\pp}} (\pp : \pprocoutty{\q}{}{\probn{1}}{\labelselli{j}}{\Basic}{\T_{\pp,j}'}, \q : \T_\q, \Delta')$ as:
        \begin{equation}
        \inferrule*[left=\rulename{ct-sc}]
        {
            \inferrule*[left=\rulename{ct-nd}]
            {
                (\pprocoutty{\q}{}{\probn{1}}{\labelselli{j}}{\Basic}{\T_{\pp,j}'}) \in \T_\pp
                \quad
                |M| + 1 > 1
            }
            {
                \pp : \T_{\pp} \predcontnd{\actionnd{\pp}} \pp : \pprocoutty{\q}{}{\probn{1}}{\labelselli{j}}{\Basic}{\T_{\pp,j}'}
            }
        }
        {
            \pp : \T_{\pp}, \q : \T_{\q}, \Delta'\predcontnd{\actionnd{\pp}} (\pp : \pprocoutty{\q}{}{\probn{1}}{\labelselli{j}}{\Basic}{\T_{\pp,j}'}, \q : \T_\q, \Delta')
        }
        \end{equation}

        We note that as $\safe{\Delta}$ and \textit{safety} propagates (\cref{def:safety}) we have $\safe{\pp :\pprocoutty{\q}{}{\probn{1}}{\labelselli{j}}{\Basic}{\T_{\pp,j}'}, \q : \T_\q, \Delta'}$.
        Next we show that:
        \begin{equation}
          (\pp : \pprocoutty{\q}{}{\probn{1}}{\labelselli{j}}{\Basic}{\T_{\pp,j}'}, \q : \T_\q, \Delta') \predcont{\actiontau{\pp}{\q}{\labelselli{j}}}{\probn{1}} \Delta_1
        \end{equation}
         (which also covers the case when $\Delta \predcont{\actiontau{\pp}{\q}{\labelselli{j}}}{\probn{1}} \Delta_1$ directly). The output action is enabled \emph{at the singleton selection}:
        \begin{equation}
            \pp : {\pprocoutty{\q}{}{\probn{1}}{\labelselli{j}}{\Basic}{\T_{\pp,j}'}} \predcont{\actionprocout{\pp}{\q}{\labelselli{j}}{\Basic_j}}{\probn{1}}
            {\pp : \T_{\pp, j}'}\quad{\small \text{ by } \rulename{ct-out}}
        \end{equation}
        and, as we require a non-empty index-set $K\neq\emptyset$ for \textit{branching} to be well-formed, the shape of $\T_{\q} = \pprocinnew{\pp_k}{k\in K}{\ell_k}{\Basic_k}{\T_{\q,k}'}$ from \cref{eq:23} must include some label $\labelselli{k}$, and thus a transition by \rulename{ct-in} is enabled:
        \begin{equation}
            \q : {\T_{\q}}
            \predcont{\actionprocin{\q}{\pp}{\labelselli{k}}{\Basic_k}}{\probn{1}}
            {\q : \T_{\q, k}'}
            \quad \text{ by }\rulename{ct-in}
        \end{equation}

        With both an output action from $\pp$ to $\q$ and an input action at $\q$ from $\pp$ enabled, \textit{safety} (\cref{def:safety}) gives $j \in K$, so \rulename{ct-in} fires at $\labelselli{j}$:
        \begin{equation}
            \label{eq:29}
            \q : {\T_{\q}}
            \predcont{\actionprocin{\q}{\pp}{\labelselli{j}}{\Basic_j}}{\probn{1}}
            {\q : \T_{\q, j}'}
            \quad \text{ by }\rulename{ct-in}
        \end{equation}

        giving derivation:
        \begin{equation}
        \inferrule*[left=\rulename{ct-sc}]
        {
            \inferrule*[left=\rulename{ct-$\tau$}]
            {
            \inferrule*[left=\rulename{ct-out}]
            { }
            {
              \pp : \pprocoutty{\q}{}{\probn{1}}{\labelselli{j}}{\Basic}{\T_{\pp,j}'}
              \predcont{\actionprocout{\pp}{\q}{\ell_j}{\Basic_j}}{\probn{1}}
              \pp : \T_{\pp,j}'
            }
              \quad
              \inferrule*[left=\rulename{ct-in}]
              {j \in K}
              {
              \q : \T_{\q} \predcont{\actionprocin{\q}{\pp}{\ell_j}{\Basic_j}}{\probn{1}}
              \q : \T_{\q,j}'
              }
            }
            {
            \pp : \pprocoutty{\q}{}{\probn{1}}{\labelselli{j}}{\Basic}{\T_{\pp,j}'}, \q : \T_\q
            \predcont{\actiontau{\pp}{\q}{\labelselli{j}}}{\probn{1}}
            \pp : \T_{\pp,j}',\q : \T_{\q,j}'
            }
        }
        {
            \pp : \pprocoutty{\q}{}{\probn{1}}{\labelselli{j}}{\Basic}{\T_{\pp,j}'}, \q : \T_{\q}, \Delta'
            \predcont{\actiontau{\pp}{\q}{\labelselli{j}}}{\probn{1}}
            \pp : \T_{\pp,j}',\q : \T_{\q,j}',\Delta'
        }
        \end{equation}

        Again, as \textit{safety} propagates we have $\safe{\Delta',{\pp : \T_{\pp, j}'},{\q : \T_{\q, j}'}}$ by \cref{def:safety}.

        The types $\T_{\pp,j}'$ and $\T_{\q,j}'$ can be used to type the continuations of $\pp$ and $\q$:

        \begin{itemize}
            \item $\der{}{\PP}{\T_{\pp,j}'}$: since from \cref{eq:25} we have:
        $\der{}{\PP}{\T_{\pp,j}}$ with $\T_{\pp,j} \subt \T_{\pp,j}'$ thus by subtyping:
        \begin{equation}
        \label{eq:212}
        \inferrule*[left=\rulename{t-sub}]
        {
        \der{}{\PP}{\T_{\pp,j}}
        \quad
        \T_{\pp,j}\subt \T_{\pp,j}'
        }
        {
        \der{}{\PP}{\T_{\pp,j}'}
        }
        \end{equation}

    \item $\der{}{\Q_j\subst{v}{\x_j}}{\T_{\q,j}'}$: since from \cref{eq:23} we have: $\der{\x_j : \Basic_j}{\Q_j}{\T_{\q,j}}$ with $\T_{\q,j} \subt \T_{\q,j}'$ thus by subtyping:
        \begin{equation}
        \label{eq:213}
        \inferrule*[left=\rulename{t-sub}]
        {
            \der{\x_j : \Basic_j}{\Q_j}{\T_{\q,j}}
            \quad
            \T_{\q,j}\subt \T_{\q,j}'
        }
        {
            \der{\x_j : \Basic_j}{\Q_j}{\T_{\q,j}'}
        }
        \end{equation}
        \end{itemize}

        and since $\eval{\e}{\val}$ with $\der{}{\val}{\Basic_j}$ we can apply \cref{lemma:subst} to get $\der{}{\Q_j\subst{\val}{\x_j}}{\T_{\q,j}'}$

        Thus we can use this $\Delta_1 = \Delta',{\pp : \T_{\pp, j}'},{\q : \T_{\q, j}'}$ to type $\N_1$:
        \begin{equation}
        \label{eq:214}
      \inferrule*[left=\rulename{b-sess}]
      {
      \forall \pr \in \Delta': \;\der{ \Gamma }{\PP_{\pr}}{\T_{\pr}}
      \quad
      \der{}{\PP}{\T_{\pp, j}'}
      \quad
      \der{}{\Q_j}{\T_{\q, j}'}
      \quad
      \safe{\Delta_1}
      }
      {\der{}{\N_1}\Delta_1
      }
      \end{equation}
      \mbox{} \hfill {\small by $\rulename{b-sess}$}

    \item \rulename{r-t-if}:
    \setcounter{equation}{0}
        by $\rulename{r-t-if}$ we have:
        \begin{align}
        &\N = \pa\pp{\cond{\e}{\PP_1}{\PP_2}}\pc \N'\\
        &\N_1 = \pa\pp\PP_1 \pc \N'
        \end{align}
        As $\N \predcont{}{\probn{1}}\N_1$ we will show $\der{}{\N_1}{\Delta}$.
        Let $\Delta = \pp : \T, \Delta'$, then by case (6) of type inversion (\cref{lemma:typeinversion}):

        \begin{equation}
      \inferrule*[left=\rulename{b-sess}]
      {
      \der{}{\cond{\e}{\PP_1}{\PP_2}}{\T}
      \quad
      \safe{\Delta}
      \quad
      \forall (\pr : \T_{\pr}) \in \Delta': \;\der{}{\PP_{\pr}}{\T_{\pr}}
      }
      {
      \der{}{\N}\Delta
      }
      \end{equation}

    Next we apply case (5) of type inversion (\cref{lemma:typeinversion}) to recover the typing of $\pp$:

            \begin{equation}
            \inferrule*[left=\rulename{t-sub}]
            {
            \inferrule*[left=\rulename{t-cond}]
            {
            \der{\varnothing}\e\tbool\quad
            \der{\varnothing}{\PP_1}{\T'}\quad
            \der{\varnothing}{\PP_2}{\T'}\quad
            }
            {
            \der
            {\varnothing}
            {{\cond{\e}{\PP_1}{\PP_2}}}
            {\T'}
            }
            \T' \subt \T
            }
            {
            \der
            {\varnothing}
            {{\cond{\e}{\PP_1}{\PP_2}}}
            {\T}
            }
            \end{equation}

            as $\der{\Gamma }{\PP_1}{\T'}$ with $\T' \subt \T$ it's easy to see that $\der{\Gamma }{\PP_1}{\T}$ by \rulename{t-sub} and thus:

        \begin{equation}
        \inferrule*[left=\rulename{b-sess}]
        {
        \der{ \varnothing }{\PP_1}{\T}
        \quad
        \forall (\pr : \T_{\pr}) \in \Delta': \;\der{ \varnothing }{\PP_{\pr}}{\T_{\pr}}
        \quad
        \safe{\Delta}
        }
        {\der{}{\N_1}\Delta
        }
        \end{equation}
    \item \rulename{r-f-if} analogous to \rulename{r-t-if}.

    \item \rulename{r-sc}

    \setcounter{equation}{0}

    First we invert \rulename{r-sc} to get $\exists \N', \N''$ s.t:\\
            \begin{equation}
            \inferrule*[left=\rulename{r-sc}]
             {
              \N \spcon \N'
              \quad
              \N' \predcont{}{\probn{p}} \N''
              \quad
              \N'' \spcon \N_1
             }
             {
              \N \predcont{}{\probn{p}} \N_1
             }
             \end{equation}

        Since $\der{}{\N}{\Delta}$ and $\N \spcon \N'$ we can apply \cref{lemma:precong} to get:
        \begin{equation}
        \der{}{\N'}{\Delta}
        \end{equation}
        The remainder of this case depends on the top-level rule used to derive $\N' \predcont{}{\prob}\N''$:
        \begin{itemize}
            \item by \rulename{r-t-if}, \rulename{r-f-if}:

            Since $\der{}{\N'}{\Delta}$ we can apply $\IH$ on $\N' \predcont{}{\probn{1}}\N''$ to get:
                \begin{equation}
                \der{}{\N''}{\Delta}
                \end{equation}

        Finally, since $\der{}{\N''}{\Delta}$ and $\N'' \spcon \N_1$ we can apply \cref{lemma:precong} to get:
        \begin{equation}
        \der{}{\N_1}{\Delta}
        \end{equation}
                And the statement holds by case (1)

            \item by \rulename{r-comm}:

            Since $\der{}{\N'}{\Delta}$ we can apply $\IH$ on $\N' \predcont{}{\probn{\prob}}\N''$ to get:
                \begin{align}
                    \exists \Delta',\Delta_1\st& \Delta \predcontnd{}\Delta'\predcont{}{\probn{1}}\Delta_1\quad (\text{or } \exists \Delta_1 \st \Delta \predcont{}{\probn{1}}\Delta_1)\\
                \text{with }&\der{}{\N''}{\Delta_1}
                \end{align}
                finally by the \textit{subject precongruence} \cref{lemma:precong}:
                $\der{}{\N_1}{\Delta_1}$
                And the statement holds by case (2)

    \item by \rulename{r-head}, \rulename{r-tail}:

            Since $\der{}{\N'}{\Delta}$ we can apply $\IH$ on $\N' \predcont{}{\probn{\prob}}\N''$ to get:
                    \begin{align}
                      &\exists \pp,\N_2,\T_1, \T_2, \Delta' \st  \Delta = \pp : \T, \Delta'\\
                      &\N'\predcont{}{\probn{1}-\prob} \N_2\\
                      &\der{}{\N''}{(\pp{:}\T_1, \Delta')}\\
                      &\der{}{\N_2}{(\pp{:}\T_2, \Delta')}\\
                      &\pp{:}\T_1 \probmerge{\prob}\T_2,\Delta' \subt \Delta
                    \end{align}

                Finally, we apply \cref{lemma:precong} to get:
                $\der{}{\N_1}{(\pp{:}\T_1, \Delta')}$
                And the statement holds by case (3)
    \end{itemize}
\end{itemize}
\end{proof}

\begin{remark}[Path conventions for~\cref{lemma:processinversion,thm:sessfidelity}]
We distinguish between the head and tail outcomes of a flip even when their
targets coincide; structural unfolding does not create distinct paths.
Path endpoints are read up to head unfolding, including for the empty
path, whose weight is $\probn{1}$.
\end{remark}

\begin{restatable}[Process Inversion]{lemma}{processinversion}
\label{lemma:processinversion}
Let $\der{}{\PP}{\T}$.
\begin{enumerate}
    \item If
    $\unf{\T}=\probpprocoutty{i\in I}{\q_j}{j\in J_i}
        {\probi{i,j}}{\ell_j}{\Basic_j}{\T_{i,j}}$,
    then there exists $i\in I$ such that, for every $j\in J_i$, letting
    \[
        \Rpath_j=\left\{\rpath\;\middle|\;
            \pa\pp\PP\predcontstar{\rpath}{\probn{p_\rpath}}
            \pa\pp{\procout{\q_j}{\ell_j}{\e_\rpath}{\Q_\rpath}}
        \right\},
    \]
    we have
    \begin{enumerate}
        \item $\sum_{\rpath\in\Rpath_j}\probn{p_\rpath}
            =\probi{i,j}$;
        \item $\der{}{\e_\rpath}{\Basic_j}$ and
            $\der{}{\Q_\rpath}{\T_{i,j}}$
            for every $\rpath\in\Rpath_j$.
    \end{enumerate}

    \item If
    $\unf{\T}=\pprocinnew{\q_i}{i\in I}{\ell_i}{\Basic_i}{\T_i}$,
    let
    \[
        \Rpath=\left\{\rpath\;\middle|\;
        \begin{array}{l}
            \pa\pp\PP\predcontstar{\rpath}{\probn{p_\rpath}}
            \pa\pp{\psyntin{j\in J_\rpath}{\q_j}{j}
                {\x_{j,\rpath}}{\Q_{j,\rpath}}},\\
            J_\rpath\supseteq I
        \end{array}
        \right\}.
    \]
    Then
    \begin{enumerate}
        \item $\sum_{\rpath\in\Rpath}\probn{p_\rpath}=\probn{1}$;
        \item $\der{\x_{i,\rpath}:\Basic_i}{\Q_{i,\rpath}}{\T_i}$
            for every $i\in I$ and $\rpath\in\Rpath$.
    \end{enumerate}

    \item If $\unf{\T}=\tend$, let
    \[
        \Rpath=\left\{\rpath\;\middle|\;
            \pa\pp\PP\predcontstar{\rpath}{\probn{p_\rpath}}
            \pa\pp\inact
        \right\}.
    \]
    Then $\sum_{\rpath\in\Rpath}\probn{p_\rpath}=\probn{1}$.
\end{enumerate}
\end{restatable}
\begin{proof}
By lexicographic induction on the number of process constructors not
beneath an input or output prefix and on the derivation height.
The same induction gives finitely many maximal paths, all finite
and ending at a send in the chosen $J_i$ in case~(1), a branching
offering $I$ in case~(2), or $\inact$ in case~(3).

\begin{enumerate}
    \item
    $\unf{\T}=
    \probpprocoutty{i\in I}{\q_j}{j\in J_i}
        {\probi{i,j}}{\ell_j}{\Basic_j}{\T_{i,j}}$.
        We define for each relevant $j$
        \begin{equation}
            \Pi_{j}(\PP) =\left\{
               \rpath
                \;\middle|\;
                    \pa\pp\PP \predcontstar{\rpath}{p_{\rpath}} \pa\pp \procout{\pchan{q_j}}{\ell_j}{\e_\rpath}{\Q_\rpath}
            \right\}
        \end{equation}
        the set of reduction paths from $\pa\pp\PP$ that lead to a $\q_{\pchan{j}}!\labelselli{j}$ prefixed send, where each $\rpath \in \Pi_j(\PP)$ determines a continuation $\Q_\rpath$, a payload $\e_\rpath$, and a path probability $\probn{p_\rpath}$. %
    \begin{itemize}
        \item Case \rulename{t-sel}.
        \[
            \inferrule*[left=\rulename{t-sel}]
            {\der{}{\e}{\Basic}\quad\der{}{\Q}{\T'}}
            {\der{}{\procout{\q}{\ell}{\e}{\Q}}
                {\pprocoutty{\q}{}{\probn{1}}{\ell}{\Basic}{\T'}}}.
        \]
        Here $I=J_1=\{1\}$ up to reindexing; the only maximal
        path is empty, of weight $\probn{1}$. All claims follow
        from the premises.

        \item Case \rulename{t-sub}.
        We have $\der{}{\PP}{\T'}$, $\T'\subt\T$, and
        \[
            \unf{\T'}=
            \probpprocoutty{k\in K}{\q_j}{j\in J_k'}
                {\probip{k,j}}{\ell_j}{\Basic_j}{\T_{k,j}'}.
        \]
        For the witness $k$ given by $\IH$,
        \rulename{sub-$\Sigma$} and \rulename{sub-$\oplus$}
        give $i\in I$ with
        \[
            J_k'=J_i,\qquad
            \probip{k,j}=\probi{i,j},\qquad
            \T_{k,j}'\subt\T_{i,j}\quad(j\in J_i).
        \]
        Thus
        \[
            \sum_{\rpath\in\Pi_j(\PP)}\probn{p_\rpath}
                =\probip{k,j}=\probi{i,j},
        \]
        and, for every $\rpath\in\Pi_j(\PP)$,
        \[
            \inferrule*[left=\rulename{t-sub}]
            {\der{}{\Q_\rpath}{\T_{k,j}'}
                \quad\T_{k,j}'\subt\T_{i,j}}
            {\der{}{\Q_\rpath}{\T_{i,j}}}.
        \]
        Support and payload typing are unchanged.

        \item Case \rulename{t-flip}.
        We have $\PP=\flip{\prob}{\PP_1}{\PP_2}$,
        $\der{}{\PP_b}{\T_b}$ for $b=1,2$, and
        $\T_1\probmerge{\prob}\T_2=\T$, where
        \[
        \begin{aligned}
            \unf{\T_1}
                &=\probpprocoutty{k\in I_1}{\q_j}{j\in J_k'}
                    {\probip{k,j}}{\ell_j}{\Basic_j}{\T_{k,j}'},\\
            \unf{\T_2}
                &=\probpprocoutty{l\in I_2}{\q_j}{j\in J_l''}
                    {\probipp{l,j}}{\ell_j}{\Basic_j}{\T_{l,j}''}.
        \end{aligned}
        \]
        By $\IH$, choose $k\in I_1$ and $l\in I_2$ with
        \[
        \begin{aligned}
            \sum_{\rpath_1\in\Pi_j(\PP_1)}
                \probn{p_{\rpath_1}}&=\probip{k,j},
            &\der{}{\Q_{\rpath_1}}{\T_{k,j}'}
            &&(j\in J_k'),\\
            \sum_{\rpath_2\in\Pi_j(\PP_2)}
                \probn{p_{\rpath_2}}&=\probipp{l,j},
            &\der{}{\Q_{\rpath_2}}{\T_{l,j}''}
            &&(j\in J_l''),
        \end{aligned}
        \]
        for every corresponding path; outside these supports the
        path sets are empty. By the merge rules, take $i=(k,l)$, with
        $J_i=J_k'\cup J_l''$ and
        \[
        (\probi{i,j},\T_{i,j})=
        \begin{cases}
            (\prob\cdot\probip{k,j},\T_{k,j}'),
                &j\in J_k'\setminus J_l'',\\
            (\prob\cdot\probip{k,j}
                +(\probn{1}-\prob)\cdot\probipp{l,j},
                \T_{k,j}'\sqcup\T_{l,j}''),
                &j\in J_k'\cap J_l'',\\
            ((\probn{1}-\prob)\cdot\probipp{l,j},\T_{l,j}''),
                &j\in J_l''\setminus J_k'.
        \end{cases}
        \]
        Splitting paths by their first flip outcome gives
        \[
            \sum_{\rpath\in\Pi_j(\PP)}\probn{p_\rpath}
            =\prob\sum_{\rpath_1\in\Pi_j(\PP_1)}
                \probn{p_{\rpath_1}}
            +(\probn{1}-\prob)
                \sum_{\rpath_2\in\Pi_j(\PP_2)}
                    \probn{p_{\rpath_2}}
            =\probi{i,j}.
        \]
        Moreover,
        \[
            \T_{k,j}'\subt\T_{i,j}\quad(j\in J_k'),
            \qquad
            \T_{l,j}''\subt\T_{i,j}\quad(j\in J_l''),
        \]
        so $\der{}{\Q_\rpath}{\T_{i,j}}$ by $\IH$ and
        \rulename{t-sub}. Payload typing follows from $\IH$,
        and every maximal path ends at a send in $J_i$.

        \item Case \rulename{t-cond}.
        For $\PP=\cond{\e}{\PP_1}{\PP_2}$, the closed guard
        selects $\PP_b$ with probability $\probn{1}$ by
        \rulename{r-t-if} or \rulename{r-f-if}.
        Prefixing this step preserves path weights and endpoints;
        apply $\IH$ to $\der{}{\PP_b}{\T}$.

        \item Case \rulename{t-rec}.
        We have $\PP=\mu X.\PP'$ and
        $\der{X:\T}{\PP'}{\T}$. By \cref{lemma:subst},
        \[
            \der{}{\PP'\subst{\mu X.\PP'}{X}}{\T}.
        \]
        Guardedness makes the first induction component strictly
        smaller after unfolding. Apply $\IH$ and the weight-preserving
        correspondence of paths under head unfolding.
    \end{itemize}

    \item
    $\unf{\T}=\pprocinnew{\q_i}{i\in I}{\ell_i}{\Basic_i}{\T_i}$.
    Write $\Pi$ for the path set in the statement.
    \begin{itemize}
        \item Case \rulename{t-bra}.
        We have
        \[
            \PP=\psyntin{i\in I}{\q_i}{i}{\x_i}{\Q_i},
            \qquad
            \der{\x_i:\Basic_i}{\Q_i}{\T_i}\quad(i\in I).
        \]
        The only maximal path is empty, of weight $\probn{1}$.

        \item Case \rulename{t-sub}.
        We have $\der{}{\PP}{\T'}$, $\T'\subt\T$, and, by
        \rulename{sub-$\&$},
        \[
            \unf{\T'}=
            \pprocinnew{\q_k}{k\in K}{\ell_k}{\Basic_k}{\T_k'},
            \qquad K\supseteq I,
            \qquad\T_i'\subt\T_i\quad(i\in I).
        \]
        By $\IH$, every maximal path offers $K$, so the path set
        for $K$ is exactly $\Pi$. Hence
        \[
            \sum_{\rpath\in\Pi}\probn{p_\rpath}=\probn{1},
            \qquad
            \inferrule*[left=\rulename{t-sub}]
            {\der{\x_{i,\rpath}:\Basic_i}
                {\Q_{i,\rpath}}{\T_i'}\quad\T_i'\subt\T_i}
            {\der{\x_{i,\rpath}:\Basic_i}
                {\Q_{i,\rpath}}{\T_i}}
            \quad(i\in I,\ \rpath\in\Pi).
        \]

        \item Case \rulename{t-flip}.
        We have $\PP=\flip{\prob}{\PP_1}{\PP_2}$,
        $\der{}{\PP_b}{\T_b}$ for $b=1,2$, and
        $\T_1\probmerge{\prob}\T_2=\T$. By \rulename{m-$\&$},
        \[
        \begin{aligned}
            \unf{\T_1}
                &=\pprocinnew{\q_k}{k\in I_1}
                    {\ell_k}{\Basic_k}{\T_k'},\\
            \unf{\T_2}
                &=\pprocinnew{\q_l}{l\in I_2}
                    {\ell_l}{\Basic_l}{\T_l''},\\
            I&=\{i\in I_1\cap I_2\mid
                    \T_i'\sqcup\T_i''\text{ is defined}\},
            \qquad\T_i=\T_i'\sqcup\T_i''.
        \end{aligned}
        \]
        Let $\Pi_b$ be the path set from $\IH$ for $\PP_b$.
        Every maximal path of $\PP_b$ offers $I_b\supseteq I$;
        splitting $\Pi$ by its first flip outcome gives
        \[
            \sum_{\rpath\in\Pi}\probn{p_\rpath}
            =\prob\sum_{\rpath_1\in\Pi_1}\probn{p_{\rpath_1}}
            +(\probn{1}-\prob)
                \sum_{\rpath_2\in\Pi_2}\probn{p_{\rpath_2}}
            =\prob+(\probn{1}-\prob)=\probn{1}.
        \]
        For every $i\in I$,
        \[
            \T_i'\subt\T_i'\sqcup\T_i''=\T_i,
            \qquad
            \T_i''\subt\T_i'\sqcup\T_i''=\T_i.
        \]
        Thus $\der{\x_{i,\rpath}:\Basic_i}{\Q_{i,\rpath}}{\T_i}$
        on either kind of path, by $\IH$ and \rulename{t-sub}.

        \item Cases \rulename{t-cond} and \rulename{t-rec}.
        As in case~(1).
    \end{itemize}

    \item
    $\unf{\T}=\tend$.
    Write $\Pi_\tend(\PP)$ for the path set in the statement.
    \begin{itemize}
        \item Case \rulename{t-$\inact$}.
        We have $\PP=\inact$; the only maximal path is empty,
        of weight $\probn{1}$.

        \item Case \rulename{t-sub}.
        From $\der{}{\PP}{\T'}$ and $\T'\subt\T$, we obtain
        $\unf{\T'}=\tend$ by \rulename{sub-end}; apply $\IH$.

        \item Case \rulename{t-flip}.
        We have $\PP=\flip{\prob}{\PP_1}{\PP_2}$,
        $\der{}{\PP_b}{\T_b}$ for $b=1,2$, and
        $\unf{\T_1}=\unf{\T_2}=\tend$ by \rulename{m-end}.
        By $\IH$, every maximal path ends at $\inact$, and
        \[
        \begin{aligned}
            \sum_{\rpath\in\Pi_\tend(\PP)}\probn{p_\rpath}
            &=\prob\sum_{\rpath_1\in\Pi_\tend(\PP_1)}
                \probn{p_{\rpath_1}}
              +(\probn{1}-\prob)
                \sum_{\rpath_2\in\Pi_\tend(\PP_2)}
                    \probn{p_{\rpath_2}}\\
            &=\prob+(\probn{1}-\prob)=\probn{1}.
        \end{aligned}
        \]

        \item Cases \rulename{t-cond} and \rulename{t-rec}.
        As in case~(1).
    \end{itemize}
\end{enumerate}
\end{proof}

\begin{restatable}[Session Fidelity full]{theorem}{sessfidelity}
\label{thm:sessfidelity}
Let $\der{}{\Multi}{\Delta}$, writing
$\Multi=\pa\pp\PP\pc\pa\q\Q\pc\N'$ in~(1) and
$\Multi=\pa\pp\PP\pc\N'$ in~(2--3).
\begin{enumerate}
    \item If $\Delta=\pp:\T_\pp,\q:\T_\q,\Delta'$ and
    $\Delta\predcont{\actiontau{\pp}{\q}{\ell}}{\probn{1}}
        \pp:\T_\pp',\q:\T_\q',\Delta'$,
    then there exist path sets $\Rpath_\pp,\Rpath_\q$ and
    corresponding processes $\PP_\rpath,\Q_{\rpath'}$ and
    sessions $\N_{\rpath,\rpath'}$ such that
    \begin{enumerate}
        \item For every
        $(\rpath,\rpath')\in\Rpath_\pp\times\Rpath_\q$,
        $\pa\pp\PP\predcontstar{\rpath}{\probn{p_\rpath}}
            \pa\pp\PP_\rpath$ and
        $\pa\q\Q\predcontstar{\rpath'}{\probn{p_{\rpath'}}}
            \pa\q\Q_{\rpath'}$, with
        $\pa\pp\PP_\rpath\pc\pa\q\Q_{\rpath'}\pc\N'
            \predcont{\actiontau{\pp}{\q}{\ell}}{\probn{1}}
            \N_{\rpath,\rpath'}$
        and
        $\der{}{\N_{\rpath,\rpath'}}
            {\pp:\T_\pp',\q:\T_\q',\Delta'}$;
        \item $\sum_{\rpath\in\Rpath_\pp}\probn{p_\rpath}
            =\sum_{\rpath'\in\Rpath_\q}\probn{p_{\rpath'}}
            =\probn{1}$.
    \end{enumerate}

    \item If $\Delta=\pp:\T,\Delta'$ and
    $\pp:\T\predcontnd{\actionnd{\pp}}$, then:
    \begin{enumerate}
        \item If $\nexists\PP',\probp\st
            \Multi\pred{\probp}(\pa\pp\PP'\pc\N')$,
        there exists $\T'$ such that
        $\pp:\T\predcontnd{\actionnd{\pp}}\pp:\T'$
        and $\der{}{\Multi}{\pp:\T',\Delta'}$;
        \item Otherwise, there exist $\Rpath$,
        $\{(\PP_\rpath,\T_\rpath)\}_{\rpath\in\Rpath}$,
        and $\T'$ such that
        \begin{enumerate*}
            \item for every $\rpath\in\Rpath$,
            $\pa\pp\PP\predcontstar{\rpath}{\probn{p_\rpath}}
                \pa\pp\PP_\rpath$, with
            $\pa\pp\PP_\rpath\nrightarrow$ and
            $\der{}{\pa\pp\PP_\rpath\pc\N'}
                {\pp:\T_\rpath,\Delta'}$;
            \item $\sum_{\rpath\in\Rpath}\probn{p_\rpath}
                =\probn{1}$;
            \item $\pp:\T\predcontnd{\actionnd{\pp}}\pp:\T'$;
            \item $\probmergebig{\rpath\in\Rpath}
                {\T_\rpath}{\probn{p_\rpath}}\subt\T'$.
        \end{enumerate*}
    \end{enumerate}

    \item If $\Delta=\pp:\T,\Delta'$ and
    $\pp:\T\predcont{\actionprob{\pp}}{\prob}
        \pp:\pprocoutty{\q}{}{\probn{1}}{\ell}{\Basic}{\T'}$,
    then there exist $\Rpath$ and
    $\{(\e_\rpath,\PP_\rpath)\}_{\rpath\in\Rpath}$ such that
    \begin{enumerate*}
        \item for every $\rpath\in\Rpath$,
        $\pa\pp\PP\predcontstar{\rpath}{\probn{p_\rpath}}
            \pa\pp{\procout{\q}{\ell}{\e_\rpath}{\PP_\rpath}}$,
        with
        $\der{}{\pa\pp{\procout{\q}{\ell}{\e_\rpath}{\PP_\rpath}}
            \pc\N'}
            {\pp:\pprocoutty{\q}{}{\probn{1}}
                {\ell}{\Basic}{\T'},\Delta'}$;
        \item $\sum_{\rpath\in\Rpath}\probn{p_\rpath}=\prob$.
    \end{enumerate*}
\end{enumerate}
\end{restatable}
\begin{proof}
By induction on the context-transition derivation, carrying an
arbitrary unchanged context and its session components as a frame.
By \rulename{b-sess},
\[
    \safe{\Delta},\qquad
    \der{}{\Multi(\pr)}{\Delta(\pr)}
        \quad(\pr\in\dom{\Delta}).
\]
\begin{itemize}
    \item Case \rulename{ct-$\tau$}.
    Inverting its premises through \rulename{ct-rec} and
    \rulename{ct-sc} gives
    \[
    \begin{gathered}
    \begin{aligned}
        \unf{\T_\pp}
            &=\pprocoutty{\q}{}{\probn{1}}
                {\ell_k}{\Basic_k}{\T_\pp'},\\
        \unf{\T_\q}
            &=\pprocinnew{\q_i}{i\in I}
                {\ell_i}{\Basic_i}{\T_{\q,i}'},
    \end{aligned}\\
    k\in I,\quad\q_k=\pp,\quad
    \ell=\ell_k,\quad\T_\q'=\T_{\q,k}'.
    \end{gathered}
    \]
    By \cref{lemma:processinversion}, there are path sets
    $\Rpath_\pp,\Rpath_\q$ with
    \[
    \begin{gathered}
        \pa\pp\PP\predcontstar{\rpath}{\probn{p_\rpath}}
            \pa\pp\PP_\rpath,
        \qquad
        \PP_\rpath=\procout{\q}{\ell_k}{\e_\rpath}{\PP_\rpath'},\\
        \pa\q\Q\predcontstar{\rpath'}{\probn{p_{\rpath'}}}
            \pa\q\Q_{\rpath'},
        \qquad
        \Q_{\rpath'}=\psyntin{l\in L_{\rpath'}}{\q_l}{l}
            {\x_{l,\rpath'}}{\Q_{l,\rpath'}},
        \quad L_{\rpath'}\supseteq I,\\
        \sum_{\rpath\in\Rpath_\pp}\probn{p_\rpath}
            =\sum_{\rpath'\in\Rpath_\q}\probn{p_{\rpath'}}
            =\probn{1},\\
        \der{}{\e_\rpath}{\Basic_k},\qquad
        \der{}{\PP_\rpath'}{\T_\pp'},\qquad
        \der{\x_{k,\rpath'}:\Basic_k}
            {\Q_{k,\rpath'}}{\T_{\q,k}'}.
    \end{gathered}
    \]
    For each $(\rpath,\rpath')\in\Rpath_\pp\times\Rpath_\q$,
    evaluation and \cref{lemma:subst} give
    \[
        \eval{\e_\rpath}{\val_\rpath},\qquad
        \der{}{\val_\rpath}{\Basic_k},\qquad
        \der{}{\Q_{k,\rpath'}
            \subst{\val_\rpath}{\x_{k,\rpath'}}}{\T_{\q,k}'}.
    \]
    Hence, by \rulename{r-comm},
    \[
    \begin{aligned}
        \pa\pp\PP_\rpath\pc\pa\q\Q_{\rpath'}\pc\N'
            &\predcont{\actiontau{\pp}{\q}{\ell_k}}{\probn{1}}
                \N_{\rpath,\rpath'},\\
        \N_{\rpath,\rpath'}
            &:=\pa\pp{\PP_\rpath'}
                \pc\pa\q{\Q_{k,\rpath'}
                    \subst{\val_\rpath}{\x_{k,\rpath'}}}\pc\N'.
    \end{aligned}
    \]
    The given transition preserves safety by \cref{def:safety};
    therefore \rulename{b-sess} gives
    \[
        \der{}{\N_{\rpath,\rpath'}}
            {\pp:\T_\pp',\q:\T_{\q,k}',\Delta'}.
    \]

    \item Case \rulename{ct-prob}.
    Here
    \[
    \begin{gathered}
        \pp:\underbrace{\pprocoutty{\q_j}{j\in J}{\probi{j}}
            {\ell_j}{\Basic_j}{\T_j}}_{\T}
        \predcont{\actionprob{\pp}}{\probi{k}}
        \pp:\pprocoutty{\q_k}{}{\probn{1}}
            {\ell_k}{\Basic_k}{\T_k},\\
        k\in J,\qquad\probi{k}\in(0,1).
    \end{gathered}
    \]
    The $k$-th path set from \cref{lemma:processinversion} satisfies
    \[
    \begin{gathered}
        \pa\pp\PP\predcontstar{\rpath}{\probn{p_\rpath}}
            \pa\pp{\procout{\q_k}{\ell_k}{\e_\rpath}{\PP_\rpath}},
        \qquad
        \sum_{\rpath\in\Rpath}\probn{p_\rpath}=\probi{k},\\
        \der{}{\e_\rpath}{\Basic_k},\qquad
        \der{}{\PP_\rpath}{\T_k}
        \quad(\rpath\in\Rpath).
    \end{gathered}
    \]
    By \rulename{t-sel}, safety of the framed target, and
    \rulename{b-sess},
    \[
        \der{}{\pa\pp{\procout{\q_k}{\ell_k}{\e_\rpath}{\PP_\rpath}}
            \pc\N'}
            {\pp:\pprocoutty{\q_k}{}{\probn{1}}
                {\ell_k}{\Basic_k}{\T_k},\Delta'}.
    \]
    This is clause~(3) with
    $(\q,\ell,\prob,\T')=(\q_k,\ell_k,\probi{k},\T_k)$.

    \item Case \rulename{ct-nd}.
    Write
    \[
    \begin{gathered}
        \T=\probpprocoutty{i\in I}{\q_j}{j\in J_i}
            {\probi{i,j}}{\ell_j}{\Basic_j}{\T_{i,j}},
        \qquad |I|>1,\\
        \T^{(i)}:=\pprocoutty{\q_j}{j\in J_i}
            {\probi{i,j}}{\ell_j}{\Basic_j}{\T_{i,j}}.
    \end{gathered}
    \]
    \begin{enumerate}
        \item If clause~(2a)'s hypothesis holds, then
        $\pa\pp\PP\nrightarrow$.
        A locally irreducible process typed by a sum of selections
        is typable by one summand. This follows by induction on
        its typing derivation, allowing environments containing
        only process variables whose free occurrences are guarded.
        The \rulename{t-sel} case is immediate; \rulename{t-sub}
        uses \rulename{sub-$\Sigma$} and \rulename{sub-$\oplus$}.
        For \rulename{t-rec}, guardedness preserves local
        irreducibility of the body, and the induction hypothesis gives
        \[
            \der{\Gamma,X:\T_0}{\PP_0}{\T_*},
            \qquad\T_*\subt\T_0,
        \]
        where $\T_*$ is a summand of $\unf{\T_0}$.
        Replacing uses of $X:\T_0$ by \rulename{t-var}
        and \rulename{t-sub} yields
        \[
            \der{\Gamma,X:\T_*}{\PP_0}{\T_*}
            \quad\Longrightarrow\quad
            \der{\Gamma}{\mu X.\PP_0}{\T_*}
            \quad\text{by \rulename{t-rec}}.
        \]
        All other rules are excluded by the type shape,
        guardedness, or local irreducibility.

        Thus some $i\in I$ satisfies $\der{}{\PP}{\T^{(i)}}$.
        Taking $\T'=\T^{(i)}$, \rulename{ct-nd},
        \cref{def:safety}, and \rulename{b-sess} give
        \[
            \pp:\T\predcontnd{\actionnd{\pp}}\pp:\T',
            \qquad\der{}{\Multi}{\pp:\T',\Delta'}.
        \]

        \item Otherwise, \cref{lemma:processinversion} gives
        $i\in I$ and disjoint path sets $\Rpath_j$ such that
        \[
        \begin{gathered}
            \pa\pp\PP\predcontstar{\rpath}{\probn{p_\rpath}}
                \pa\pp{\procout{\q_j}{\ell_j}{\e_\rpath}{\Q_\rpath}},
            \qquad
            \sum_{\rpath\in\Rpath_j}\probn{p_\rpath}
                =\probi{i,j},\\
            \der{}{\e_\rpath}{\Basic_j},\qquad
            \der{}{\Q_\rpath}{\T_{i,j}}
            \quad(j\in J_i,\ \rpath\in\Rpath_j).
        \end{gathered}
        \]
        Set
        \[
        \begin{gathered}
            \Rpath:=\biguplus_{j\in J_i}\Rpath_j,
            \qquad\T':=\T^{(i)},\\
            \PP_\rpath:=\procout{\q_j}{\ell_j}{\e_\rpath}{\Q_\rpath},
            \qquad
            \T_\rpath:=\pprocoutty{\q_j}{}{\probn{1}}
                {\ell_j}{\Basic_j}{\T_{i,j}}
            \quad(\rpath\in\Rpath_j).
        \end{gathered}
        \]
        The four claims follow:
        \begin{enumerate}
            \item By \rulename{t-sel},
            $\der{}{\PP_\rpath}{\T_\rpath}$.
            The full context $\pp:\T_\rpath,\Delta'$ is reachable
            from $\Delta$ by \rulename{ct-nd}, followed by
            \rulename{ct-prob} if $\probi{i,j}<1$;
            if $\probi{i,j}=1$, then $J_i=\{j\}$.
            Hence \cref{def:safety} and \rulename{b-sess} give
            \[
                \der{}{\pa\pp\PP_\rpath\pc\N'}
                    {\pp:\T_\rpath,\Delta'}.
            \]

            \item
            $\displaystyle
                \sum_{\rpath\in\Rpath}\probn{p_\rpath}
                =\sum_{j\in J_i}\sum_{\rpath\in\Rpath_j}
                    \probn{p_\rpath}
                =\sum_{j\in J_i}\probi{i,j}=\probn{1}.$

            \item
            $\pp:\T\predcontnd{\actionnd{\pp}}\pp:\T'$
            by \rulename{ct-nd}.

            \item By \rulename{m-$\oplus$} and idempotence
            of $\sqcup$ up to mutual subtyping,
            \[
            \begin{aligned}
                \probmergebig{\rpath\in\Rpath}
                    {\T_\rpath}{\probn{p_\rpath}}
                &\subt\pprocoutty{\q_j}{j\in J_i}
                    {\sum_{\rpath\in\Rpath_j}\probn{p_\rpath}}
                    {\ell_j}{\Basic_j}{\T_{i,j}}\\
                &=\pprocoutty{\q_j}{j\in J_i}{\probi{i,j}}
                    {\ell_j}{\Basic_j}{\T_{i,j}}=\T'.
            \end{aligned}
            \]
            The path set is finite and its weights are positive,
            so the multi-way merge is well-defined.
        \end{enumerate}
    \end{enumerate}

    \item Case \rulename{ct-rec}.
    The premise and conclusion are
    \[
        \pp:\tunroll\predcont{\alpha}{t}\pp:\T_0,
        \qquad
        \pp:\mut\predcont{\alpha}{t}\pp:\T_0.
    \]
    Folded and unfolded types are mutually subtypes; thus
    \rulename{t-sub}, \cref{lemma:safeproperties}, and
    \rulename{b-sess} give
    \[
        \der{}{\PP}{\tunroll},\qquad
        \safe{\pp:\tunroll,\Delta'},\qquad
        \der{}{\Multi}{\pp:\tunroll,\Delta'}.
    \]
    Apply $\IH$ to the premise with the same frame.
    In clause~(2), lift its witness by \rulename{ct-rec}:
    \[
        \pp:\tunroll\predcontnd{\actionnd{\pp}}\pp:\T'
        \quad\Longrightarrow\quad
        \pp:\mut\predcontnd{\actionnd{\pp}}\pp:\T'.
    \]
    All other conclusions, including clause~(3), are unchanged.

    \item Case \rulename{ct-sc}.
    Apply $\IH$ to the premise with the additional unchanged
    participant included in the frame. It gives the required
    paths and typings for the same full session and context.
\end{itemize}
\end{proof}

\sessionfidelitysmall*
\begin{proof}
Corollary of~\cref{thm:sessfidelity}
\end{proof}

\subsection{Proofs for~\Cref{sec:typingcontextproperties} (Downward closure of properties)}
\label{app:path}

\begin{restatable}[Downward closure of $\sqcup$ definedness]{lemma}{lubclosure}
\label{lemma:lubclosure}
\[
    \T_1\subt \T_2,\quad \T_1'\subt \T_2',\quad
    \T_2\sqcup \T_2'\text{ defined}
    \quad\Longrightarrow\quad
    \T_1\sqcup \T_1'\text{ defined}.
\]
\end{restatable}
\begin{proof}
By the proof of~\cref{lem:supdecidable}, two types admit a join whenever they have a common upper bound, the result follows by transitivity of $\subt$.
\end{proof}

\begin{restatable}[Downward closure of merge definedness]{lemma}{definednessclosure}
\label{lemma:definednessclosure}
    Suppose $\T_1 \subt \T_2$ and $\T_1' \subt \T_2'$ with $\T_2 \probmerge{\prob}\T_2'$ defined, then $\T_1 \probmerge{\prob}\T_1'$ defined
\end{restatable}
\begin{proof}
We proceed by induction on the finite derivation of
$\T_2\probmerge{\prob}\T_2'$.

\begin{itemize}
    \item case \rulename{m-end}:
    By \rulename{sub-end}, $\T_1=\T_1'=\tend$, so
    $\T_1\probmerge{\prob}\T_1'=\tend$ by \rulename{m-end}.

    \item case \rulename{m-$\oplus$}:
    Inverting \rulename{sub-$\oplus$} or
    \rulename{sub-$\Sigma$}, write
    \[
        \T_1=\sum_{a\in A}\T_{1,a},
        \qquad
        \T_1'=\sum_{b\in B}\T_{1,b}',
    \]
    where the summands are selection distributions, singleton sums are allowed, and
    \[
        \forall a\in A:\ \T_{1,a}\subt\T_2,
        \qquad
        \forall b\in B:\ \T_{1,b}'\subt\T_2'.
    \]
    These comparisons use \rulename{sub-$\oplus$}, preserving the respective prefixes and probabilities. Let $I,J$ index the branches of $\T_2,\T_2'$. For every $(a,b,k)\in A\times B\times(I\cap J)$,
    \[
        \T_{1,a,k}\subt\T_{2,k},
        \qquad
        \T_{1,b,k}'\subt\T_{2,k}',
        \qquad
        \T_{2,k}\sqcup\T_{2,k}'\text{ defined}.
    \]
    Thus $\T_{1,a,k}\sqcup\T_{1,b,k}'$ is defined by
    \cref{lemma:lubclosure}, and
    \rulename{m-$\oplus$} gives
    \[
        \forall(a,b)\in A\times B:\quad
        \T_{1,a}\probmerge{\prob}\T_{1,b}'\text{ defined}.
    \]
    Apply \rulename{m-$\Sigma$} if $|A|\cdot|B|>1$; otherwise the unique component is the required merge.

    \item case \rulename{m-$\&$}:
    Let $I_r,J_r$ index the branches of $\T_r,\T_r'$ for $r\in\{1,2\}$. Inverting \rulename{sub-$\&$} gives
    \[
    \begin{gathered}
        I_2\subseteq I_1,\qquad J_2\subseteq J_1,\\
        \forall i\in I_2:\ \T_{1,i}\subt\T_{2,i},
        \qquad
        \forall j\in J_2:\ \T_{1,j}'\subt\T_{2,j}',\\
        \{\q_i\}_{i\in I_1}=\{\q_i\}_{i\in I_2},
        \qquad
        \{\q_j\}_{j\in J_1}=\{\q_j\}_{j\in J_2}.
    \end{gathered}
    \]
    Put
    \[
        K_r=
        \{\,k\in I_r\cap J_r
        \mid \T_{r,k}\sqcup\T_{r,k}'\text{ defined}\,\}.
    \]
    By~\cref{lemma:lubclosure},
    $K_2\subseteq K_1$. The participant conditions of the
    upper merge and the subtypings therefore give
    \[
        \{\q_k\}_{k\in K_2}
        \subseteq \{\q_k\}_{k\in K_1}
        \subseteq \{\q_i\}_{i\in I_1}
        = \{\q_j\}_{j\in J_1}
        = \{\q_k\}_{k\in K_2}.
    \]
    Hence $\{\q_k\}_{k\in K_1}
    =\{\q_i\}_{i\in I_1}
    =\{\q_j\}_{j\in J_1}$, so \rulename{m-$\&$} applies.

    \item case \rulename{m-$\Sigma$}:
    Write
    \[
        \T_r=\sum_{i\in I_r}\T_{r,i},
        \qquad
        \T_r'=\sum_{j\in J_r}\T_{r,j}'
        \qquad(r\in\{1,2\}),
    \]
    allowing singleton sums.
    Inverting \rulename{sub-$\Sigma$} or
    \rulename{sub-$\oplus$} gives
    \[
    \begin{aligned}
        &\forall a\in I_1:\ \exists i\in I_2:
            \T_{1,a}\subt\T_{2,i},\\
        &\forall b\in J_1:\ \exists j\in J_2:
            \T_{1,b}'\subt\T_{2,j}'.
    \end{aligned}
    \]
    For each $(a,b)\in I_1\times J_1$, choose such $i,j$. Since $\T_{2,i}\probmerge{\prob}\T_{2,j}'$ is defined by a premise of the given derivation, the induction hypothesis yields $\T_{1,a}\probmerge{\prob}\T_{1,b}'$ defined. Apply \rulename{m-$\Sigma$} if $|I_1|\cdot|J_1|>1$; otherwise the unique component is the required merge.
\end{itemize}
\end{proof}

\monotonicmerge*
\begin{proof}
By \cref{lemma:definednessclosure},
$\T_1\probmerge{\prob}\T_2$ is defined.
We use the following consequence of~\cref{lemma:lubclosure}, transitivity and the
least-upper-bound property:
\begin{equation}
\label{eq:merge-join-monotonicity}
\begin{gathered}
    \T_a\subt\T_a',\qquad
    \T_b\subt\T_b',\qquad
    \T_a'\sqcup\T_b'\text{ defined}
    \\
    \Longrightarrow\quad
    \T_a\sqcup\T_b\text{ defined}
    \;\land\;
    \T_a\sqcup\T_b\subt\T_a'\sqcup\T_b'.
\end{gathered}
\end{equation}
Indeed, $\T_a'\sqcup\T_b'$ is a common upper bound of
$\T_a,\T_b$.

We prove
$\T_1\probmerge{\prob}\T_2
 \subt \T_1'\probmerge{\prob}\T_2'$
by induction on the finite derivation of
$\T_1'\probmerge{\prob}\T_2'$.

\begin{itemize}
    \item case \rulename{m-end}:
    $\T_1'=\T_2'=\tend$, hence
    $\T_1=\T_2=\tend$ by \rulename{sub-end}.
    The claim follows by \rulename{m-end} and reflexivity.

    \item case \rulename{m-$\oplus$}:
    Write
    \[
    \begin{aligned}
        \T_1'
        &=
        \pprocoutty{\pp_i}{i\in I}{\probi{i}}
            {\ell_i}{\Basic_i}{\T_{1,i}'},
        \\
        \T_2'
        &=
        \pprocoutty{\pp_j}{j\in J}{\probip{j}}
            {\ell_j}{\Basic_j}{\T_{2,j}'}.
    \end{aligned}
    \]
    Inverting \rulename{sub-$\oplus$} or
    \rulename{sub-$\Sigma$} gives
    \[
    \begin{gathered}
        \T_1=\sum_{a\in A}\T_{1,a},
        \qquad
        \T_2=\sum_{b\in B}\T_{2,b},
        \\
        \forall a\in A:\ \T_{1,a}\subt\T_1',
        \qquad
        \forall b\in B:\ \T_{2,b}\subt\T_2',
    \end{gathered}
    \]
    where all summands are selection distributions. Each summand comparison uses \rulename{sub-$\oplus$},
    preserving the respective prefixes and probabilities. Write $\T_{1,a,i},\T_{2,b,j}$ for their continuations.

    Fix $(a,b)\in A\times B$.
    The two distribution merges have the same prefixes and probabilities, and their continuations satisfy
    \[
    \begin{array}{ll}
        k\in I\setminus J:
        &\T_{1,a,k}\subt\T_{1,k}',
        \\[1mm]
        k\in J\setminus I:
        &\T_{2,b,k}\subt\T_{2,k}',
        \\[1mm]
        k\in I\cap J:
        &\T_{1,a,k}\sqcup\T_{2,b,k}
         \subt
         \T_{1,k}'\sqcup\T_{2,k}'.
    \end{array}
    \]
    The first two comparisons follow from
    \rulename{sub-$\oplus$}; the third, including existence
    of the lower join, follows from
    \eqref{eq:merge-join-monotonicity}.
    Thus \rulename{m-$\oplus$} and
    \rulename{sub-$\oplus$} give
    \[
        \T_{1,a}\probmerge{\prob}\T_{2,b}
        \subt
        \T_1'\probmerge{\prob}\T_2'.
    \]
    If $|A|\cdot|B|>1$, \rulename{m-$\Sigma$} and
    \rulename{sub-$\Sigma$} therefore yield
    \[
        \T_1\probmerge{\prob}\T_2
        =
        \sum_{(a,b)\in A\times B}
            \bigl(\T_{1,a}\probmerge{\prob}\T_{2,b}\bigr)
        \subt
        \T_1'\probmerge{\prob}\T_2'.
    \]
    Otherwise, the unique component comparison is already
    the required inequality.

    \item case \rulename{m-$\&$}:
    Let $I,J$ and $I',J'$ be the respective branch sets of
    $\T_1,\T_2$ and $\T_1',\T_2'$.
    Inverting \rulename{sub-$\&$} gives
    \[
    \begin{gathered}
        I'\subseteq I,\qquad J'\subseteq J,
        \\
        \forall i\in I':\ \T_{1,i}\subt\T_{1,i}',
        \qquad
        \forall j\in J':\ \T_{2,j}\subt\T_{2,j}'.
    \end{gathered}
    \]
    Put
    \[
    \begin{aligned}
        K
        &=
        \{\,k\in I\cap J
           \mid \T_{1,k}\sqcup\T_{2,k}\text{ defined}\,\},
        \\
        K'
        &=
        \{\,k\in I'\cap J'
           \mid \T_{1,k}'\sqcup\T_{2,k}'\text{ defined}\,\}.
    \end{aligned}
    \]
    By \eqref{eq:merge-join-monotonicity},
    \[
        K'\subseteq K,
        \qquad
        \forall k\in K':\
        \T_{1,k}\sqcup\T_{2,k}
        \subt
        \T_{1,k}'\sqcup\T_{2,k}'.
    \]
    Since both merges are defined, the side conditions of
    \rulename{m-$\&$} and \rulename{sub-$\&$} give
    \[
        \{\pp_k\}_{k\in K}
        =
        \{\pp_i\}_{i\in I}
        =
        \{\pp_i\}_{i\in I'}
        =
        \{\pp_k\}_{k\in K'}.
    \]
    Hence \rulename{sub-$\&$} yields
    \[
        \pprocinnew{\pp_k}{k\in K}{\ell_k}{\Basic_k}
            {(\T_{1,k}\sqcup\T_{2,k})}
        \subt
        \pprocinnew{\pp_k}{k\in K'}{\ell_k}{\Basic_k}
            {(\T_{1,k}'\sqcup\T_{2,k}')},
    \]
    which is precisely the required inequality by
    \rulename{m-$\&$}.

    \item case \rulename{m-$\Sigma$}:
    Write
    \[
    \begin{aligned}
        \T_1&=\sum_{a\in A}\T_{1,a},
        &\qquad
        \T_2&=\sum_{b\in B}\T_{2,b},
        \\
        \T_1'&=\sum_{i\in I}\T_{1,i}',
        &
        \T_2'&=\sum_{j\in J}\T_{2,j}',
    \end{aligned}
    \]
    where $|I|\cdot|J|>1$.
    Inverting \rulename{sub-$\Sigma$} or
    \rulename{sub-$\oplus$} gives
    \[
    \begin{aligned}
        &\forall a\in A:\ \exists i\in I:\
            \T_{1,a}\subt\T_{1,i}',
        \\
        &\forall b\in B:\ \exists j\in J:\
            \T_{2,b}\subt\T_{2,j}'.
    \end{aligned}
    \]
    Fix $(a,b)\in A\times B$ and choose such $i,j$.
    The merge
    $\T_{1,i}'\probmerge{\prob}\T_{2,j}'$
    is defined by a premise of the given derivation.
    Hence the induction hypothesis gives
    \[
        \T_{1,a}\probmerge{\prob}\T_{2,b}
        \subt
        \T_{1,i}'\probmerge{\prob}\T_{2,j}'.
    \]
    Thus every component of the lower merge is a subtype
    of some component of the upper merge.
    Since
    $(|A|\cdot|B|)\cdot(|I|\cdot|J|)>1$,
    \rulename{sub-$\Sigma$} yields
    \[
        \sum_{(a,b)\in A\times B}
            \bigl(\T_{1,a}\probmerge{\prob}\T_{2,b}\bigr)
        \subt
        \sum_{(i,j)\in I\times J}
            \bigl(\T_{1,i}'\probmerge{\prob}\T_{2,j}'\bigr).
    \]
    By \rulename{m-$\Sigma$}, these are the required merges.
    If $|A|\cdot|B|=1$, the lower sum instead denotes its
    unique selection merge, derived by
    \rulename{m-$\oplus$}; the same
    \rulename{sub-$\Sigma$} application still applies.
\end{itemize}
\end{proof}

\begin{notation}[Property Abbreviations]
In the following we write $\sdf{\Delta}$ if $\safe{\Delta}$ and $\df{\Delta}$; $\saslive{\Delta}$ if $\safe{\Delta}$ and $\aslive{\Delta}$; and $\slive{\Delta}$ if $\safe{\Delta}$ and $\live{\Delta}$.
\end{notation}

\begin{remark}[Factored sums]
\label{rem:factoring}
\rulename{sub-$\Sigma$} requires every summand $\T_i$ in $\sum_{i \in I}\T_i$ of the subtype to be a subtype of \emph{some} summand $\T_j'$ in $\sum_{j \in J}\T_j'$ of the supertype, but the map need not be injective: several $\T_i$ sharing a common prefix may all be subtype of a single supertype $\T_j'$ that \emph{factors} that prefix. For instance
\begin{equation}
    {\begin{array}{l}
    \sum\Big\{\,
    \pprocoutty{\q}{}{\probn{1}}{\ell_1}{\Basic}{\pprocoutty{\q}{}{\probn{1}}{\ell_2}{\Basic}{\tend}}
    \;,\;
    \pprocoutty{\q}{}{\probn{1}}{\ell_1}{\Basic}{\pprocoutty{\q}{}{\probn{1}}{\ell_3}{\Basic}{\tend}}
    \,\Big\}\; \subt \\
    \hphantom{xxx}\pprocoutty{\q}{}{\probn{1}}{\ell_1}{\Basic}{\sum\Big\{\,
    \pprocoutty{\q}{}{\probn{1}}{\ell_2}{\Basic}{\tend}
    \,,\,
    \pprocoutty{\q}{}{\probn{1}}{\ell_3}{\Basic}{\tend}
    \,\Big\}}
    \end{array}}
\end{equation}
Consequently a subtype context $\Delta \subt \Delta'$ may take a non-deterministic step $\Delta\ndstep$ that $\Delta'$ can't mirror $\Delta' \not\ndstep$: the supertype $\Delta'$ is then already at the factored form, lying above \emph{every} outcome of the subtype's choice i.e. $\forall \Delta'' \st \Delta \ndstep \Delta''$ we have $\Delta'' \subt \Delta'$ --- no supertype step is needed (clause 2(a) of \cref{lemma:deltasimulation}).
\end{remark}

\begin{restatable}[Type context simulation]{lemma}{deltasimulation}
\label{lemma:deltasimulation}
If $\Delta_0 \subt \Delta_0'$ and $\safe{\Delta_0'}$ then $\Delta_0'$ can simulate $\Delta_0$:
\begin{enumerate}
    \item $\Delta_0 \predcont{\actiontau{\pp}{\q}{\labelselli{}}}{\probn{1}} \Delta_1$ implies
        \begin{enumerate}
            \item if $\Delta_0'(\pp) \not\ndstep$ then $\exists\Delta_1'\supt \Delta_1$ such that $\Delta_0'\predcont{\actiontau{\pp}{\q}{\labelselli{}}}{\probn{1}} \Delta_1'$;
            \item if $\Delta_0'(\pp) \ndstep$ then $\exists\Delta''\supt \Delta_0$ and $\exists \Delta_1' \supt \Delta_1$ such that $\Delta_0' \predcontnd{\actionnd{\pp}}\Delta''\predcont{\actiontau{\pp}{\q}{\labelselli{}}}{\probn{1}} \Delta_1'$;
        \end{enumerate}
    \item $\Delta_0 \predcontnd{\actionnd{\pp}}\Delta_1$ implies
        \begin{enumerate}
            \item if $\Delta_0'(\pp) \not\ndstep$ then $\Delta_0' \supt \Delta_1$
            \item if $\Delta_0'(\pp) \ndstep$ then $\exists \Delta_1' \supt \Delta_1$ such that $\Delta_0' \predcontnd{\actionnd{\pp}}\Delta_1'$.
        \end{enumerate}
    \item $\Delta_0 \predcont{\actionprob{\pp}}{\prob} \Delta_1$ implies
        \begin{enumerate}
            \item if $\Delta_0'(\pp) \not\ndstep$ then $\exists\Delta_1'\supt \Delta_1$ such that $\Delta_0' \predcont{\actionprob{\pp}}{\prob} \Delta_1'$;
            \item if $\Delta_0'(\pp) \ndstep$ then $\exists\Delta''\supt \Delta_0$ and $\exists \Delta_1' \supt \Delta_1$ such that $\Delta_0' \predcontnd{\actionnd{\pp}}\Delta''\predcont{\actionprob{\pp}}{\prob} \Delta_1'$.
        \end{enumerate}
\end{enumerate}
\end{restatable}

\begin{proof}
\setcounter{equation}{0}
By induction on the derivation of $\Delta_0 \predcont{\alpha}{t} \Delta_1$, with base cases \rulename{ct-nd}, \rulename{ct-prob} and \rulename{ct-$\tau$} and inductive cases \rulename{ct-sc} and \rulename{ct-rec}:
\begin{itemize}
    \item case \rulename{ct-nd}: $\Delta_0 \predcontnd{\actionnd{\pp}} \Delta_1$ by:
        \begin{equation}
         \inferrule*[left=\rulename{ct-nd}]
        {
          k \in I \quad |I| > 1
        }
        {
            \pp :  \sum_{i \in I}\T_i
          \predcontnd{\actionnd{\pp}}{
              {\pp}:\T_k
          }
        }
        \end{equation}
        thus $\Delta_0 = \pp : \sum_{i \in I}\T_i$ with $|I| > 1$, and as $\Delta_0 \subt \Delta_0'$ the only applicable rule is \rulename{sub-$\Sigma$}, so $\Delta_0' = \pp : \sum_{j \in J}\T_j'$ where $\forall i \in I\st \exists j \in J\st \T_i \subt \T_j'$ --- note $J$ may well be a singleton, several summands of $\Delta_0(\pp)$ sharing the same supertype summand (\cref{rem:factoring}).
        \begin{itemize}
            \item if $\Delta_0'(\pp) \not\ndstep$ (case 2a) then $J = \{l\}$ is a singleton, and in particular $\T_k \subt \T_l'$, so $\Delta_1 = \pp:\T_k \subt \pp:\T_l' = \Delta_0'$ and no step is needed;
            \item if $\Delta_0'(\pp) \ndstep$ (case 2b) then $|J| > 1$; picking $l\in J$ with $\T_k \subt \T_l'$,
            \begin{equation}
             \inferrule*[left=\rulename{ct-nd}]
            {
              l \in J \quad |J| > 1
            }
            {
                \pp :  \sum_{j \in J}\T_j'
              \predcontnd{\actionnd{\pp}}{
                  {\pp}:\T_l'
              }
            }
            \end{equation}
            with $\Delta_1 = \pp:\T_k \subt \pp:\T_l' = \Delta_1'$.
        \end{itemize}
    \item case \rulename{ct-prob}: $\Delta_0 \predcont{\actionprob{\pp}}{\probi{k}} \Delta_1$ by:
        \begin{equation}
            \label{eq:dsprob1}
        \inferrule*[left=\rulename{ct-prob}]
        {
          k \in J \quad \probi{k} \in (0;1)
        }
        {
            \pp : \pprocoutty{\q_j}{j \in J}{\probi{j}}{\ell_j}{\Basic_j}{\T_j}
          \predcont{\actionprob{\pp}}{\probi{k}}
            {\pp}:{\pprocoutty{\q_k}{}{\probn{1}}{\ell_k}{\Basic_k}{\T_k}}
        }
        \end{equation}
        thus $\Delta_0 = \pp : \pprocoutty{\q_j}{j \in J}{\probi{j}}{\ell_j}{\Basic_j}{\T_j}$ with $|J| > 1$ (as $\probi{k} < 1$ and the branch probabilities sum to $\probn{1}$). From $\Delta_0(\pp) \subt \Delta_0'(\pp)$, by either \rulename{sub-$\oplus$} or \rulename{sub-$\Sigma$}:
        \begin{itemize}
            \item by \rulename{sub-$\oplus$} (case 3a): $\Delta_0'(\pp) = \pprocoutty{\q_j}{j \in J}{\probi{j}}{\ell_j}{\Basic_j}{\T_j'''}$ with $\T_j \subt \T_j'''$ for all $j \in J$ --- \emph{same} labels, partners and probabilities --- and $\Delta_0'(\pp) \not\ndstep$. Then \rulename{ct-prob} at the same branch $k$ gives
            \begin{equation}
                \Delta_0' \predcont{\actionprob{\pp}}{\probi{k}} \pp : \pprocoutty{\q_k}{}{\probn{1}}{\ell_k}{\Basic_k}{\T_k'''} =: \Delta_1'
            \end{equation}
            with $\Delta_1 \subt \Delta_1'$ by \rulename{sub-$\oplus$} on the committed singletons, since $\T_k \subt \T_k'''$;
            \item by \rulename{sub-$\Sigma$} (case 3b): as in the \rulename{ct-nd} case, $\Delta_0'$ first commits by \rulename{ct-nd} to the summand $\Delta''(\pp)$ with $\Delta_0(\pp) \subt \Delta''(\pp)$ by \rulename{sub-$\oplus$}, so $\Delta_0 \subt \Delta''$, and the previous bullet applies to $\Delta''$, giving $\Delta'' \predcont{\actionprob{\pp}}{\probi{k}} \Delta_1' \supt \Delta_1$.
        \end{itemize}
        Note that no probability slack arises: \rulename{sub-$\oplus$} preserves the distribution exactly, so the supertype commits at precisely $\probi{k}$.
    \item case \rulename{ct-$\tau$}:
        \begin{equation}\footnotesize
            \label{eq:ds3}
        \inferrule*[left=\rulename{ct-$\tau$}]
        {
            \inferrule*[left=\rulename{ct-out}]
            { }
            {
                \pp :  \pprocoutty{\q}{}{\probn{1}}{\ell_k}{\Basic_k}{\T_k}
                \predcont{\actionprocout{\pp}{\q}{\ell_k}{\Basic_k}}{\probn{1}}{{\pp}:{\T_k}}
            }
            \inferrule*[left=\rulename{ct-in}]
            {
              k \in I
            }
            {
                \q : \pprocinnew{\pp_i}{i \in I}{\ell_i}{\Basic_i}{\T_i'}
                \predcont{\actionprocin{\q}{\pp_k}{\ell_k}{\Basic_k}}{1}{\q : \T_k'}
            }
        }
        {
            \pp :  \pprocoutty{\q}{}{\probn{1}}{\ell_k}{\Basic_k}{\T_k},\q : \pprocinnew{\pp_i}{i \in I}{\ell_i}{\Basic_i}{\T_i'}
            \predcont{\actiontau{\pp}{\q}{\labelselli{k}}}{\probn{1}} \pp : \T_k, \q : \T_k'
        }
        \end{equation}
        We have from $\Delta_0 \subt \Delta_0'$, componentwise:
        \begin{itemize}
            \item $\Delta_0(\q) \subt \Delta_0'(\q)$: by \rulename{sub-$\&$} then $\Delta_0'(\q) = \pprocinnew{\pp_i}{i \in I'}{\ell_i}{\Basic_i}{\T_i''}$ with $I' \subseteq I$ and $\T_i' \subt \T_i''$ for all $i \in I'$
            \item $\Delta_0(\pp) \subt \Delta_0'(\pp)$: the subtype is a committed singleton, so by \rulename{sub-$\oplus$} the matching (summand of the) supertype is the \emph{same} singleton $\pprocoutty{\q}{}{\probn{1}}{\ell_k}{\Basic_k}{\T_k'''}$ with $\T_k \subt \T_k'''$ --- a proper distribution is never a supertype of a singleton, as \rulename{sub-$\oplus$} preserves probabilities exactly.
        \end{itemize}
        By either \rulename{sub-$\oplus$} directly or \rulename{sub-$\Sigma$} embedding the singleton into a sum, we consider each in turn:
        \begin{itemize}
            \item by \rulename{sub-$\oplus$} (case 1a): $\Delta_0'(\pp) = \pprocoutty{\q}{}{\probn{1}}{\ell_k}{\Basic_k}{\T_k'''}$ with $\Delta_0'(\pp) \not\ndstep$, then as $\safe{\Delta_0'}$ and $\pp$'s (sole) output $\labelselli{k}$ towards $\q$ is enabled, $\q$ must be able to receive $\labelselli{k}$, i.e.\ $k \in I'$:
            \begin{equation}
                \Delta_0' \predcont{\actiontau{\pp}{\q}{\labelselli{k}}}{\probn{1}} \pp : \T_k''',\; \q : \T_k'' =: \Delta_1'
            \end{equation}
            with $\Delta_1 = \pp:\T_k, \q:\T_k' \subt \Delta_1'$ since $\T_k \subt \T_k'''$ and $\T_k' \subt \T_k''$; this is clause (1a) with a lone synchronisation ($\prob = \probn{1}$);
            \item by \rulename{sub-$\Sigma$} (case 1b): first commit by \rulename{ct-nd} to the summand with $\pprocoutty{\q}{}{\probn{1}}{\ell_k}{\Basic_k}{\T_k} \subt \pprocoutty{\q}{}{\probn{1}}{\ell_k}{\Basic_k}{\T_k'''}$, giving $\Delta_0' \predcontnd{\actionnd{\pp}} \Delta''$ with $\Delta_0 \subt \Delta''$. As safety is preserved by $\ndstep$ (\cref{def:safety}), the previous bullet applies to $\Delta''$, giving $\Delta'' \predcont{\actiontau{\pp}{\q}{\labelselli{k}}}{\probn{1}} \Delta_1' \supt \Delta_1$.
        \end{itemize}

    \item case \rulename{ct-sc}:
        \begin{equation}
        \inferrule*[left=\rulename{ct-sc}]
        { \Delta_s \predcont{\alpha}{t} \Delta_s' \quad t \in \{\prob, \nd\} }
        { \Delta_s, \pr : \T_\pr \predcont{\alpha}{t} \Delta_s', \pr : \T_\pr }
        \end{equation}
        so $\Delta_0 = \Delta_s, \pr:\T_\pr \predcont{\alpha}{t} \Delta_s', \pr:\T_\pr = \Delta_1$. As $\subt$ is componentwise, $\Delta_0' = \Delta_s'', \pr:\T_\pr'$ with $\T_\pr \subt \T_\pr'$ and $\Delta_s \subt \Delta_s''$ and by case 2 of \cref{lemma:safeproperties} $\safe{\Delta_s''}$.
        Since $\pr$ is not a participant of $\alpha$ (now one of $\actiontau{\pp}{\q}{\labelselli{}}$, $\actionnd{\pp}$, $\actionprob{\pp}$), the role $\pp$ lies in $\dom{\Delta_s}$ and the case split on $\Delta_0'(\pp) = \Delta_s''(\pp)$ is that of the $\IH$ applied to $\Delta_s \subt \Delta_s''$; re-framing each resulting transition and subtyping with $\pr:\T_\pr$ ($\pr:\T_\pr'$ resp.) preserves the statement, as $\T_\pr \subt \T_\pr'$.

    \item case \rulename{ct-rec}:
        \begin{equation}
        \inferrule*[left=\rulename{ct-rec}]
        { \pp: \T\subst{\mut}{\ty} \predcont{\alpha}{t} \pp:\T' \quad t \in \{\prob, \nd\} }
        { \pp: \mut \predcont{\alpha}{t} \pp:\T' }
        \end{equation}
        i.e. $\Delta_0 = \pp:\mut$ and $\Delta_1 = \pp:\T'$. Since $\subt$ is preserved under unfolding, $\pp:\T\subst{\mut}{\ty} \subt \Delta_0'$, and the $\IH$ on the premise --- same $\Delta_0'$, same target $\Delta_1$ --- gives the claim verbatim.
\end{itemize}
\end{proof}

\begin{definition}[Synchronised subtyping]
\label{def:synct}
We write $\Delta \synct \hat{\Delta}$ when $\Delta \subt \hat{\Delta}$, $\safe{\hat{\Delta}}$, and the supertype is committed wherever the subtype is:
\begin{equation*}
    \forall \pr \in \dom{\Delta}\st \Delta(\pr) \not\ndstep \text{ implies } \hat{\Delta}(\pr) \not\ndstep.
\end{equation*}
\end{definition}

\noindent Intuitively, $\hat\Delta$ has resolved its non-determinism at least as far as $\Delta$ has; the converse implication can fail, as a committed $\hat\Delta(\pr)$ may factor a sum of $\Delta(\pr)$ (\cref{rem:factoring}). Synchronised contexts enable exactly the same communications:

\begin{lemma}[Enabledness under synchronised subtyping (no phantoms)]
\label{lemma:enabledness}
Let $\Delta \synct \hat{\Delta}$. Then:
\begin{enumerate}
    \item $\forall \pp,\q,\labelsell\st$ $\Delta(\pp) \not\ndstep$ implies
        $\;\Delta \predcont{\actiontau{\pp}{\q}{\labelselli{}}}{\probn{1}} \;\text{ iff }\; \hat{\Delta} \predcont{\actiontau{\pp}{\q}{\labelselli{}}}{\probn{1}}$
    \item $\forall \pp,\prob\st$ $\Delta(\pp) \not\ndstep$ implies
        $\;\Delta \predcont{\actionprob{\pp}}{\prob} \;\text{ iff }\; \hat{\Delta} \predcont{\actionprob{\pp}}{\prob}$
    \item $\forall \pr\st \hat{\Delta}(\pr) \ndstep$ implies $\Delta(\pr) \ndstep$.
\end{enumerate}
\end{lemma}
\begin{proof}
    (3) is the contrapositive of the condition in \cref{def:synct}. (1), (2): as $\Delta(\pp)\not\ndstep$, by that condition $\hat{\Delta}(\pp) \not\ndstep$ too, so $\Delta(\pp) \subt \hat{\Delta}(\pp)$ is by \rulename{sub-$\oplus$}: both carry the \emph{same} branches with the \emph{same} probabilities towards the same partners. (2) is immediate: one side is a proper distribution iff the other is, resolving the corresponding branch at the same $\prob$; no partner is involved. For (1), both $\pp$-components are the same singleton selection, and by \rulename{sub-$\&$} the branches of $\hat{\Delta}(\q)$ are a subset of those of $\Delta(\q)$:
    \begin{itemize}
        \item ($\Rightarrow$) If $\Delta \predcont{\actiontau{\pp}{\q}{\labelselli{}}}{\probn{1}}$ then $\hat{\Delta}(\pp)\predcont{\actionprocout{\pp}{\q}{\ell}{\Basic}}{\probn{1}}$, and $\safe{\hat{\Delta}}$ (\cref{def:safety}) guarantees $\hat{\Delta}(\q)\predcont{\actionprocin{\q}{\pp}{\ell}{\Basic}}{\probn{1}}$.
        \item ($\Leftarrow$) If $\hat{\Delta} \predcont{\actiontau{\pp}{\q}{\labelselli{}}}{\probn{1}}$ then $\Delta(\pp)\predcont{\actionprocout{\pp}{\q}{\ell}{\Basic}}{\probn{1}}$, and by contravariance of \rulename{sub-$\&$}, $\ell$ is offered by $\q$ too: $\Delta(\q)\predcont{\actionprocin{\q}{\pp}{\ell}{\Basic}}{\probn{1}}$. \qedhere
    \end{itemize}
\end{proof}

\begin{lemma}[Synchronisation]
\label{lemma:sync}
If $\Delta \subt \Delta'$ and $\safe{\Delta'}$ then $\Delta' \predcontndstar{}{} \hat{\Delta}$ with $\Delta \synct \hat{\Delta}$.
\end{lemma}
\begin{proof}
    Consider any $\pp$ with $\Delta(\pp) \not \ndstep$ but $\Delta'(\pp) \ndstep$. We must have $\Delta'(\pp) = \sum_{i \in I}\T_i$ with $|I|>1$ (as $\Delta'(\pp)\ndstep$), from \rulename{sub-$\Sigma$} we know $\exists i \in I \st \T_i \supt \Delta(\pp)$, commiting to this $\T_i$ through a transition on \rulename{ct-nd} gives: $\Delta' \predcontnd{\actionnd{\pp}}\Delta''$ where $\Delta \subt \Delta''$, $\safe{\Delta''}$ (by \cref{def:safety}) and $\Delta''(\pp) \not \ndstep$. After at most $|\dom{\Delta}| - 1$ more steps $\Delta''\predcontndstar{}{} \hat{\Delta}$ every such $\pp$ is committed, giving $\Delta \synct \hat{\Delta}$.
\end{proof}

\begin{lemma}[One-step preservation]
\label{lemma:syncstep}
Let $\Delta \synct \hat{\Delta}$ and $\Delta \predcont{\alpha}{} \Delta_1$. Then there is $\hat{\Delta}_1$ with $\Delta_1 \synct \hat{\Delta}_1$, where:
\begin{enumerate}
    \item if $\alpha = \actiontau{\pp}{\q}{\labelselli{}}$, then $\hat{\Delta} \predcont{\actiontau{\pp}{\q}{\labelselli{}}}{\probn{1}} \cdot \predcontndstar{}{} \hat{\Delta}_1$: the \emph{same} communication, followed by nd-steps;
    \item if $\alpha = \actionnd{\pr}$, then either $\hat{\Delta} \predcontnd{\actionnd{\pr}} \hat{\Delta}_1$ or $\hat{\Delta}_1 = \hat{\Delta}$ (no step; the factored situation of \cref{rem:factoring});
    \item if $\alpha = \actionprob{\pp}$ at probability $\prob$, then $\hat{\Delta} \predcont{\actionprob{\pp}}{\prob} \hat{\Delta}_1$: the \emph{same} prob-step, at the \emph{same} probability.
\end{enumerate}
\end{lemma}
\begin{proof}
    (1) A synchronisation fires only from a singleton selection, so $\Delta(\pp) \not\ndstep$ and, by \cref{def:synct}, $\hat{\Delta}(\pp) \not\ndstep$; \cref{lemma:deltasimulation}(1a) gives $\hat{\Delta} \predcont{\actiontau{\pp}{\q}{\labelselli{}}}{\probn{1}} \Delta_1'$ with $\Delta_1 \subt \Delta_1'$, and \cref{lemma:sync} re-synchronises $\Delta_1'$ to $\hat{\Delta}_1$ with $\Delta_1 \synct \hat{\Delta}_1$ (only the continuations at $\pp$ and $\q$ may need committing).
    (2) By \cref{lemma:deltasimulation}(2): if $\hat{\Delta}(\pr) \ndstep$, case (2b) gives $\hat{\Delta} \predcontnd{\actionnd{\pr}} \hat{\Delta}_1$ with $\Delta_1 \subt \hat{\Delta}_1$; otherwise case (2a) gives $\Delta_1 \subt \hat{\Delta} =: \hat{\Delta}_1$. Synchronisation holds without further steps: the only component newly committed in $\Delta_1$ is $\pr$, and $\hat{\Delta}_1(\pr)$ is committed (in (2b) by the step just taken, in (2a) by hypothesis); all other components are unchanged on both sides.
    (3) A prob-step $\pp : p$ fires only from a ($\Sigma$-committed) proper distribution, so as in (1), $\hat{\Delta}(\pp) \not\ndstep$, and \cref{lemma:deltasimulation}(3a) gives $\hat{\Delta} \predcont{\actionprob{\pp}}{\prob} \hat{\Delta}_1$ with $\Delta_1 \subt \hat{\Delta}_1$. Synchronisation needs no further steps: only $\pp$ changes, both sides becoming singleton selections, hence $\not\ndstep$; all else is unchanged.
\end{proof}

\begin{corollary}[Lock-step path]
\label{corollary:lockstep}
Let $\Delta_0 \subt \Delta_0'$ with $\safe{\Delta_0'}$, and let $\tpath = \Delta_0 \predcont{\alpha_1}{} \Delta_1 \predcont{\alpha_2}{} \cdots$ be a (finite or infinite) path. Then there is a path $\tpath'$ from $\Delta_0'$ and states $\hat{\Delta}_0, \hat{\Delta}_1, \ldots$ along it --- the \emph{stages} of $\tpath'$ --- with $\Delta_i \synct \hat{\Delta}_i$ for all $i$ and $\Delta_0' \predcontndstar{}{} \hat{\Delta}_0$, such that the segment of $\tpath'$ from $\hat{\Delta}_i$ to $\hat{\Delta}_{i+1}$ is:
\begin{itemize}
    \item the same communication $\alpha_{i+1}$ (at probability $\probn{1}$), followed by finitely many $\ndstep$;
    \item the same prob-step $\alpha_{i+1} = \actionprob{\pp}$, at the same probability;
    \item at most one nd-step $\ndstep$, if $\alpha_{i+1} = \actionnd{\pr}$ (taken iff $\hat{\Delta}_i(\pr) \ndstep$).
\end{itemize}
In particular $\tpath$ and $\tpath'$ perform the same sequence of communications and prob-steps, with the same labels and probabilities; as all padding is by nd-steps, corresponding finite prefixes carry the same product of transition probabilities.
\end{corollary}
\begin{proof}
    \cref{lemma:sync} gives $\Delta_0' \predcontndstar{}{} \hat{\Delta}_0$ with $\Delta_0 \synct \hat{\Delta}_0$; iterating \cref{lemma:syncstep} along $\tpath$ (cases (1)/(3)/(2) respectively) gives each segment and stage.
\end{proof}

\begin{lemma}[Scheduler transfer along subtyping]
\label{lem:scheduler-transfer}
Let $\Delta \leqslant \Delta'$ with $\safe{\Delta'}$, and let
$\phi \in \Phi_F^{\Delta}$ be a fair scheduler.  Then there exists a
fair scheduler $\phi' \in \Phi_F^{\Delta'}$ and a mapping
$(\cdot)^\dagger : \Tpath_{\phi}(\Delta) \to \Tpath_{\phi'}(\Delta')$
such that:
\begin{enumerate}
  \item \emph{(measure-isomorphic)} $(\cdot)^\dagger$ is a bijection
        of $\Tpath_{\phi}(\Delta)$ onto $\Tpath_{\phi'}(\Delta')$,
        and $E \subseteq \Tpath_{\phi}(\Delta)$ is an event iff
        $E^\dagger$ is, with $\probpathmeasure{\phi}{\Delta}(E) =
        \probpathmeasure{\phi'}{\Delta'}(E^\dagger)$;
  \item \emph{(property-reflecting)} $\tpath$ reaches a deadlocked
        context iff $\tpath^\dagger$ does; and if $\tpath^\dagger$
        is live, then so is $\tpath$.
\end{enumerate}
\end{lemma}
\begin{proof}
    Take any path of the $\dtmc$ induced by $\phi$: $\Delta \predcont{\alpha_1}{} \Delta_1 \predcont{\alpha_2}{} \cdots \in \Tpath_\phi(\Delta)$.
    We map it to its \emph{lock-step path} $\tpath'$ with stages $\hat{\Delta}_i \st \Delta_i \synct \hat{\Delta}_i$ (\cref{corollary:lockstep}), made unique by fixing a total order on participants and committing $\predcontnd{\actionnd{\pp}}$ before $\predcontnd{\actionnd{\q}}$ if $\pp \prec \q$; we write $\tpath^\dagger = \tpath'$ for this mapping.
    The scheduler $\phi'$ then simply mirrors $\phi$'s decisions through the correspondence, performing the canonical padding steps in between.

    \myparagraph{$\tpath'$ is fair.} We discharge the three fairness requirements on $\tpath'$ from those on $\tpath$. Throughout we use that internal enabledness is \emph{persistent}: a participant at a proper sum (resp.\ proper distribution) can only move by its own nd-step (resp.\ prob-step), so it stays enabled until it moves.
    \begin{enumerate}[label=\roman*)]
        \item (\emph{internal steps}) Suppose towards a contradiction that an internal step of $\pr$ ($\actionnd{\pr}$ or $\actionprob{\pr}$) is enabled infinitely often but taken finitely often in $\tpath'$; by persistence, from some stage $m$ on it is continuously enabled and $\pr$ never moves in $\tpath'$.
            Then $\pr$'s step is eventually continuously enabled in $\tpath$ too (\cref{lemma:enabledness}(2,3)) --- for a prob-step, $\Delta_i(\pr)$ may by \rulename{sub-$\Sigma$} first be a factored proper sum (\cref{rem:factoring}), whose resolution fairness of $\tpath$ provides, absorbed in $\tpath'$ without movement (\cref{lemma:syncstep}(2)).
            By fairness of $\tpath$, $\pr$ then takes its step at some $j \geq m$, and \cref{lemma:syncstep}(2,3) mirrors it
in $\tpath'$ --- contradiction.
        \item (\emph{communication}) Suppose $a = \actiontau{\pp}{\q}{\labelselli{}}$ is enabled infinitely often in $\tpath'$; as $\tpath$ and $\tpath'$ share their communication sequence, it suffices that $\pp$ communicates infinitely often in $\tpath$. Suppose not: after its last communication, fairness of $\tpath$ eventually resolves $\pp$'s nd- and prob-steps, so from some stage $N$ on $\Delta_i(\pp) = \Delta_N(\pp)$ is a singleton selection, a branching, or $\tend$.
            There are infinitely many $i \geq N$ with $\hat{\Delta}_i \predcont{a}{\probn{1}}$, where $\hat{\Delta}_i(\pp)$ is a singleton selection; being its subtype, $\Delta_N(\pp)$ is by \rulename{sub-$\oplus$} the \emph{same} singleton, so \cref{lemma:enabledness}(1) gives $\Delta_i \predcont{a}{\probn{1}}$. Thus $a$ is enabled infinitely often in $\tpath$, and fairness makes $\pp$ communicate infinitely often --- contradiction.
    \end{enumerate}

Finally, we note that \emph{images are complete paths} --- let $\tpath \in \Xi_\phi(\Delta)$. If $\tpath$ is finite with $\Delta_{|\tpath|} \not\rightarrow$, then its final stage $\hat{\Delta}$ satisfies $\Delta_{|\tpath|} \ltimes \hat{\Delta}$, and $\hat{\Delta} \not\rightarrow$: no nd-step by \cref{def:synct} (the supertype is committed wherever the subtype is, and $\Delta_{|\tpath|}$ enables no nd-step), and no probabilistic step or communication by \cref{lemma:enabledness}(1,2).
Hence $\tpath^\dagger$ is maximal. If $\tpath$ is infinite, we claim $\tpath^\dagger$ is infinite: $\tpath$ can contain no more than $|\dom{\Delta}|$ consecutive nd-steps, so $\tpath$ performs infinitely many communications and probabilistic steps, each mirrored in $\tpath^\dagger$ by \cref{corollary:lockstep}.
Finally, every complete path induced by $\phi'$ from $\Delta'$ is by construction the companion of a (unique) $\phi$-path, so $(\cdot)^\dagger$ maps $\Xi_\phi(\Delta)$ onto $\Xi_{\phi'}(\Delta')$.

    The 2 claims of the lemma now follow:
    \begin{enumerate}
        \item By the closing paragraph above, $(\cdot)^\dagger$ maps $\Tpath_\phi(\Delta)$ onto $\Tpath_{\phi'}(\Delta')$; moreover, the cylinder sets of the two \textsf{DTMC}s correspond: for a finite $\phi$-prefix $\tpath$ with companion $\tpath^\dagger$, $\cylsig{\tpath}^\dagger = \cylsigp{\tpath^\dagger}$ at equal probability, since the interleaved nd-padding carries probability $\probn{1}$. We show injectivity as follows: Consider two $\phi$-induced paths $\tpath_1$ and $\tpath_2$, they share the scheduler's selection along their common prefix, so their first difference lies between outcomes of a single scheduled action --- necessarily a prob-step, every other action having probability $\probn{1}$ --- and by the lock-step correspondence (\cref{corollary:lockstep}) their images $\tpath_1^\dagger$ and $\tpath_2^\dagger$ perform the same probabilistic step, resolved the same way, at the corresponding stage, and hence differ as well.

        \item \emph{Deadlock.} If $\xi$ ends in a deadlocked $\Delta_{|\xi|}$, then as shown above $\xi^\dagger$ ends in the stuck stage $\hat{\Delta} \geqslant \Delta_{|\xi|}$, and the deadlock witness transfers: $\Delta_{|\xi|}(\pp) \neq \tend$ gives $\hat{\Delta}(\pp) \neq \tend$, as \rulename{sub-end} relates $\tend$ only to $\tend$.
            Conversely, suppose $\tpath^\dagger$ reaches a deadlocked context; being stuck, it is the endpoint of $\tpath^\dagger$, i.e.\ its final stage $\hat{\Delta}$.
            Then $\tpath$ is finite: otherwise it performs infinitely many communications and prob-steps (bounded nd-runs, as above), each extending $\tpath^\dagger$ beyond $\hat{\Delta}$ (\cref{corollary:lockstep}) --- contradiction.
            Being complete, $\Delta_{|\tpath|} \not\rightarrow$, with $\Delta_{|\tpath|} \synct \hat{\Delta}$; and the witness transfers down: $\hat{\Delta}(\pp) \neq \tend$ forces $\Delta_{|\tpath|}(\pp) \neq \tend$ by \rulename{sub-end}. Hence $\tpath$ ends deadlocked.

    \emph{Liveness.} Suppose $\tpath^\dagger$ is live; we show every pending action along $\xi$ is eventually resolved.
    If $\Delta_i(\pp) \predcont{\actionprocout{\pp}{\q}{\ell}{\Basic}}{\probn{1}}$, then $\Delta_i(\pp)$ is a singleton selection, so by $\synct$ and \rulename{sub-$\oplus$} the stage $\hat{\Delta}_i(\pp)$ is the \emph{same} singleton and the output is pending in $\xi^\dagger$; liveness of $\xi^\dagger$ provides a communication of $\pp$ at some later point, which lies in the shared communication sequence of the lock-step sequence, and hence occurs in $\xi$.
    If $\Delta_i(\q) \predcont{\actionprocin{\q}{\pp}{\ell}{\Basic}}{\probn{1}}$, then by contravariance of \rulename{sub-$\&$} the stage $\hat{\Delta}_i(\q)$ offers a (subset of $\q$'s) input, pending in $\xi^\dagger$; liveness of $\xi^\dagger$ provides a communication involving $\q$ as receiver, shared with $\xi$ likewise. Hence $\xi$
is live.
    \end{enumerate}
\end{proof}

\begin{corollary}[deadlock-freedom is downwards-closed]
\label{corollary:dfclosed}
If $\Delta \subt \Delta'$ with $\sdf{\Delta'}$ then $\df{\Delta}$.
\end{corollary}
\begin{proof}
Suppose $\Delta \predcontstar{}{} \Delta_n$ with $\deadlock{\Delta_n}$; extend this finite path to a complete path $\tpath$, induced by a fair scheduler.
By \cref{lem:scheduler-transfer}(2), $\tpath^\dagger$ reaches a deadlocked context, contradicting $\df{\Delta'}$.
\end{proof}

\begin{corollary}[liveness is downwards-closed]
\label{corollary:livenessclosed}
If $\Delta \subt \Delta'$ with $\slive{\Delta'}$
then $\live{\Delta}$.
\end{corollary}
\begin{proof}
Take any fair complete path $\tpath \in \pathcont{\Delta}$, induced by a fair scheduler. Then $\tpath^\dagger \in \Xi_{\phi'}(\Delta')$ is fair (\cref{lem:scheduler-transfer}), hence live by $\live{\Delta'}$, hence $\tpath$ is live by \cref{lem:scheduler-transfer}(2). As $\tpath$ was arbitrary, $\live{\Delta}$.
\end{proof}

\submono*
\begin{proof}
    Each of (1) and (2) are proved in turn:
    \begin{enumerate}
        \item \mbox{}
        \begin{enumerate}[label=\roman*)]
            \item $\safe{\Delta'} \implies \safe{\Delta}$ by~\cref{lemma:safeproperties}(3)
            \item $\df{\Delta'} \implies \df{\Delta}$ by~\cref{corollary:dfclosed}
            \item $\live{\Delta'} \implies \live{\Delta}$ by~\cref{corollary:livenessclosed}
        \end{enumerate}
    \item \mbox{}
        \begin{enumerate}[label=\roman*)]
            \item $\probdfmin{\Delta'} \leq \probdfmin{\Delta}$: take any fair $\phi \in \Phi_F^{\Delta}$ and let $\phi'$, $(\cdot)^\dagger$ be as in \cref{lem:scheduler-transfer}. By case (2), a path reaches a deadlocked context iff its image does; with bijectivity (case (1)), the deadlock-reaching events on the two sides are therefore each other's images, and case (1) gives
        $\probpathmeasure{\phi'}{\Delta'}(\deadlockk) = \probpathmeasure{\phi}{\Delta}(\deadlockk)$.
        As $\phi' \in \Phi_F^{\Delta'}$,
  \[
    \max_{\phi'' \in \Phi_F^{\Delta'}}
      \probpathmeasure{\phi''}{\Delta'}(\deadlockk)
    \;\geq\;
    \probpathmeasure{\phi'}{\Delta'}(\deadlockk)
    \;=\;
    \probpathmeasure{\phi}{\Delta}(\deadlockk),
  \]
  and as $\phi \in \Phi_F^{\Delta}$ was arbitrary,
  $\max_{\psi \in \Phi_F^{\Delta'}}
   \probpathmeasure{\psi}{\Delta'}(\deadlockk) \geq
   \max_{\phi \in \Phi_F^{\Delta}}
   \probpathmeasure{\phi}{\Delta}(\deadlockk)$; subtracting from
  $\probn{1}$ flips the inequality, giving
  $\probdfmin{\Delta'} \leq \probdfmin{\Delta}$.

  \item $\problivemin{\Delta'} \leq \problivemin{\Delta}$:
              take any fair $\phi \in \Phi_F^{\Delta}$ and let $\phi'$, $(\cdot)^\dagger$ be as in \cref{lem:scheduler-transfer}.  Write $L = \{\tpath \in \Xi_\phi(\Delta) \SEP \live{\tpath}\}$, $L' = \{\tpath' \in \Xi_{\phi'}(\Delta') \SEP \live{\tpath'}\}$, and $A = \{\tpath \in \Xi_\phi(\Delta) \SEP \live{\tpath^\dagger}\}$.  Since $(\cdot)^\dagger$ is a bijection of $\Xi_\phi(\Delta)$ onto $\Xi_{\phi'}(\Delta')$ (case (1)), $A^\dagger = L'$; hence $A$ is an event iff $L'$ is --- and $L'$ is --- with $\probpathmeasure{\phi}{\Delta}(A) = \probpathmeasure{\phi'}{\Delta'}(L')$ by case (1).
              By case (2), $\live{\tpath^\dagger}$ implies $\live{\tpath}$, i.e.\ $A \subseteq L$.  As $\phi'$ is one particular fair scheduler of $\Delta'$ while $\problivemin{\Delta'}$ is an infimum over all of them:
              \[
                \problivemin{\Delta'}
                \;\leq\; \probpathmeasure{\phi'}{\Delta'}(L')
                \;=\; \probpathmeasure{\phi}{\Delta}(A)
                \;\leq\; \probpathmeasure{\phi}{\Delta}(L).
              \]
              As $\phi \in \Phi_F^{\Delta}$ was arbitrary, taking the infimum over $\phi$ on the right yields $\problivemin{\Delta'} \leq \problivemin{\Delta}$.
        \end{enumerate}
    \end{enumerate}
\end{proof}

\subsection{Proofs for \Cref{sec:processproperties} (Lower bounds on properties checked on $\Delta$ transfer to $\N$)}
\label{app:sec44proofs}

\begin{definition}[Distribution reduction semantics]
The internal distribution reductions on sessions and typing contexts,
their lifting to distributions, and their reflexive--transitive
closure and normalization are defined in Fig.~\ref{fig:intred}.
We denote by $\distN$ and $\distD$ distributions over sessions $\N$ and typing contexts $\Delta$ respectively, and lift singleton states $\N$ or $\Delta$ to distributions $\dist$ in the expected way, as $\{\N \mapsto \probn{1}\}$ and $\{\Delta \mapsto \probn{1}\}$. We use $\dist$ to refer to either of $\distN$ and $\distD$ when the statement is general enough to warrant it, and $\config$ to either of $\N$ or $\Delta$.
We denote by $\dom{\dist}$ the finite positive support (domain) of $\dist$.

In the figure, $\dist \not\intred{}$ means $\not\exists \dist',\alpha \st \dist\intred{}\dist'$ and we say that $\dist$ is in ``communication normal-form'' $\dist \models \intnf{}$ iff $\dist \not\intred{}$ and note that:
\[
    \dist \models \intnf{}
    \iff
    \forall \config\in\dom{\dist}: \config\models\intnf{}.
\]
\begin{figure}[t]
\centering
\small
\begin{minipage}{\linewidth}
\setlength{\abovedisplayskip}{4pt}
\setlength{\belowdisplayskip}{4pt}
\setlength{\abovedisplayshortskip}{4pt}
\setlength{\belowdisplayshortskip}{4pt}
\noindent\textbf{(a) Session reductions ($\N$)}
\[
\begin{gathered}
    \inferrule[\rulename{i-flip-diff}]
    {
        \N\spcon
        \pa{\pp}{\flip{\prob}{\PP}{\Q}}\pc\N_0
        \\
        \pa{\pp}{\PP}\pc\N_0\spcon\N'
        \quad
        \pa{\pp}{\Q}\pc\N_0\spcon\N''
        \quad
        \N'\not\equiv\N''
    }
    {
        \N\intred{}
        \{\N'\mapsto\prob,\,
          \N''\mapsto\probn{1}-\prob\}
    }
    \\[1mm]
    \inferrule[\rulename{i-flip-id}]
    {
        \N\spcon
        \pa{\pp}{\flip{\prob}{\PP}{\Q}}\pc\N_0
        \\
        \pa{\pp}{\PP}\pc\N_0\spcon\N'
        \quad
        \pa{\pp}{\Q}\pc\N_0\spcon\N'
    }
    {
        \N\intred{}\{\N'\mapsto\probn{1}\}
    }
    \\[1mm]
    \inferrule[\rulename{i-if}]
    {
        \N\spcon
        \pa{\pp}{\cond{\e}{\PP_{\true}}{\PP_{\false}}}
        \pc\N_0
        \\
        b\in\{\true,\false\}
        \quad
        \peval{\e}{b}{}
        \quad
        \pa{\pp}{\PP_b}\pc\N_0\spcon\N'
    }
    {
        \N\intred{}\{\N'\mapsto\probn{1}\}
    }
\end{gathered}
\]
\par\hrule height 0.4pt\relax
\vspace{1mm}
\noindent\textbf{(b) Typing-context reductions ($\Delta$)}
\[
    \inferrule[\rulename{i-prob}]
    {
        \{\Delta
          \predcont{\actionprob{\pp}}{\probi{i}}
          \Delta_i\}_{i\in I}
        \quad
        \Sigma_{i\in I}\probi{i}=\probn{1}
    }
    {
        \Delta \intred{}
        \{\Delta_i\mapsto\probi{i}\}_{i\in I}
    }
    \qquad
    \inferrule[\rulename{i-nd}]
    {
        \Delta \predcontnd{\actionnd{\pp}} \Delta'
    }
    {
        \Delta \intred{}
        \{\Delta'\mapsto\probn{1}\}
    }
\]

\par\hrule height 0.4pt\relax
\vspace{1mm}
\noindent\textbf{(c) Distribution lifting, closure and normal-form}
\[
\begin{gathered}
    \inferrule[\rulename{i-dist}]
    {
        \prob + \Sigma_{i \in I} \probi{i} = \probn{1}
        \quad
        \config \intred{}
        \{\config_j\mapsto\probip{j}\}_{j\in J}
    }
    {
        \{\config\mapsto\prob\}
        \uplus
        \{\config_i\mapsto\probi{i}\}_{i\in I}
        \intred{}
        \{\config_i\mapsto\probi{i}\}_{i\in I\setminus J}
        \uplus
        \{\config_i\mapsto
          \prob\cdot\probip{i}\}_{i\in J\setminus I}
        \uplus
        \{\config_i\mapsto
          \probi{i}+\prob\cdot\probip{i}\}_{i\in I\cap J}
    }
    \\[2mm]
    \inferrule[\rulename{i-refl}]
    { }
    {
        \dist \intredstar{} \dist
    }
    \qquad
    \inferrule[\rulename{i-trans}]
    {
        \dist \intred{} \dist'
        \quad
        \dist' \intredstar{} \dist''
    }
    {
        \dist
        \intredstar{}
        \dist''
    }
    \qquad
    \inferrule[\rulename{i-nf}]
    {
        \dist \intredstar{} \dist'
        \quad
        \dist' \not\intred{}
    }
    {
        \dist \intrednf{} \dist'
    }
\end{gathered}
\]
\end{minipage}

\caption{Internal distribution reductions of sessions and typing
contexts, their lifting to distributions, and normalization.
Here $\config$ ranges over $\N$ and $\Delta$, and $\dist$ over
$\distN$ and $\distD$, respectively.
}
\label{fig:intred}
\label{fig:intrednf}
\end{figure}

\end{definition}

\begin{definition}[Typing coupling]
\label{def:typing-coupling}
We lift typing derivations $\der{}{\N}{\Delta}$ to distributions on sessions and contexts as $\dercoup{}{\distN}{\distD}{\gamma}$ iff the \emph{coupling} $\gamma : \dom{\distN} \times \dom{\distD} \rightarrow [0;1]$ satisfies:
    \begin{enumerate}
        \item $\displaystyle\forall\N' \in \dom{\distN} :\sum_{\Delta' \in \dom{\distD}} \gamma(\N',\Delta') = \distN(\N')$
        \item $\displaystyle\forall\Delta' \in \dom{\distD}:\sum_{\N' \in \dom{\distN}} \gamma(\N',\Delta') = \distD(\Delta')$
        \item $\forall\N' \in \dom{\distN}:\forall \Delta'\in\dom{\distD}: \gamma(\N',\Delta')>0 $ implies $\der{}{\N'}{\Delta'}$
    \end{enumerate}
\end{definition}

\begin{definition}[Internal-first schedulers]
\label{def:internal-first-schedulers}
We define the set of \emph{internal-first} schedulers over sessions or contexts $\config$;
\[
\begin{aligned}
    \Phi_I^{\config}
    =
    \bigl\{
        \phi\in\Phi^{\config}
        \;\bigm|\;&
        \forall\tpath\in\Tpath_\phi(\config):
        \forall i<|\tpath|\;
        \forall\pp,\q,\ell:
        \tpath[i]
        \predcont{\actiontau{\pp}{\q}{\ell}}{\probn{1}}
        \tpath[i+1]
        \implies
        \tpath[i]\models\intnf{}
    \bigr\},
\end{aligned}
\]
i.e. schedulers that perform internal reductions before any communication.
The set of \emph{fair internal-first} schedulers is then: $\Phi_{I,F}^{\config} = \Phi_I^{\config}\cap\Phi_F^{\config}$.
\end{definition}

\begin{notation}
\mbox{}
\begin{enumerate}
    \item For a scheduler $\phi \in \Phi_I^\config$, we write $\config \intrednfsched{}{\phi} \dist$ for the distribution $\dist$ induced by $\phi$ reducing $\config$ to normal-form.
    \item For a scheduler $\phi \in \Phi^\config$, we write $\phi_\tpath$ for the residual scheduler after executing $\tpath$ i.e. $\phi_\tpath = \phi(\tpath;\; \cdot)$.
\end{enumerate}
\end{notation}

\begin{restatable}[Internal normalization] {lemma}{internalNormalization}
\label{lemma:internal-normalization}
\mbox{}
\begin{enumerate}
    \item For all sessions $\N$:
        \begin{enumerate}
            \item $\forall \phi \in \Phi_I^\N:\exists\distN: \N \intrednfsched{}{\phi} \distN$
            \item $\forall \phi, \phi' \in \Phi_I^\N: \left( \N \intrednfsched{}{\phi} \distN \land \N \intrednfsched{}{\phi'} \distN' \right) \implies \distN = \distN'$
        \end{enumerate}
    \item For all contexts $\Delta$:
        \begin{enumerate}
            \item $\forall \phi \in \Phi_I^\Delta:\exists\distD: \Delta \intrednfsched{}{\phi} \distD$
        \end{enumerate}
\end{enumerate}
\end{restatable}
\begin{proof}
\begin{enumerate*}
    \item
        \begin{enumerate*}
            \item follows from guardedness of recursion (all recursive variables $X$ must occur following some communication in its prefix). Hence since our types are regular, firing internal actions must eventually lead to a \emph{non-internal} action.
            \item Since both of $\phi$ and $\phi'$ are \emph{internal-first}, and there's no \emph{internal nondeterminism} at the process-level, they can only differ up to permutation in their scheduling of the same set of internally enabled actions.
        \end{enumerate*}
    \item
        \begin{enumerate*}
            \item Follows by an argument analogous to (1)(a) and from the fact that recursion in types are guarded.
        \end{enumerate*}
\end{enumerate*}
\end{proof}

\begin{restatable}[Process inversion under internal normalization]
    {lemma}{processInversionNF}
\label{lemma:process-inversion-nf}
For every participant $\pp$,
\[
    \der{}{\PP}{T}
    \;\land\;
    \pa\pp{\PP}\intrednf{}\distN
    \;\implies\;
    \exists \gamma, \phi \in \Phi_I^{\pp{:}\T},\distD:
    \left(
        (\pp:T)\intrednfsched{}{\phi}\distD
        \;\land\;
        \dercoup{}{\distN}{\distD}{\gamma}
    \right).
\]
\end{restatable}
\begin{proof}
Put $\Delta_0=\pp:\T$. Singleton contexts are safe, so
$\der{}{\pa\pp\PP}{\Delta_0}$ by \rulename{b-sess}.

\begin{enumerate}
    \item $\Delta_0\models\intnf{}$.
    Then $\unf{\T}$ is $\tend$, a branching, or a singleton
    selection. The merge and subtyping rules give
    \[
        \T_1\probmerge{\prob}\T_2\subt\T
        \quad\Longrightarrow\quad
        \T_b\subt\T\qquad(b=1,2).
    \]
    Hence \cref{thm:subred}(1,3) and \rulename{t-sub}
    preserve typing at $\Delta_0$ through every internal step.
    By \rulename{i-dist},
    \[
        \forall\N'\in\dom{\distN}:\qquad
        \der{}{\N'}{\Delta_0}.
    \]
    Choose any $\phi\in\Phi_I^{\Delta_0}$ and set
    \[
        \distD=\{\Delta_0\mapsto\probn{1}\},
        \qquad
        \gamma(\N',\Delta_0)=\distN(\N').
    \]
    Then $\Delta_0\intrednfsched{}{\phi}\distD$ by
    \rulename{i-refl} and \rulename{i-nf}, and both coupling
    marginals are immediate.

    \item $\Delta_0\not\models\intnf{}$.
    Then
    \[
        \unf{\T}=
        \probpprocoutty{i\in I}{\q_j}{j\in J_i}
            {\probi{i,j}}{\ell_j}{\Basic_j}{\T_{i,j}}.
    \]
    Reduce the session into its finite tree of normalized
    outcomes. Write $\Rpath$ for the multiset of leaf occurrences (indexed by the unique path $\rpath$ that lead there), $\N_\rpath=\pa\pp{\PP_\rpath}$ for their endpoints , and $\probn{p_\rpath}$ for their weights. Thus
    \[
        \distN(\N')
        =\sum_{\substack{\rpath\in\Rpath\\
                         \N_\rpath=\N'}}
            \probn{p_\rpath}.
    \]
    By \cref{lemma:processinversion}, some $i\in I$ gives
    \[
        \Rpath=\biguplus_{j\in J_i}\Rpath_j,
        \qquad
        \sum_{\rpath\in\Rpath_j}\probn{p_\rpath}
            =\probi{i,j},
    \]
    where $\Rpath_j$ contains the leaves exposing
    $\q_j!\ell_j$. Put
    \[
        \T_j^*:=
        \pprocoutty{\q_j}{}{\probn{1}}
            {\ell_j}{\Basic_j}{\T_{i,j}},
        \qquad
        \Delta_j:=\pp:\T_j^*.
    \]

    The same inversion induction also gives the whole-endpoint
    typings
    \[
        \der{}{\PP_\rpath}{\T_j^*}
        \qquad(j\in J_i,\ \rpath\in\Rpath_j).
    \]
    By \rulename{b-sess},
    \[
        \der{}{\N_\rpath}{\Delta_j}
        \qquad(\rpath\in\Rpath_j).
    \]

    Let $\phi$ select summand $i$ when $|I|>1$, then schedule
    its probabilistic step when $|J_i|>1$, extending it
    internal-first elsewhere. By \rulename{ct-rec},
    \rulename{i-nd}, and \rulename{i-prob},
    \[
        \Delta_0\intrednfsched{}{\phi}\distD,
        \qquad
        \distD=\{\Delta_j\mapsto\probi{i,j}\}_{j\in J_i}.
    \]
    Define
    \[
        \gamma(\N',\Delta_j)
        :=
        \sum_{\substack{\rpath\in\Rpath_j\\
                         \N_\rpath=\N'}}
            \probn{p_\rpath}.
    \]
    Its marginals are
    \[
    \begin{aligned}
        \sum_{j\in J_i}\gamma(\N',\Delta_j)
            &=\sum_{\substack{\rpath\in\Rpath\\
                              \N_\rpath=\N'}}
                \probn{p_\rpath}
              =\distN(\N'),\\
        \sum_{\N'\in\dom{\distN}}\gamma(\N',\Delta_j)
            &=\sum_{\rpath\in\Rpath_j}\probn{p_\rpath}
              =\probi{i,j}
              =\distD(\Delta_j).
    \end{aligned}
    \]
    Moreover,
    $\gamma(\N',\Delta_j)>0$ implies
    $\der{}{\N'}{\Delta_j}$ by the endpoint typings.
    Hence $\dercoup{}{\distN}{\distD}{\gamma}$.
\end{enumerate}
\end{proof}

\begin{restatable}[Session normalization correspondence]{corollary}{sessionNormalizationCorrespondence}
\label{corollary:session-normalization}
\[
    \der{}{\N}{\Delta}
    \;\land\;
    \N\intrednf{}\distN
    \;\implies\;
    \exists \gamma, \phi \in \Phi_I^{\Delta},\distD:
    \left(
        \Delta\intrednfsched{}{\phi}\distD
        \;\land\;
        \dercoup{}{\distN}{\distD}{\gamma}
    \right).
\]
\end{restatable}

\begin{restatable}[Internal-first session schedulers]{lemma}{internalFirstSessionSchedulers}
\label{lemma:intredsched}
\[
    \forall\phi\in\Phi_F^\N:
    \exists\phi'\in\Phi_{I,F}^\N:
    \forall\mathsf{prop}\in\{\livee,\dff\}:
    \probpathmeasure{\phi}{\N}(\mathsf{prop})
    =
    \probpathmeasure{\phi'}{\N}(\mathsf{prop}).
\]
Hence, for any property $\mathsf{prop}\in\{\livee,\dff\}$: $\min_{\phi\in\Phi_F^\N} \probpathmeasure{\phi}{\N}(\mathsf{prop}) = \min_{\phi\in\Phi_{I,F}^\N}
\probpathmeasure{\phi}{\N}(\mathsf{prop})$
\end{restatable}
\begin{proof}
Fix $\phi\in\Phi_F^\N$.
We show how to construct an ``internal-first'' scheduler $\phi'\in\Phi_I^\N$ with the same bounds as $\phi$ as follows:
Simulate $\phi$, performing internal reductions early in a
fixed participant order until normal form, but revealing
their outcomes to $\phi$ only when it requests those steps (essentially $\phi'$ is a bit ahead of $\phi$).
Retain $\phi$'s communication and error choices.
Internal reductions commute with unrelated communications
and preserve enabled errors; by
\cref{lemma:internal-normalization}, normalization is finite.

Couple $\xi,\xi'$ using the same outcome for each participant's successive flips.
Revealing outcomes only upon replay preserves both marginal laws.
Almost surely, $\xi$ is fair, so every early internal step is eventually replayed.
Thus the paths have the same local reductions (just shifted), communication sequence, and terminal outcome.

Normalization ensures internal fairness of $\xi'$.
If a pair $(\pp,\q)$ were enabled infinitely often on $\xi'$
but communicated only finitely often, $\pp$'s output would
eventually remain pending.
That output is eventually exposed also in $\xi$, where the
corresponding input prefixes of $\q$ recur or persist,
contradicting fairness.
Hence $\phi'\in\Phi_{I,F}^\N$.

The shared reductions preserve liveness and deadlock-freedom, giving the probability equalities.
The equality of infima follows from $\Phi_{I,F}^\N\subseteq\Phi_F^\N$.
\end{proof}

\begin{restatable}[Internal persistence]{lemma}{internalLocality}
\label{lemma:internal-locality}
Suppose $\config\intredstar{}\dist$.
For every $\alpha=\actiontau{\pp}{\q}{\ell}$,
\[
    \config\predcont{\alpha}{\probn{1}}
    \quad\Longrightarrow\quad
    \forall\config'\in\dom{\dist}:\;
    \config'\predcont{\alpha}{\probn{1}}.
\]
\end{restatable}
\begin{proof}
An enabled communication exposes matching output and input
prefixes at $\pp,\q$ (a singleton selection and a branching
for contexts).
Neither component admits an internal reduction, so internal
steps leave both unchanged, modulo unfolding.
The claim follows by induction through \rulename{i-dist}
and the reflexive--transitive closure.
\end{proof}

\begin{definition}[1-step blocks]
    We say that a participant $\pp$ is 1-step blocked in $\Delta$ (denoted $\blkp{\pp}{\Delta}$) iff
\[
\Delta(\pp) \neq \tend \land
\Delta \not\predcontnd{\actionnd{\pp}} \land \not\exists \prob \st\Delta \predcont{\actionprob{\pp}}{\prob} \land \not\exists \q, \ell \st \left( \Delta \predcont{\actiontau{\pp}{\q}{\ell}}{\probn{1}} \lor \Delta \predcont{\actiontau{\q}{\pp}{\ell}}{\probn{1}}\right)
\]
analogously, $\blkp{\pp}{\N}$ iff
\[
({\N(\pp)}\models\intnf{}) \;\land\;
\N(\pp) \neq \inact \land
\not\exists \q, \ell \st \left( \N \predcont{\actiontau{\pp}{\q}{\ell}}{\probn{1}} \lor \N \predcont{\actiontau{\q}{\pp}{\ell}}{\probn{1}}\right)
\]
\end{definition}

\begin{restatable}[Communication correspondence at normal forms]{lemma}{normalFormCommunication}
\label{lemma:normal-form-communication}
\mbox{}\\
Suppose $\der{}{\N}{\Delta}$ with $\N \models \intnf{}$ and $\Delta \models \intnf{}$, then:
\begin{enumerate}
    \item For every $\alpha=\actiontau{\pp}{\q}{\ell}$:
    \[
        \begin{array}{rl}
        & \N\predcont{\alpha}{\probn{1}}
        \;\text{ iff }\;
        \Delta\predcont{\alpha}{\probn{1}}\\
        \forall \N',\Delta': &
            \N\predcont{\alpha}{\probn{1}}\N'
            \;\land\;
            \Delta\predcont{\alpha}{\probn{1}}\Delta'
            \;\implies
            \der{}{\N'}{\Delta'}.
        \end{array}
    \]

    \item
        For every $\pp\in\dom{\Delta}$: $\N(\pp) \equiv 0 \;\text{ iff } \unf{\Delta(\pp)}=\mathsf{end}$.
\end{enumerate}
\end{restatable}
\begin{proof}
By inversion of \rulename{b-sess}, $\safemp{\Delta}$ and
each local process is typed by its corresponding entry.
We work up to head unfolding, using \cref{lemma:precong}.

\emph{(1)}
Fix $\alpha=\actiontau{\pp}{\q}{\ell}$.
If either side enables $\alpha$, inversion of
\rulename{t-sel}, \rulename{t-bra}, and \rulename{t-sub} gives
\[
\begin{aligned}
    \N
    &\equiv
    \pa\pp{\procout{\q}{\ell}{\e}{\PP}}
    \pc
    \pa\q{\psyntin{i\in I}{\pp_i}{i}{x_i}{\Q_i}}
    \pc \N_0,
    \\
    \unf{\Delta(\pp)}
    &=
    \pprocoutty{\q}{}{\probn{1}}{\ell}{\Basic}{\T},
    \\
    \unf{\Delta(\q)}
    &=
    \pprocinnew{\pp_j}{j\in J}{\ell_j}{\Basic_j}{\T_j},
\end{aligned}
\]
where, by \rulename{sub-$\oplus$} and \rulename{sub-$\&$},
\[
\begin{gathered}
    J\subseteq I,
    \qquad
    \{\pp_j\}_{j\in J}=\{\pp_i\}_{i\in I},
    \\
    \der{}{\e}{\Basic},
    \qquad
    \der{}{\PP}{\T},
    \qquad
    \forall j\in J:\quad
    \der{x_j:\Basic_j}{\Q_j}{\T_j}.
\end{gathered}
\]

If $\N$ enables $\alpha$, then
$\pp\in\{\pp_i\}_{i\in I}=\{\pp_j\}_{j\in J}$.
Thus $\Delta$ enables an input at $\q$ from $\pp$;
together with its output at $\pp$, safety then gives
\[
    \exists j\in J:\qquad
    (\pp_j,\ell_j,\Basic_j)=(\pp,\ell,\Basic).
\]
Hence $\Delta$ enables $\alpha$ by \rulename{ct-$\tau$}.
Conversely, a context $\alpha$-step supplies such a
$j\in J\subseteq I$, so \rulename{r-comm} gives a session
$\alpha$-step.

For any matching pair of steps, the branch $j$ is unique
by well-formedness, and the successors have the forms
\[
\begin{aligned}
    \N'
    &\equiv
    \pa\pp{\PP}
    \pc
    \pa\q{\Q_j\subst{\val}{x_j}}
    \pc \N_0,
    \\
    \Delta'
    &=\Delta[\pp\mapsto\T,\q\mapsto\T_j].
\end{aligned}
\]
By \cref{lemma:subst}(2),
$\der{}{\Q_j\subst{\val}{x_j}}{\T_j}$.
Safety is preserved by $\Delta\predcont{\alpha}{\probn{1}}\Delta'$,
and all other components retain their typings.
Thus \rulename{b-sess} and \cref{lemma:precong} give
$\der{}{\N'}{\Delta'}$.

\par\noindent
\emph{(2)}
Since $\N\models\intnf{}$, each local process unfolds to
$\inact$, an output, or an input.
Typing inversion and the subtyping rules match these
respectively with $\tend$, a singleton selection, or a
branching. Therefore
\[
    \N(\pp)\equiv\inact
    \quad\Longleftrightarrow\quad
    \unf{\Delta(\pp)}=\tend.
\]
\end{proof}

\begin{restatable}[1-step block correspondence]{corollary}{deadlockcorrespondence}
\label{cor:deadlockcorrespondence}
\mbox{}\\
Suppose $\der{}{\N}{\Delta}$ with $\N \models \intnf{}$ and $\Delta \models \intnf{}$ then: $\blkp{\pp}{\N}$ iff $\blkp{\pp}{\Delta}$
\end{restatable}

\begin{restatable}[Non-liveness as eventual permanent blocking]
    {lemma}{nonLivenessBlockedSuffix}
\label{lemma:nonliveness-blocked-suffix}
Let $\tpath$ be a complete fair path from
$\config$ (for $\config$ either a session $\N$ or context $\Delta$).

Then
\[
    \tpath\not\models\livee
    \;\text{ iff }\;
    \exists\pp\in\dom{\config}:
    \exists m\geq0:
    \forall k\geq m:
    \blkp{\pp}{\tpath[k]}.
\]
\end{restatable}
\begin{proof}
For finite $\tpath$, interpret $\tpath[k]=\last{\tpath}$
whenever $k>|\tpath|$; these repetitions perform no reductions.

$(\Rightarrow)$
Suppose $\tpath\not\models\livee$. There exist $\pp$ and $i$ such that an input or output pending at $\tpath[i]$ is never consumed. Its prefix therefore persists: at every subsequent position, $\pp$ is nonterminated and has no internal reduction. Define:
\[
    E=
    \left\{
        k\geq i
        \;\middle|\;
        \exists\q,\ell:
        \left(
            \tpath[k]
            \predcont{\actiontau{\pp}{\q}{\ell}}{\probn{1}}
            \;\lor\;
            \tpath[k]
            \predcont{\actiontau{\q}{\pp}{\ell}}{\probn{1}}
        \right)
    \right\}.
\]
The fixed pending prefix admits finitely many partners and labels. If $E$ were infinite, some serving communication label (respectively, participant pair for sessions) would be enabled infinitely often.
Fairness would force a communication consuming that prefix, a contradiction. Hence $E$ is finite. Choosing $m>\max(E\cup\{i\})$ gives
\[
    \forall k\geq m:\qquad
    \blkp{\pp}{\tpath[k]}.
\]
This also covers finite paths, since their terminal states enable no reductions.

\par\noindent
$(\Leftarrow)$
Suppose $\exists\pp,m:\forall k\geq m: \blkp{\pp}{\tpath[k]}$. By the definition of blockedness and the process/type syntax, $\pp$ has a pending input or output at $\tpath[m]$. If that prefix were consumed at some step $k\geq m$, a communication involving $\pp$ would be enabled at $\tpath[k]$, contradicting $\blkp{\pp}{\tpath[k]}$.
Thus the prefix is never consumed, and $\tpath\not\models\livee$.
\end{proof}

\begin{definition}[Normalization paths]
\label{def:normalization-paths}
For an internal-first scheduler $\phi$, define
\[
    \fpaths{\phi}{\config}
    =
    \left\{
        \xi[1..n]
        \;\middle|\;
        \begin{array}{l}
            \xi\in\Tpath_\phi(\config),
            \xi[n]\models\intnf{},
            \;\;
            \forall i<n:
                \tpath[i]\not\models\intnf{}
        \end{array}
    \right\}.
\]
Thus $\fpaths{\phi}{\config}$ contains precisely the finite
prefixes stopping at their first communication normal form.
If $\config\models\intnf{}$, its only member is the
one-state path $\config$. We note that $\forall \phi \in \Phi_I^{\config}:\forall \tpath \in \Tpath_\phi(\config): \exists \tpath' \in \fpaths{\phi}{\config} \st \tpath' \preceq \tpath$.
\end{definition}

\begin{restatable}[Fair scheduler comparison]
    {lemma}{fairSchedulerComparison}
\label{lemma:fair-scheduler-comparison}
Suppose $\der{}{\N}{\Delta}$. Then
\[
\begin{gathered}
    \forall\phi\in\Phi_{I,F}^{\N}:
    \forall\mathsf{prop}\in\{\livee,\dff\}:
    \exists\phi'\in\Phi_{I,F}^{\Delta}:
    \\[1mm]
    \probpathmeasure{\phi'}{\Delta}(\neg\mathsf{prop})
    \geq
    \probpathmeasure{\phi}{\N}(\neg\mathsf{prop}).
\end{gathered}
\]
Here $\neg\dff$ denotes the event of reaching a deadlocked state.
\end{restatable}

\begin{proof}
Fix $\phi\in\Phi_{I,F}^{\N}$ and
$\mathsf{prop}\in\{\livee,\dff\}$.

\par\noindent\emph{Coupling.}
Construct paired histories $h,g$ in rounds, starting from
$(h,g)=(\N,\Delta)$ and maintaining
$\der{}{\last{h}}{\last{g}}$.
For each such pair, \cref{lemma:internal-normalization,corollary:session-normalization} give witnesses $\gamma$ and $\psi$ such that:
\[
    \last{h}\intrednfsched{}{\phi_h}\distN,
    \qquad
    \last{g}\intrednfsched{}{\psi}\distD,
    \qquad
    \dercoup{}{\distN}{\distD}{\gamma}.
\]
A context scheduler resolves nondeterminism but cannot choose
the outcomes of probabilistic reductions.
We therefore normalize the context first under $\psi$, obtaining
\[
    s\in\fpaths{\psi}{\last{g}},
    \qquad
    \Delta'=\last{s}\in\dom{\distD},
\]
then sample a compatible session path
$r\in\fpaths{\phi_h}{\last{h}}$ according to
\[
    \Pr_{h,g}(r\mid\Delta')
    =
    \frac{\gamma(\last{r},\Delta')}{\distD(\Delta')}
    \frac{
        \probpathmeasure{\phi_h}{\last{h}}(\cyl{r})
    }{
        \distN(\last{r})
    }.
\]
The coupling marginals and the partition into normalization
cylinders give
\[
\begin{aligned}
    \sum_{r\in\fpaths{\phi_h}{\last{h}}}
        \Pr_{h,g}(r\mid\Delta')
    &=1,
    \\
    \sum_{\Delta'\in\dom{\distD}}
        \distD(\Delta')\Pr_{h,g}(r\mid\Delta')
    &=
        \probpathmeasure{\phi_h}{\last{h}}(\cyl{r}).
\end{aligned}
\]
Append both normalization paths, identifying shared endpoints:
\[
    h'=h;r,
    \qquad
    g'=g;s.
\]
Since $\gamma(\last{h'},\last{g'})>0$,
we have $\der{}{\last{h'}}{\last{g'}}$.
If a communication is enabled, append the step selected by
$\phi(h')$ and its matching context step, then repeat;
\cref{lemma:normal-form-communication} preserves typing.
Otherwise both paths terminate, since communication safety
excludes error reductions.

The sampled history guides later context choices.
This procedure gives a randomized context scheduler, whose choices
are distributions over possible next actions (each weighted by the sum of $\Pr_{h,g}(r\mid\Delta')$ over compatible session paths $r$ that prescribe that action).

To get rid of this auxiliary randomness and extract a deterministic scheduler, let $U_0,U_1,\ldots$
be independent random variables, each uniform on $[0,1)$
and independent of the context's own random draws;
write $U=(U_0,U_1,\ldots)$.
In round $n$, enumerate the possible session paths
$r_1,\ldots,r_k$ and set
\[
    p_i=\Pr_{h,g}(r_i\mid\Delta'),
    \qquad
    I_i=
    \left[
        \sum_{j<i}p_j,\,
        \sum_{j\leq i}p_j
    \right).
\]
Select $r=r_i$ when $U_n\in I_i$;
uniformity gives each path its required probability $p_i$.

For fixed $u=(u_0,u_1,\ldots)$, reconstruct $h$ from the observed context history, starting at $h=\N$. At each context normal-form endpoint $\Delta'$, the known $h,g,\Delta'$ determine the intervals, and $u_n$ selects the subsequent $r$.
Then $\phi(h;r)$ determines the communication, after which the fixed witnesses determine the next normalization policy. This procedure defines a deterministic $\psi_u\in\Phi_I^\Delta$.

Each normalization is finite and each nonterminal round adds
a communication, so iteration yields complete paired paths
$(\xi_\N,\xi_\Delta)$.
The session marginal is $\probpathmeasure{\phi}{\N}$;
conditional on $U=u$, the context marginal is
$\probpathmeasure{\psi_u}{\Delta}$.

\par\noindent\emph{Fairness and failure.}
Let $(\N_n,\Delta_n)$ be the corresponding normal-form
boundaries, repeating a terminal pair only for indexing.
By construction and \cref{cor:deadlockcorrespondence},
\[
    \der{}{\N_n}{\Delta_n},
    \qquad
    \forall\pp:\quad
    \blkp{\pp}{\N_n}
    \iff
    \blkp{\pp}{\Delta_n}.
\]

Assume $\xi_\N$ is fair.
Finite normalization satisfies the context's internal
fairness obligations.
If $\alpha=\actiontau{\pp}{\q}{\ell}$ were enabled infinitely
often on $\xi_\Delta$ but performed only finitely often,
\cref{lemma:internal-locality,lemma:normal-form-communication}
would make $\alpha$ enabled at infinitely many $\N_n$.
Choose such a boundary $n_0$ after all $\alpha$-steps. Then
\[
    \unf{\Delta_{n_0}(\pp)}
    =
    \pprocoutty{\q}{}{\probn{1}}{\ell}{\Basic}{\T}.
\]
Only $\alpha$ can consume this prefix.
Session pair-fairness forces a later $(\pp,\q)$-communication,
whose matching context step must therefore be $\alpha$,
a contradiction.
Thus $\xi_\Delta$ is fair.

Corresponding terminal states are deadlocked together by
\cref{lemma:normal-form-communication}.
For liveness, \cref{lemma:nonliveness-blocked-suffix}
and the boundary invariant give
\[
    \xi_\N\not\models\livee
    \quad\Longrightarrow\quad
    \exists\pp,n_0:\ \forall n\geq n_0:
    \blkp{\pp}{\Delta_n}.
\]
No later communication involves $\pp$, and internal steps
preserve its pending prefix, so $\xi_\Delta\not\models\livee$.

\par\noindent\emph{Deterministic extraction.}
Since $\phi$ is fair, the marginal identities and the
pathwise implications above yield
\[
\begin{aligned}
    \mathbb E_U\!\left[
        \probpathmeasure{\psi_U}{\Delta}(\fairr)
    \right]
    &=1,
    \\
    \mathbb E_U\!\left[
        \probpathmeasure{\psi_U}{\Delta}(\neg\mathsf{prop})
    \right]
    &\geq
        \probpathmeasure{\phi}{\N}(\neg\mathsf{prop}).
\end{aligned}
\]
These expectations average over the deterministic schedulers
$\psi_u$ as $u$ varies.
Since probabilities are bounded by $1$, the first equality
implies that $\psi_u$ is fair for almost every $u$
by \cref{def:fairscheduler}.
Among these fair schedulers, some has failure probability
at least their average, hence at least the session's.
Taking that scheduler as $\phi'=\psi_u$ proves the claim.
\end{proof}

\procdf*
\begin{proof}
\emph{(1)}
Suppose $\df{\Delta}$ and
$\Multi\predcontstar{}{}\N'$ with $\deadlock{\N'}$.
Commute independent internal steps to make this finite
execution internal-first (possible by~\cref{lemma:intredsched}).

At each normalization phase, the coupling supplied by
\cref{corollary:session-normalization} satisfies
\[
    0<\distNsub{k}(\N_k)
    =\sum_{\Gamma\in\dom{\distDsub{k}}}
        \gamma_k(\N_k,\Gamma).
\]
Choose $\Delta_k$ with $\gamma_k(\N_k,\Delta_k)>0$.
Then $\der{}{\N_k}{\Delta_k}$, and
\cref{lemma:normal-form-communication} matches the next
communication. Induction over these blocks yields
\[
    \Delta\predcontstar{}{}\Delta',
    \qquad
    \der{}{\N'}{\Delta'},
    \qquad
    \N',\Delta'\models\intnf{}.
\]
By \cref{lemma:normal-form-communication},
\[
\begin{aligned}
    \N'\nrightarrow
    &\Longrightarrow \Delta'\nrightarrow,\\
    \exists\pp:\N'(\pp)\not\equiv\inact
    &\Longrightarrow \exists\pp:\Delta'(\pp)\neq\tend.
\end{aligned}
\]
Thus $\deadlock{\Delta'}$, contradicting $\df{\Delta}$.

\par\noindent\emph{(2)}
Taking complements in
\cref{lemma:fair-scheduler-comparison}, and using
\cref{lemma:intredsched}, gives
\[
\begin{aligned}
    \probdfmin{\Delta}
    &\leq
        \min_{\psi\in\Phi_{I,F}^{\Delta}}
        \probpathmeasure{\psi}{\Delta}(\dff)\\
    &\leq
        \min_{\phi\in\Phi_{I,F}^{\Multi}}
        \probpathmeasure{\phi}{\Multi}(\dff)
     =\probdfmin{\Multi}.
\end{aligned}
\]
\end{proof}

\livenss*
\begin{proof}
\emph{(1)}
Suppose $\live{\Delta}$ and $\neg\live{\Multi}$.
Then the pathwise rearrangement in the proof of
\cref{lemma:intredsched} gives a complete, fair,
internal-first path $\tpath$ with $\tpath\not\models\livee$ from $\N$.

Lift $\tpath$ by the support construction in
\cref{thm:proc:df}(1).
The pathwise fairness argument in the proof of
\cref{lemma:fair-scheduler-comparison} gives a complete
fair context path $\tpath'$ with the same communications
and corresponding normal-form boundaries
\[
    \der{}{\N_k}{\Delta_k},
    \qquad
    \N_k,\Delta_k\models\intnf{}.
\]
Repeat terminal boundaries when necessary.
By \cref{lemma:nonliveness-blocked-suffix,cor:deadlockcorrespondence},
\[
    \exists\pp,k_0:\ \forall k\geq k_0:\qquad
    \blkp{\pp}{\N_k}
    \;\land\;
    \blkp{\pp}{\Delta_k}.
\]
No subsequent communication involves $\pp$, and internal
steps leave its pending input or output unchanged.
Hence $\tpath'\not\models\livee$, contradicting
$\live{\Delta}$.

\par\noindent\emph{(2)}
Taking complements in
\cref{lemma:fair-scheduler-comparison}, and using
\cref{lemma:intredsched}, gives
\[
\begin{aligned}
    \problivemin{\Delta}
    &\leq
        \min_{\psi\in\Phi_{I,F}^{\Delta}}
        \probpathmeasure{\psi}{\Delta}(\livee)\\
    &\leq
        \min_{\phi\in\Phi_{I,F}^{\Multi}}
        \probpathmeasure{\phi}{\Multi}(\livee)
     =\problivemin{\Multi}.
\end{aligned}
\]
\end{proof}

\section{Appendix for Section \ref{sec:translation}}
\label{app:translation}

\subsection{Full proofs for soundness and completeness of \prism{} translation}
\label{app:sec5proofs}
\subsubsection{Soundness}
Bisimulation allows us to write a \emph{coinductive} proof.
\begin{definition}
\label{def:bisimulation}
  Let $(M_1, I_1)$ and $(M_2, I_2)$ be model-state pairs.
  A binary relation $\rel$ is a \emph{bisimulation} if:
  \begin{enumerate*}
    \item $(I_1, I_2) \in \rel$, and
    \item For all $(S_1, S_2) \in \rel$, action $\alpha$ and
  $\prob \in [0, 1]$,\\
\hspace*{1cm}$(M_1, S_1) \prismreduce{\alpha}{\prob} (M_1, S_1') \implies \exists S_2'$
      s.t. $(M_2, S_2)  \prismreduce{\alpha}{\prob} (M_2, S_2')$
      and $(S_1', S_2') \in \rel$; and \\
\hspace*{1cm}$(M_2, S_2) \prismreduce{\alpha}{\prob}(M_2, S_2')
\implies \exists S_1'$
      s.t. $(M_1, S_1) \prismreduce{\alpha}{\prob} (M_1, S_1')$ and $ (S_1', S_2') \in \rel$.
  \end{enumerate*}
We say $(M_1, I_1)$ and $(M_2, I_2)$ are \emph{bisimilar} if such a relation $\rel$
  exists, and write $(M_1, I_1) \sim (M_2, I_2)$.
\end{definition}
To show two
model-state pairs are bisimilar, we define a relation $\mathcal{R}$
and verify it
satisfies the bisimulation conditions.

Defining $\mathcal{R}$ is challenging as we do not have the necessary
mechanism to reason
about the reductions available from a given state, nor are we able to relate
states together according to this mechanism.

To solve the first issue, we introduce a \emph{source function}
(\Cref{def:source}).
This, combined with \Cref{prop:source}, precisely identifies the location of a
state within its encoding, and identifies the reductions available from it.
For the second issue, we introduce a novel \emph{unfold correspondence} relation
(\Cref{def:unfoldcor}) that relates equivalent partial types of $\mut$ and
$\tunroll$. Using these, we define a binary relation between states of
$\pmuttran$ and $\ptunrolltran$, and check using
\Cref{prop:source} that it is indeed a bisimulation relation. See
\Cref{lemma:soundrecur} for details.

The rest of the proof proceeds as follows. We show that model-state pairs
that enter the state space for a closed type behaves exactly like the closed
type (\Cref{lemma:shift}). We then use this to prove soundness of
$\trantwo{\cdot}$ (\Cref{lemma:soundcomm}). Next, we define what it means to
\emph{combine} states (\Cref{def:stateunion}), using this to relate
composition and bisimulation (\Cref{lemma:lsimcombine}). This allows us to use
\Cref{lemma:soundcomm} to prove soundness of
$\tranone{\cdot}$.

\myparagraph{Detailed proofs}
Taking a compositional approach, we first show that the inner encoding
$\trantwo{\Delta}$ is sound with respect to the reductions of a single
participant: the nondeterministic reductions $\predcontnd{}$, the
probabilistic reductions (with action $\actionprob{\pp}$) and the
send/receive actions
(\Cref{lemma:soundcomm}). We then use this result to demonstrate the soundness
of $\tranone{\Delta}$.

\begin{figure}
    \small
    \centering
    \begin{align*}
          \sourcenobrackets{\pp}{\T} &= \sourcep{\T}{\pp}{0}{\varnothing}
                              \ \cup \ \{ \ \{S_\pp \mapsto \StateS{\T}\} \mapsto (\tend, \varnothing) \ \}
          \\
          \sourcepp{\tend} &= \varnothing
          \\
          \sourcepp{\ty} &= \varnothing
          \\
          \sourcepp{\mu \ty . \T} &= \sourcep{\T}{\pp}{n}{f \cup \{\ty \mapsto \{S_\pp \mapsto n\}\}}
          \\
          \sourcepp{\pprocinnew{\pp_i}{i \in I}{\ell_i}{\Basic_i}{\T_i}}
                             &= \{\ \{S_\pp \mapsto n\} \mapsto
                               (\pprocinnew{\pp_i}{i \in I}{\ell_i}{\Basic_i}{\T_i}, f)\ \} \\
                             & \cup \textstyle\bigcup_{i \in I}
                               \sourcep{\T_i}{\pp}{\nexts{i}}{f} \\
            \text{where } \nexts{i} &= n + 1
                                           + \textstyle\sum_{ \{ i' \in I | i' < i \} } \StateS{\T_{i'}} \\
          \\
          \sourcepp{\textstyle\pprocoutty{\q}{}{\probn{1}}{\ell}{\Basic}{\T}}
                             &= \{\ \{S_\pp \mapsto n\} \mapsto
                               (\textstyle\pprocoutty{\q}{}{\probn{1}}{\ell}{\Basic}{\T}, f)\ \} \\
                             & \cup \sourcep{\T}{\pp}{n + 1}{f}
          \\[2mm]
          \sourcepp{\textstyle\pprocoutty{\pp_j}{j \in J}{\probi{j}}{\ell_j}{\Basic_j}{\T_{j}}}
                             &= \{\ \{S_\pp \mapsto n\} \mapsto
                               (\textstyle\pprocoutty{\pp_j}{j \in J}{\probi{j}}{\ell_j}{\Basic_j}{\T_{j}}, f)\ \} \\
                             & \cup \textstyle\bigcup_{j \in J}
                               \sourcep{\comsel{j}}{\pp}{\come{j}}{f}
                               \qquad \qquad (|J| \geq 2)\\
          \text{where } \comsel{j} &= \pprocoutty{\pp_j}{}{\probn{1}}{\ell_j}{\Basic_j}{\T_{j}} \\
          \come{j} &= n + 1 + \textstyle\sum_{j' < j}\StateS{\comsel{j'}}
          \\[2mm]
          \sourcepp{\textstyle\probpprocoutty{i \in I}{\pp_j}{j \in J_i}{\probi{(i,j)}}{\ell_j}{\Basic_j}{\T_{(i,j)}}}
                             &= \{\ \{S_\pp \mapsto n\} \mapsto
                               (\textstyle\probpprocoutty{i \in I}{\pp_j}{j \in J_i}{\probi{(i,j)}}{\ell_j}{\Basic_j}{\T_{(i,j)}}, f)\ \} \\
                             & \cup \textstyle\bigcup_{i \in I}
                               \sourcep{\selsummand{i}}{\pp}{\sume{i}}{f}
                               \qquad \qquad (|I| \geq 2)\\
          \text{where } \selsummand{i} &= \pprocoutty{\pp_j}{j \in J_i}{\probi{(i,j)}}{\ell_j}{\Basic_j}{\T_{(i,j)}} \\
          \sume{i} &= n + 1 + \textstyle\sum_{i' < i}\StateS{\selsummand{i'}}
    \end{align*}
  \caption{Definition of $\source{\pp}{\T}{S}$.}
    \label{fig:source}
\end{figure}

\begin{definition}
  \label{def:source}
  The \emph{source} of a state $S$ within the module $\trantwo{\pp : \T}$,
  written $\source{\pp}{\T}{S}$, is given in \Cref{fig:source}.
\end{definition}

The function outputs a pair $\source{\pp}{\T}{S} = (\T', f)$.
The first element $\T'$ is an $\tend$, a proper sum $\sum\oplus$ (with $|I|
\geq 2$), a proper selection $\oplus$ (with $|J| \geq 2$), a committed
selection (a single branch of probability $\probn{1}$), or a
branching $\&$ occurring as a
\emph{partial type} of $\T$, representing the substructure of $\T$ associated
with the state: every state of the encoding is the entry point of such a
partial type. In particular, the state reached by nondeterministically
picking a summand $i$ of a proper sum is the entry point of the singleton
selection $\selsummand{i}$, and the state reached by probabilistically
committing to a branch $j$ of a proper selection is the entry point of the
committed selection $\comsel{j}$. Finally,
$f$ maps variables to their corresponding states.

The following proposition says that $\source{\pp}{\T}{S}$ really does what we expect.

\begin{restatable}{proposition}{propsource}
  \label{prop:source}
  Let $\T$ be a type and $S$ a state for $\pttran$. Then:
  \begin{enumerate}
    \item If $\source{\pp}{\T}{S} = (\tend, \varnothing)$, then $(\pttran, S) \not
      \rightarrow$.
    \item \label{prop:source:bra} If $\source{\pp}{\T}{S} = (\pprocinnew{\pp_i}{i \in I}{\labelselli{i}}{\Basic_i}{\T_i},
      f)$, then $\forall i \in I : (\pttran, S)
      \prismreduce{\ppi{i}\conc\pp\conc\labelselli{i}}{\probn{1}} (\pttran, S')$
      where $S' =
      f(\ty)$ if $\T_i = \ty$, and otherwise
      \[
        \source{\pp}{\T}{S'} =
        \begin{cases}
          (\tend, \varnothing) &\text{if } \T_i = \tend \\
          (\T', f \cup \bigcup_{k}\{\ty_k \mapsto S'\}) &\text{if } \T_i = \mu\ty_1\cdots\mu\ty_n. \T' \\
          (\T_i, f) & \text{otherwise.}
        \end{cases}
      \]
    \item \label{prop:source:sum} If $\source{\pp}{\T}{S} = (\probpprocoutty{i \in I}{\pp_j}{j \in J_i}{\probi{(i,j)}}{\labelselli{j}}{\Basic_j}{\T_{(i,j)}},
      f)$ with $|I| \geq 2$, then $\forall i \in I : (\pttran, S) \prismreduce{\pp\conc\tau}{\probn{1}}
      (\pttran, S')$ with
      $\source{\pp}{\T}{S'} = (\selsummand{i}, f)$, where $\selsummand{i} =
      \pprocoutty{\pp_j}{j \in J_i}{\probi{(i,j)}}{\labelselli{j}}{\Basic_j}{\T_{(i,j)}}$.
    \item \label{prop:source:selentry} If $\source{\pp}{\T}{S} = (\pprocoutty{\pp_j}{j \in J}{\probi{j}}{\labelselli{j}}{\Basic_j}{\T_{j}},
      f)$ with $|J| \geq 2$, then $\forall j \in J : (\pttran, S) \prismreduce{\pp\conc p}{\probi{j}}
      (\pttran, S')$ with
      $\source{\pp}{\T}{S'} = (\comsel{j}, f)$, where $\comsel{j} =
      \pprocoutty{\pp_j}{}{\probn{1}}{\labelselli{j}}{\Basic_j}{\T_{j}}$.
    \item \label{prop:source:selint} If $\source{\pp}{\T}{S} = (\pprocoutty{\q}{}{\probn{1}}{\labelsell}{\Basic}{\T''},
      f)$, then $(\pttran, S)
      \prismreduce{\pp\conc\q\conc\labelsell}{\probn{1}} (\pttran, S')$ where $S' =
      f(\ty)$ if $\T'' = \ty$, and otherwise
      \[
        \source{\pp}{\T}{S'} =
        \begin{cases}
          (\tend, \varnothing) &\text{if } \T'' = \tend \\
          (\T', f \cup \bigcup_{k}\{\ty_k \mapsto S'\}) &\text{if } \T'' = \mu \ty_1\cdots\mu\ty_n. \T' \\
          (\T'', f) & \text{otherwise.}
        \end{cases}
      \]
  \end{enumerate}
Moreover, $(\trantwo{\pp : \T}, S)$ cannot make any other transitions.
\end{restatable}
\begin{proof}
  We walk through the recursive calls of $\sourcepp{\T}$ and $\oxtrantwo{\T}{\pp}{n}{m}{f'}$ in parallel.
  In particular, by induction on the structure of $\T'$, we show that we
  maintain the invariant that
  $\sourcepp{\T'}$ is called iff $\oxtrantwo{\T'}{\pp}{n}{m}{f'}$ is called,
  where $f(\ty) = \{S_\pp \mapsto f'(\ty)\}$ for all $\ty$. The result then
  follows as the transitions that $(\trantwo{\pp : \T}, \{S_\pp \mapsto n\})$ can
  make is fully determined by the commands introduced in
  $\oxtrantwo{\T'}{\pp}{n'}{\StateS{\T}}{f'}$ for some $\T'$ a partial type of
  $\T$, and $n' < n$. By the invariant, the corresponding output of
  $\source{\pp}{\T}{S}$ is
  determined by the mappings added in $\sourcep{\T'}{\pp}{n'}{f}$,
  which adds the correct transitions since $\T'$ is the same on both sides.

  \case{$\T' = \tend$ or $\ty$}
  Then both $\sourcepp{\T'}$ and $\oxtrantwo{\T'}{\pp}{n}{m}{f'}$ does not make
  any further recursive calls.

  \case{$\T' = \mu \ty . \T''$}
  Then $\sourcepp{\mu \ty . \T''}$ calls $\sourcep{\T''}{\pp}{n}{f \cup \{\ty
    \mapsto {S_\pp \mapsto n}\}}$ and $\oxtrantwo{\mu \ty. \T''}{\pp}{n}{m}{f'}$ calls
  $\oxtrantwo{\T''}{\pp}{n}{m}{f' \cup \{\ty \mapsto n\}}$, and the invariant is maintained by
  the induction hypothesis.

  \case{$\T' = \pprocoutty{\q}{}{\probn{1}}{\labelsell}{\Basic}{\T''}$}
  Then
  $\sourcepp{\T'}$ calls
  $\sourcep{\T''}{\pp}{n + 1}{f}$ and
  $\oxtrantwo{\T'}{\pp}{n}{m}{f}$ calls
  $\oxtrantwo{\T''}{\pp}{n + 1}{m}{f}$. Moreover, the entry mapping
  $\{S_\pp \mapsto n\}$ added by $\sourcepp{\T'}$ corresponds exactly to
  the guard $(S_\pp = n)$
  of the $\pp\conc\q\conc\labelsell$ command. The
  invariant is therefore maintained by the induction hypothesis.

  \case{$\T' = \pprocoutty{\pp_j}{j \in J}{\probi{j}}{\labelselli{j}}{\Basic_j}{\T''_{j}}$ with $|J| \geq 2$}
  Then
  $\sourcepp{\T'}$ calls
  $\sourcep{\comsel{j}}{\pp}{\come{j}}{f}$ and
  $\oxtrantwo{\T'}{\pp}{n}{m}{f}$ calls
  $\oxtrantwo{\comsel{j}}{\pp}{\come{j}}{m}{f}$, with the same
  $\comsel{j} = \pprocoutty{\pp_j}{}{\probn{1}}{\labelselli{j}}{\Basic_j}{\T''_{j}}$
  and
  $\come{j} = n + 1
  + \sum_{j'<j}\StateS{\comsel{j'}}$ on both sides, for all $j \in J$. Moreover, the entry mapping
  $\{S_\pp \mapsto n\}$ added by $\sourcepp{\T'}$ corresponds exactly to
  the guard $(S_\pp = n)$ of the $\pp\conc p$ command, whose updates
  $(S'_\pp = \come{j})$ target the entry mappings added by
  $\sourcep{\comsel{j}}{\pp}{\come{j}}{f}$. The
  invariant is therefore maintained by the induction hypothesis.

  \case{$\T' = \probpprocoutty{i \in I}{\pp_j}{j \in J_i}{\probi{(i,j)}}{\labelselli{j}}{\Basic_j}{\T''_{(i,j)}}$ with $|I| \geq 2$}
  Then
  $\sourcepp{\T'}$ calls
  $\sourcep{\selsummand{i}}{\pp}{\sume{i}}{f}$ and
  $\oxtrantwo{\T'}{\pp}{n}{m}{f}$ calls
  $\oxtrantwo{\selsummand{i}}{\pp}{\sume{i}}{m}{f}$, with the same
  $\selsummand{i} = \pprocoutty{\pp_j}{j \in J_i}{\probi{(i,j)}}{\labelselli{j}}{\Basic_j}{\T''_{(i,j)}}$
  and
  $\sume{i} = n + 1 + \sum_{i'<i}\StateS{\selsummand{i'}}$ on both sides, for all $i \in I$.
  Moreover, the entry mapping
  $\{S_\pp \mapsto n\}$ added by $\sourcepp{\T'}$ corresponds exactly to
  the guard $(S_\pp = n)$ of the $|I|$ $\pp\conc\tau$ commands, whose updates
  $(S'_\pp = \sume{i})$ target the entry mappings added by
  $\sourcep{\selsummand{i}}{\pp}{\sume{i}}{f}$. The
  invariant is therefore maintained by the induction hypothesis.

  \case{$\T' = \pprocinnew{\pp_i}{i \in I}{\labelselli{i}}{\Basic_i}{\T''_i}$}
  Similar to the committed selection case: the only
  mapping added by $\sourcepp{\T'}$ for the node itself is $\{S_\pp \mapsto n\}$,
  corresponding to the guard $(S_\pp = n)$ of the $|I|$ branching commands, and
  the recursive calls $\sourcep{\T''_i}{\pp}{\nexts{i}}{f}$ and
  $\oxtrantwo{\T''_i}{\pp}{\nexts{i}}{m}{f}$ share the same
  $\nexts{i} = n + 1 + \textstyle\sum_{ \{ i' \in I \SEP i' < i \} }
  \StateS{\T''_{i'}}$.

  Finally, we conclude by noting that $\pttran$ calls
  $\oxtrantwo{\T}{\pp}{0}{\StateS{\T}}{\varnothing}$ and
  $\sourcenobrackets{\pp}{\T}$ calls $\sourcep{\T}{\pp}{0}{\varnothing}$, so the
  invariant holds in the beginning.
\end{proof}

As mentioned above, we define a relation to relate partial types of $\mut$ and
$\tunroll$. We do this through a series of rules that build up equivalent types
modulo unrolling of $\ty$.

\begin{definition}
  \label{def:unfoldcor}
    The \emph{unfold correspondence} of two partial types \emph{with respect to $\T$ and $\ty$}, written
    $\sconf$, is the least relation satisfying the following rules.

    \begin{mathpar}
    \inferrule[\rulename{uc-refl}]{ }{\T' \sconf \T'}

    \inferrule[\rulename{uc-t}]{ }{\ty \sconf \mut}

    \inferrule[\rulename{uc-rec}]{\ty' \neq \ty \\ \T_1 \sconf \T_2}{\mu \ty' . \T_1 \sconf \mu \ty' . \T_2}\\

    \inferrule[\rulename{uc-out}]{\forall i \in I, j \in J_i : \T_{(i,j)} \sconf \T'_{(i,j)}}{\probpprocoutty{i \in I}{\q_j}{j \in
    J_i}{\probi{(i,j)}}{\labelselli{j}}{\Basic_j}{\T_{(i,j)}} \sconf \probpprocoutty{i \in I}{\q_j}{j \in
    J_i}{\probi{(i,j)}}{\labelselli{j}}{\Basic_j}{\T'_{(i,j)}} }

    \inferrule[\rulename{uc-in}]{\forall i \in I : \T_i \sconf \T'_i}{\pprocinnew{\q_i}{i \in
    I}{\labelselli{i}}{\Basic_i}{\T_i} \sconf \pprocinnew{\q_i}{i \in
    I}{\labelselli{i}}{\Basic_i}{\T'_i} }
\end{mathpar}
\end{definition}

\begin{restatable}{lemma}{unrollsconf}
  \label{lemma:unrollsconf}
  For every type $\T, \T'$ and variable $\ty$, $\T' \sconf \tprimeunroll$.
\end{restatable}
\begin{proof}
  We prove by induction on the structure of $\T'$.

  \case{$\T' = \tend$}
  Follows directly from \rulename{uc-refl}.

  \case{$\T' = \ty'$}
  There are two further cases. If $\ty' = \ty$, then
  $\tprimeunroll = \mut$, so we can apply \rulename{uc-t}. If
  $\ty' \neq \ty$, then $\tprimeunroll = \ty'$, so we can apply
  \rulename{uc-refl}.

  \case{$\T' = \mu \ty' . \T''$}
  By the inductive hypothesis, we have
  $\T'' \sconf \T''\subst{\mu \ty . \T}{\ty}$. Similarly to the variable case, there
  are two cases. If $\ty' = \ty$, then $\tprimeunroll = \mu \ty . \T''$ since $\ty$ is
  rebound, and therefore we can apply \rulename{uc-refl} immediately. If $\ty' \neq \ty$,
  then we have $\tprimeunroll = \mu \ty' . (\T''\subst{\mut}{\ty})$, and using the
  inductive hypothesis we can apply \rulename{uc-rec}.

  \case{$\T' = \probpprocoutty{i \in I}{\q_j}{j \in
      J_i}{\probi{(i,j)}}{\labelselli{j}}{\Basic_j}{\T''_{(i,j)}}$}
  By the inductive hypothesis, we have $\T''_{(i,j)} \sconf \T''_{(i,j)}\subst{\mut}{\ty}$ for
  all $i \in I$ and $j \in J_i$. Thus $\tprimeunroll = \probpprocoutty{i \in I}{\q_j}{j \in
    J_i}{\probi{(i,j)}}{\labelselli{j}}{\Basic_j}{(\T''_{(i,j)} \subst{\mut}{\ty})}$, so we can apply
  \rulename{uc-out}.

  \case{$\T' = \pprocinnew{\q_i}{i \in
      I}{\labelselli{i}}{\Basic_i}{\T''_i}$}
  Similarly to the above case, by the inductive hypothesis, we have $\T''_i
  \sconf \T''_i\subst{\mut}{\ty}$ for all $i \in I$.
  Note $\tprimeunroll = \pprocinnew{\q_i}{i \in
    I}{\labelselli{i}}{\Basic_i}{(\T''_i\subst{\mut}{\ty})}$, so we can apply
  \rulename{uc-in}.
\end{proof}

\begin{restatable}{lemma}{soundrecur}\label{lemma:soundrecur}
  For every type $\T$, $\pmuttran \sim \ptunrolltran$.
\end{restatable}
\begin{proof}
  Suppose $S_1$ and $S_2$ are states for $\pmuttran$ and $\ptunrolltran$
  respectively, and let $(\T_1, f_1) = \mutsource{S_1}$ and
  similarly $(\T_2, f_2) = \tunrollsource{S_2}$.

  Define a binary relation $\rel$ to be the greatest relation such that if $(S_1, S_2) \in \rel$,
  then all of the following hold:
  \begin{enumerate}
    \item $\T_1 \sconf \T_2$;
    \item for all $\ty' \in \dom{f_1} \cap \dom{f_2}$, $(f_1(\ty'), f_2(\ty')) \in \rel$.
  \end{enumerate}
  We will argue that $\rel$ is a bisimulation relation.

  Take any $(S_1, S_2) \in \rel$, and again let $(\T_1, f_1) = \mutsource{S_1}$ and
  $(\T_2, f_2) = \tunrollsource{S_2}$. By our definition of $\rel$, both $\T_1$
  and $\T_2$ must be
  $\tend$, a proper sum, a proper selection, a committed selection, or a
  branching (the latter four with the
  same participants,
  labels and basic types, and the same probabilities for sums and selections).
  In the first
  case, both $(\pmuttran, S_1)$  and $(\ptunrolltran, S_2)$
  cannot make any transitions, so there is nothing to check.

  For the sum case, suppose
  \[ \T_1 = \probpprocoutty{i \in I}{\q_j}{j \in
    J_i}{\probi{(i,j)}}{\labelselli{j}}{\Basic_j}{\T'_{(i,j)}}, \qquad
     \T_2 = \probpprocoutty{i \in I}{\q_j}{j \in
    J_i}{\probi{(i,j)}}{\labelselli{j}}{\Basic_j}{\T''_{(i,j)}},\]
  with $|I| \geq 2$. By
  \Cref{prop:source}~(\ref{prop:source:sum}), for every $i \in I$,
  \[(\pmuttran, S_1) \prismreduce{\pp\conc\tau}{\probn{1}} (\pmuttran, S'_1)\]
  with $\mutsource{S'_1} = (\pprocoutty{\q_j}{j \in
    J_i}{\probi{(i,j)}}{\labelselli{j}}{\Basic_j}{\T'_{(i,j)}}, f_1)$, and
  $S_2$ is able to make the matching transition
  \[(\ptunrolltran, S_2) \prismreduce{\pp\conc\tau}{\probn{1}} (\ptunrolltran, S'_2)\]
  with $\tunrollsource{S'_2} = (\pprocoutty{\q_j}{j \in
    J_i}{\probi{(i,j)}}{\labelselli{j}}{\Basic_j}{\T''_{(i,j)}}, f_2)$.
  Since $(S_1, S_2) \in \rel$, we have $\T_1 \sconf \T_2$, which must be derived
  by \rulename{uc-refl} or \rulename{uc-out}; in either case $\T'_{(i,j)}
  \sconf \T''_{(i,j)}$ for all $j \in J_i$, so applying \rulename{uc-out}
  relates the two singleton selections above. Therefore
  $(S'_1, S'_2) \in \rel$. The opposite direction is symmetrical.

  For the proper selection case, suppose
  \[ \T_1 = \pprocoutty{\q_j}{j \in
    J}{\probi{j}}{\labelselli{j}}{\Basic_j}{\T'_{j}}, \qquad
     \T_2 = \pprocoutty{\q_j}{j \in
    J}{\probi{j}}{\labelselli{j}}{\Basic_j}{\T''_{j}},\]
  with $|J| \geq 2$.
  By \Cref{prop:source}~(\ref{prop:source:selentry}), for
      every $j \in J$,
      \[(\pmuttran, S_1) \prismreduce{\pp\conc p}{\probi{j}} (\pmuttran, S'_1)\]
      with $\mutsource{S'_1} = (\pprocoutty{\q_j}{}{\probn{1}}{\labelselli{j}}{\Basic_j}{\T'_{j}}, f_1)$, and $S_2$ is able to make the
      matching transition
      \[(\ptunrolltran, S_2) \prismreduce{\pp\conc p}{\probi{j}} (\ptunrolltran, S'_2)\]
      with $\tunrollsource{S'_2} = (\pprocoutty{\q_j}{}{\probn{1}}{\labelselli{j}}{\Basic_j}{\T''_{j}}, f_2)$. Since $(S_1, S_2) \in \rel$,
      we have $\T_1 \sconf \T_2$, which must be derived by \rulename{uc-refl}
      or \rulename{uc-out}; in either case $\T'_{j} \sconf \T''_{j}$, so
      applying \rulename{uc-out} relates the two committed selections above.
      Therefore, $(S'_1, S'_2) \in \rel$ also.
      The opposite direction is symmetrical.

  For the committed selection case, suppose
  \[ \T_1 = \pprocoutty{\q_j}{}{\probn{1}}{\labelselli{j}}{\Basic_j}{\T'_{j}}, \qquad
     \T_2 = \pprocoutty{\q_j}{}{\probn{1}}{\labelselli{j}}{\Basic_j}{\T''_{j}}.\]
  By \Cref{prop:source}~(\ref{prop:source:selint}),
      the only possible transitions are
      \begin{align*}
        (\pmuttran, S_1) &\prismreduce{\pp::\qi{j}::\labelselli{j}}{\probn{1}} (\pmuttran, S'_1) \text{, and} \\
        (\ptunrolltran, S_2) &\prismreduce{\pp::\qi{j}::\labelselli{j}}{\probn{1}} (\ptunrolltran, S'_2).
      \end{align*}
    We consider the possible cases of $\T'_{j}$ and $\T''_{j}$. Again by \Cref{prop:source}:
    \begin{enumerate}
      \label{lemma:tcases}
      \item
      If $\T'_{j} = \T''_{j} = \tend$,
        $\mutsource{S'_1} = \tunrollsource{S'_2} = (\tend, \varnothing)$.
      \item
      If $\T'_{j} = \T''_{j} = \ty'$ (potentially $\ty' = \ty$),
      we have $S'_1 = f_1(\ty')$ and $S'_2 = f_2(\ty')$.
      Hence by definition, $(S'_1, S'_2) \in \rel$.
      \item If $\T'_{j} = \mu \ty'_1 \cdots \mu \ty'_n . \T'$ and $\T''_{j} = \mu
        \ty'_1 \cdots \mu \ty'_n . \T''$, then
        $\mutsource{S'_1} = (\T', f_1 \cup \bigcup_{k} \{\ty'_k
        \mapsto S'_1\})$ and $\tunrollsource{S'_2} = (\T'', f_2 \cup \bigcup_k
        \{\ty'_k \mapsto S'_2\})$.
      \item If $\T'_{j} = \ty$ and $\T''_{j} = \mut$, then $\mutsource{S'_1} = (\T,
        \{\ty \mapsto S'_1\})$ and $\tunrollsource{S'_2} = (\T, f_2 \cup
        \{\ty \mapsto S'_2\}).$
      \item Otherwise, we have $\mutsource{S'_1} = (\T'_{j}, f_1)$ and
        $\tunrollsource{S'_2} = (\T''_{j}, f_2)$, with $\T'_{j} \sconf \T''_{j}$ by the
        definition of $\sconf$.
    \end{enumerate}
    Thus, in all cases, $(S'_1, S'_2) \in \rel$.

  For the branching case, suppose $\T_1 = \pprocinnew{\q_i}{i \in
  I}{\labelselli{i}}{\Basic_i}{\T'_i}$, and $\T_2 = \pprocinnew{\q_i}{i \in
  I}{\labelselli{i}}{\Basic_i}{\T''_i}$. Here,
  by \Cref{prop:source}~(\ref{prop:source:bra}), for each $i \in I$ both modules
  make the single transition
  \begin{gather*}
    (\pmuttran, S_1) \prismreduce{\qi{i} :: \pp :: \labelselli{i}}{\probn{1}} (\pmuttran, S'_1) \\
    (\ptunrolltran, S_2) \prismreduce{\qi{i} :: \pp :: \labelselli{i}}{\probn{1}} (\ptunrolltran,
    S'_2)
  \end{gather*}
  since the senders $\q_i$, choices $I$, labels and basic types are the same on
  both sides. We then consider the different cases of $\T'_i$ and $\T''_i$ in
  exactly the same way as the committed selection case~(\ref{lemma:tcases}).
  Therefore we have $(S'_1, S'_2) \in \rel$, and the opposite direction is symmetrical.

  Finally, from the definition
  of $\emptytran$, we have $\mutsource{\varnothing} = (T, \{\ty \mapsto \varnothing\})$ and
  $\tunrollsource{\varnothing} = (\tunroll, \varnothing)$, so applying \Cref{lemma:unrollsconf}
  again, $(\varnothing, \varnothing) \in \rel$. Hence $\rel$ is a bisimulation relation, and we can
  conclude $\pmuttran \sim \ptunrolltran$.
\end{proof}

\begin{restatable}{lemma}{lemmashift}
  \label{lemma:shift}
  Let $\T, \T'$ be (closed) types, and $S$ a state for $\pttran$ such that
  $\source{\pp}{\T}{S} = (\T', f)$. Then $(\pttran, S) \sim \ptptran$.
\end{restatable}
\begin{proof}
  Let $S_1, S_2$ be states for $\pttran$ and $\ptptran$, respectively. Let
  $(\T_1, f_1) = \source{\pp}{\T}{S_1}$ and $(\T_2, f_2) =
  \source{\pp}{\T'}{S_2}$.
  We construct the bisimulation relation $\rel$ to be the greatest relation such that
  if $(S_1, S_2) \in \rel$:
  \begin{enumerate}
    \item $\T_1 = \T_2$; and
    \item for all $\ty' \in \dom{f_1} \cap \dom{f_2}$, $(f_1(\ty'), f_2(\ty')) \in \rel$.
  \end{enumerate}
  We then use a similar structure as the proof of
  \Cref{lemma:soundrecur} to argue that $\rel$ is a bisimulation relation. Notably,
  $(S, \varnothing) \in \rel$.
\end{proof}

\begin{restatable}{lemma}{lemmasoundcomm}
  \label{lemma:soundcomm}
  ~\begin{enumerate}
    \item \label{lemma:soundnd} For every type $\T$ with $\pp : \T
      \predcontnd{\actionnd{\pp}} \pp : \T'$, there exists a state $S_1$ such
      that \[\trantwo{\pp : \T} \prismreduce{\pp\conc\tau}{\probn{1}}
      (\trantwo{\pp : \T}, S_1) \sim \trantwo{\pp : \T'}.\]
    \item \label{lemma:soundprob} For every type $\T$ and probability $\prob$ with $\pp : \T
      \predcont{\actionprob{\pp}}{\prob} \pp : \T'$, there
      exists a state $S_1$ such that \[\trantwo{\pp : \T} \prismreduce{\pp\conc p}{\prob}
      (\trantwo{\pp : \T}, S_1) \sim \trantwo{\pp : \T'}.\]
    \item \label{lemma:soundsend} For every type $\T$ with $\pp : \T
      \predcont{\pp : \q!\labelsell\langle\Basic\rangle}{\probn{1}} \pp : \T'$, there
      exists a state $S_1$ such that \[\trantwo{\pp : \T}
      \prismreduce{\pp::\q::\labelsell}{\probn{1}}
      (\trantwo{\pp : \T}, S_1) \sim \trantwo{\pp : \T'}.\]
    \item \label{lemma:soundrecv} For every type $\T$ with $\pp : \T
      \predcont{\pp : \q?\labelsell(\Basic)}{1} \pp : \T'$, there exists a state $S_1$
      such that \[\trantwo{\pp : \T} \prismreduce{\q::\pp::\labelsell}{\probn{1}}
        (\trantwo{\pp : \T}, S_1) \sim \trantwo{\pp :
          \T'}.
        \]
  \end{enumerate}
\end{restatable}
\begin{proof}
  ~\begin{enumerate}
     \item\label{proof:soundnd}
       We proceed by induction on the derivation of the reduction.
       \case{\rulename{ct-nd}}
       Let $\T = \probpprocoutty{i \in I}{\q_j}{j \in J_i}{\probi{(i,j)}}{\labelselli{j}}{\Basic_j}{\T_{(i,j)}}$
       with $|I| > 1$.
       The reduction selects a summand $k \in I$, so that
       $\T' = \selsummand{k} = \pprocoutty{\q_j}{j \in
         J_k}{\probi{(k,j)}}{\labelselli{j}}{\Basic_j}{\T_{(k,j)}}$. By the
       definition of $\emptytran$, we have the summand step \[\trantwo{\pp : \T}
       \prismreduce{\pp\conc\tau}{\probn{1}} (\trantwo{\pp : \T}, S_1)\] with
       $S_1 = \{S_\pp \mapsto \sume{k}\}$, so
       $\source{\pp}{\T}{S_1} = (\selsummand{k}, \varnothing)$. Since $\T$
       is closed, so is $\selsummand{k}$, and we can apply \Cref{lemma:shift}
       to conclude $(\trantwo{\pp : \T}, S_1) \sim \trantwo{\pp : \selsummand{k}}$.

       \case{\rulename{ct-rec}}
       Suppose $\pp : \tunroll \predcontnd{\actionnd{\pp}} \pp : \T'$. By the
       inductive hypothesis, there exists a state $S_1$ such that
       \[
         \ptunrolltran \prismreduce{\pp\conc\tau}{\probn{1}} (\ptunrolltran, S_1)
         \sim \trantwo{\pp : \T'}.
       \]
       We have $\pmuttran \sim \ptunrolltran$ from \Cref{lemma:soundrecur},
       and hence there must indeed exist a state $S'_1$ such that
       \[\pmuttran \prismreduce{\pp\conc\tau}{\probn{1}} (\pmuttran, S'_1)
       \sim \trantwo{\pp : \T'}.\]

     \item\label{proof:soundprob}
       We proceed by induction on the derivation of the reduction.
       \case{\rulename{ct-prob}}
       Let $\T = \pprocoutty{\q_j}{j \in J}{\probi{j}}{\labelselli{j}}{\Basic_j}{\T_{j}}$.
       The reduction selects a branch $l \in J$ with $\probi{l} \in (0;1)$
       (hence $|J| \geq 2$, i.e.\ $\T$ is a proper selection), so that
       $\prob = \probi{l}$ and $\T' = \comsel{l} =
       \pprocoutty{\q_l}{}{\probn{1}}{\labelselli{l}}{\Basic_l}{\T_{l}}$.
       We can assume $\T_{l}$ is
       closed (otherwise, instances of \rulename{ct-rec} can be applied to
       unfold the recursion). Then by the definition of
       $\emptytran$, we have the commitment step \[\trantwo{\pp : \T}
       \prismreduce{\pp\conc p}{\probi{l}} (\trantwo{\pp : \T}, S_1)\]
       with $S_1 = \{S_\pp \mapsto \come{l}\}$, so
       $\source{\pp}{\T}{S_1} = (\comsel{l}, \varnothing)$. Since
       $\comsel{l}$ is closed, we can apply \Cref{lemma:shift} to conclude
       $(\trantwo{\pp : \T}, S_1) \sim \trantwo{\pp : \comsel{l}}$.

       \case{\rulename{ct-rec}}
       Suppose $\pp : \tunroll \predcont{\actionprob{\pp}}{\prob}
       \pp : \T'$. By the inductive hypothesis, there exists a state $S_1$
       such that
       \[
         \ptunrolltran \prismreduce{\pp\conc p}{\prob} (\ptunrolltran, S_1)
         \sim \trantwo{\pp : \T'}.
       \]
       We have $\pmuttran \sim \ptunrolltran$ from \Cref{lemma:soundrecur},
       and hence there must indeed exist a state $S'_1$ such
       that \[\pmuttran \prismreduce{\pp\conc p}{\prob} (\pmuttran, S'_1)
       \sim \trantwo{\pp : \T'}.\]

     \item\label{proof:soundsend}
       We proceed by induction on the derivation of the reduction.
       \case{\rulename{ct-out}}
       Let $\T = \pprocoutty{\q}{}{\probn{1}}{\labelsell}{\Basic}{\T'}$ be a
       committed selection, so that the reduction emits $\labelsell$ to $\q$
       with $\T'$ the continuation. We can assume $\T'$ is
       closed (otherwise, instances of \rulename{ct-rec} can be applied to
       unfold the recursion). Then by the definition of
       $\emptytran$, we have the synchronisation \[\trantwo{\pp : \T}
       \prismreduce{\pp::\q::\labelsell}{\probn{1}} (\trantwo{\pp : \T}, S_1)\] for
       a state $S_1$.

       It therefore suffices to show that
       $(\trantwo{\pp : \T}, S_1) \sim \trantwo{\pp : \T'}$.
       We consider the two possible cases for $\T'$.

       \begin{enumerate}
         \item If $\T' = \tend$, then $S_1 = \{S_\pp \mapsto \StateS{\T}\}$. By the
           encoding, both $(\trantwo{\pp : \T}, S_1)$ and $\trantwo{\pp
           : \tend}$ cannot make any transitions, and hence they are
           bisimilar.
         \item Otherwise, $S_1 = \{S_\pp \mapsto 1\}$, so
           $\source{\pp}{\T}{S_1} = (\T', \varnothing)$. Since
           $\T'$ is closed, we can apply \Cref{lemma:shift} to conclude
           $(\trantwo{\pp : \T}, S_1) \sim \trantwo{\pp :
           \T'}$.
       \end{enumerate}

       \case{\rulename{ct-rec}}
       Suppose $\pp : \tunroll \predcont{\pp : \q!\labelsell\langle\Basic\rangle}{\probn{1}}
       \pp : \T'$. By the inductive hypothesis, there exists a state $S_1$
       such that
       \[
         \ptunrolltran \prismreduce{\pp::\q::\labelsell}{\probn{1}} (\ptunrolltran, S_1)
         \sim \trantwo{\pp : \T'}.
       \]
       We have $\pmuttran \sim \ptunrolltran$ from \Cref{lemma:soundrecur},
       and hence there must indeed exist a state
       $S'_1$ such
       that \[\pmuttran
       \prismreduce{\pp::\q::\labelsell}{\probn{1}}(\pmuttran, S'_1) \sim \trantwo{\pp :
       \T'}.\]

     \item\label{proof:soundrecv}
       Analogous to (\ref{proof:soundsend}), using \rulename{ct-in} in place of
       \rulename{ct-out}. Here $\T = \pprocinnew{\q_i}{i \in I}{\labelselli{i}}{\Basic_i}{\T_i}$
       and the reduction selects a branch $k \in I$ with $\q = \q_k$,
       $\labelsell = \labelselli{k}$ and $\T' = \T_k$. By the definition of
       $\emptytran$ the receiver makes the single transition
       \[\trantwo{\pp : \T} \prismreduce{\q_k::\pp::\labelselli{k}}{\probn{1}} (\trantwo{\pp : \T}, S_1),\]
       and $(\trantwo{\pp : \T}, S_1) \sim \trantwo{\pp : \T_k}$ follows exactly as
       in the send case: directly when $\T_k = \tend$, and otherwise by
       \Cref{lemma:shift} since $\source{\pp}{\T}{S_1} = (\T_k, \varnothing)$.
       The \rulename{ct-rec} case is identical, using \Cref{lemma:soundrecur}.
     \end{enumerate}
\end{proof}

\begin{definition}
  \label{def:stateunion}
    Let $S_1$ and $S_2$ be states for a \prism{} module $M$. Their \emph{union}
    is given by
    \[
        (S_1 \cup S_2)(x) =
        \begin{cases}
            S_1(x) & \text{if } x \in \dom{S_1}, x \notin \dom{S_2} \\
            S_2(x) & \text{if } x \in \dom{S_2}, x \notin \dom{S_1} \\
            S_1(x) & \text{if } S_1(x) = S_2(x) \\
            \text{undefined} & \text{otherwise.}
        \end{cases}
    \]
\end{definition}

\begin{restatable}{lemma}{lemmacomposition}
  \label{lemma:composition}
  Let $\Delta$ and $\Delta'$ be contexts with $\dom{\Delta} \cap \dom{\Delta'} = \varnothing$, and suppose $(\dtrantwo, S_1) \prismreduce{\alpha}{\prob} (\dtrantwo,
  S'_1)$ for some states $S_1$ and $S'_1$. Let $S_2$ be a state for $\dptrantwo$.
  \begin{enumerate}
    \item If $\alpha = \pp\conc\q\conc\labelsell$ and
      $(\dptrantwo, S_2) \prismreduce{\alpha}{\probp} (\dptrantwo, S'_2)$
      for some state $S'_2$, then
      \[(\ddptrantwo, S_1 \cup S_2) \prismreduce{\alpha}{\prob \cdot \probp}
        (\ddptrantwo, S'_1 \cup S'_2).\]

    \item If $\alpha$ is not present in $\dptrantwo$, then
      \[(\ddptrantwo, S_1 \cup S_2) \prismreduce{\alpha}{\prob} (\ddptrantwo,
        S'_1 \cup S_2).\]
  \end{enumerate}
\end{restatable}
\begin{proof}
  ~\begin{enumerate}
    \item Since $\alpha = \pp\conc\q\conc\labelsell$, by construction only
    participants $\pp$ and $\q$ can have transitions with
    this action. Furthermore, by assumption, there are no overlapping
    participants in $\Delta$ and $\Delta'$. Thus, without loss of generality, we
    assume $\Delta$ contains $\pp$ and $\Delta'$ contains $\q$.

    By the definition of $\emptytran$, each such command is deterministic, so we have
    \begin{align*}
      \dtrantwo &\prismhas \prismcommand{\alpha}{(S_\pp = n)}{\probn{1} : (S'_\pp = n')} \\
      \dptrantwo &\prismhas \prismcommand{\alpha}{(S_\q = m)}{\probn{1} : (S'_\q = m')},
    \end{align*}
    for $n = S_1(S_\pp), m = S_2(S_\q)$ with
    $S_1[(S'_\pp = n')] = S'_1$ and $S_2[(S'_\q = m')] = S'_2$. Note from this we
    have $\prob = \probp = 1$.

    Applying \rulename{M-sync}, we have
    \begin{align*}
      \ddptrantwo \prismhas \prismcommand{\alpha}{(S_\pp = n) \land (S_\q = m)}{\probn{1} : (S'_\pp = n') \land
        (S'_\q = m')}.
    \end{align*}
    Hence by applying \rulename{S-step}, we conclude
    \[
      (\ddptrantwo, S_1 \cup S_2) \prismreduce{\pp::\q::\labelsell}{\probn{1}} (\ddptrantwo,
      S'_1 \cup S'_2),
    \]
    and $\prob \cdot \probp = 1$.

    \item By definition of $\emptytran$, this transition does not
      access or modify variables in $S_2$. So the result is immediate by
      applying \rulename{M-move}
      followed by \rulename{S-step}.
  \end{enumerate}
\end{proof}

\begin{restatable}{lemma}{comptran}
  \label{lemma:comptran}
  Let $\Delta = \pp : \T, \q : \T', \Delta'$ be a context, and $\alpha = \pp::\q::\labelsell$. If $(\pttran, S_1) \prismreduce{\alpha}{\prob} (\pttran, S'_1)$ and
  $(\qttran, S_2) \prismreduce{\alpha}{\probn{1}} (\qttran, S'_2)$,
  \[(\dtran, S_1 \cup S_2 \cup S) \prismreduce{\alpha}{\prob} (\dtran, S'_1 \cup
    S'_2 \cup S).\]
\end{restatable}
\begin{proof}
  Let $u_1$ and $u_2$ be the updates performed in the two transitions, so that
  $S'_1 = S_1[u_1]$ and $S'_2 = S_2[u_2]$.
  Examining the definition of $\emptytran$, the updates must have the form
  \[
    u_1 = (S'_\pp = n') \qquad u_2 = (S'_\q = m').
  \]
  These are distinct variables, so $u_1 \wedge u_2$ is well-defined.
  Applying \rulename{M-sync} and \rulename{S-step}, we have
  \[(\dtran, S_1 \cup S_2 \cup S) \prismreduce{\alpha}{\prob} (\dtran, (S_1
    \cup S_2 \cup S)[u_1 \wedge u_2]).\]
  Note $S_1$ and $S_2$ must be of the form
  \[
    S_1 = \{S_\pp \mapsto n\} \qquad S_2 = \{S_\q \mapsto m\}
  \]
  and hence
  \begin{align*}
    (S_1 \cup S_2 \cup S)[u_1 \wedge u_2] &=
    \{S_\pp \mapsto n', S_\q \mapsto m'\} \cup S \\
    &= S_1[u_1] \cup S_2[u_2] \cup S \\ &= S'_1 \cup S'_2 \cup S,
  \end{align*}
  as required.
\end{proof}

\begin{restatable}{lemma}{lemmaclosure}
  \label{lemma:closure}
  Let $\Delta = \pp : \T, \q : \T', \Delta'$ be a context. Further let $S_1$ and
  $S'_1$ be states of $\pttran$, $S_2$ and $S'_2$ states of $\qttran$, and $S$ a state
  of $\dptrantwo$.
  If \[(\dtran, S_1 \cup S_2 \cup S) \prismreduce{\alpha}{\prob} (\dtran, S'_1
  \cup S'_2 \cup S),\]
  with $\alpha = \pp::\q::\labelsell$, then
  \[(\pttran, S_1) \prismreduce{\alpha}{\prob} (\pttran, S'_1) \text{ and }
    (\qttran, S_2) \prismreduce{\alpha}{\probn{1}} (\qttran, S'_2).\]
\end{restatable}
\begin{proof}
  We first argue that $\alpha$
  must appear as an action in both $\pttran$ and $\qttran$.

  Suppose $\alpha \in \actions{\Delta}{\pp}$ but
  $\alpha \notin \actions{\Delta}{\q}$. Note \[\dtran = \pttran \cmp \qttran \cmp
    \dptrantwo \cmp \closure{\Delta},\] and according to
  \Cref{def:closure},
  \[
    \closure{\Delta} \prismhas \prismcommand{\alpha}{\false}{\probn{1} : ()},
  \]
  Hence the only commands of $\dtran$ with action $\alpha$ are of the form
  \[
    \dtran \prismhas \prismcommand{\alpha}{\false \wedge g}{\probn{1} : u},
  \]
  and since $S \not \vdash \false \wedge g$ for any state $S$, $\dtran
  \not \prismreduce{\alpha}{ }$, which is a contradiction. The case where $\alpha
  \notin \actions{\Delta}{\pp}$ and $\alpha \in \actions{\Delta}{\q}$ is identical.

  Hence $\alpha$ appears in both modules, and by the definition of $\emptytran$
  each such command is deterministic:
  \begin{align*}
    \pttran &\prismhas \prismcommand{\alpha}{(S_\pp = n')}{\probn{1} : (S'_\pp = n'')} \\
    \qttran &\prismhas \prismcommand{\alpha}{(S_\q = m')}{\probn{1} : (S'_\q = m'')},
  \end{align*}
  and furthermore by \rulename{M-sync} and \rulename{S-step},
  \begin{gather}
    \dtran \prismhas \prismcommand{\alpha}{(S_\pp = n') \wedge (S_\q =
      m')}{\probn{1} : (S_\pp '= n'') \wedge (S_\q '=
      m'')}, \nonumber\\
    \label{eq:sosts}
    S_1 \cup S_2 \cup S \vdash (S_\pp = n') \wedge (S_\q = m'), \\
    S'_1 \cup S'_2 \cup S = (S_1 \cup S_2 \cup S)[(S_\pp '= n'') \wedge (S_\q
    '= m'')]. \nonumber
  \end{gather}
  In particular $\prob = \probn{1}$.

  Note that states are of the form
  \begin{alignat*}{3}
    &S_1 = \{S_\pp \mapsto n\} \qquad &&S_2 = \{S_\q \mapsto m\}
  \end{alignat*}
  and by (\ref{eq:sosts}), in fact $n = n'$ and $m
  = m'$. Hence we can apply
  \rulename{S-step} to each of $(\pttran, S_1)$ and $(\qttran, S_2)$ to arrive at
  the conclusion.
\end{proof}

\begin{restatable}{lemma}{lsimcombine}
  \label{lemma:lsimcombine}
  Let $\Delta = \deltaupto{n}$ and $\Delta' = \deltaprimeupto{n}$.
  If $(\ptran{i}, S_i) \sim \ptranprime{i}$ for all $i$, then $(\dtran, S_1 \cup
  \cdots \cup S_n) \sim \dptran$.
\end{restatable}
\begin{proof}
   For each $i$, let $\rel_i$ be the
   bisimulation relation of $(\ptran{i}, S_i) \sim \ptranprime{i}$.
   We construct a binary relation
  \[
    \rel = \{((\allstatest), (\allstatespt)) \ | \ \forall i \cdot (T_i, T'_i)
    \in \rel_i\}.
  \]
  where each $S_i$ is a state for modules with participant $\ppi{i}$.
  We will show that $\rel$ is a bisimulation relation for $(\dtran, A)$ and
  $\dptran$.

  Let us take an arbitrary $(\allstatest, \allstatespt) \in \rel$.
  Taking the shorthand $T = \allstatest$,
  now suppose $(\dtran, T) \prismreduce{\alpha}{\prob} (\dtran, T')$ for some
  $T'$.
  For the purposes of this proof, we use the notation
  \[
    T[i \mapsto U] = T_1 \cup \cdots \cup T_{i-1} \cup U \cup T_{i+1} \cup \cdots \cup T_n
  \]
  (i.e. $T$ with the $i$-th union being changed to $U$), and using the obvious
  generalisation to more mappings (e.g $T[i \mapsto U, j \mapsto U']$).
  We now consider the possible cases of $\alpha$.

  \case{$\alpha = \ppi{i}\conc\tau$ or $\alpha = \ppi{i}\conc p$ for some $i$}
  The transitions with these labels
  in the encoding are the internal steps (summand and
  commitment steps, respectively) of participant $\ppi{i}$, which
  do not synchronise (as participant names are unique, these actions appear
  only in $\ppi{i}$'s module). So exactly one
  component caused the transition:
  $(\ptran{i}, T_i) \prismreduce{\alpha}{\prob} (\ptran{i}, U_i)$ and
  $T' = T[i \mapsto U_i]$. Since $(T_i, T'_i) \in \rel_i$, there exists a state
  $U'_i$ with $(\ptranprime{i}, T'_i) \prismreduce{\alpha}{\prob}
  (\ptranprime{i}, U'_i)$ and $(U_i, U'_i) \in \rel_i$. By
  \Cref{lemma:composition}, we have both $(\dtran, T)
  \prismreduce{\alpha}{\prob} (\dtran, T[i \mapsto U_i])$ and
  $(\dptran, \allstatespt) \prismreduce{\alpha}{\prob} (\dptran, T'[i \mapsto U'_i])$,
  with $(T[i \mapsto U_i], T'[i \mapsto U'_i]) \in \rel$.

  \case{$\alpha = \ppi{i} :: \ppi{j} :: \labelsell$}
  Firstly, applying \Cref{lemma:closure}, there must exist
  states $U_i$ and $U_j$ such that
  \begin{align*}
    (\ptran{i}, T_i) &\prismreduce{\alpha}{\prob} (\ptran{i}, U_i) \\
    (\ptran{j}, T_j) &\prismreduce{\alpha}{\pone} (\ptran{j}, U_j).
  \end{align*}
  Since $(T_i, T'_i) \in \rel_i$ and similarly $(T_j, T'_j) \in \rel_j$, there must
  also exist states $U'_i, U'_j$ such that
  \begin{gather*}
    (\ptranprime{i}, T'_i) \prismreduce{\alpha}{\prob}
    (\ptranprime{i}, U'_i) \\
    (\ptranprime{j}, T'_j) \prismreduce{\alpha}{\pone} (\ptranprime{j}, U'_j),
  \end{gather*}
  with $(U_i, U'_i) \in \rel_i$ and $(U_j, U'_j) \in \rel_j$.
  Applying \Cref{lemma:comptran}, we have
  \begin{align*}
    (\dtran, \allstatest) &\prismreduce{\alpha}{\prob} (\dtran, T[i \mapsto U_i, j \mapsto U_j]) \\
    (\dptran, \allstatespt) &\prismreduce{\alpha}{\prob} (\dptran, T'[i \mapsto U'_i, j \mapsto U'_j]),
  \end{align*}
  with $(T[i \mapsto U_i, j \mapsto U_j], T'[i \mapsto U'_i, j \mapsto U'_j]) \in \rel$ as required.

  Hence, any transition in $(\dtran, \allstatest)$ can be matched with
  a transition in $(\dptran, \allstatespt)$, with the resulting pair
  of states still in $\rel$. The other direction is completely symmetrical.
  We conclude by observing that $(\allstates, \varnothing) \in \rel$, so $\rel$ is
  indeed a bisimulation
  relation, and $(\dtran, \allstates) \sim \dptran.$
\end{proof}

\soundness*
\begin{proof}
  Let $\Delta = \deltaupto{n}.$

  \begin{enumerate}
    \item
  Without loss of generality, we assume $\pp = \ppi{1}$, so that
  $\Delta' = \ppi{1} : \T'_1, \ppi{2} : \T_2, \cdots,
  \ppi{n} : \T_n$ with $\ppi{1} : \T_1 \predcontnd{\actionnd{\ppi{1}}} \ppi{1}
  : \T'_1$.
  Applying \Cref{lemma:soundcomm}~(\ref{lemma:soundnd}), there exists a state
  $S_1$ such that
  \[
    \ptran{1} \prismreduce{\ppi{1}\conc\tau}{\probn{1}} (\ptran{1}, S_1) \sim
    \ptranprime{1}.
  \]
  By \Cref{lemma:composition}, this summand step does not
  involve the other components and lifts to
  \[
    \dtran \prismreduce{\ppi{1}\conc\tau}{\probn{1}} (\dtran, S_1).
  \]
  Since $(\ptran{i}, \varnothing) \sim \ptran{i} = \trantwo{\ppi{i} : \T'_i}$
  trivially for $i \geq 2$, we appeal to \Cref{lemma:lsimcombine} to conclude
  $(\dtran, S_1) \sim \dptran$.

  \item
  Without loss of generality, we assume $\pp = \ppi{1}$, so that
  $\Delta' = \ppi{1} : \T'_1, \ppi{2} : \T_2, \cdots, \ppi{n} : \T_n$
  with $\ppi{1} : \T_1 \predcont{\actionprob{\ppi{1}}}{\prob} \ppi{1} :
  \T'_1$.
  Applying \Cref{lemma:soundcomm}~(\ref{lemma:soundprob}), there exists a
  state $S_1$ such that
  \[
    \ptran{1} \prismreduce{\ppi{1}\conc p}{\prob} (\ptran{1}, S_1) \sim
    \ptranprime{1}.
  \]
  By \Cref{lemma:composition}, this commitment step does not
  involve the other components and lifts to
  \[
    \dtran \prismreduce{\ppi{1}\conc p}{\prob} (\dtran, S_1).
  \]
  Since $(\ptran{i}, \varnothing) \sim \ptran{i} = \trantwo{\ppi{i} : \T'_i}$
  trivially for $i \geq 2$, we appeal to \Cref{lemma:lsimcombine} to conclude
  $(\dtran, S_1) \sim \dptran$.

  \item
  Without loss of generality, we assume $\pp = \ppi{1}$ and $\q = \ppi{2}$,
  so that $\Delta' = \deltaprimeuptotwo{n}$, and we have
  \begin{align*}
    \tranone{\Delta} &= \tranupto{n} \cmp \closure{\Delta} \text{ and} \\
    \tranone{\Delta'} &= \tranprimeuptotwo{n} \cmp \closure{\Delta'},
  \end{align*}
  where $\closure{\Delta}$ and $\closure{\Delta'}$ are the closure terms
  introduced in \Cref{def:closure}.

  By \rulename{ct-$\tau$}, we must have $\ppi{1}: \T_1
  \predcont{\ppi{1} : \ppi{2}!\labelsell\langle\Basic\rangle}{\probn{1}} \ppi{1}: \T'_1$
  and $\ppi{2}: \T_2 \predcont{\ppi{2} : \ppi{1}?\labelsell(\Basic)}{1} \ppi{2}: \T'_2$.
  Applying \Cref{lemma:soundcomm}~(\ref{lemma:soundsend},~\ref{lemma:soundrecv})
  to these reductions, there exist states $S_1, S_2$ such that
  \begin{gather*}
    \ptran{1} \prismreduce{\poptl}{\probn{1}} (\ptran{1}, S_1) \sim \ptranprime{1}, \\
    \ptran{2} \prismreduce{\poptl}{\pone} (\ptran{2}, S_2) \sim \ptranprime{2}.
  \end{gather*}
  We combine the sender's and receiver's synchronising steps through
  \Cref{lemma:comptran} to get
  \[
    \dtran \prismreduce{\poptl}{\pone} (\dtran, S_1 \cup S_2).
  \]
  Finally, we appeal to \Cref{lemma:lsimcombine} to conclude $(\dtran, S_1 \cup S_2) \sim
  \dptran$.
  \end{enumerate}
\end{proof}

\subsubsection{Completeness}
We follow the approach for soundness by first considering $\trantwo{\cdot}$
(\Cref{lemma:completenesslitepart}) then generalising to $\tranone{\cdot}$ (\Cref{thm:completeness}).

\begin{restatable}{lemma}{completenesslitepart}
  \label{lemma:completenesslitepart}
  ~\begin{enumerate}
    \item \label{lemma:completenesslitepartnd} For every type $\T$ and state
      $S$ with
      \[\pttran \prismreduce{\pp\conc\tau}{\probn{1}} (\pttran, S),\]
      there exists $\T'$
      such that $\pp : \T \predcontnd{\actionnd{\pp}} \pp : \T'$ and $(\pttran, S)
      \sim \ptptran$.
    \item \label{lemma:completenesslitepartprob} For every type $\T$,
      state $S$ and probability $\prob$ with
      \[\pttran \prismreduce{\pp\conc p}{\prob} (\pttran, S),\] there exists $\T'$
      such that $\pp : \T \predcont{\actionprob{\pp}}{\prob} \pp : \T'$ and $(\pttran, S)
      \sim \ptptran$.
    \item \label{lemma:completenesslitepartsend} For every type $\T$ and state $S'$ with
      \[\pttran \prismreduce{\pp::\q::\labelsell}{\pone} (\pttran, S'),\] there exists $\T'$
      such that $\pp : \T \predcont{\pqsend}{\pone} \pp : \T'$ and $(\pttran, S')
      \sim \ptptran$.
    \item \label{lemma:completenesslitepartrecv} For every type $\T$ and state $S'$ with
      \[
        \pttran \prismreduce{\q::\pp::\labelsell}{\pone}
        (\pttran, S'),
        \]
      there exists $\T'$ such that $\pp : \T \predcont{\pqrecv}{\pone} \pp : \T'$
      and $(\pttran, S') \sim \ptptran$.
    \end{enumerate}
\end{restatable}
\begin{proof}
  ~\begin{enumerate}
    \item
    By \Cref{prop:source}, since $\pttran$ makes a summand step (labelled
    $\pp\conc\tau$) in the initial state, $\T$ must be of the form
    \[
      \T = \mu \ty_1 . \mu \ty_2 . \cdots \mu \ty_n . \probpprocoutty{i \in I}{\q_j}{j \in
      J_i}{\probi{(i,j)}}{\labelselli{j}}{\Basic_j}{\T'_{(i,j)}}
    \]
    with $|I| > 1$.
    We proceed by induction on $n$, the number of $\mu$s before the sum (we
    will call this the $\mu$-depth).

    \case{$n = 0$}
    Then $\T = \probpprocoutty{i \in I}{\q_j}{j \in J_i}{\probi{(i,j)}}{\labelselli{j}}{\Basic_j}{\T'_{(i,j)}}$.
    The summand step corresponds to some summand $k \in I$, so
    $S = \{S_\pp \mapsto \sume{k}\}$ and
    $\source{\pp}{\T}{S} = (\selsummand{k}, \varnothing)$, where
    $\selsummand{k} = \pprocoutty{\q_j}{j \in
      J_k}{\probi{(k,j)}}{\labelselli{j}}{\Basic_j}{\T'_{(k,j)}}$.
    By \rulename{ct-nd}, we have
    $\pp : \T \predcontnd{\actionnd{\pp}} \pp : \selsummand{k}$, and since
    $\selsummand{k}$ is closed, we can apply \Cref{lemma:shift} to conclude
    $(\pttran, S) \sim \trantwo{\pp : \selsummand{k}}$.

    \case{$n = k + 1$}
    Then $\T = \mu \ty . \T'$, where $\T'$ has a $\mu$-depth of $k$. By
    \Cref{lemma:soundrecur} and the definition of $\sim$, there must exist a
    state $S_1$ such that
    \[
      \ptprimeunrolltran \prismreduce{\pp\conc\tau}{\probn{1}} (\ptprimeunrolltran, S_1)
      \sim (\pttran, S).
    \]
    Note $\ptprimeunroll$ has a $\mu$-depth of $k$, so we can apply the
    induction hypothesis to obtain a type $\T''$ such that
    \[
      \ptprimeunroll \predcontnd{\actionnd{\pp}} \pp : \T'',
      \qquad (\ptprimeunrolltran, S_1) \sim \trantwo{\pp : \T''}.
    \]
    By \rulename{ct-rec}, $\pp : \T = \pmutprime \predcontnd{\actionnd{\pp}} \pp :
    \T''$ also, and since $\sim$ is transitive, $(\pttran, S) \sim
    \trantwo{\pp : \T''}$ as required.

    \item
    By \Cref{prop:source}, since $\pttran$ makes a commitment step (labelled
    $\pp\conc p$) in the initial state, $\T$ must be of the form
    \[
      \T = \mu \ty_1 . \mu \ty_2 . \cdots \mu \ty_n . \pprocoutty{\q_j}{j \in
      J}{\probi{j}}{\labelselli{j}}{\Basic_j}{\T'_{j}}
    \]
    with $|J| \geq 2$.
    We proceed by induction on the $\mu$-depth $n$.

    \case{$n = 0$}
    Then $\T = \pprocoutty{\q_j}{j \in J}{\probi{j}}{\labelselli{j}}{\Basic_j}{\T'_{j}}$.
    The commitment step corresponds to some branch
    $l \in J$, so $S = \{S_\pp \mapsto \come{l}\}$, $\prob = \probi{l}$ and
    $\source{\pp}{\T}{S} = (\comsel{l}, \varnothing)$, where
    $\comsel{l} = \pprocoutty{\q_l}{}{\probn{1}}{\labelselli{l}}{\Basic_l}{\T'_{l}}$.
    Since $|J| \geq 2$ and the probabilities are positive and sum to
    $\probn{1}$, we have $\probi{l} \in (0;1)$, so by \rulename{ct-prob},
    $\pp : \T \predcont{\actionprob{\pp}}{\probi{l}} \pp : \comsel{l}$; and since
    $\comsel{l}$ is closed, we can apply \Cref{lemma:shift} to conclude
    $(\pttran, S) \sim \trantwo{\pp : \comsel{l}}$.

    \case{$n = k + 1$}
    Then $\T = \mu \ty . \T'$, where $\T'$ has a $\mu$-depth of $k$. By
    \Cref{lemma:soundrecur} and the definition of $\sim$, there must exist a
    state $S_1$ such that
    \[
      \ptprimeunrolltran \prismreduce{\pp\conc p}{\prob} (\ptprimeunrolltran, S_1)
      \sim (\pttran, S).
    \]
    Note $\ptprimeunroll$ has a $\mu$-depth of $k$, so we can apply the
    induction hypothesis to obtain a type $\T''$ such that
    \[
      \ptprimeunroll \predcont{\actionprob{\pp}}{\prob} \pp : \T'',
      \qquad (\ptprimeunrolltran, S_1) \sim \trantwo{\pp : \T''}.
    \]
    By \rulename{ct-rec}, $\pp : \T = \pmutprime \predcont{\actionprob{\pp}}{\prob} \pp :
    \T''$ also, and since $\sim$ is transitive, $(\pttran, S) \sim
    \trantwo{\pp : \T''}$ as required.

    \item
    By \Cref{prop:source}, since $\pttran$ makes a synchronisation labelled
    $\pp\conc\q\conc\labelsell$ (with $\pp$ the sender) in the initial state,
    $\T$ must be of the form
    \[
      \T = \mu \ty_1 . \mu \ty_2 . \cdots \mu \ty_n .
      \pprocoutty{\q}{}{\probn{1}}{\labelsell}{\Basic}{\T''},
    \]
    a committed selection under $n$ binders.
    We proceed by induction on the $\mu$-depth $n$.

    \case{$n = 0$}
    Then $\T = \pprocoutty{\q}{}{\probn{1}}{\labelsell}{\Basic}{\T''}$.
    By \rulename{ct-out}, we have
    $\pp : \T \predcont{\pqsend}{\pone} \pp : \T''$, and it is
    sufficient to show that $(\pttran, S') \sim \trantwo{\pp : \T''}$. We
    consider the two possible cases of $\T''$.
    \begin{enumerate}
      \item If $\T'' = \tend$, then $(\pttran, S') \not \prismreduce{ }{ }$ and
        $\trantwo{\pp : \tend} \not \prismreduce{ }{ }$.
        Hence $(\pttran, S') \sim \trantwo{\pp : \T''}$ as
        required.
     \item Otherwise, $S' = \{S_\pp \mapsto 1\}$, and thus
       $\source{\pp}{\T}{S'} = (\T'', \varnothing)$. Hence, we
        can apply \Cref{lemma:shift} to obtain the desired result.
    \end{enumerate}

    \case{$n = k + 1$}
    Then $\T = \mu \ty . \T'$, where $\T'$ has a $\mu$-depth of $k$. By
    \Cref{lemma:soundrecur} and the definition of $\sim$, there must exist a
    state $S_1$ such that
    \[
      \ptprimeunrolltran \prismreduce{\pp::\q::\labelsell}{\pone} (\ptprimeunrolltran, S_1)
      \sim (\pttran, S').
    \]
    Note $\ptprimeunroll$ has a $\mu$-depth of $k$, so we can apply the
    induction hypothesis to obtain a type $\T''$ such that
    \[
      \ptprimeunroll \predcont{\pqsend}{\pone} \pp : \T'',
      \qquad (\ptprimeunrolltran, S_1) \sim \trantwo{\pp : \T''}.
    \]
    By \rulename{ct-rec}, $\pp : \T = \pmutprime \predcont{\pqsend}{\pone} \pp :
    \T''$ also, and since $\sim$ is transitive, $(\pttran, S') \sim
    \trantwo{\pp : \T''}$ as required.

    \item Analogous to the above, with branching in place of the committed
    selection. Here
    $\T = \mu \ty_1 \cdots \mu \ty_n . \pprocinnew{\q_i}{i \in I}{\labelselli{i}}{\Basic_i}{\T'_i}$,
    and the single $\q\conc\pp\conc\labelsell$ transition selects a branch
    $k \in I$ with $\q = \q_k$ and $\labelsell = \labelselli{k}$. By
    \rulename{ct-in}, $\pp : \T \predcont{\pqrecv}{\pone} \pp : \T'_k$, and
    $(\pttran, S') \sim \trantwo{\pp : \T'_k}$ follows as in the send case
    (directly when $\T'_k = \tend$, otherwise by \Cref{lemma:shift} since
    $\source{\pp}{\T}{S'} = (\T'_k, \varnothing)$). The $\mu$-depth induction
    uses \Cref{lemma:soundrecur} exactly as above.

  \end{enumerate}
\end{proof}

\completeness*
\begin{proof}
  Let $\Delta = \deltaupto{n}$.
  Then, we have
  \begin{align*}
    \tranone{\Delta} &= \tranupto{n} \cmp \closure{\Delta},
  \end{align*}
  where $\closure{\Delta}$ is the closure term introduced in \Cref{def:closure}.
  Since internal steps do not synchronise (as participant names are unique) and
  each internal command labelled $\ppi{i}\conc\tau$ or $\ppi{i}\conc p$
  modifies only the variable $S_{\ppi{i}}$ of the acting participant, every
  internal step is made by the single component of its acting participant.

  \begin{enumerate}
     \item
  Without loss of generality, we let $\pp = \ppi{1}$.
  The summand step is labelled $\ppi{1}\conc\tau$, and is therefore made by
  the component $\ptran{1}$:
  \[
    \ptran{1} \prismreduce{\ppi{1}\conc\tau}{\probn{1}} (\ptran{1}, S_1),
  \]
  with the other components unchanged, so $S = S_1$.

  Applying
  \Cref{lemma:completenesslitepart}~(\ref{lemma:completenesslitepartnd}) to
  this step, there must exist a type $\T'_1$ such that
  \[
    \ppi{1} : \T_1 \predcontnd{\actionnd{\ppi{1}}} \ppi{1} : \T'_1, \qquad
    (\ptran{1}, S_1) \sim \ptranprime{1}.
  \]

  Let $\Delta' = \ppi{1} : \T'_1, \ppi{2} : \T_2, \cdots, \ppi{n} : \T_n$.
  Then $\Delta \predcontnd{\actionnd{\ppi{1}}} \Delta'$ by \rulename{ct-sc},
  and since $(\ptran{i}, \varnothing) \sim \ptran{i} = \trantwo{\ppi{i} :
    \T'_i}$ trivially for $i \geq 2$, we appeal to \Cref{lemma:lsimcombine} to
  conclude $(\dtran, S) = (\dtran, S_1) \sim \dptran$, as required.

  \item
  Without loss of generality, we let $\pp = \ppi{1}$.
  The commitment step is labelled $\ppi{1}\conc p$, and is therefore made by
  the component $\ptran{1}$:
  \[
    \ptran{1} \prismreduce{\ppi{1}\conc p}{\prob} (\ptran{1}, S_1),
  \]
  with the other components unchanged, so $S = S_1$.

  Applying
  \Cref{lemma:completenesslitepart}~(\ref{lemma:completenesslitepartprob}) to
  this step, there must exist a type $\T'_1$ such that
  \[
    \ppi{1} : \T_1 \predcont{\actionprob{\ppi{1}}}{\prob} \ppi{1} : \T'_1, \qquad
    (\ptran{1}, S_1) \sim \ptranprime{1}.
  \]

  Let $\Delta' = \ppi{1} : \T'_1, \ppi{2} : \T_2, \cdots, \ppi{n} : \T_n$.
  Then $\Delta \predcont{\actionprob{\ppi{1}}}{\prob} \Delta'$ by \rulename{ct-sc},
  and since $(\ptran{i}, \varnothing) \sim \ptran{i} = \trantwo{\ppi{i} :
    \T'_i}$ trivially for $i \geq 2$, we appeal to \Cref{lemma:lsimcombine} to
  conclude $(\dtran, S) = (\dtran, S_1) \sim \dptran$, as required.

  \item
  Without loss of generality, we let $\pp =
  \ppi{1}$ and $\q = \ppi{2}$.
  The synchronising step
  $\dtran \prismreduce{\ppi{1}::\ppi{2}::\labelsell}{\pone} (\dtran, S)$
  decomposes, by \Cref{lemma:closure}, into
  \begin{gather*}
    \ptran{1} \prismreduce{\ppi{1}::\ppi{2}::\labelsell}{\pone} (\ptran{1}, S'_1), \qquad
    \ptran{2} \prismreduce{\ppi{1}::\ppi{2}::\labelsell}{\pone} (\ptran{2}, S'_2),
  \end{gather*}
  where $S = S'_1 \cup S'_2$, with the sender $\ptran{1}$ and the receiver
  $\ptran{2}$ each transitioning from their initial states.

  Applying \Cref{lemma:completenesslitepart}~(\ref{lemma:completenesslitepartsend})
  to the sender's step and~(\ref{lemma:completenesslitepartrecv}) to the receiver's
  step, there must exist types $\T'_1$ and
  $\T'_2$ such that
  \begin{gather*}
    \ppi{1} : \T_1 \predcont{\potsend}{\pone} \ppi{1} : \T'_1,
      \qquad \ppi{2} : \T_2 \predcont{\ptorecv}{1} \ppi{2} : \T'_2, \\
    (\ptran{1}, S'_1) \sim \ptranprime{1}, \qquad (\ptran{2}, S'_2) \sim \ptranprime{2}.
  \end{gather*}

  Let $\Delta' = \deltaprimeuptotwo{n}$. Then $\Delta \predcont{(\ppi{1}, \ppi{2})
    \labelsell}{\pone} \Delta'$ by \rulename{ct-$\tau$} and \rulename{ct-sc}, and by
  \Cref{lemma:lsimcombine}, we have $(\dtran, S) = (\dtran, S'_1 \cup S'_2)
  \sim \dptran$, as required.
  \end{enumerate}
\end{proof}

\begin{restatable}{lemma}{lemmastatespace}
  \label{lemma:statespace}
  For every command of the form
  \[
    \oxtrantwo{\T}{\pp}{n}{m}{f} \prismhas \prismcommandgeneric{\alpha}{(S_\pp = k) \wedge g}{i}{\probi{i}}{u_i},
  \]
  we must have $n \leq k < n + \StateS{T}$.
\end{restatable}
\begin{proof}
  We proceed by induction of the structure of $\T$.

  \case{$\T = \tend$ or $\ty$}
  Then $\parttran = \varnothing$, so the statement
  is trivial.

  \case{$\T = \mu \ty . \T'$}
  The commands in $\parttran$ are exactly the commands in $\parttranp$, and
  $\StateS{\T} = \StateS{\T'}$. Thus, we can apply the induction hypothesis on $\T'$.

  \case{$\T = \pprocoutty{\q}{}{\probn{1}}{\labelsell}{\Basic}{\T'}$}
  Then we have $\StateS{T} = 1 + \StateS{\T'}$. By the
  definition of $\emptytran$, the commands in
  $\parttran$ are of the form
  \begin{align*}
    \parttran &\prismhas \prismcommand{\pp::\q::\labelsell}{S_\pp = n}{1 : u},
  \end{align*}
  along with the commands in
  $\oxtrantwo{\T'}{\pp}{n + 1}{m}{f}$.
  Obviously $n \in [n, n + \StateS{\T})$, and by the induction hypothesis,
  every guard $(S_\pp = k)$ in $\oxtrantwo{\T'}{\pp}{n + 1}{m}{f}$ has
  \[
    n + 1 \leq k < n + 1 + \StateS{\T'} = n + \StateS{\T},
  \]
  and so $k \in [n, n + \StateS{\T})$ as required.

  \case{$\T = \pprocoutty{\q_j}{j \in
      J}{\probi{j}}{\labelselli{j}}{\Basic_j}{\T'_{j}}$ with $|J| \geq 2$}
  Then we have $\StateS{T} = 1 + \sum_{j \in J} \StateS{\comsel{j}}$, where
  $\comsel{j} = \pprocoutty{\q_j}{}{\probn{1}}{\labelselli{j}}{\Basic_j}{\T'_{j}}$. By the
  definition of $\emptytran$, the commands in
  $\parttran$ are of the form
  \begin{align*}
    \parttran &\prismhas \prismcommandgeneric{\pp\conc p}{S_\pp = n}{j \in J}{\probi{j}}{(S'_\pp = \come{j})},
  \end{align*}
  along with the commands in each of
  $\oxtrantwo{\comsel{j}}{\pp}{\come{j}}{m}{f}$,
  where $\come{j} = n + 1 + \textstyle\sum_{j' < j}\StateS{\comsel{j'}}$.
  Obviously $n \in [n, n + \StateS{\T})$. Furthermore, by the induction
  hypothesis, every guard $(S_\pp =
  k)$ in $\oxtrantwo{\comsel{j}}{\pp}{\come{j}}{m}{f}$ has
  \[
    \come{j} \leq k < \come{j} + \StateS{\comsel{j}}.
  \]
  Note $\come{j} \geq n$ and, since the prefix sum covers all branches
  $j'$ preceding $j$,
  \[
    \come{j} + \StateS{\comsel{j}} \leq n + 1 + \textstyle\sum_{j \in J}\StateS{\comsel{j}} = n + \StateS{\T},
  \]
  and so $k \in [n, n + \StateS{\T})$ as required.

  \case{$\T = \probpprocoutty{i \in I}{\q_j}{j \in
      J_i}{\probi{(i,j)}}{\labelselli{j}}{\Basic_j}{\T'_{(i,j)}}$ with $|I| \geq 2$}
  Then we have $\StateS{T} = 1 + \sum_{i \in I} \StateS{\selsummand{i}}$, where
  $\selsummand{i} = \pprocoutty{\q_j}{j \in
    J_i}{\probi{(i,j)}}{\labelselli{j}}{\Basic_j}{\T'_{(i,j)}}$. By the
  definition of $\emptytran$, the commands in
  $\parttran$ are of the form
  \begin{align*}
    \parttran &\prismhas \prismcommand{\pp\conc\tau}{S_\pp = n}{1 : (S'_\pp = \sume{i})},
  \end{align*}
  along with the commands in each of
  $\oxtrantwo{\selsummand{i}}{\pp}{\sume{i}}{m}{f}$,
  where $\sume{i} = n + 1 + \textstyle\sum_{i' < i}\StateS{\selsummand{i'}}$.
  Obviously $n \in [n, n + \StateS{\T})$. Furthermore, by the induction
  hypothesis, every guard $(S_\pp = k)$ in
  $\oxtrantwo{\selsummand{i}}{\pp}{\sume{i}}{m}{f}$ has
  \[
    \sume{i} \leq k < \sume{i} + \StateS{\selsummand{i}}.
  \]
  Note $\sume{i} \geq n$ and, since the prefix sum covers all summands $i'$
  preceding $i$,
  \[
    \sume{i} + \StateS{\selsummand{i}} \leq n + 1 + \textstyle\sum_{i \in I}\StateS{\selsummand{i}} = n + \StateS{\T},
  \]
  and so $k \in [n, n + \StateS{\T})$ as required.

  \case{$\T = \pprocinnew{\q_i}{i \in
      I}{\labelselli{i}}{\Basic_i}{\T'_i}$}
  Then $\StateS{\T} = 1 + \sum_{i \in I}\StateS{\T'_i}$. The
  new commands have the single guard clause $(S_\pp = n) \in [n, n + \StateS{\T})$,
  and the recursive commands in $\oxtrantwo{\T'_i}{\pp}{\nexts{i}}{m}{f}$, with
  $\nexts{i} = n + 1 + \sum_{i' < i}\StateS{\T'_{i'}}$, are bounded into
  $[n, n + \StateS{\T})$ in exactly the same way as the proper selection case.
\end{proof}

\section{Appendix for Section \ref{sec:modelchecking}}
\label{app:modelchecking}

\subsection{Computing the Weak Almost-Surely Livelocked Region}
\label{app:wasll-alg}
Throughout this section we fix $\Delta$ to be the ``initial'' context, and participant $\pp$ to be the one whose livelock region we are computing. As a refresher,~\cref{def:wasll} places $\Delta'$ in $\wasllp{\pp}{\Delta}$ when
\[\exists \phi \in \Phi_F^{\Delta'}: \forall \tpath \in \Tpath_\phi(\Delta'): \visited{\tpath} \subseteq \pend{\pp}{\Delta}
\]
This section shows how to compute, over the finite $\ReachD{\Delta}$, the region $\wasllp{\pp}{\Delta}$.

Before outlining the approach we briefly outline the structure of the \textsf{MDP} we are dealing with, and define the notion of an \emph{end-component} (shorthand; ``EC'') as a tuple (S,A).

\begin{definition}[Scheduler-available Actions]
\label{def:actions}
The \emph{scheduler-available actions} (or just \emph{actions}) enabled at a state $\Delta'$ are the transition labels of Fig.~\ref{fig:pmpst_tct}, with nondeterministic labels paired with their resolution state ``$\Delta_\nd$'' (as these are determined by the scheduler):
\[
\ActD(\Delta') \;:=\;
\{\, \actiontau{\q}{\pr}{\ell} \SEP
   \Delta' \predcont{\actiontau{\q}{\pr}{\ell}}{\probn{1}} \,\}
\;\cup\;
\{\, (\actionnd{\q}{:}\Delta_{\nd}) \SEP
   \Delta' \predcontnd{\actionnd{\q}}{} \Delta_\nd \,\}
\;\cup\;
\{\, \actionprob{\q} \SEP
   \Delta' \predcont{\actionprob{\q}}{\probi{k}} \,\}.
\]
\end{definition}

\begin{definition}[step function]
\label{def:stepfunc}
The function $\stepd{\cdot}{\cdot}$ takes a state and a scheduler-available action and returns a function from states to distributions over states:
\begin{equation*}
\begin{array}{l@{\quad}l}
    \stepd{\Delta'}{(\actionnd{\q}{:}\Delta_{\nd})}(\Delta'') = \probn{1}
  & \iff\; \Delta' \predcontnd{\actionnd{\q}}{} \Delta''
    \,\wedge\, \Delta'' = \Delta_{\nd} \\[1mm]
\stepd{\Delta'}{\actionprob{\q}}(\Delta'') = \prob
  & \iff\; \Delta' \predcont{\actionprob{\q}}{\probi{k}} \Delta''
    \\[1mm]
\stepd{\Delta'}{\actiontau{\q}{\pr}{\ell}}(\Delta'') = \probn{1}
  & \iff\; \Delta' \predcont{\actiontau{\q}{\pr}{\ell}}{\probn{1}}
    \Delta''
\end{array}
\end{equation*}
and $\stepd{\Delta'}{\alpha}(\Delta'') = \probn{0}$ otherwise, for
$\alpha \in \ActD(\Delta')$.  Each $\stepd{\Delta'}{\alpha}$ is a
probability distribution: both communication and internal nondeterminism reach their target state with probability $\probn{1}$, while a probabilistic action may reach a state at $\prob < \probn{1}$.
\end{definition}

Thus $(S, \ActD, \mathsf{step})$ is a finite MDP (where $S=\ReachD{\Delta}$), whose schedulers select one enabled action per history, inducing the measures of~\Cref{def:cylinder}.

\begin{definition}[End components]
An \emph{end component} (EC) of $(S, \ActD, \mathsf{step})$ is a pair $(S',A')$
with the requirements:
\begin{enumerate}
    \item $\emptyset \neq S' \subseteq S$ and for all $\Delta' \in S'$: $\emptyset \neq A'(\Delta')  \subseteq \ActD\restrict{S'}(\Delta')$ where $\ActD\restrict{S'}(\Delta') = \{\alpha \in \ActD(\Delta') | \suppd(\stepd{\Delta'}{\alpha}) \subseteq S' \}$.
    \item The induced graph $(S', A', \mathsf{step})$ is strongly connected.
\end{enumerate}
Unions of overlapping ECs are ECs, so \emph{maximal} ECs (MECs) partition $S$.
\end{definition}

\noindent Using this notion of \emph{end-components}, we outline the approach in two stages,
(all lemmas and the correctness proof \Cref{prop:wasll-correct} follow in \S\ref{app:wasll-proofs});

\myparagraph{Stage 1: closure}\\[1mm]
\label{def:stage1}
Let $S^0 := \pend{\pp}{\Delta}$ and
$S^{i+1} := \{\Delta' \in S^i \SEP \ActD(\Delta') = \emptyset \vee
\exists \alpha \in \ActD(\Delta') \st
\suppd(\stepd{\Delta'}{\alpha}) \subseteq S^i\}$; write $\Sclos$ for
the limit of this decreasing sequence and $\Aclos(\Delta') :=
\{\alpha \SEP \suppd(\stepd{\Delta'}{\alpha}) \subseteq \Sclos\}$.
Thus $S^* \subseteq \pend{\pp}{\Delta}$ is the largest set
whose every state is action-less or has an action whose reachable states are fully supported in $S^*$.\\[1mm]
The result $S^*$ consists of the union over all \emph{end-components} $(S_i, A_i)$ whose states are all pending, i.e. $\bigcup_{i \in I}S_i$ where $S_i \subseteq \pend{\pp}{\Delta}$ --- (an actionless \emph{deadlocked} state may be considered an EC, e.g. if extended with a \emph{self-loop}, but in the following we treat them separately), along with any states $S_{\neg\text{SCC}} \subseteq \pend{\pp}{\Delta} $ that can reach $\bigcup_{i \in I}(S_i, A_i)$ with probability $\probn{1}$ (but are not strongly-connected), i.e. $S^* = S_{\neg\text{SCC}} \uplus \bigcup_{i \in I}(S_i, A_i)$.

\begin{remark}
\label{rem:exclusion}
We note that excluding any state in the not strongly-connected part $S_{\neg\text{SCC}}$ from $\wasllp{\pp}{\Delta}$ does not change the probability of reaching in $\wasllp{\pp}{\Delta}$, hence it will be excluded from the computed region as PRISM's reachability will handle it. This will be proved in~\cref{lem:absorb}. We will call the final computed region (that excludes $S_{\neg\text{SCC}}$) ``$\GoodP{\pp}$'', and show (\cref{prop:wasll-correct}):
\[
\max_{\phi \in \Phi^{\Delta}}
\probpathmeasure{\phi}{\Delta}\Bigl(\textstyle\bigcup_{\pp \in \dom{\Delta}}
  \GoodP{\pp}\Bigr)
\;=\;
\max_{\phi \in \Phi^{\Delta}}
    \probpathmeasure{\phi}{\Delta}\bigl(\wasll{\Delta}\bigr).
\]
\end{remark}

\myparagraph{Stage 2: pruning}\\[1mm]
\label{def:stage2}
Let $(S_1, A_1), \ldots, (S_m, A_m)$ be the \emph{maximal end-components} (MECs) of
$(S_0^*, \Aclos{\big|}_{S_0^*})$ where $S_0^* := \Sclos \setminus \{\Delta'
\SEP \ActD(\Delta') = \emptyset\}$, we then compute $\GoodP{\pp}{\Delta}$ as\\[1mm]
\centerline{$\GoodP{\pp}{\Delta} \;:=\; \bigcup_{i \leq m}
\mathsf{prune}(S_i, A_i) \;\cup\; \{\Delta' \in \Sclos \SEP
\ActD(\Delta') = \emptyset\}$.}
the set of states that lie in $\pendd{\pp}$, and which may be closed under \emph{fair} scheduling ($\mathsf{prune}(\cdot, \cdot)$ to be introduced soon). Note that deadlocked states bypass the fairness pass and join $\GoodP{\pp}$ directly as any path to a deadlocked state is trivially fair.

 The algorithm $\mathsf{prune}(\cdot,\cdot)$ run on each MEC $(S_i, A_i)$ recursively removes states that can't be visited infinitely often under a fair scheduler, and either re-decomposes the resulting states into new MECs and runs prune on these, or, returns $S_i$ if every state of $S_i$ can be visited infinitely often by a fair scheduler:
\[
\begin{array}{l}
\mathsf{prune}(S, A) :=\\[0.5mm]
\quad \mathit{nd} := \emptyset;\quad \mathit{prob} := \emptyset;\quad
  \mathit{comm} := \emptyset\\[0.5mm]
\quad \mathbf{forall}\ \Delta' \in S,\ \alpha \in \ActD(\Delta'){:}
  \qquad \text{(fairness obligations to discharge)}\\
\qquad \mathbf{if}\ \alpha = (\actionnd{\q}, \cdot)\
  \mathbf{then}\ \mathit{nd} := \mathit{nd} \cup \{\q\}
  \qquad\qquad \text{(clause 2 of Def.~\ref{def:fairpath})}\\
\qquad \mathbf{if}\ \alpha = \actionprob{\q}\
  \mathbf{then}\ \mathit{prob} := \mathit{prob} \cup \{\q\}
  \qquad\qquad \text{(clause 3)}\\
\qquad \mathbf{if}\ \alpha = \actiontau{\q}{\pr}{\ell}\
  \mathbf{then}\ \mathit{comm} := \mathit{comm} \cup \{(\q, \pr, \labelsell)\}
  \qquad \text{(clause 1)}\\[0.5mm]
\quad \mathbf{forall}\ \Delta' \in S,\ \alpha \in A(\Delta'){:}
  \qquad\qquad \text{(obligations dischargeable within the component)}\\
\qquad \mathbf{if}\ \alpha = (\actionnd{\q}, \cdot)\
  \mathbf{then}\ \mathit{nd} := \mathit{nd} \setminus \{\q\}\\
\qquad \mathbf{if}\ \alpha = \actionprob{\q}\
  \mathbf{then}\ \mathit{prob} := \mathit{prob} \setminus \{\q\}\\
\qquad \mathbf{if}\ \alpha = \actiontau{\q}{\pr}{\ell}\
  \mathbf{then}\ \mathit{comm} := \mathit{comm} \setminus \{(\q,\pr,\labelsell)\}\\[0.5mm]
\quad D := \{\, \Delta' \in S \SEP \exists \alpha \in
    \ActD(\Delta') \st \big(\alpha = (\actionnd{\q}, \_) \,\wedge\, \q \in \mathit{nd}
  \,\big)\ \lor\\
\;\qquad\qquad\qquad\qquad\qquad\qquad\quad \big(\alpha = \actionprob{\q} \,\wedge\, \q \in \mathit{prob} \,\big)\ \lor\\
\;\qquad\qquad\qquad\qquad\qquad\qquad\quad  \big(\alpha = \actiontau{\q}{\pr}{\ell} \,\wedge\, \alpha \in \mathit{comm}\big) \,\}\\
\quad \mathbf{if}\ D = \emptyset\ \mathbf{then\ return}\ S\\
\quad \mathbf{else\ return}\
  \bigcup\,\{\, \mathsf{prune}(S', A') \SEP (S', A')\ \text{a MEC
  of}\ (S \setminus D,\ A{\big|}_{S \setminus D}) \,\}
\end{array}
\]

The first loop records the \emph{fairness} obligations that the actions $\{\alpha \in \ActD(\Delta') \SEP \Delta' \in S\}$ incur --- the second loop then checks that each of these obligations can be discharged by some action in $\{\alpha \in A(\Delta') \SEP \Delta' \in S\}$, pruning state set $D$ whose members incur unfulfilled obligations. The resulting states $S\setminus D$ may no longer be a single MEC, so we re-decompose it into MECs and rerun \emph{prune} on each. Each recursive call strictly shrinks $S$, so $\mathsf{prune}$ terminates.

\subsubsection{Correctness of the Computation}\mbox{}
\label{app:wasll-proofs}
We use the following standard facts for finite MDPs:
\begin{enumerate}
    \item Every EC is contained in a unique MEC \cite[Notation~10.124]{model-checking-book}.
    \item Every EC admits a scheduler that stays inside it on every path and visits every retained (state, action) pair infinitely often almost surely \cite[Lemma~10.119 and its proof]{model-checking-book}.
    \item Under every scheduler, the recurrent (state, action) pairs of almost every infinite path form an EC \cite[Theorem~10.120]{model-checking-book}.
    \item Finally, maximal reachability probabilities are attained by deterministic memoryless schedulers \cite[Lemma~10.102]{model-checking-book}. Actionless states are handled separately; for reachability, they may equivalently be made absorbing.
\end{enumerate}

\begin{definition}[Fair EC]
An EC $(S,A)$ is \emph{fair} if there exists an infinite fair path using only actions in $A$ and visiting every state of $S$ infinitely often, or equivalently; every fairness obligation enabled somewhere in $S$, evaluated using the original $\ActD$, has a discharging action somewhere in $A$.
\end{definition}

Consequently, the recurrent scheduler above is almost surely fair whenever its EC is fair, since every enabled fairness obligation has a discharging action that recurs infinitely often.

\begin{lemma}[Pruning is sound and lossless]
\label{lem:prune}
\mbox{}
\begin{enumerate}
    \item Every EC returned by $\mathsf{prune}$ is fair, and
    \item $S \subseteq \GoodP{\pp}$ for every fair EC $(S,A)$ of the context MDP with $S \subseteq \pend{\pp}{\Delta}$.
\end{enumerate}
\end{lemma}
\begin{proof}
    \mbox{}
    \begin{enumerate}
        \item $\mathsf{prune}$ returns a component only when $D=\emptyset$, so every obligation it incurs has a discharging action in the component. Hence it is fair.

        \item Let $(S',A')$ be a fair EC with $S'\subseteq\pend{\pp}{\Delta}$. Since every state of $S'$ has an action in $A'$ supported in $S'$, closure preserves $(S',A')$ in $(\Sclos,\Aclos)$. Its states have actions, so it avoids the deadlocked carve-out and is contained in an initial MEC.
        Suppose a recursive call $\mathsf{prune}(S,A)$ contains $(S',A')$, in both states and actions. If $\Delta'\in S'\cap D$, then $\Delta'$ incurs an obligation with no discharge in $A$, whereas fairness supplies a discharging action in $A'(\Delta'')\subseteq A(\Delta'')$ for some $\Delta''\in S'$, a contradiction. Thus $S'\cap D=\emptyset$. All actions of $A'$ remain supported in $S'$, so $(S',A')$ remains an EC after restriction to $S\setminus D$. If $D=\emptyset$, the call returns $S\supseteq S'$. Otherwise, MEC containment places $(S',A')$ in a child MEC, and following that child preserves the invariant. Since each nonterminal call strictly decreases $S$ while retaining the nonempty set $S'$, this branch eventually returns a component containing $S'$. Hence $S'\subseteq\GoodP{\pp}$.
    \end{enumerate}
\end{proof}

\begin{lemma}[Absorption]
\label{lem:absorb}
If $T \subseteq U \subseteq \ReachD{\Delta}$ and
$\max_{\phi \in \Phi^{\Delta'}} \probpathmeasure{\phi}{\Delta'}(T) = \probn{1}$ for every
$\Delta' \in U$, then
$\max_{\phi \in \Phi^\Delta} \probpathmeasure{\phi}{\Delta}(U) = \max_{\phi \in \Phi^\Delta}
\probpathmeasure{\phi}{\Delta}(T)$.
\end{lemma}
\begin{proof}
We apply the argument in the proof of~\cite[Theorem~10.15]{model-checking-book} to the Markov chain induced by $\phi$. Since $T\subseteq U$ and $T$ is reached almost surely after entering $U$, this gives the desired result.
\end{proof}

\begin{proposition}[Correctness of the computation]
\label{prop:wasll-correct}
For every $\Delta$,
\[
\max_{\phi \in \Phi^{\Delta}}
\probpathmeasure{\phi}{\Delta}\Bigl(\textstyle\bigcup_{\pp \in \dom{\Delta}}
  \GoodP{\pp}\Bigr)
\;=\;
\max_{\phi \in \Phi^{\Delta}}
    \probpathmeasure{\phi}{\Delta}\bigl(\wasll{\Delta}\bigr).
\]
\end{proposition}

\begin{proof}
By~\cref{lem:absorb} with $T:=\bigcup_{\pp\in\dom{\Delta}}\GoodP{\pp}$ and $U:=\wasll{\Delta}$, it suffices to show, for every $\pp\in\dom{\Delta}$:
\begin{enumerate}
    \item $\GoodP{\pp}\subseteq\wasllp{\pp}{\Delta}$
    \item every $\Delta'\in\wasllp{\pp}{\Delta}$ admits a scheduler reaching $\GoodP{\pp}$ almost surely.
\end{enumerate}
(1) Every state of $\GoodP{\pp}$ lies in $\Sclos\subseteq\pend{\pp}{\Delta}$ and is either actionless or belongs to an EC returned by $\mathsf{prune}$. An actionless state satisfies Def.~\ref{def:wasll} immediately. Otherwise, by~\cref{lem:prune}(1) the returned EC is fair, and the EC recurrence property provides a fair scheduler whose every path remains in that EC. Thus $\GoodP{\pp}\subseteq\wasllp{\pp}{\Delta}$.

(2) Let $\Delta'\in\wasllp{\pp}{\Delta}$, and take its witnessing scheduler $\phi\in\Phi_F^{\Delta'}$ from Def.~\ref{def:wasll}. Every induced path remains in $\pend{\pp}{\Delta}$. By the EC limit property and fairness of $\phi$, almost every infinite induced path is fair and has a suffix using only the states and actions of its limit EC $(S_{\lim},A_{\lim})$. This suffix is fair and visits every state of $S_{\lim}$ infinitely often, so the limit EC is fair. Since $S_{\lim}\subseteq\pend{\pp}{\Delta}$, \cref{lem:prune}(2) gives $S_{\lim}\subseteq\GoodP{\pp}$. Every finite complete induced path ends at an actionless pending state, which is included directly.
Therefore
\[
    \probpathmeasure{\phi}{\Delta'}(\GoodP{\pp})
    = \probn{1}.
\]
Since $\Phi_F^{\Delta'}\subseteq\Phi^{\Delta'}$, this establishes (ii), and~\cref{lem:absorb} yields the claimed equality.
\end{proof}

\begin{remark}[Implementation]
\label{rem:impl}
By Prop.~\ref{prop:wasll-correct},~\cref{thm:wasll} may be model checked with the label of \S E.2 marking $\bigcup_{\pp} \GoodP{\pp}$ directly.  The marked set may be a strict subset of $\wasll{\Delta}$, but the two sets are reached with the same probability, which is all the theorem requires.
\end{remark}

\subsection{Proofs for~\cref{sec:spec:pctl}}
\label{app:spec:pctl}

\begin{lemma}[Reachability insensitive to fairness]
\label{prop:reachabilityfairness}
For any predicate $f$ of typing contexts,
\[
  \max_{\phi \in \Phi_F^{\Delta}} \probpathsig{\Delta}{\eventually f}
  \;=\;
  \max_{\phi \in \Phi^{\Delta}} \probpathsig{\Delta}{\eventually f}.
\]
\end{lemma}
\begin{proof}
Apply \cite[Lemma~10.131]{model-checking-book}.
The cited lemma gives the equality and a fair scheduler attaining the maximum.
\end{proof}

\dffair*
\begin{proof}
    Apply~\cref{prop:reachabilityfairness} with predicate $f = \deadlockk$.
\end{proof}

For the arguments below, finite complete paths are extended by stuttering at their final state (self-loops); without introducing any new fairness obligation. We write $\mathsf{Fair}$ for the LTL encoding of our fairness conditions.

An enabled internal step persists until the corresponding participant takes it. Thus, together with communication fairness, this makes $\mathsf{Fair}$ a limit property (meaning only the set of states visited \emph{infinitely often} determine its truth value). In the following proofs we will appeal to a few results from~\cite{model-checking-book} that relies on \emph{limit properties} of \textsf{MDPs}.%

\begin{restatable}[Non-live path characterisation]{lemma}{nonlivepath}
\label{lem:non-live-path}
For every fair complete path $\tpath$ from $\Delta$,\\[1mm]
\centerline{$
    \neg \live{\tpath}
    \;\text{ iff }\;
    \tpath \models
    \bigvee_{\pp \in \dom{\Delta}}
        \eventually\always \blkpp{\pp}
$
}

Moreover, for each $\pp \in \dom{\Delta}$:
$
    \tpath \models \eventually\always \blkpp{\pp}
    \;\text{ iff }\;
    \exists j \geq 0 \st
        \visited{\tpath[j\ldots]}
        \subseteq \pend{\pp}{\Delta}.
$
\end{restatable}

\begin{proof}
Suppose that $\tpath$ is not live. Some participant $\pp$ has an input or committed output that is never discharged. Since $\pp$ is nonterminated and has no internal step at this input or committed output, $\blkpp{\pp}$ holds forever from that point onwards.

Conversely, if $\blkpp{\pp}$ holds permanently from some point, then $\pp$ remains at a nonterminated input or committed output that is never discharged. Hence $\tpath$ is not live.

For the second equivalence, $\blkpp{\pp}(\Delta')$ implies $\Delta' \in \pend{\pp}{\Delta}$ directly. Conversely, suppose a fair suffix stays in $\pend{\pp}{\Delta}$. The participant $\pp$ cannot communicate on this suffix. Fairness resolves any remaining internal choices: at most a nondeterministic resolution followed by a probabilistic selection. Thereafter, $\pp$ is at an input or committed output and is permanently one-step blocked.
\end{proof}

\wasllthm*
\begin{proof}
Let
\[
    G :=
    \bigcup_{\pp \in \dom{\Delta}} \GoodP{\pp},
    \qquad
    \Psi :=
    \left(
        \mathsf{Fair}
        \rightarrow
        \bigwedge_{\pp \in \dom{\Delta}}
            \always\eventually\neg\blkpp{\pp}
    \right).
\]
The formula $\Psi$ is a limit property.

We first identify its rejecting end components. For any fair EC $(S,A)$ and participant $\pp$,
\[
    S \subseteq \pend{\pp}{\Delta}
    \quad\Longleftrightarrow\quad
    \forall \Delta' \in S:\; \blkpp{\pp}(\Delta').
\]
The right-to-left implication follows directly from the definitions. For the converse, take a fair path confined to $(S,A)$ that visits every state of $S$ infinitely often.
Since $S \subseteq \pend{\pp}{\Delta}$, Lemma~\ref{lem:non-live-path} gives $\eventually\always\blkpp{\pp}$ on this path. Every state of $S$ recurs, so every state of $S$ must satisfy $\blkpp{\pp}$.

Now
\[
    \neg\Psi
    \equiv
    \mathsf{Fair}
    \land
    \bigvee_{\pp \in \dom{\Delta}}
        \eventually\always\blkpp{\pp}.
\]
Consequently, the nonterminal ECs rejecting $\Psi$ are precisely the fair ECs $(S,A)$ with $S \subseteq \pend{\pp}{\Delta}$ for some $\pp \in \dom{\Delta}$. By Lemma~\ref{lem:prune}(1--2), the union of their state sets, together with the actionless pending states, is exactly $G$.

It follows that
\[
\begin{aligned}
    \min_{\phi \in \Phi_F^\Delta}
        \probpathmeasure{\phi}{\Delta}
        \bigl(
            \{\tpath \in \Tpath_\phi(\Delta)
                \SEP \live{\tpath}\}
        \bigr)
    &=
    \min_{\phi \in \Phi^\Delta}
        \probpathmeasure{\phi}{\Delta}(\Psi)
    \\[1mm]
    &=
    \probn{1}
    -
    \max_{\phi \in \Phi^\Delta}
        \probpathmeasure{\phi}{\Delta}(G)
    \\[1mm]
    &=
    \probn{1}
    -
    \max_{\phi \in \Phi^\Delta}
        \probpathmeasure{\phi}{\Delta}(\wasll{\Delta}).
\end{aligned}
\]
The first equality follows from
Lemma~\ref{lem:non-live-path} and
\cite[Theorem~10.134]{model-checking-book};
the second follows from
\cite[Theorem~10.122(b)]{model-checking-book};
and the third is Proposition~\ref{prop:wasll-correct}.
The cited results also ensure that the minimum is attained.
\end{proof}

\subsubsection{Complexity results}
\label{app:complexity}

We write $|\T|$ for the number of nodes in the syntax tree of $\T$, counting
each branch as a node, and $|\Delta| = \sum_{\pp \in \dom{\Delta}}
|\Delta(\pp)|$. Following~\cite[Notation~10.88]{model-checking-book}, the size
of an MDP is its number of transitions $(s, \alpha, t)$ with positive
probability.

\complexitythm*
\begin{proof}
Let $\mathcal{M}$ be the MDP of $\tranone{\Delta}$ restricted to its reachable
states. At each state, each of the at most $|\Delta|$ participants enables at
most $|\Delta|$ actions, each with at most $|\Delta|$ successors, so
$\mathcal{M}$ has size at most $N|\Delta|^3$ and is built by exploration from
the initial state.
The context MDP $\mathcal{C} = (\ReachD{\Delta}, \ActD, \mathsf{step})$ (used in
the livelock computation in \cref{app:wasll-alg}) has $R$ states and thus
size at most $R|\Delta|^3$.

\emph{Livelock precomputation.} We compute $\GoodP{\pp}$ for each participant
$\pp$. Membership of $\pend{\pp}{\Delta}$ is a local check at each state, and
the closure sequence of Stage~1 stabilises after at most $R$ steps. An MEC
decomposition takes time $O(R \cdot (R + |\mathcal{C}|))$
\cite[Lemma~10.126]{model-checking-book}. Apart from its recursive calls, each
call of $\mathsf{prune}$ performs two loops and one MEC decomposition. The
children of a call are MECs of $S \setminus D$ with $D \neq \emptyset$, so
they are pairwise disjoint \cite[Notation~10.124]{model-checking-book} and
strictly smaller than $S$. Hence there are at most $R$ calls at each recursion
depth and at most $R$ depths. The resulting label $\livelocklabel$ is a
disjunction of at most $N$ valuations.

\emph{Model checking.} Each label is evaluated at every state in time
polynomial in $N$ and $|\Delta|$. Safety requires $\pmax{\eventually
\neg\psi} = \probn{0}$ for the safety formula $\psi$, which holds iff no
reachable state violates $\psi$ \cite[Theorem~10.100]{model-checking-book},
and is decided by graph search. Deadlock-freedom and liveness are both
maximum reachability queries, which are known to be computable in time polynomial in
$|\mathcal{M}|$ \cite[Corollary~10.107]{model-checking-book}.
\end{proof}

\complexitycor*
\begin{proof}
Each $S_\pp$ ranges over $[0, \StateS{\Delta(\pp)}]$, and
$\StateS{\T} \leq |\T|$ by induction on $\T$. Hence $N \leq \prod_{\pp \in
\dom{\Delta}} (\StateS{\Delta(\pp)} + 1) \leq (|\Delta| + 1)^k$. Moreover, by
induction on reductions, every type reachable from $\Delta(\pp)$ is a
continuation, committed selection, summand or $\mu$-subterm of $\Delta(\pp)$
whose free variables are replaced by their (likewise closed) $\mu$-binders, so
$\pp$ has at most $|\Delta(\pp)|$ reachable types and, with $R$ as in
\cref{thm:complexity}, $R \leq |\Delta|^k$.
\Cref{thm:complexity} then gives time polynomial in $(|\Delta| + 1)^k$.
\end{proof}

\subsection{Definitions of \prism{} Labels}
We first define several functions that tell us the states in which certain
transitions are available.

\begin{definition}[Enabled Output Labels]
  \begin{align*}
    \EOL{\pp}{\q}{\labelsell}{\Delta} & = \EOLT{\pp}{\q}{\labelsell}{\Delta(\pp)}{0} \\
    \EOLT{\pp}{\q}{\labelsell}{\tend}{n} & = \varnothing \qquad
    \EOLT{\pp}{\q}{\labelsell}{\ty}{n} = \varnothing \\
    \EOLT{\pp}{\q}{\labelsell}{ \mu \ty . \T }{n} & = \EOLT{\pp}{\q}{\labelsell}{\T}{n} \\
    \EOLT{\pp}{\q}{\labelsell}{\pprocinnew{\ppi{i}}{i \in I}{\labelselli{i}}{\Basic_i}{\T_i}}{n} &=
        \textstyle \bigcup_{i \in I} \EOLT{\pp}{\q}{\labelsell}{ {\T_i} }{\nexts{i}}
    \\
    \EOLT{\pp}{\q}{\labelsell}{\pprocoutty{\q'}{}{\probn{1}}{\labelsell'}{\Basic}{\T}}{n} & =
        \{ n \SEP \q' = \q, \ \labelsell' = \labelsell \}
        \cup \EOLT{\pp}{\q}{\labelsell}{\T}{n + 1} \\
    \EOLT{\pp}{\q}{\labelsell}{\pprocoutty{\ppi{j}}{j \in J}{\probi{j}}{\labelselli{j}}{\Basic_j}{\T_j}}{n} & =
        \textstyle \bigcup_{j \in J} \EOLT{\pp}{\q}{\labelsell}{\comsel{j}}{\come{j}}
        \qquad (|J| \geq 2) \\
    \EOLT{\pp}{\q}{\labelsell}{\probpprocoutty{i \in I}{\ppi{j}}{j \in J_i}{\probi{(i,j)}}{\labelselli{j}}{\Basic_j}{\T_{(i,j)}}}{n} & =
        \textstyle \bigcup_{i \in I} \EOLT{\pp}{\q}{\labelsell}{\selsummand{i}}{\sume{i}}
        \qquad (|I| \geq 2)
  \end{align*}
\end{definition}

\begin{definition}[Enabled Branching]
  \begin{align*}
    \EB{\pp}{\q}{\Delta} & = \EBT{\pp}{\q}{\Delta(\q)}{0} \\
    \EBT{\pp}{\q}{ \tend }{n} & = \varnothing \qquad
    \EBT{\pp}{\q}{ \ty }{n} = \varnothing \qquad
    \EBT{\pp}{\q}{ \mu \ty . \T }{n} = \EBT{\pp}{\q}{\T}{n} \\
    \EBT{\pp}{\q}{\pprocinnew{\ppi{i}}{i \in I}{\labelselli{i}}{\Basic_i}{\T_i}}{n} & =
       \begin{cases}
         \{n\} \cup
         \bigcup_{i \in I} \EBT{\pp}{\q}{ {\T_i} }{\nexts{i}} & \text{if } \exists i \in I : \ppi{i} = \pp \\
         \bigcup_{i \in I} \EBT{\pp}{\q}{ {\T_i} }{\nexts{i}} & \text{otherwise}
       \end{cases}  \\
    \EBT{\pp}{\q}{\pprocoutty{\q'}{}{\probn{1}}{\labelsell'}{\Basic}{\T}}{n} & =
        \EBT{\pp}{\q}{\T}{n + 1} \\
    \EBT{\pp}{\q}{\pprocoutty{\ppi{j}}{j \in J}{\probi{j}}{\labelselli{j}}{\Basic_j}{\T_j}}{n} & =
        \textstyle \bigcup_{j \in J} \EBT{\pp}{\q}{\comsel{j}}{\come{j}}
        \qquad (|J| \geq 2) \\
    \EBT{\pp}{\q}{\probpprocoutty{i \in I}{\ppi{j}}{j \in J_i}{\probi{(i,j)}}{\labelselli{j}}{\Basic_j}{\T_{(i,j)}}}{n} & =
        \textstyle \bigcup_{i \in I} \EBT{\pp}{\q}{\selsummand{i}}{\sume{i}}
        \qquad (|I| \geq 2)
  \end{align*}
\end{definition}

\begin{definition}[Enabled Branching Labels]
  \begin{align*}
    \EBL{\pp}{\q}{\labelsell}{\Delta} & = \EBLT{\pp}{\q}{\labelsell}{\Delta(\q)}{0} \\
    \EBLT{\pp}{\q}{\labelsell}{\tend}{n} & = \varnothing \qquad
    \EBLT{\pp}{\q}{\labelsell}{\ty}{n} = \varnothing \\
    \EBLT{\pp}{\q}{\labelsell}{ \mu \ty . \T }{n} & = \EBLT{\pp}{\q}{\labelsell}{\T}{n} \\
    \EBLT{\pp}{\q}{\labelsell}{\pprocinnew{\ppi{i}}{i \in I}{\labelselli{i}}{\Basic_i}{\T_i}}{n} &=
       \begin{cases}
         \{n\} \cup
         \bigcup_{i \in I} \EBLT{\pp}{\q}{\labelsell}{ {\T_i} }{\nexts{i}} & \text{if } \exists i : \ppi{i} = \pp \land \labelselli{i} = \labelsell \\
         \bigcup_{i \in I} \EBLT{\pp}{\q}{\labelsell}{ {\T_i} }{\nexts{i}} & \text{otherwise}
       \end{cases}  \\
    \EBLT{\pp}{\q}{\labelsell}{\pprocoutty{\q'}{}{\probn{1}}{\labelsell'}{\Basic}{\T}}{n} & =
        \EBLT{\pp}{\q}{\labelsell}{\T}{n + 1} \\
    \EBLT{\pp}{\q}{\labelsell}{\pprocoutty{\ppi{j}}{j \in J}{\probi{j}}{\labelselli{j}}{\Basic_j}{\T_j}}{n} & =
        \textstyle \bigcup_{j \in J} \EBLT{\pp}{\q}{\labelsell}{\comsel{j}}{\come{j}}
        \qquad (|J| \geq 2) \\
    \EBLT{\pp}{\q}{\labelsell}{\probpprocoutty{i \in I}{\ppi{j}}{j \in J_i}{\probi{(i,j)}}{\labelselli{j}}{\Basic_j}{\T_{(i,j)}}}{n} & =
        \textstyle \bigcup_{i \in I} \EBLT{\pp}{\q}{\labelsell}{\selsummand{i}}{\sume{i}}
        \qquad (|I| \geq 2)
  \end{align*}
\end{definition}

Given these functions, we define the following labels, adopting the
convention that an empty disjunction is $\false$:

\newcommand{\prismlabel}[2]{\texttt{label}\text{ }#1 = #2}
\begin{gather*}
  \prismlabel{\labelend}{\bigwedge_{\pp \in \dom{\Delta}} (S_\pp =
    \StateS{\Delta(\pp)})} \\
  \forall \pp, \q \in \dom{\Delta} :
  \prismlabel{\receiveanylabel{\pp}{\q}}{\bigvee_{n \in \EB{\pp}{\q}{\Delta}}
    (S_\q = n)} \\
\begin{aligned}
  \forall \pp \conc \q \conc \labelsell \in \actions{}{\Delta} : \
  &\prismlabel{\receivelabel{\pp}{\q}{\labelsell}}{\bigvee_{n \in
    \EBL{\pp}{\q}{\labelsell}{\Delta}} (S_\q = n)} \\
  &\prismlabel{\sendlabel{\pp}{\q}{\labelsell}}{\bigvee_{n \in
    \EOL{\pp}{\q}{\labelsell}{\Delta}} (S_\pp = n)}
\end{aligned}
\end{gather*}

By definition, these meet the specification given in \Cref{def:labels}.

\subsection{Typing Contexts Used for Benchmarking}
\label{sec:benchmarkingcontexts}
\foreach \fname in {
sessions/auth.promt,
sessions/dice.promt,
sessions/dining.promt,
sessions/example-4-16.promt,
sessions/gamblers-ruin.promt,
sessions/leader-election.promt,
sessions/monty-hall-change.promt,
sessions/monty-hall-stay.promt,
sessions/multiparty-workers.promt,
sessions/rec-map-reduce.promt,
sessions/rec-two-buyers.promt,
sessions/von-neumann-coin.promt}{
 \filename{\fname}
 \lstinputlisting{\fname}
}

\subsection{Results of Running \prose{} on Examples}
\label{app:proseresults}
\begin{table}[]
    \centering
    \caption{\prose{} outputs on the typing contexts inferred from
      (\Cref{sec:benchmarkingcontexts}), used for benchmarking. \textsf{DF} and \textsf{Liveness} are
      the lower-bound probabilities of deadlock-freedom and liveness
      (minimised over all fair schedulers), rounded to two decimal points.}
    \small
    \begin{tabular}{l|@{ }c@{ }|@{ }c@{ }|@{ }c}
      \textsf{\textbf{Typing context}} & \textsf{\textbf{Safety}} & \textsf{\textbf{DF}} & \textsf{\textbf{Liveness}} \\
      \hline
      \textsf{auth} & yes & 0.76 & 0.36 \\
      \textsf{dice} & yes & 0.17 & 0 \\
      \textsf{dining} & yes & 1 & 0 \\
      \textsf{leader-election} & yes & 1 & 1 \\
      \textsf{monty-hall-change} & no & 0.66 & 0.66 \\
      \textsf{monty-hall-stay} & no & 0.34 & 0.34 \\
      \textsf{multiparty-workers} & yes & 1 & 1 \\
      \textsf{gamblers-ruin} & yes & 1 & 1 \\
      \textsf{rec-map-reduce} & yes & 1 & 1 \\
      \textsf{rec-two-buyers} & yes & 1 & 1 \\
      \textsf{example-4-16} & yes & 0.69 & 0.53 \\
      \textsf{von-neumann-coin} & yes & 1 & 1 \\
    \end{tabular}\label{tab:proseresults}
\end{table}

\Cref{tab:proseresults} shows the results of running \prose{} on the typing contexts inferred from the examples given in \Cref{sec:benchmarkingcontexts}.
All probabilistic results are rounded to two decimal points.

\subsection{Comparison With Model checking Sessions}
\label{app:modelcheckingprocesses}
Consider the following session $\Multi_n$. For a given $n$, workers $\wrk{i}$ $(0 \leq i < n)$ each compute the factorial of $i$
by querying worker $\wrk{i-1}$ (except for $\wrk{0}$, which always outputs $1$) and multiplying the result by $i$.
However, each worker may probabilistically produce an error; moreover, the probability of each worker producing an error is larger
if the value received is at least $100$. If a worker receives an error, it simply propagates it. Finally, the worker $\wrk{n}$ requests
the value of $(n-1)$ factorial. If $\wrk{n}$ receives a result, it sends a message to the dummy worker $\dum$ forever; if it receives an error,
then it terminates, leaving the remaining workers waiting for a request that never arrives. Thus, the probability of
deadlock-freedom $\probdfmin{\Multi_n}$ gives the probability of an execution where $\wrk{n}$
receives a result.
    \begin{align*}
      &P_0 = \mu X . \procin{\wrk{1}}{\req}{\flip{0.7}{\procout{\wrk{1}}{\ress}{1}{X}}{\procoutu{\wrk{1}}{\err}{X}}} \\
      &\begin{aligned}
        P_{i \in [1,n-1]} = \mu X . &\procin{\wrk{i+1}}{\req}{\procoutu{\wrk{i-1}}{\req}{ \\
        & (\procin{\wrk{i-1}}{\err}{\procoutu{\wrk{i+1}}{\err}{X}} \\
        & + \procin{\wrk{i-1}}{\ress(x)}{\\
          & \quad \cond{x < 100}{\flip{0.7}{\procout{\wrk{i+1}}{\ress}{x * i}{X}}{\procoutu{\wrk{i+1}}{\err}{X}} \\
          & \ \ \ \ }{\flip{0.5}{\procout{\wrk{i+1}}{\ress}{x * i}{X}}{\procoutu{\wrk{i+1}}{\err}{X}}}}
        )}}
      \end{aligned} \\
      &P_n = \procoutu{\wrk{n-1}}{\req}{
        (\procin{\wrk{n-1}}{\ress(x)}{\mu X . \procoutu{\dum}{\donel}{X}} + \procin{\wrk{n-1}}{\err}{\mpNil})
      } \\
      &P_d = \mu X . \procin{\wrk{n}}{\labels{\donel}}{X} \\
      &\Multi_n = \pa{\wrk{0}}{P_0} \pc \cdots \pc \pa{\wrk{n}}{P_n} \pc \pa{\dum}{P_d}
    \end{align*}
This session is typable, with $P_{i \in [1, n-1]}$ having the
following type $\T_i$.
    {\small
    \begin{align*}
      \T_i = \mu\ty.\;\&\wrk{i+1}?\req.\;\oplus\wrk{i-1}!\req.\;\&\big\{\,
        &\wrk{i-1}?\err.\;\oplus\wrk{i+1}!\err.\,\ty, \\
        &\wrk{i-1}?\ress(\ints).\;{\textstyle\sum}\big\{\,
          \oplus\{\,\probn{0.7}\!:\wrk{i+1}!\ress\langle\ints\rangle.\,\ty,\;
                   \probn{0.3}\!:\wrk{i+1}!\err.\,\ty\,\}, \\
        &\phantom{\wrk{i-1}?\ress(\ints).\;{\textstyle\sum}\big\{\,}
          \oplus\{\,\probn{0.5}\!:\wrk{i+1}!\ress\langle\ints\rangle.\,\ty,\;
                   \probn{0.5}\!:\wrk{i+1}!\err.\,\ty\,\}
        \,\big\}\,\big\}
    \end{align*}}%
The non-deterministic sum $\sum$ originates from the use of subtyping (\rulename{sub-$\Sigma$}), which is necessary to type
the conditional using \rulename{t-cond}: the test $x < 100$ is abstracted to a non-deterministic choice
between the two output distributions (cf.~\Cref{exam:runningtype}).
The types for other participants are obvious and are omitted.

We translate the sessions and their typing contexts into \prism{} modules, allowing us to compare
their efficiency and evaluate the quality of the lower bound obtained from the typing context. The
\prism{} definitions for the translated processes are presented in \Cref{app:factorialprism}. We invoke \prism{}
with the explicit engine (\texttt{-ex} flag) for the session verification, since unlike the default
hybrid engine it does not require integer variables to be bounded.

\Cref{fig:factorial:a} shows a comparison of the time taken to verify deadlock-freedom probabilities of
$\Multi_n$ and its typing context, with varying values of $n$.
We observe that as $n$ increases, the time taken for direct verification
of sessions increases significantly faster than that of typing contexts. In fact, for $n \geq 14$, the
\prism{} invocations for session verifications ran out of the allocated Java heap
space\footnote{We used the default allocated heap size of 1 GB.}, and consequently
did not finish execution.

This is due to the increased
state space of processes: any integer variables used by processes need to be present in the state space
for the module corresponding to that participant. In general, each integer variable can lie within a
large range\footnote{Theoretically infinite, but would be bounded in practice as in our example.} of values,
and each new variable contributes to the product state of variables. By contrast, the typing system
treats each variable simply as a member of the $\ints$ sort, collapsing the state space required.

More generally, if we extend our calculus with a richer language of expressions $e$, checking
properties for processes becomes undecidable
due to the presence of conditionals and recursion (\Cref{prop:procsafeundec}). However, checking safety, termination and
deadlock-freedom for typing contexts has been proven decidable in the
non-probabilistic setting~\cite{Padovani16,Udomsrirungruang2024a} (which implies our probabilistic case).

The trade-off of verifying typing contexts is that we compute a lower bound
(\Cref{thm:proc:df}) of the deadlock-freedom probability, rather than the exact probability.
The difference is due to the conditional $x < 100$ being abstracted to a non-deterministic
choice in $\T_i$: the worst-case scheduler always resolves it towards the distribution with
the larger error probability $\probn{0.5}$, whereas the session only takes this branch once
the received value reaches $100$.
\Cref{fig:factorial:b} shows a comparison of the deadlock-freedom probability obtained from $\Multi_n$ and its
corresponding context: both tend to $\probn{0}$ as $n$ grows, since every worker in the chain must succeed,
with the lower bound from the typing context lying below the exact probability of the session.

\subsection{\prism{} Models for \Cref{app:modelcheckingprocesses}}
\label{app:factorialprism}
An example model for $n = 3$ is given here; others are straightforward extensions. Our artifact includes scripts
to generate the sessions and typing contexts (and hence these models) for other values of $n$.

\begin{lstlisting}
module w0
  w0 : int init 0;
  w0_w1_err_unit : int init 0;
  w0_w1_res_unit : int init 0;

  [w1_w0] w0=0 -> 1:(w0'=1);
  [w1_w0_req_unit] w0=1 -> 1:(w0'=2);
  [] w0=2 -> 0.7:(w0'=3) + 0.3:(w0'=6);
  [w0_w1] w0=3 -> 1:(w0'=4)&(w0_w1_res_unit'=1);
  [w0_w1_res_unit] w0=4 -> 1:(w0'=5);
  [] w0=5 -> 1:(w0'=0);
  [w0_w1] w0=6 -> 1:(w0'=7)&(w0_w1_err_unit'=0);
  [w0_w1_err_unit] w0=7 -> 1:(w0'=8);
  [] w0=8 -> 1:(w0'=0);
endmodule

module w1
  w1 : int init 0;
  w1_w2_err_unit : int init 0;
  w1_w2_res_unit : int init 0;
  w1_x : int init 0;

  [w2_w1] w1=0 -> 1:(w1'=1);
  [w2_w1_req_unit] w1=1 -> 1:(w1'=2);
  [w1_w0] w1=2 -> 1:(w1'=3);
  [w1_w0_req_unit] w1=3 -> 1:(w1'=4);
  [w0_w1] w1=4 -> 1:(w1'=5);
  [w0_w1_res_unit] w1=5 -> 1:(w1'=10)&(w1_x'=w0_w1_res_unit);
  [] (w1=10) & ((w1_x)<(100)) -> 1:(w1'=11);
  [] (w1=10) & (!((w1_x)<(100))) -> 1:(w1'=18);
  [] w1=11 -> 0.7:(w1'=12) + 0.3:(w1'=15);
  [w1_w2] w1=12 -> 1:(w1'=13)&(w1_w2_res_unit'=(w1_x)*(1));
  [w1_w2_res_unit] w1=13 -> 1:(w1'=14);
  [] w1=14 -> 1:(w1'=0);
  [w1_w2] w1=15 -> 1:(w1'=16)&(w1_w2_err_unit'=0);
  [w1_w2_err_unit] w1=16 -> 1:(w1'=17);
  [] w1=17 -> 1:(w1'=0);
  [] w1=18 -> 0.5:(w1'=19) + 0.5:(w1'=22);
  [w1_w2] w1=19 -> 1:(w1'=20)&(w1_w2_res_unit'=(w1_x)*(1));
  [w1_w2_res_unit] w1=20 -> 1:(w1'=21);
  [] w1=21 -> 1:(w1'=0);
  [w1_w2] w1=22 -> 1:(w1'=23)&(w1_w2_err_unit'=0);
  [w1_w2_err_unit] w1=23 -> 1:(w1'=24);
  [] w1=24 -> 1:(w1'=0);
  [w0_w1_err_unit] w1=5 -> 1:(w1'=6)&(w1_x'=w0_w1_err_unit);
  [w1_w2] w1=6 -> 1:(w1'=7)&(w1_w2_err_unit'=0);
  [w1_w2_err_unit] w1=7 -> 1:(w1'=8);
  [] w1=8 -> 1:(w1'=0);
endmodule

module w2
  w2 : int init 0;
  w2_w3_err_unit : int init 0;
  w2_w3_res_unit : int init 0;
  w2_x : int init 0;

  [w3_w2] w2=0 -> 1:(w2'=1);
  [w3_w2_req_unit] w2=1 -> 1:(w2'=2);
  [w2_w1] w2=2 -> 1:(w2'=3);
  [w2_w1_req_unit] w2=3 -> 1:(w2'=4);
  [w1_w2] w2=4 -> 1:(w2'=5);
  [w1_w2_res_unit] w2=5 -> 1:(w2'=10)&(w2_x'=w1_w2_res_unit);
  [] (w2=10) & ((w2_x)<(100)) -> 1:(w2'=11);
  [] (w2=10) & (!((w2_x)<(100))) -> 1:(w2'=18);
  [] w2=11 -> 0.7:(w2'=12) + 0.3:(w2'=15);
  [w2_w3] w2=12 -> 1:(w2'=13)&(w2_w3_res_unit'=(w2_x)*(2));
  [w2_w3_res_unit] w2=13 -> 1:(w2'=14);
  [] w2=14 -> 1:(w2'=0);
  [w2_w3] w2=15 -> 1:(w2'=16)&(w2_w3_err_unit'=0);
  [w2_w3_err_unit] w2=16 -> 1:(w2'=17);
  [] w2=17 -> 1:(w2'=0);
  [] w2=18 -> 0.5:(w2'=19) + 0.5:(w2'=22);
  [w2_w3] w2=19 -> 1:(w2'=20)&(w2_w3_res_unit'=(w2_x)*(2));
  [w2_w3_res_unit] w2=20 -> 1:(w2'=21);
  [] w2=21 -> 1:(w2'=0);
  [w2_w3] w2=22 -> 1:(w2'=23)&(w2_w3_err_unit'=0);
  [w2_w3_err_unit] w2=23 -> 1:(w2'=24);
  [] w2=24 -> 1:(w2'=0);
  [w1_w2_err_unit] w2=5 -> 1:(w2'=6)&(w2_x'=w1_w2_err_unit);
  [w2_w3] w2=6 -> 1:(w2'=7)&(w2_w3_err_unit'=0);
  [w2_w3_err_unit] w2=7 -> 1:(w2'=8);
  [] w2=8 -> 1:(w2'=0);
endmodule

module w3
  w3 : int init 0;
  w3_x : int init 0;

  [w3_w2] w3=0 -> 1:(w3'=1);
  [w3_w2_req_unit] w3=1 -> 1:(w3'=2);
  [w2_w3] w3=2 -> 1:(w3'=3);
  [w2_w3_res_unit] w3=3 -> 1:(w3'=6)&(w3_x'=w2_w3_res_unit);
  [w3_dummy] w3=6 -> 1:(w3'=7);
  [w3_dummy_done_unit] w3=7 -> 1:(w3'=8);
  [] w3=8 -> 1:(w3'=6);
  [w2_w3_err_unit] w3=3 -> 1:(w3'=4)&(w3_x'=w2_w3_err_unit);
endmodule

module dummy
  dummy : int init 0;

  [w3_dummy] dummy=0 -> 1:(dummy'=1);
  [w3_dummy_done_unit] dummy=1 -> 1:(dummy'=2);
  [] dummy=2 -> 1:(dummy'=0);
endmodule

label "end" = (false) & (false) & (false) & (w3=4) & (false);
\end{lstlisting}

We check the probability of deadlock-freedom using the PLTL formula
$\pmin{\always (\deadlocklabel \implies \labelend)}$, i.e.~the same specification
as for typing contexts in \Cref{sec:spec:pctl}.

\end{document}